\documentclass{ruthesis}
\usepackage{graphicx}
\usepackage[english]{babel}
\usepackage{amsmath}

\newcommand{\beq}{\begin{equation}}
\newcommand{\eeq}{\end{equation}}
\newcommand{\beqa}{\begin{eqnarray}}
\newcommand{\eeqa}{\end{eqnarray}}

\def\Im{\hbox{Im}}
\def\Re{\hbox{Re}}

\def\latt{\text{latt}}

\begin{document}

\phd

\title{Investigation of Strongly Correlated Electron Systems with Cellular Dynamical Mean Field Theory}

\author{Marcello Civelli}

\program{Physics}

\director{Prof. B. Gabriel Kotliar}

\approvals{5}

\submissionyear{2006}

\submissionmonth{May}

\abstract{In this thesis we study the strongly correlated
electron physics in the framework of the longstanding H-T$_{C}$
superconductivity problem using a non-perturbative method, the
Dynamical Mean Field Theory (DMFT), capable to go beyond standard
perturbation theory techniques. DMFT is by construction a local
theory which neglects spatial correlation. The latter is however
shown in experiments to be a fundamental property of cuprate
materials. In a first step, we approach the problem of the spatial
correlation in the normal state of cuprate materials using a
phenomenological Fermi-Liquid-Boltzmann model. We then introduce
and develop in detail an extension to DMFT, the Cellular Dynamical
Mean Field Theory (CDMFT), capable of considering short-ranged
spatial correlation in a system and implemented it with the exact
diagonalization algorithm . After testing CDMFT in an exact
limiting case, we apply it to study the density-driven Mott
metal-insulator transition in the two-dimensional Hubbard Model
with particular attention to the anomalous properties of the
normal state as the Mott insulator is approached. We finally
study the superconducting state. We show that within CDMFT the
one-band Hubbard Model supports a d-wave superconductive state,
which strongly departs from the standard BCS theory. We
conjecture a link between the instabilities found in the normal
state and the onset of superconductivity.}
\beforepreface
\acknowledgements{This thesis owns its birth to the ideas and
long-term frontier research of Prof. Gabriel Kotliar. I am most
grateful to him for providing the chance of studying very
interesting and challenging problems. He has guided me trough
these graduate-school years teaching the way to do research and
living with me either the moments of difficulties and of great
exaltation.

I thank the people that worked with me in this project (in order
of appearance): Andrea Perali,  Srivenkateswara Sarma Kancharla,
Massimo Capone, Olivier Parcollet, Tudor Stanescu. From them I
learned a lot. In this work I also enjoyed the collaboration and
the discussions with Bumsoo Kyung and Andr\'e-Marie Trambley.

I am very grateful to the many people in the big Prof. Kotliar's
"family" and to the many friends in graduate school. With them I
shared a lot of things: offices, computers, lunch and dinner
times, seminars, long nights (never mornings however), difficult
and joyful moments (at the departments and outside). From them I
got help, courage, advise and support.

I am grateful to the members of the committee, from whom I always
had a great help: Prof. Eva Andrei, Prof. Paul Leath and Prof.
Claude Lovelace. I own gratitude to many professors of the
Physics and Astronomy Department, in particular to the three
graduate directors and the professors of the Condensed Matter
group, whose advise and suggestions have always been present. I
enjoyed teaching with Prof. John Hughes, Prof. Gordon Thomson and
Gabriel Alba. Special thanks to Kathy Di Meo, the first person I
met in the Rutgers Physics Department, and to Francesca De Lucia.

I own the decision to pursue studies in physics to Alberto Parola
"un fuoriclasse". He is my first teacher and a friend.

To my parents, who always supported me in any decision, all my
gratitude. I thank you the constant encouragement from all my
family (in Italy and USA). V\'eronique, to her my love.

Finally I thank you all the friends and people that I had the
luck to meet in Rutgers all over these years. Many of them have
been like a family for me, that leaving Rutgers is like leaving
home a second time.}

\dedication{\begin{center}To my grand-parents, Ai miei nonni, A
mis aubuelos. \end{center}} \afterpreface


\addcontentsline{toc}{chapter}{Introduction} 
\pagenumbering{arabic} \setcounter{page}{1}



\chapter*{Introduction}

In 1986 J. G. Bednorz e K. A. Muller announced the discovery of a
superconductor material with the record high critical temperature
of 30 $K$. This was most unexpected, as this superconductive
material was a cuprate ceramic, which is known to be a good
insulator at room temperature. In the light of the well tested
classical theory of superconductivity, the
Bardeen-Cooper-Schrifferer (BCS) \cite{BCS}, it was not possible
to explain the origin of the superconductivity in these materials.
Even more mysterious was the fact that their critical temperature
was an order of magnitude higher than that of the classical BCS
superconductors.

These extraordinary results were successively re-confirmed by
other experimental groups (Takagi et alteri \cite{Takagi}). Since
then many high-T$_{C}$ superconductors have been discovered. (So
far, the record critical temperature is from
mercury-thallium-barium-calcium-copper-oxygen that becomes
superconducting at 138 K [-135 C or -211 F]).
Since 77 K is the boiling point of nitrogen, new technological
applications are expected, such as superconducting quantum
interference devices (SQUID), Josephson's effect based
electronics and magnetic levitation of super-fast trains. This
exciting technological scenario however looks still far in the
future. Up to now the search for new materials has been an
empirical procedure, since the key-mechanism the of high-T$_{c}$
superconductivity remains theoretically obscure. Many theories
have been proposed, but none of them have given a definitive
answer. On one hand, because of the complexity of the cuprates
materials, it is very difficult to interpret the experimental
results. Antiferromagnetism, disorder, phonons and strong
anisotropy conspire together to hide the key ingredients. On the
other hand, the theoretical tools to face the longstanding
problem of strongly correlated systems, where standard
perturbation theory cannot be applied, are still scarcely
available. The theoretical model capable of explaining
high-T$_{c}$ superconductivity may have been already identified,
but we are not able to extract its properties even approximately.

In this thesis we face the problem of high-T$_{C}$ superconducting
materials from a theoretical point of view by studying the
two-dimensional Hubbard Model on a simple square lattice. This
model, proposed by J. Hubbard in 1964 \cite{hubbard:64}, is
universally considered the simplest minimal description of
cuprate materials. In spite of this model's simplicity however,
its properties have been well determined only in the
one-dimensional limit, where an exact Bethe Ansatz (BA) solution
exists (1968 \cite{Lieb-Wu}), and more recently in the
infinite-dimensional limit (1989 \cite{Metzner:89}), where the
exact solution has been determined by Dynamical Mean Field Theory
(DMFT). However, no general consensus on its properties has been
reached for finite dimensions (the realm of real materials), so it
is not clear wheather this model embodies the physics necessary
to understand the high-T$_{C}$ superconducting mechanism.

In the first chapter of this thesis we introduce the high-T$_{C}$
superconducting materials and the two-dimensional Hubbard Model
and review some of the results which have been obtained in the
past using standard techniques. We show that the Hubbard Model is
able to describe some of the physical properties of the cuprates
materials, especially in the insulating state, and answers the
fundamental question of the origin of the insulating (Mott) gap in
half-filled-band systems. This is not however sufficient to
understand the metallic and superconductive phases. To approach
this problem we start from the infinite-dimensional-limit
viewpoint and adopt an extension of the DMFT, the Cellular
Dynamical Mean Field Theory (CDMFT, \cite{cdmft}), as an
approximate tool to approach finite-dimensional systems. To
support this approach, in chapter 2, we present a
Boltzmann-Fermi-liquid study of the transport properties of the
cuprate materials in the normal state. In this framework, we
emphasize the importance of considering momentum-dependent
scattering to describe the physics of these systems. DMFT, by
construction, is not able to describe momentum-dependent
(spatial-correlated) properties. It is therefore at its
foundations a poor approximation for understanding cuprates.
CDMFT, instead, is designed to introduce short-range spatial
correlations into the system, allowing for the description of
momentum-dependent quantities. A presentation of the building of
CDMFT from DMFT is given in chapter 3. In this work the Lanczos
algorithm \cite{Lanczos} is adopted to solve the associated
quantum impurity problem (see also APPENDIX B). In chapter 4 we
test CDMFT a the one dimensional exact Bethe-Ansatz solution, the
worst case scenario for a mean field theory, and with previous
Quantum Monte Carlo results. These tests provide a benchmark for
the method. We then present details on the implementation of
CDMFT and the schemes adopted to extract physical quantities in
real and momentum space. In chapter 5 we apply CDMFT to the
normal state of the two-dimensional Hubbard model, with attention
to experimental results on cuprate materials and we study
hole-doped as well as electron-doped cases. A complete
description of the density-driven Mott metal-insulator transition
is presented for the two-dimensional Hubbard Model. The creation
of an anomalous Fermi-liquid state (pseudogap phase) is presented
as a reliable possibility in the region which preludes a Mott
transition. Finally in chapter 6 the superconducting state is
studied. We find that the Hubbard Model in two dimensions and
zero temperature supports a d-wave superconducting state, which
clearly departs from the standard BCS theory. Its properties are
described in light of the proximity to a parent antiferromagnetic
Mott-insulator. A connection between the anomalous
superconducting properties and the anomalies of the normal state
are inferred.

\chapter{High Temperature Superconductors}

\section{Structure and Phase Diagram}

A characteristic feature of cuprate superconductors is a
multilayered structure of CuO$_{2}$ planes, separated by layers
of other atoms (La, O, Ba,...). There are strong hints that the
high-T$_{c}$ superconductivity is a two-dimensional phenomenon
taking place within the CuO$_{2}$ planes, while the intermediate
layers are simply reservoirs for charges. Another fundamental
property common to all these materials is an antiferromagnetic
insulating state. The insulator is of the {\it Mott} kind, with
the electronic band half-filled. Upon doping, the long-range
antiferromagnetic order is destroyed, and the system becomes a
paramagnetic metal which superconducts at temperatures below a
critical value. Short-range antiferromagnetic correlations may,
however, remain relevant and play a fundamental role in
determining the superconductive mechanism.

Many high-T$_{c}$ superconducting materials have been synthesized:
it is sufficient to vary the number of layers in the unit cell or
the kind of intra-layer atoms to obtain an enormous variety. In
the following table we report some of them along with their
critical temperatures. For comparison, we also report the
critical temperatures of the "classical" superconductors Nb, Pb
and Nb$_{3}$Ge, which
had the highest T$_{c}$ before 1986. \bigskip \\
\begin{center}
\begin{tabular}[position]{rcl}
Material&      &T$_{c}$(K)\\ \hline
&&\\
H$_{g}$Ba$_{2}$Ca$_{2}$Cu$_{3}$O$_{8+\delta}$ &     & 133  \\
Tl$_{2}$Ca$_{2}$Ba$_{2}$Cu$_{3}$O$_{10}$ &     & 125       \\
YBa$_{2}$Cu$_{3}$O$_{7}$ &     & 92\\
Bi$_{2}$Sr$_{2}$CaCu$_{2}$O$_{8}$ &     & 89\\
La$_{1.85}$Sr$_{0.15}$CuO$_{4}$ &     & 39\\
Nd$_{1.85}$Ce$_{0.15}$CuO$_{4}$ &     &24\\
Nb$_{3}$Ge &     & 23.2\\
Nb &     & 9.25\\
Pb &     & 7.20\\ \hline
\end{tabular}
\end{center}
\vspace{1cm} Let's analyze in detail the structure of the most
studied high-T$_{c}$ superconductor, the
La$_{2-x}$Sr$_{x}$CuO$_{4}$, as shown in Fig \ref{intro1b} below.

\subsection{La$_{2-x}$Sr$_{x}$CuO$_{4}$}

\begin{figure}[!htbp]
\begin{center}
\includegraphics[width=6cm,height=8cm] {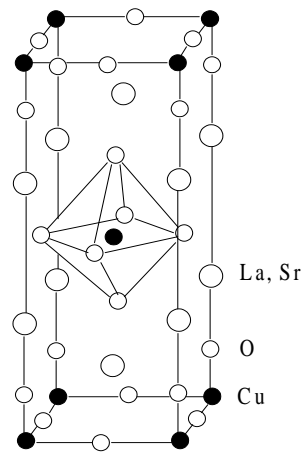}
\caption[entry for the LOF]{Structure of La$_{2-x}$Sr$_{x}$CuO$_{4}$}
\label{intro1b}
\end{center}
\end{figure}

This material, one of the first discovered, gives a good example
of the properties common to all H-T$_{c}$ superconductors. Let us
consider first the insulating state. La$_{2}$CuO$_{4}$ has a
structure body-centered-tetragonal (bct), well known from studies
on K$_{2}$NiF$_{4}$, which is described in Fig.\ref{intro1b}. This
structure can be seen as a pile of CuO$_{2}$ planes 6.6 \AA\
apart and separated by two planes of LaO, which play the role of
charge reservoirs for the CuO$_{2}$ planes, once doping has been
added into the system. The Cu atoms in the CuO$_{2}$ planes are
each of them surrounded by four coplanar O atoms, plus two other
O atoms, denoted O$_{z}$, in the upward and downward direction
perpendicular to the CuO$_{2}$ planes. Each Cu atom is thus
surrounded by an octahedron of oxygens; however the Cu-O$_{z}$
distance is roughly 2.4 \AA, considerably greater than the
distance Cu-O in the two-dimensional conducting planes, which is
roughly 1.9 \AA. The latter therefore are the dominant bonds,
although the importance of the O$_{z}$ atoms in the
superconductive mechanism is strongly debated. The electronic
configurations are Cu:[Ar](3d)$^{10}$(4s), O:[He](2s)$^{2}$,
La:[Xe](5d)(6s)$^{2}$ e Sr:[Kr](5s)$^{2}$. To first approximation,
it is possible to schematize the electronic structure with the
oxygen in a O$^{2-}$ state, which completes its p shell, and the
lanthanium in a La$^{3+}$ state, which has the more stable
electronic configuration of Xe. In order to maintain electrical
neutrality, the copper atoms must be in the Cu$^{2+}$ state,
having lost an electron from the 4s level and one from the 3d
level, where a hole is formed with a total spin of $\frac{1}{2}$.
Upon doping, by substituting some percentage of La$^{3+}$ with
Sr$^{2+}$, fewer electrons remain for the oxygens in the CuO$_{2}$
planes. Therefore, in these planes more holes are formed. Either
the oxygen completes again the p shell, getting a further electron
from a Cu atom, where a hole is formed, or it remains in the
state O$^{-}$. The hole in this case localizes on the oxygen. We
will see that the latter is the situation energetically favored.
If the hole were localized on the Cu atom it would experience a
strong Coulomb repulsion from the hole already present. Since the
conduction band in these materials is very narrow, the effective
mass of the carriers is very large, and thus their mobility
greatly reduced and interactions between them are thereby
amplified. The strong on-site coulomb repulsion at the Cu atoms
is also the origin of the insulating state even when the
electronic band is half-filled. The holes, which localize at the
Cu atom in the insulating state, have the tendency to freeze at
their own site and thus do not find other hole-free Cu-sites to
occupy. This is a classical case in which the free electron
approximation fails completely. It is indeed the strong
interaction between electrons which determines the macroscopic
properties of the matter. With only a small percentage of doping
(x$\geq$ 0.04), the system becomes metallic. The sign of the Hall
coefficient shows that the electronic carriers are holes, as
expected.
\subsection{Phase Diagram}
\begin{figure}[!htbp]
\begin{center}
\includegraphics[width=10cm,height=7cm] {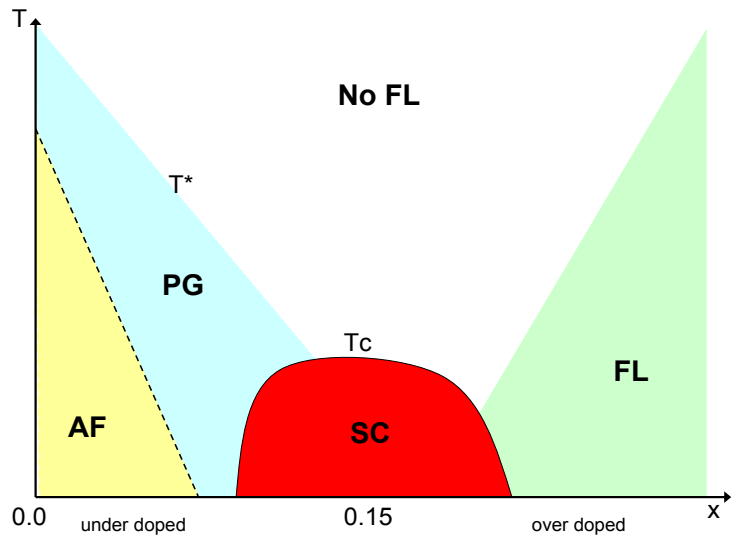}
\caption[entry for the LOF]{Schematic phase diagram of cuprate
superconductors. The phases: Antiferomagnetic (AF), pseudogap
(PG), superconductive (SC), Fermi-Liquid (FL) and
Non-Fermi-Liquid (No FL).} \label{intro2}
\end{center}
\end{figure}
\begin{figure}[!htbp]
\begin{center}
\includegraphics[width=10cm,height=7cm] {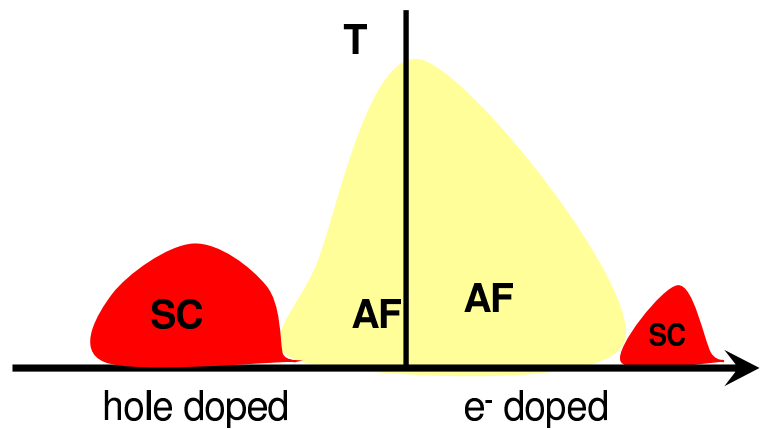}\\
\caption[entry for the LOF]{Schematic phase diagram of the
hole-electron-doping asymmetry in cuprate superconductor
materials. In the electron doped case (right hand side), the AF
region spans a wider region of doping and the SC phase is strongly
reduced with a much lower critical temperature. } \label{intro2b}
\end{center}
\end{figure}
The generic phase diagram of the cuprates shows a wide variety of
behavior at different temperatures and levels of doping (Fig.
\ref{intro2}). All the cuprate compounds investigated so far show
similar characteristic changes in their thermodynamic and
transport properties as the temperature or the number of holes
per unit cell of CuO$_{2}$ is varied. The number of holes per
CuO$_{2}$ unit, x, is a convenient parameter that can be used to
compare the different cuprates. The physical properties of the
cuprates change abruptly at the superconducting transition (and
also at the antiferromagnetic transition). In the other regions
of the phase diagram, however, the properties change gradually
and there is a "cross-over" rather than a well defined phase
transition. Understanding the phase diagram in Fig. \ref{intro2}
is tantamount to understanding the cuprates and their puzzling
behaviors, including high-temperature superconductivity. We are
interested in the thermodynamic, magnetic and transport
properties of these materials, but it is a challenge to develop a
microscopic theory that predicts all of these properties.
The antiferromagnetic region (AF) is the best understood region in
the phase diagram. At zero doping the cuprates are all insulators,
that, below a few hundred Kelvin, are also antiferromagnets (i.e.
the electron spins on neighbouring copper ions point in opposite
directions). However, when the doping $x$ is increased above a
critical value ( about 5\%, although this varies from compound to
compound), the antiferromagnetic state disappears and we enter
the so-called {\it pseudogap} (PG), or underdoped region, which
will be discussed in chapter 5. This region is called "underdoped"
because the level of doping is less than that which maximizes the
superconducting transition temperature. Some of the most unusual
behavior observed in the cuprates occurs in this region. The {\it
Fermi-liquid} (FL) region of the phase diagram (at high doping)
is also well understood. One of the central concepts in
condensed-matter physics, introduced by Lev Landau, is the
"quasiparticle" concept. In a so-called Landau-Fermi liquid the
properties of single free electrons are "renormalized" by
interactions with other electrons to form "quasiparticles". The
properties of the material can then be understood in terms of the
weak residual interactions between the quasiparticles and their
excitations. A key feature of the quasiparticle concept is that
low-energy single-particle excitations have very narrow
line-widths: $\Delta \omega \simeq \omega^{2}$ where $\omega$ is
the energy of the excitation. When the quasiparticle approach is
valid, there is a well defined boundary between particles and
holes in both energy and momentum space at zero temperature. This
boundary occurs at the Fermi energy and defines the "Fermi
surface" in momentum space. However, the Landau quasiparticle
model can only explain part of the phase diagram of the cuprates.
We shall return on the concept of Fermi-liquid in the next chapter
where we will attempt to approach the normal state properties of
H-T$_{C}$ superconductors from a phenomenological point of view.
We will consider the part of the phase diagram between the
underdoped and Fermi-liquid regions, and above the area with the
highest superconducting transition temperatures, which is called
the {\it non-Fermi-liquid region}(No FL). The thermodynamic
properties in this region are un-exceptional and, within
experimental uncertainties, are similar to the behavior of a
Fermi liquid. However, this region is characterized by simple but
unusual power laws in all of its transport properties as a
function of temperature. These transport properties include
resistivity, optical conductivity, electronic Raman-scattering
intensity, thermal conductivity, various nuclear relaxation
rates, Hall conductivity and magnetoresistance. Because of these
unusual transport properties this part of the phase diagram is
called No FL region.

Another important property of these materials is the strong
asymmetry in the phase diagram for hole-doped and electron-doped
systems (Fig. \ref{intro2b}). Compared with the hole-doped case,
the electron-doped phase diagram displays a more stable AF state
which extends to a wider region of doping (up to 15\%) and a
reduced SC state with a much lower critical temperature. We will
investigate this asymmetry in the normal state (chapter 5),
enlightening the physical mechanism underlying its formation, in
conjunction to the proximity to the Mott metal-insulator
transition and the set up of a d-wave superconductive state
(chapter 6).

\section{Multi-band Hamiltonian}

\begin{figure}[!htbp]
\begin{center}
\includegraphics[width=7cm,height=8.5cm] {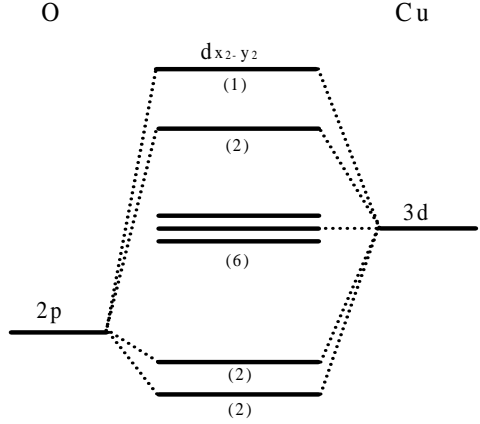}
\caption[entry for the LOF]{Hybridization of the $Cu^{2+}$ and  $O^{2-}$ orbitals}
\label{intro3}
\end{center}
\end{figure}

In order to find a Hamiltonian that is able to describe the
complex structure of high-T$_{c}$ cuprate materials, it is
unavoidable that we make some simplifications. First we consider
only the electrons in the two-dimensional CuO$_{2}$ planes where
the Cu-O bond is stronger so that a conduction band is formed. By
ignoring the interaction between planes we neglect some effects
that are present in real systems, such as the existence of a
non-zero N\'eel critical temperature for the AF state, which
quantum fluctuations would reduce to zero in a pure
two-dimensional system. Nevertheless, we hope to be able to
re-inject this feature in the system {\it a posteriori}, once the
physics that rules the CuO$_{2}$ two-dimensional electronic has
been fully understood. However, in spite of this first necessary
approximation the problem still remains extremely complex. There
are nine electrons in the five {\it d} orbitals of the Cu$^{2+}$
ions, while three {\it p} orbitals of the O$^{2-}$ are occupied.
So, in order to have a model from which we can extract important
results it is necessary to introduce further simplifications. In
the cuprates, every Cu atom is surrounded by oxygen ions: for
example in the La$_{2-x}$Sr$_{x}$CuO$_{4}$ there is a stretched
octahedron around every Cu$^{2+}$. This structure breaks the
rotational degeneracy in the {\it d} orbitals of an isolated
copper atom: the {\it d} orbitals of the Cu and {\it p} orbitals
of the O hybridize and split into separate levels (Fig.
\ref{intro3}).

The state with the highest energy, where a hole remains, is of the
{\it d$_{x^{2}-y^{2}}$} symmetry, hence a spin $\frac{1}{2}$
localizes here. So, in the insulating state, there is a hole on
each Cu ion of the CuO$_{2}$ planes, and the system is well
described by a model of localized spins with an AF groundstate.
The orbitals at lower energy are all occupied and, to first
approximation, can be disregarded in the model. But once doping is
introduced into the system, electrons are removed from the
CuO$_{2}$ planes which is equivalent to adding holes in more
energy levels of the scheme of Fig. \ref{intro3}. Where will the
holes be added? A first guess might be to place another hole in
the {\it d$_{x^{2}-y^{2}}$} level, but we must take into
consideration the strong Coulomb repulsion between holes in the
same orbital. So, we introduce a {\it multi-band} Hamiltonian in
an hole notation, with the vacuum state defined in such a way that
all the orbitals in Fig. \ref{intro3} are electron-occupied.

$$ H= -t_{pd} \sum_{<ij>}^{} p^{+}_{j}d_{i}+h.c.-t_{pp}\sum_{<jj^{'}>}^{} p^{+}_{j}p_{j^{'}}+
\varepsilon_{d} \sum_{i}^{} n_{i}^{d}+\varepsilon_{p} \sum_{j}^{} n_{j}^{p}+$$
$$+ U_{d} \sum_{i}^{} n_{i \uparrow}^{d}n_{i \downarrow}^{d}+U_{p} \sum_{j}^{} n_{j \uparrow}^{p}n_{j \downarrow}^{p}+
U_{dp} \sum_{<ij>}^{} n_{i}^{d}n_{j}^{p}$$  \bigskip \\
where $p_{j}$ is the fermionic operator which destroys a hole
localized on the O of site j, $d_{i}$ is the fermionic operator
which destroys the hole on the Cu of site i, $\langle ij \rangle$
refers to the nearest neighbor ions O in j and Cu in i. The
parameter $t_{pd}$ is the overlapping integral between the O and
the Cu atoms. For completeness, we have also introduced the term
$t_{pp}$ for hopping between neighboring oxygens. $U_{d}$ and
$U_{p}$ are positive constants which represent the Coulomb
repulsion between two holes on the same orbital {\it p} or {\it
d}. $U_{pd}$ has a similar meaning for two holes occupying
adjacent copper and oxygen ions. Generally distances longer than
nearest neighbors should also be taken into account, but we
assume here that the screening effect of the electrons cuts down
the Coulomb repulsion beyond near neighbors. Finally, the
$\varepsilon_{d}$ and $\varepsilon_{p}$ represent the different
hole-occupation energies of the orbitals O and Cu respectively.
In the insulating halfilled-band case, there is a hole on each Cu
ion which is energetically favored since
$\varepsilon_{p}-\varepsilon_{d}=\Delta
>0$. When another hole
is added to the unit cell, with $U_{d}> \Delta$, it prefers to
occupy an oxygen orbital, because this is again energetically
favored. This is in agreement with
electron-energy-loss-spectroscopy (EELS) experiments
\cite{Nucker}. Theoretical band structure calculations
\cite{Hybertsen} have been used to fix the universally recognized
numerical values of the parameters in {\it H}, as shown in the
following table:

\begin{center}
\begin{tabular}[c]{cccccc} \hline

$U_{d}   $& $U_{p}   $& $U_{pd}  $ &
$\varepsilon_{p}-\varepsilon_{d}  $ &$ t_{pd} $  & $t_{pp} $\\
\hline
                                                                                       \\
10.5    &    4    &     1.2  &      3.6                          &     1.3  &     0.65 \\
\hline
\end{tabular}\\ \vspace{0.5cm}
Parameters of the multi-band Hamiltonian in eV
\end{center} \bigskip
These data are consistent with the hypothesis presented above on
the structure of the cuprate superconductors.

\section{The Hubbard Model}

\begin{figure}[!htbp]
\begin{center}
\includegraphics[width=10cm,height=15cm] {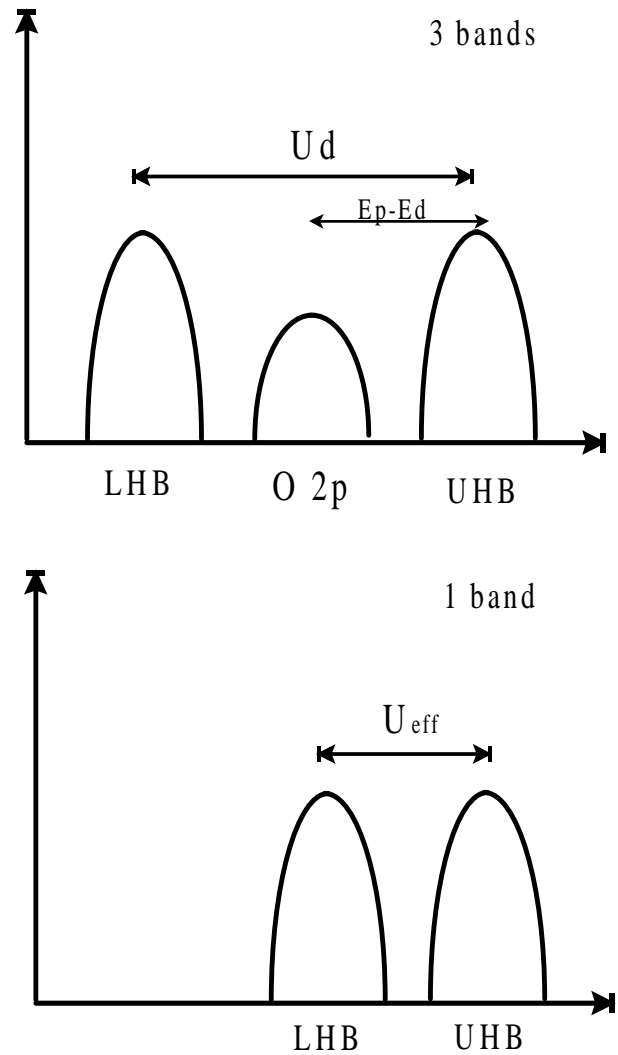}
\caption[entry for the LOF]{ Effective band structure of the
$CuO_{2}$ planes.}\label{intro4}
\end{center}
\end{figure}

The multi-band Hamiltonian presented in the previous section,
though a plausible first-order-approximate description
forcuprates, is still too complicated to perform calculations. We
are therefore forced to introduce an even simpler Hamiltonian,
which nevertheless is still able to describe the low energy
properties of the system. Since the early days of H-T$_{c}$
superconductivity a simple model yet useful to describe the
properties of the cuprates is the Hubbard Model, proposed by
Hubbard in 1963 \cite{hubbard:64} to study the dynamics of the
electrons in the transition metals. It is defined by this
Hamiltonian:

$$ H= -t \sum_{<ij>,\sigma}^{} (c^{+}_{i\sigma}c_{j\sigma}+ h.c.) +U \sum_{i}^{} n_{i\uparrow}n_{i\downarrow}$$
\bigskip  \\
where the fermionic operator $c^{+}_{i\sigma}$ creates an electron
with spin $\sigma$ on site i of a square lattice, {\it U} is the
on-site Coulomb repulsion, and {\it t} is the hopping amplitude.
Cuprate superconductors are better described by a structure with
at least three main bands, as explained in the previous section,
but the Hubbard model tries to simulate the presence of the gap
$\Delta= \varepsilon_{p}-\varepsilon_{d}$ through an effective
value of the Coulomb repulsion U. In this way only two bands are
involved, with the oxygen band being at lower energy (as
schematically shown in the bottom panel of Fig. \ref{intro4}).
%
This Hamiltonian however does not consider the presence of the O,
and the Coulomb interactions are reduced to on-site interactions
only. There is not in fact a clear justifications {\it a priori}
for the choice of this model. However an {\it a posteriori}
justification is that many of its properties qualitatively
describe the cuprate superconductors in the normal state.
Unfortunately there is not conclusive evidence concerning the
superconductivity. This may be due to the limitations of the
model, which misses some fundamental ingredients, or a failure of
the techniques used to solve it. As in many cases when a
theoretical model tries to explain a more complex reality, its
introduction can be justified {\it a posteriori} only after its
properties are completely understood.

\subsection{Weak Interaction Limit U $<<$t}

To gain some understanding of the properties of the Hubbard Model
we study the weak coupling limit U $<<$t. Let us first consider
the trivial U $=$0 case, which describes a simple non-interacting
Fermi gas. Its ground state is given by:

$$ |\Phi_{0}\rangle= \prod_{k<k_{f},\sigma}^{} \,  c^{+}_{{\bf k},\sigma}|0\rangle $$  \bigskip \\
where $|0\rangle$ is the vacuum state of the fermionic
destruction operator $c^{}_{{\bf k},\sigma}$. The creation
operator $c^{+}_{{\bf k},\sigma}$ is the Fourier transform of the
site creation operator, namely
\begin{equation}
c^{+}_{{\bf k},\sigma}= \frac{1}{\sqrt{N}} \sum_{i}^{} e^{i{\bf
kR_{i}}} \hspace{2pt} c^{+}_{i,\sigma} \label{1.1}
\end{equation}
where the quasimomentum $k$ is given by periodic boundary
conditions ${\bf k}= \frac{2\pi}{\sqrt{N}a}(n_{x}n_{y})$, N is
the total number of sites in the system, {\it a} the lattice
constant and $ \frac{\sqrt{N}}{2}+1\leq
n_{x,y}\leq\frac{\sqrt{N}}{2}$. We start from the Hamiltonian
$H_{o}$:
$$H_{o}=-t\sum_{\langle ij\rangle,\sigma}^{}(c^{+}_{i\sigma}c_{j\sigma}+c^{+}_{j\sigma}c_{i\sigma}),$$
which becomes:
$$H_{o}=-t\sum_{{\bf n}, i}^{}\frac{1}{N}\sum_{{\bf k,k^{'}}\sigma}^{} e^{i({\bf k-k^{'} })
{\bf R_{i}}} e^{i{\bf kn}a} c^{+}_{{\bf k\sigma}} c^{+}_{{\bf
k'\sigma}}$$ where the sum $\langle ij\rangle$ over nearest
neighbors has become a sum over i and ${\bf n}$ with ${\bf
R_{j}}={\bf R_{i}}+ a{\bf n}$, where ${\bf n}$ is the versor
indicating the four possible directions in a two-dimensional
square lattice. Thus, we find:
$$H_{o}= -2t\sum_{{\bf k,\sigma}}^{} \varepsilon_{{\bf k}} c^{+}_{{\bf k}\sigma}c_{{\bf k}\sigma}$$
where the dispersion is given by $\varepsilon_{{\bf k}}= -2t\,(\cos k_{x}a+\cos k_{y}a)$.

The ground state is obtained by filling up the band to the Fermi
level. As each single-particle state, labeled by the
quasi-momentum vector {\bf k}, is spin-doubly degenerate, the
total number of states for a system of $N$ sites is $2N$. If (like
in the case of the un-doped cuprates) there are also $N$
electrons, one for each site, the band is half-filled and the
Fermi level is given by: $\cos(k_{x}a)+\cos(k_{y}a)=0$, i.e.
$|k_{x}|+|k_{y}|=\frac{\pi}{a}$. According to single-band theory
such system is a good metal. We have seen, however, that the
un-doped cuprates superconductors materials are instead good
insulators. This clearly shows the failure of standard band
theory to describe such materials. It is therefore necessary to
add an interaction U $>$t into the system. However, adding a small
perturbation to a non-interacting electron Hamiltonian only
re-normalizes the free electron gas, affecting only the energy
levels $\varepsilon_{{\bf k}}$ close to the Fermi energy. In
order to radically change the physics of the system we must
consider an interaction $U$ strong enough to break the free-electron
energy band and produce a new kind of fermionic gas. 

\subsection{Hartree-Fock Approximation}

Let us now apply to the un-perturbed Hamiltonian $H_{0}$ a
mean-field Hartree-Fock interaction term $H_{1}$:
$H_{HF}=H_{0}+H_{1}$ with
$$H_{1}= U\sum_{i}^{} \, ( \langle n_{i\uparrow}\rangle
n_{i\downarrow} + n_{i\uparrow}\langle n_{i\downarrow}\rangle )$$
where $\langle n_{i\uparrow}\rangle$ is the ground-state
expectation value of the density operator $n_{i\uparrow}$, which
has to be determined self-consistently from $H_{HF}$. We can in
fact write $n_{i\sigma}= \langle n_{i\sigma}\rangle + \Theta$ to
good approximation if $\langle \Theta^{2}\rangle<<\langle
n_{i\sigma}\rangle^{2}$. Since $(n_{i\uparrow}-\langle
n_{i\uparrow}\rangle) (n_{i\downarrow}-\langle
n_{i\downarrow}\rangle) \sim \Theta^{2}$, we have to first order
in perturbation theory:
$$n_{i\uparrow}n_{i\downarrow}\cong \langle n_{i\uparrow}\rangle n_{i\downarrow}
+n_{i\uparrow}\langle n_{i\downarrow}\rangle-\langle
n_{i\uparrow}\rangle\langle n_{i\downarrow}\rangle .$$ In $H_{1}$
we have omitted the last term which is a constant. We once again
consider the case of the half-filled band ($\rho$= 1) materials.
For this case we anticipate antiferromagnetic (AF) properties to
be relevant. So, to describe them we divide the lattice into two
equivalent sub-lattices, each describing the two different spin
orientations in a symmetry-broken AF state.
\begin{figure}[!htbp]
\begin{center}
\includegraphics[width=10cm,height=15cm] {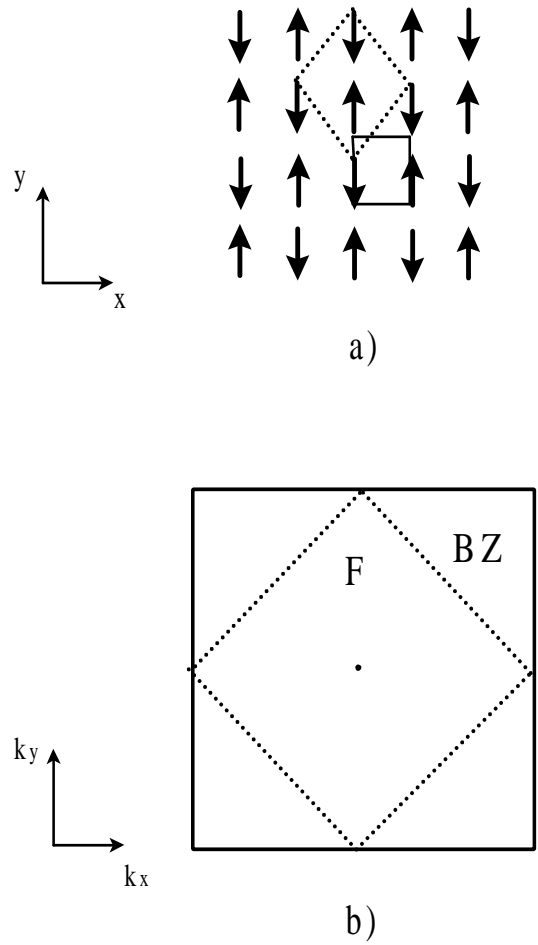}
\caption[entry for the LOF]{Unitary cell and BZ in an AF lattice.
} \label{intro5}
\end{center}
\end{figure}
In Fig. \ref{intro5}a we indicate the AF ground-state, with the
spins polarized along opposite $z$-directions on the two different
sublattices. In this representation the unit cell (dashed line in
{\it a}), is twice the unit cell of the original lattice. Hence
the Brillouin Zone (BZ) in the reciprocal space F (bottom
drawing) is half of the original BZ, reflecting the periodicity
of the AF state. The unit vectors in the reciprocal lattice
change to:
$$ \left\{ \begin{matrix}
\frac{2\pi}{a}(1,0) \\
\frac{2\pi}{a}(0,1)
 \end{matrix} \right.
\rightarrow  \qquad \left \{
\begin{matrix}
 \frac{\pi}{a}(1,1) \hfill \\
 \frac{\pi}{a}(1,-1)
\end{matrix} \right. $$
If ${\bf R_{j}}=(ra,sa)$ is the vector localizing site j in the
lattice, we parameterize: \begin{equation} \left\{
\begin{matrix}
\langle n_{j\uparrow} \rangle=  & \frac{1}{2}+(-1)^{r+s} m \\                                                                \\
\langle n_{j\downarrow} \rangle=& \frac{1}{2}-(-1)^{r+s} m
\end{matrix} \right.  \label{1.2} \end{equation}
where, $m= \frac{(-1)}{2}^{r+s} (\langle n_{j\uparrow}
\rangle-\langle n_{j\downarrow} \rangle)$
 is the on-site magnetization to be self-consistently determined.
 Notice that if $m=0$
 $ \langle n_{j\uparrow} \rangle=\langle n_{j\downarrow} \rangle=\frac{1}{2} $
 and the system is paramagnetic.
 In the opposite limit 
 $ \langle n_{j} \rangle= \langle n_{j\uparrow} \rangle+\langle n_{j\downarrow} \rangle=
 \langle n_{j\uparrow} \rangle= 1$ if $r+s$ ie even, and $ \langle n_{j} \rangle=\langle n_{j\downarrow}
 \rangle= -1$ if $r+s$ is odd, and the system is in a perfect AF state.
The Hartree-Fock Hamiltonian becomes:
$$H^{SF}_{HF}= -t \sum_{\langle ij,\sigma \rangle}^{} (c^{+}_{i\sigma}c_{j\sigma}+ h.c.)+
 \sum_{i,\sigma}^{} \frac{U}{2}(1-m\sigma e^{i{\bf qR_{i}}}) c^{+}_{i\sigma}c_{i\sigma} $$
where $ \sigma= +(-)1$ for $ \uparrow(\downarrow)$ spins, and
${\bf q}=\frac{\pi}{a}(1,1)$. $e^{i{\bf qR_{j}}} = \pm 1$
according to which of the two sub-lattices $j$ belongs. We
disregard the constant term $ \sum_{i}^{}\frac{U}{2} n_{i}=
\frac{U}{2}N$, since the number of particles is conserved, and
obtain: \beq H^{AF}_{HF}= \sum_{{\bf k},\sigma}^{}
\varepsilon_{{\bf k}}c^{+}_{{\bf k}\sigma}c_{{\bf k}\sigma}- mU
\sum_{{\bf k}\sigma}^{} \sigma c^{+}_{{\bf k+q}\sigma}c_{{\bf
k}\sigma} \label{1.3}\eeq
Notice that, owing to the periodicity of the lattice \cite{Pini}: \\
\begin{center}
\begin{tabular}[position]{lll}
$\varepsilon_{{\bf k+}2{\bf q}}= \varepsilon_{{\bf k}} $ \\ $\varepsilon_{{\bf k+q}}=
-\varepsilon_{{\bf k}}$ \\
$c_{{\bf k+}2{\bf q}}= c_{{\bf k}}$ \\
\end{tabular}
\end{center}
The Hamiltonian \ref{1.3} can be diagonalized using a Bogoliubov
transformation. The time evolution of the electron destruction
operators $c_{{\bf k}\uparrow}$ and $c_{{\bf k}\downarrow}$ in
Heisenberg notation is given by: \beq i\hbar \dot c_{{\bf
k}\sigma}= [c_{{\bf k}\sigma},H_{HF}]. \label{1.4} \eeq Thus,
using eq. \ref{1.3}, we have:
$$ i\hbar \begin{pmatrix}
\dot c_{{\bf k}\sigma} \cr \dot c_{{\bf k+q}\sigma} \cr
\end{pmatrix}
 = \begin{pmatrix} \varepsilon_{{\bf k}} & -\sigma\Delta \cr
-\sigma\Delta         & -\varepsilon_{{\bf k}} \cr \end{pmatrix}
\begin{pmatrix}
 c_{{\bf k}\sigma} \cr
 c_{{\bf k+q}\sigma} \cr
\end{pmatrix} $$ where $\Delta= mU$. For convenience, let {\bf M} be
the time-evolution matrix. The transformation {\bf
S}($\sigma$,{\bf k}) which diagonalizes $H_{HF}$ also
diagonalizes {\bf M}, so that:
$${\bf S^{-1}MS}= {\bf D}=
\begin{pmatrix} E^{1}_{{\bf k}} & 0 \cr 0                &
E^{2}_{{\bf k}} \cr \end{pmatrix}.$$ This is the diagonal matrix
with the eigenvalues: \beq
\begin{pmatrix} \alpha_{{\bf k+q}\sigma}
\cr \alpha_{{\bf k}\sigma} \cr
\end{pmatrix}=
{\bf S^{-1}} \ \begin{pmatrix}
 c_{{\bf k}\sigma} \cr
 c_{{\bf k+q}\sigma} \cr
\end{pmatrix} .\label{1.5}  \eeq
The new quasiparticles thus evolve in time simply as:
$$i\hbar \begin{pmatrix}
\dot \alpha_{{\bf k+q}\sigma} \cr \dot \alpha_{{\bf k}\sigma} \cr
\end{pmatrix} =
\begin{pmatrix}
 E^{1}_{{\bf k}}&0 \cr
0                & E^{2}_{{\bf k}} \cr \end{pmatrix}
\begin{pmatrix}
 \alpha_{{\bf k}\sigma} \cr
 \alpha_{{\bf k+q}\sigma} \cr
\end{pmatrix}. $$\\
and the Hamiltonian can be written:
 $H_{HF}= \sum_{{\bf
h}\sigma}^{} (E^{1}_{{\bf k}}\alpha^{+}_{{\bf
k+q}\sigma}\alpha_{{\bf k+q}\sigma} + E^{2}_{{\bf k}}
\alpha^{+}_{{\bf k}}\alpha_{{\bf k}\sigma})$ as required by
\ref{1.4}. The eigenvalues  $E_{{\bf k}}^{1(2)}$ are determined by
resolving the determinant for {\bf M}. We obtain $E_{{\bf
k}}^{1(2)}= \pm E_{{\bf k}}= \pm \sqrt{ \varepsilon^{2}_{{\bf
k}}+\Delta^{2}}$. The transformation {\bf S} has however to be
chosen in order to preserve the canonical commutation relations,
so that: $\alpha_{{\bf k}\sigma}^{+} \hbox{ and } \alpha_{{\bf
k}\sigma}$ are real fermionic-quasiparticle construction and
destruction operators. The matrix {\bf S} is the eigenvector
matrix of {\bf M}:
$$ \begin{pmatrix}
u_{{\bf k}}          & \sigma v_{{\bf k}} \cr -\sigma v_{{\bf k}}
& u_{{\bf k}} \cr \end{pmatrix}$$
 with the constraint \beq
\Delta v_{{\bf k}}= (E_{{\bf k}}-\varepsilon_{{\bf k}})u_{{\bf
k}} . \label{1.6} \eeq If we assume $u_{{\bf k}}$ and $v_{{\bf
k}}$ real, using (\ref{1.5}) we find \cite{Pini}:
$$ \{ c_{{\bf k}\sigma},c_{{\bf k'}\sigma}^{+} \}= (u_{{\bf k}}^{2}+
v_{{\bf k}}^{2})\delta_{{\bf kk^{'}}}+u_{{\bf k}}v_{{\bf
k^{'}}}\delta_{{\bf k^{'}k+q}}+ v_{{\bf k}}u_{{\bf
k^{'}}}\delta_{{\bf kk^{'}+q}}\equiv \delta_{{\bf kk^{'}}}$$ which
requires that \beq \label{1.7} \ u_{{\bf k}}^{2}+v_{{\bf k}}^{2}=
1\eeq and that the labels ${\bf k} \hbox{ and } {\bf k^{'}+q}$ are
never equal, for all couple $({\bf k},{\bf k^{'}})$, in such a way
that $\delta_{{\bf kk^{'}+q}}= \delta_{{\bf k+qk^{'}}}=0$. This
is in fact always true if we constrain the sum of the
quasi-momenta {\bf k} to the reduced BZ $\,$ F defined by the AF
lattice. Moreover, since the Hamiltonian is diagonal with respect
to $\alpha^{+}_{{\bf k}}\alpha_{{\bf k}} \hbox{ and }
\alpha^{+}_{{\bf k+q}}\alpha_{{\bf k+q}}$, where {\bf q} is a
primitive vector of the reciprocal lattice, all the possible
states are covered even if the sum is restricted to F only.
Equations (\ref{1.6}) and (\ref{1.7}) determine in this way the
transformation {\bf S}. Finally we obtain:
$$H_{HF}=\sum_{{\bf k}\in F,\sigma}^{} E_{{\bf k}}(\alpha^{+}_{{\bf k+q}
\sigma}\alpha_{{\bf k+q}\sigma}-\alpha^{+}_{{\bf
k}\sigma}\alpha_{{\bf k}\sigma}).$$ In summary, the effect of the
interaction U with respect to the un-perturbed case U=0 is to
split the band in two sub-bands of energies: $-E_{{\bf k}} \hbox{
and } +E_{{\bf k}}$ for ${\bf k} \in F$, the reduced BZ. Being
the latter half of the original BZ, the states in each sub-band
are also half of the original un-perturbed one. In the
half-filled case ($\rho=1$), the lower band is filled, and the
first accessible states are in the upper band, which is separated
by a gap of 2$\Delta$. The Hubbard model proves therefore able to
explain the insulating properties of the half-filled state of the
cuprate materials. These are the result of the strong on-site
interactions\ between particles of opposite spin, which is
described in the model by the U parameter. Systems with these
properties are called {\it Mott insulators}.

In order to implement self-consistency in the method we have to
determine the magnetization $m$, requesting that the parameterized
form of $\langle n_{i\sigma} \rangle $ assumed in \ref{1.2} is
equal to that derived from the calculated $H_{HF}$. Writing for
example $n_{j\uparrow}$ as a function of the quasiparticle
operators which diagonalize the Hamiltonian, we find \cite{Pini}
$$ m= \frac{1}{N}\sum_{{\bf k}\in F}^{} 2u_{{\bf k}}v_{{\bf k}}$$
and using the expressions for $u_{{\bf k}} \hbox{ and } v_{{\bf
k}}$ (\ref{1.6}) (\ref{1.7}), we obtain a {\it gap equation}:
$$ \frac{U}{N}\sum_{{\bf k}\in F}^{} \frac{1}{E_{{\bf k}}}=1 $$
which in the continuous limit becomes
$$ \frac{Ua^{2}}{(2\pi)^{2}} \int_{F}^{} \frac{1}{\sqrt{\varepsilon^{2}_{{\bf k}}+\Delta^{2}}} \,
d^{2}k =1 $$ Changing variables
$\frac{a^{2}d^{2}k}{(2\pi)^{2}}\rightarrow \rho(\varepsilon)
d\varepsilon$, where $\rho(\varepsilon)$ is the density of states
$\varepsilon_{{\bf k}}$ and taking into account that for ${\bf k}
\in F, \, \varepsilon \in (-4t,0)$, we find that the gap equation
becomes :
$$ U\int_{0}^{4t} \rho(\varepsilon) \frac{1}{\sqrt{\varepsilon^{2}+\Delta^{2}}} \, d\varepsilon =1 .$$
For the case of weak interaction $U<< t$, we expect a solution
close to the un-perturbed case: being $\Delta= mU$,  $\Delta
\rightarrow 0$. The integrand is dominated by a term in the
neighborhood of  $\varepsilon=0$, where it can be shown $\rho
(\varepsilon)\rightarrow \infty$. Hence expanding the integrand
near $\varepsilon=0$ \cite{Pini}, we find:
$$ \Delta= 16t \, e^{-2\pi \sqrt{t/U}}.$$
For $U\rightarrow 0$ therefore the system exponentially converges
to the free case, with the gap $\Delta$ going to zero and the
magnetization given by $m=\Delta /U$. But this convergence is less
quick than in a system with a regular density of states at the
Fermi level $\varepsilon= 0$, for which we would find an
exponential of $-t/U$ rather than that of $-\sqrt{t/U}$. AF
instabilities show therefore to be favored in the half-filled
Hubbard Model.

In the opposite limit $U>> t$ instead we have $\Delta>> t$, which
dominates the denominator in the integrand of the gap equation.
Thus we find approximately:
$$\frac{U}{\Delta} \int_{0}^{4t} \rho(\varepsilon) \, d\varepsilon =1 \qquad \hbox{ and since }
\qquad
                   \int_{-4t}^{4t} \rho(\varepsilon) \, d\varepsilon =1 $$
$\Delta= U/2$. The gap between the lower and upper bands tends to
{\it U}, in agreement with the intuitive idea that single
particle excitations are due to the double occupation of the same
sites. The magnetization becomes $m= \Delta/U= 1/2$, which
corresponds to a classical AF. This was in fact the landscape on
which we have built the Hartree-Fock construction.

Let's now add doping into the system, with $\delta$ holes for each
site. We expect a phase transition into a metallic phase taking
place. We can re-write the relations \ref{1.2}:
$$ \left\{
\begin{matrix} \langle n_{j\uparrow} \rangle=&
\frac{1-\delta}{2}+(-1)^{r+s} m  \cr
                                                                \\
\langle n_{j\downarrow} \rangle=& \frac{1-\delta}{2}-(-1)^{r+s} m
\cr \end{matrix} \right.  $$ and the gap equation becomes
$$ \frac{U}{N}\sum_{{\bf k}\in K}^{} \frac{1}{E_{{\bf k}}}=1 $$
where $K \subset F$ is the new region of the phase space occupied
by electrons that now do not fill completely the lower band.
Taking, as above the continuous limit:
$$ U\int_{\varpi_{0}}^{\varepsilon_{0}}
\rho(\varepsilon) \frac{1}{\sqrt{\varepsilon^{2}+\Delta^{2}}} \,
d\varepsilon =1 $$ where $\varpi_{0}$ is the bottom energy of the
lower Hubbard band and the Fermi energy $\varepsilon_{0}$ of the
non-interacting system is defined as
$$ \int_{\varpi_{0}}^{\varepsilon_{0}} \rho(\varepsilon) \, d\varepsilon =\frac{1-\delta}{2}.$$
As $\varepsilon_{0} < \varepsilon_{F}= 0$, is negative, the
integrand is no longer singular in the neighborhood of
$\varepsilon_{F}= 0$, and remains finite for $\Delta=0$, where it
assumes its maximum value. There exists therefore a finite U$_{c}$
which satisfies the gap equation condition and for U$<$U$_{c}$
there is no solution. Hence, for decreasing U, there is a
metal-insulator transition, corresponding to the AF-paramagnet
transition.

In conclusion, a mean-field Hartree-Fock approach to the Hubbard
model is able to qualitatively describe the behavior of the
H-T$_{c}$ superconductors in their insulating and normal metallic
states. This explains a posteriori the introduction of this model,
which embodies the essential properties of these materials.
Whether the model can explain the rich phase diagram of H-T$_{c}$
superconductors or can reproduce a H-T$_{c}$ superconductive
mechanism remains an open question.

\section{Heisenberg Hamiltonian}

The Hartree-Fock results of the previous section suggest that the
ground-state of the Hubbard Model is close to a classical AF
state. The on-site magnetization assumes alternatively a value
which tends to $\pm\frac{1}{2}$, and each electron can be thought
to be occupying a single site with a double occupation forbidden.
In this section we want to reverse this point of view from the
Hartree-Fock treatment starting instead from the strong
interacting limit. Let's assume U$>>$ t, and let us consider the
interaction part of the Hamiltonian as the un-perturbed portion:
 $H_{0}= U\sum_{\langle ij \rangle}^{} n_{i\uparrow} n_{j\downarrow}$.
 The kinetic term $H_{1}= U\sum_{\langle ij \rangle}^{}( c^{+}_{j\sigma}c_{i\sigma} + h.c.)$,
 now becomes the perturbation. Since a doubly occupied site is very expensive in
 terms of energy (with a cost of order U), we constrain the treatment to the subspace $\Xi$,
 generated by the eigenvectors $\{ \varphi_{0} \} \hbox{ of } H_{0}$
 corresponding to the eigenvalue $E=0$, where double occupation is
 forbidden.
 In the half-filled case, the groundstate of $H_{0}$, $|\Phi_{0}\rangle$ is given by
 assigning a single electron to each site so that $H_{0}|\Phi_{0}\rangle
 =0$ and $|\Phi_{0}\rangle \in \Xi$.
 Because of the spin degeneracy there are $2^{N}$ such states,
 where $N$ is the total number of sites in the system.
 An excited state is created by doubly occupying a site. In this
 case
$H_{0} |\Phi_{1}\rangle= U|\Phi_{1}\rangle$, and in general,
$|\Phi_{n}\rangle$ corresponds to $n$ doubly occupied sites,
$H_{0} |\Phi_{n}\rangle= nU|\Phi_{n}\rangle$.  The energy-gap
between the groundstate and the first excited
 state is U, and in the limit $U\rightarrow \infty$
 double occupation is forbidden. We can therefore assume that, for
sufficiently U, confining the system to the subspace $\Xi$ would
be a good approximation. Adding the kinetic perturbation removes
the $2^{N}$ degeneracy and creates an energy-band that is narrow
compared to the energy-scale U of the charge excitation.

We now apply perturbation theory to find an effective Hamiltonian
in the subspace $\Xi$. Let's consider $H=H_{0}+\lambda H_{1}$,
with the small parameter $\lambda$ introduced for convenience,
and $|\Psi\rangle$ the generic eigenstate of $H$. We can write
 $|\Psi\rangle= |\Psi_{0}\rangle +\lambda |\Psi_{1}\rangle$, where $|\Psi_{0}\rangle \in \Xi$,
 assuming without loss of generality that $|\Psi_{1}\rangle \perp |\Psi_{0}\rangle$.
 From the eigenvalue equation $H |\Psi\rangle= E |\Psi\rangle $ we obtain:
$$ \left \{
\begin{matrix}
 H_{0} |\Psi_{0}\rangle=0 \hfill &          & \hfill
(i) \cr
                                                                \\
\lambda H_{1} |\Psi_{0}\rangle+\lambda H_{0} |\Psi_{1}\rangle = \lambda E_{1} |\Psi_{0}\rangle \hfill &     & \hfill (ii)\cr
                                                                  \\
\lambda^{2} H_{1} |\Psi_{1}\rangle = \lambda^{2} E_{2}
|\Psi_{0}\rangle \hfill &   & \hfill (iii) \cr \end{matrix}
\right.  $$ \bigskip \\
where we have expanded the eigenvalue in powers of $ \lambda$, $E=
E_{0}+ \lambda E_{1}+ \lambda^{2}E_{2}$, where $E_{0}=0$. Formula
(i) is just the eigenvalue equation of the un-perturbed
Hamiltonian. Multiplying both sides of (ii) by $\langle \Psi_{0}|$
we obtain $E_{1}= \langle \Psi_{0}| H_{1} |\Psi_{0}\rangle+
\langle \Psi_{0}| H_{0}| \Psi_{1}\rangle= 0$, since the two terms
in the second side of the equation are both zero, because $H_{1}
|\Psi_{0}\rangle \perp |\Psi_{0}\rangle$ (its effect is to move
an $e^{-}$ to the next neighbor site which however is already
occupied) and  $H_{0} |\Psi_{0}\rangle \in \Xi$ is
$\perp|\Psi_{1}\rangle$. Multiplying (iii) by $\langle \Psi_{0}|$
we obtain $E_{2}= \langle \Psi_{0}| H_{1} |\Psi_{1}\rangle$. In
order to evaluate $\langle \Psi_{0}|H_{1}| \Psi_{1} \rangle$, we
multiply (ii) by $\langle \Psi_{0}|H_{1}$ to obtain $\langle
\Psi_{0}| H_{1} |\Psi_{1}\rangle=-U^{-1}\langle \Psi_{0}|
H_{1}^{2} |\Psi_{0}\rangle$, so that finally:
$$ \langle \Psi_{0}| -\frac{1}{U}H_{1}^{2} |\Psi_{0}\rangle = E_{2}.$$
This equation is equivalent to the initial eigenvalue equation to
first order in perturbation theory and shows that the first
correction to the un-perturbed energy in the subspace $\Xi$ is
$E_{2}$. The effective Hamiltonian is therefore given by: \beq
H_{eff}= -\frac{1}{U}H_{1}^{2} \label{1.8}.\eeq The appearance of
$H_{1}^{2}$ means that forbidding double occupation restricts the
system to virtual electron-jumps to neighboring sites. This
Hamiltonian can be written in this more representative form:
$$H^{2}_{1}= t^{2} \sum_{klij,\sigma \sigma^{'}}^{} c^{+}_{k\sigma^{'}}c_{l\sigma^{'}}
c^{+}_{i\sigma}c_{j\sigma}$$ and since the only jumps permitted
are virtual, the only non-zero terms in the sum are the ones for
which  $k=j \hbox{ and } l=i$, so that:
$$H_{eff}= -\frac{1}{U} \sum_{kl,\sigma \sigma^{'}}^{} c^{+}_{k\sigma^{'}}c_{l\sigma^{'}}c^{+}
_{l\sigma}c_{k\sigma}.$$ We can now use spin-operators, defined
as:
$$\begin{matrix}
c^{+}_{\sigma}c_{\sigma} \hfill & \rightarrow \hfill & \hfill \frac{1}{2}+ \sigma S^{z} \cr
                               \\
c^{+}_{\sigma}c_{-\sigma} \hfill & \rightarrow \hfill & \hfill
S^{x}+i \sigma S^{y} \cr \end{matrix}$$ where $\sigma= \pm 1$ for
spin $\uparrow \hbox{ o } \downarrow$, and satisfying the usual
angular momentum commutation relations $[S^{x},S^{y}]= iS^{z}$.
If we keep only next-neighbors in the sum:
$$H_{eff}= \frac{4t^{2}}{U} \sum_{\langle ij\rangle}^{}({\bf S_{i}S_{j}}-\frac{1}{4})$$
The effective Hamiltonian which describes the behavior of the
Hubbard Hamiltonian in the limit of strong interaction
$U\rightarrow\infty$ is naturally the Heisenberg Hamiltonian with
coupling constant $J= 4t^{2}/U> 0$.

\subsection{AF Groundstate}

\begin{figure}[!htbp]
\begin{center}
\includegraphics[width=15cm,height=20cm] {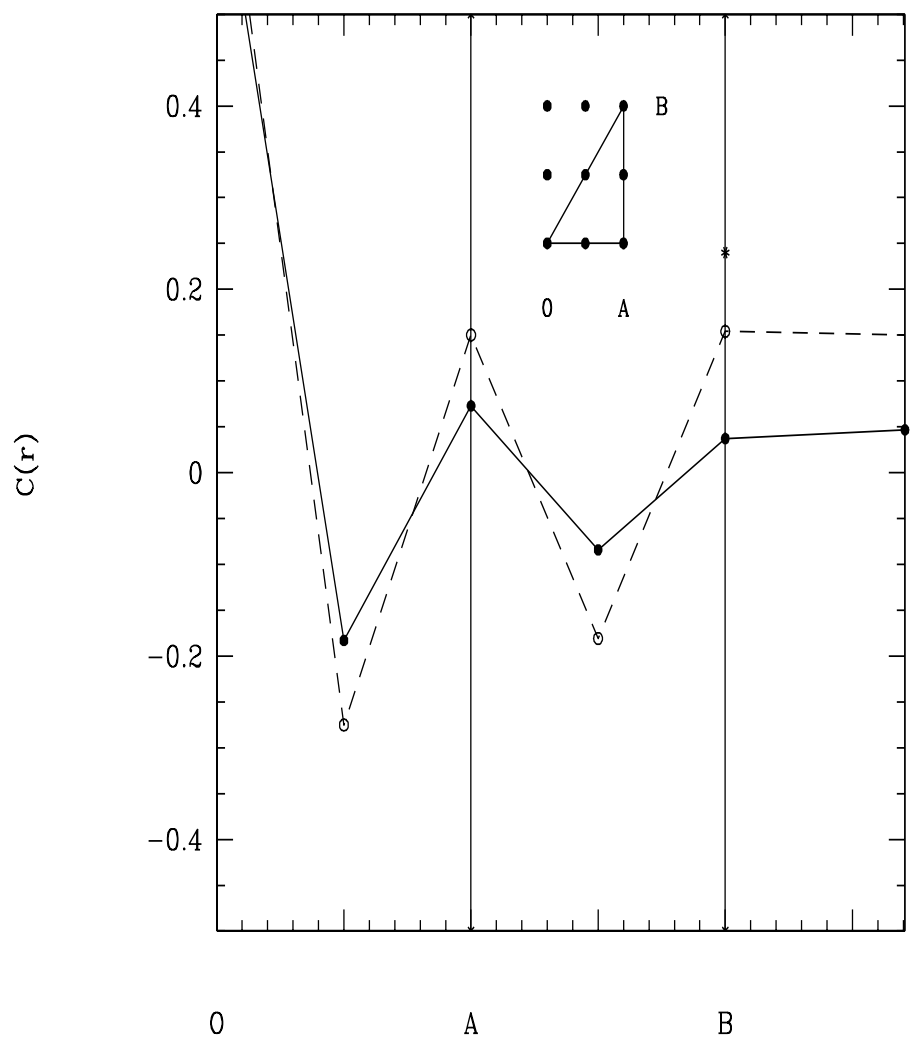}
\caption[]{Spin-spin correlation function C(r). The solid line
represents the half-filled system, the dashed line a
two-hole-doped system.L} \label{mag}
\end{center}
\end{figure}

What can we say about the groundstate? From the Hartree-Fock
treatment a natural guess would be to consider a classical AF
state where the spins are alternatively $S^{z}= \pm\frac{1}{2}$,
parting the lattice into two ferromagnetic sublattices $A$ and
$B$:
$$|\Phi_{0} \rangle= \prod_{i \in A}^{}\prod_{j \in B}^{} | \uparrow \rangle_{i} \
| \downarrow \rangle_{j}.$$ If the spins were classical vectors,
the energy of this state would be $E_{0}= -JNg/8$, where $g$ is
the number of next-neighbors. $|\Phi_{0} \rangle$ however is not
even an eigenstate of Hamiltonian, as can easily be checked by
applying it to $H_{eff}$ (the kinetic $S^{+}_{i}S^{-}_{j}$ term
in fact destroys the perfect AF arrangement in  $|\Phi_{0}
\rangle$). However, it is easily verified (\cite{AM} pg. 704) that
the real value of $E_{0}$ in the Heisenberg Hamiltonian with spin
$S$ is bounded as folllows:
$$-S(S+1)\frac{g}{2}NJ \leq E_{0} \leq -S^{2} \ \frac{g}{2}NJ.$$
Notice that in the classical limit $S\rightarrow \infty$, the
ratio of the two boundaries tends to 1 and we recover the
classical AF solution. In the present case, however, the value of
$S$ is small and a precise determination of $E_{0}$ is not
possible. For example in the exactly solvable one dimensional
case \cite{Bethe}, $E_{0}= -2NJ(ln2-1/4)= -0.886NJ$, and
$-0.5NJ\geq E_{0} \geq -1.5NJ$: far from a precise estimate of the
exact value. Eventually, if we could turn off quantum mechanics we
would recover the classical limit. Let's imagine therefore to
perform an ideal experiment in which $S\rightarrow \infty$ keeping
$JS^{2}$ finite while ${\bf S}/S\rightarrow 1$. In this way $1/S$
plays the role of Planck's constant $\hbar$. As soon as we allow
$1/S$ to grow from zero, two radical happenings may take place in
the system: i) either the AF symmetry survives until a finite
value of $1/S$, ii) or AF is immediately destroyed. The latter is
in fact what happens in the one dimensional case, while it is
likely that the first case is applicable in higher dimensions.
The correct way to realize such an ideal experiment is to
introduce bosonic operators in place of the spins trough a
Holstein-Primakoff transformation:
\bigskip $$\begin{matrix}
S^{+}= \hfill &   (2S)^{\frac{1}{2}} \sqrt{1-a^{+}a/2S} \, a \hfill \cr
                               \\
S^{-}= \hfill &   (2S)^{\frac{1}{2}} a^{+} \sqrt{1-a^{+}a/2S} \hfill \cr
                              \\
S^{z}= \hfill &   S-a^{+}a \hfill \cr \end{matrix}$$ This
transformation is exact except for the fact that the Fock's space
of the boson operators has infinite dimension, contrary to the
finite dimension of the spin's space. Its advantage is
straightforward, the bosonic operators obey to appropriate
commutation relations. Straightforward is also the disadvantage:
the transformation is not linear and it is necessary to expand in
eries the square root in every practical calculation. This
introduces non-quadratic terms in the boson operators and also
non-physical states which do not correspond to any spin state. In
order to justify the expansion of the square root the ratio
$a^{+}a/2S$ has to be small in a relevant portion of the Hilbert
space. We can then expand in powers of $1/S$. If we keep only the
leading term we find in fact the classical AF N\'eel state with
alternatively $\langle S^{z} \rangle= \pm S$ in the two
sublattices $A$ and $B$. The leading term does not have quantum
fluctuations. We start to see the quantum effects when we
consider the next term in the expansion. The Hamiltonian is in
this case quadratic in the boson operators and can be diagonalized
with a standard Bogoliubov's transformation. The result contains
two normal modes for every quasimomentum {\bf k} in the BZ, which
correspond to waves of the vector {\bf S}. These are the so called
{\it spin waves} (SW). The SW spectrum is gapless, as prescribed
by Goldstone's theorem. The expectation value of $S^{z}$ is given
by:
$$S-\langle S^{z} \rangle= \hbox{ const. }
 \int_{0}^{\Lambda} {k^{d-1}\over k} \, dk .$$
$\Lambda$ is a micro-wave cut-off, of the order of the inverse
lattice constant. In one dimension, $d=1$, the integral is
diverging: the deviation of the spin from its maximum value is
infinite because of quantum fluctuations. So the $k\rightarrow 0$
SW destroy the AF long range order, and the groundstate is a
non-degenerate singlet. There is no spontaneous symmetry breaking
with non-zero order parameter. We can suppose that if already the
first order term in the expansion is enough to kill the long range
order, terms of following order will act in the same direction,
and the groundstate in one dimension has no AF order. In bigger
dimensions the integral is convergent. For the two dimensional
case:
$$\langle S^{z}\rangle=S-0.197+o(1/S).$$
At this point we may wonder whether the $1/S$ series is
convergent. It is difficult to answer, but we can start
hypothesizing that for $d\geq 2$ quantum fluctuations are not
enough to completely destroy the N\'eel AF state. 
It is likely that the $1/S$ expansion is in fact only asymptotic.
We can however consider some numerical results, which show that
the AF order is preserved in the groundstate at zero temperature,
not only for $J\rightarrow 0$ but also for finite values of
$J=U/t$.  As an example we report a numerical study \cite{Fano},
Fig. \ref{mag}, made in a square lattice of 16 sites for
$J=U/t=4$. Using Lanczos technique, the spin-spin correlation
function
$$C(r)= \frac{1}{4} \langle
(n_{i\uparrow}-n_{i\downarrow})(n_{i\uparrow}-n_{i\downarrow})\rangle
$$ has been calculated, where the expectation value has been taken
on the groundstate of the Hubbard Model (in the notation
$i=(i_{x},i_{y})$).
The point labelled A corresponds to the site $(2,0)$ in the
lattice while B to the site $(2,2)$. The hints to a AF long range
are evident. The SW theory is able to predict characteristic
quantities of the system, and their values can be confronted with
for example Quantum Monte Carlo (QMC) calculations. Useful
quantities are for example the SW speed {\it c}:
$$c=\frac{2S\sqrt{2}ja}{\hbar} Z_{c}(S)$$ where the re-normalization factor
$Z_{c}$ is
$$ Z_{c}=1+0.158/2S+O(1/2S)^{2},$$ the magnetic susceptivity
in the direction perpendicular to the magnetization (in units
$g\mu_{B}/\hbar=1$):
$$\chi_{\perp}=\frac{\hbar^{2}}{8Ja^{2}} Z_{\chi} ,$$ where the
re-normalization factor is $Z_{\chi}=1-0.552/2S+O(1/2s)^{2}$, or
the on-site magnetization $\langle S^{z} \rangle$ itself. A
comparison of the re-normalization constants with QMC results for
$S=1/2$ \cite{Singh}:
$$\begin{matrix}
                              & SW     & &   & MC            \cr
                       \\
Z_{c}=      \hfill            & 1.158  & &   &  1.18\pm0.02   \cr
                       \\
Z_{\chi}=    \hfill           & 0.448  & &   &  0.52\pm0.03   \cr
                        \\
\langle S^{z} \rangle= \hfill & 0.303  & &   &  0.302\pm0.007 \cr
\end{matrix}.$$ \bigskip \\
appears indeed to be satisfactory and supports the SW theory and
its conclusions on a AF groundstate. The SW theory can be also
developed for temperatures higher than zero, and its predictions
confronted directly with experimental results. We find then the
H-T$_{C}$ superconductors' properties (like $La_{2}CuO_{4}$) can
be well explained on a N\'eel's ground state assumption. This
gives full legitimation to the model that has been proposed.

\begin{figure}[!htbp]
\begin{center}
\includegraphics[width=14cm,height=18cm] {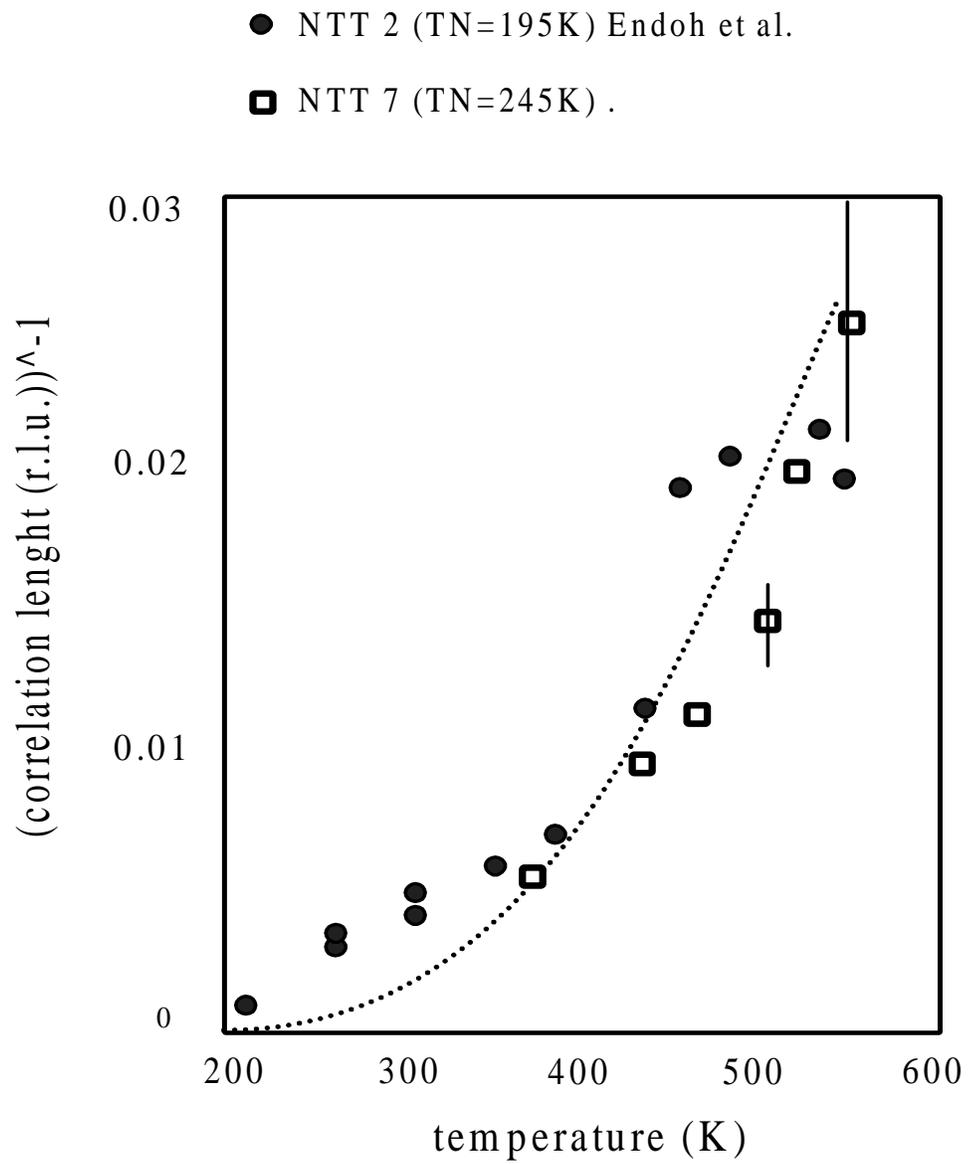}
\caption[]{Correlation length as a function of temperature}
\label{intro6}
\end{center}
\end{figure}

As an example we report in Fig. \ref{intro6} the correlation
length of the spin-spin correlation function $\xi$ as a function
of the temperature \cite{Yamada}, obtained from neutron scattering
experiments. If the correlation between the spins take place in
the $CuO_{2}$ planes, we expect the dynamical form factor $S({\bf
q_{\parallel}})$ independent from the momentum exchanged
perpendicularly to these planes. The static form factor is
therefore given:
$$ S({\bf q_{\parallel}})= \int_{-\infty}^{+\infty} S({\bf q_{\parallel}},\omega) \, d\omega.$$
The correlation length $\xi$ is given by fitting $S({\bf
q_{\parallel}})$ with a lorenzian form:
$$ S({\bf q_{\parallel}}) \sim \frac{1}{{\bf q_{\parallel}}^{2}+(1/\xi)^{2}}.$$
The dotted line is the SW-theory prediction for $S=1/2$: the agrement is in fact satisfactory.

\section{Is the Hubbard Model able to explain the key mechanisms of
cuprates?}

We end this introductory chapter with a question that is the
starting point of the thesis investigation. The answer has been
the focus of one of the most enduring open problems in condensed
matter physics. This problem is very difficult because we lack
tools capable of extracting information from the model. The
standard techniques we used in the previous sections
(perturbation theory in two opposite limits) were able to provide
some first insights in the insulating phase of these materials,
and they showed the capacity of the Hubbard Model in explaining
at least these properties. However, to answer the question of
whether the simple Hubbard Model, derived as extreme
simplification of a much more complex reality, is capable to
account for the richness of the phase diagram presented in
section 1.1 is a more difficult challenge. Many different
instabilities appear to compete: Mott metal-insulator transition,
antiferromagnetism, pseudogap, anomalous normal state, anomalous
superconductivity. Simple perturbation theory does not suffices
anymore and non-pertubative approaches are required.

In this project we tackle the problem in the frame-work of
Dynamical Mean Field Theory (DMFT) \cite{Metzner:89}. This is a
local non-perturbative method which fully takes into account the
quantum-dynamics of the system. DMFT has in recent years proved
successful in describing the longstanding problem of the Mott
metal-insulator transition of Hubbard-like systems \cite{bibble}.
For construction, however, it is a purely local theory which
becomes exact only in the limit of infinite dimension. The thesis
develops as follows:
\begin{itemize}
\item We will see that in order to describe real finite-size systems, local
correlation must be taken into account. This will be presented in
chapter 2 with a phenomenological Boltzmann-Laundau-Fermi-Liquid
approach to the cuprate properties in the normal No FL state.
\item We will then start from a microscopic approach in chapter 3
introducing an extension of DMFT, the Cellular DMFT (CDMFT)
\cite{cdmft}, capable of taking into account short-range spatial
correlations, allowing the study of the finite dimensional
Hubbard problem. We will then be able to confront its properties
with the real system ones.
\item We will benchmark CDMFT with known results and develop an
implementation of the method in chapter 4 using the Lanczos
algorithm \cite{Lanczos} to solve the associated Anderson Impurity
Model. Attention will be given to the right method of extracting
the momentum dependent quantities which are the observables
measured in real systems.
\item In chapter 5 we will apply CDMFT to study the normal state
properties of the 2-dimensional Hubbard Model, which promises to
be a metallic phase with anomalous properties, as shown in the
phase diagram of section 1.1.
\item In chapter 6 we study the superconductive phase,
 connecting its anomalous properties with those of the normal state.
\end{itemize}



\chapter{ Non-Fermi-Liquid Normal State} \vspace{-1cm}
\begin{center}
{\Large \bf A Macroscopic approach}
\end{center}

\section{Non Fermi-Liquid normal state region}

As sketched in the phase diagram picture of Fig. \ref{intro2}, in
the underdoped regime a pseudogap appears in the excitation
spectrum of the metallic state above the superconductive critical
temperature T$_{c}$ and below a doping dependent crossover
temperature $T^{*}$ (PG region in Fig. \ref{intro2}). At optimal
doping the $T^{*}$ almost coincides with T$_{c}$, and the PG
region if present is very small. The normal state of cuprates at
optimal doping however still strongly deviates from the regular
FL behavior observed in simple metals. This can be experimentally
measured in the normal state transport properties. Typical
examples are:
\begin{itemize}
  \item At optimal doping, the in-plane DC-resistivity is linear in
  temperature from T$_{c}$ to very high temperatures
  $\rho(T)\sim T$,
  \cite{Forro}\cite{Takagi2}
  \item The thermoelectric power is linear $TEP \sim -T$, \cite{McIntosh}
  \item The cotangent of the Hall angle displays a
  T$^{\gamma}$-dependence cot$\theta_{H}\sim T^{\gamma}$, with $1.6<\gamma<2$
  \cite{Chien}, \cite{Malinowski}, \cite{Konsta} and
  \cite{Ando}
  \item the magnetoresistance has approximately a T$^{\alpha}$-dependence $MR \sim T^{-\alpha}$,
  with $\alpha\simeq$ 4 \cite{Ando}
  \item The thermal Hall conductivity has approximately a T$^{\beta}$-dependence $THC\sim T^{-\beta}$,
  with $\beta\simeq$ 1.2 \cite{Zhang}
\end{itemize}
\begin{figure}[!htbp]
\begin{center}
\includegraphics[width=10.5cm,height=4.5cm] {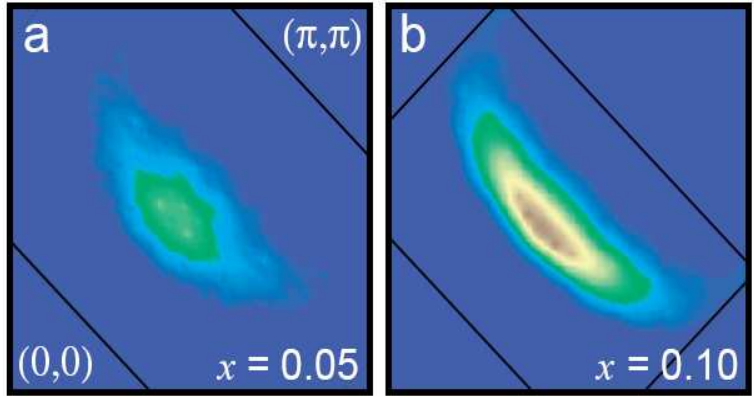}
\caption{Angle Resolved Photo-Emission Spectra $A(k,\omega=0^{+})=
-\frac{1}{\pi}\, G(k,\omega=0^{+})$ in the first quadrant of the
Brillouin Zone for the normal state of a hole-doped cuprate
superconductor material close to the Mott metal-insulator
transition. Doping is labeled $x$. (data taken form \cite{KShen})
} \label{ARPES}
\end{center}
\end{figure}

The non FL anomalies of these materials are also evident in angle
resolved photoemission experiments (ARPES)
\cite{damascelli},\cite{campuzano}. For example in Fig.
\ref{ARPES} it is shown the planar spectral function
$A(k,\omega\rightarrow 0)=\, -\frac{1}{\pi} \, \hbox{Im}
G(k,0^{+})$ in a doped cuprate superconductor \cite{KShen} in the
first quadrant of the two dimensional $(k_{x},k_{y})$-plane in
the Brillouin zone. The color scale spans from blue to bright red
for the highest spectral weight. Close to optimal doping
(right-hand panel at 10\% doping) we observe that the spectral
weight remains in the region close to the point $(\pi/2,\pi/2)$
of momentum space and almost completely disappears around
$(0,\pi)\, (\pi,0)$, indicating that in the last regions the
quasiparticles have disappeared and the Fermi Surface (FS) has
broken up. An arc remains instead close to the $(\pi/2,\pi/2)$
region. If we then look at $A(k,\omega) \hbox{vs}\, \omega$ in the
specific direction $(0,0)\rightarrow (\pi,\pi)$ of the $k$-space
(Fig. \ref{Marshall}), we observe around $(\pi/2,\pi/2)$ a
quasiparticle peak (the line-width is of the order of 0.05-0.1 eV
at $T=100$K \cite{Mesot}) and a wave-vector dispersion of this
peak together with the temperature dependence can be followed. On
the contrary, in the regions around $(0,\pi)\, (\pi,0)$ the
spectral function is very broad (the line-width is of the order
of 0.2-0.3 eV at $T=100$K \cite{Mesot}) and a quasiparticle
cannot easily be distinguished. These features are typical of
incoherent (localized) states where a very strong scattering
mechanism is dominant. The ratio of the Fermi velocities in the
two regions is $v_{F}(\pi/2,\pi/2)/v_{F}(0,\pi)\simeq 3$. The
quasiparticle states around the nodal points $(\pi/2,\pi/2)$ look
therefore coherent (delocalized states) and the scattering
mechanism is weaker and more conventional.

All these experimental observations motivate the introduction of
phenomenological semi-classical methods to describe the non-FL
normal state properties of cuprates. In this chapter we will
present a Boltzmann theory approach due to Perali, Sindel and
Kotliar \cite{PSK}. In their study they introduced a strongly
momentum dependent scattering rate, dividing the FS in {\it hot}
regions (around $(0,\pi)\, (\pi,0)$), where the scattering between
quasiparticles is strong and a the Fermi velocity low, and {\it
cold} regions (around $(\pi/2,\pi/2)$), where the scattering is
weak and the Fermi velocity large. Attributing Fermi-like
scattering properties to the cold region (a temperature $T^{2}$
dependence of the scattering rate) and insulating ones to the hot
region ($T$-independent scattering rate), we can attempt (relying
on few reasonable assumptions) to capture some of the anomalous
normal-state transport properties in the framework of a Boltzmann
Theory. We start introducing in the next section the general
formulation of the semi-classical transport Boltzmann theory for
fermion gas. We will then present the {\it multi-patch model}
parametrization of the scattering matrix of the Boltzmann equation
introduced in \cite{PSK}, where a practical {\it 2-patch} model
was studied in detail. We generalize this method to a {\it
5-patch} model and present possible conjectures that better
explain the observed experimental results on transport quantities.

\section{The Landau Fermi Liquid Theory}

\subsection{The Non Interacting Fermion Gas}

A Fermi gas of noninteracting particles, at equilibrium at some
temperature $T$, is described by the Fermi distribution function,
which expresses the probability to find a particle in a state of
energy $\epsilon= p^{2}/2m$:
\begin{equation}\label{}
  f(\epsilon)= { 1  \over 1+ \exp[(\epsilon- \mu)/kT]}
\end{equation}
where $\mu$ is the chemical potential and $k$ is the Boltzmann's
constant. For most practical purposes in studying real fermionic
systems, it is enough to consider the degenerate case $T
\rightarrow 0$, for which $f(\epsilon)$ reduces to the
Fermi-Dirac step function, which jumps from 1 to zero when
$\epsilon$ crosses the positive chemical potential $\mu$. If the
system has $N$ particles, the ground state is obtained by filling
the $N$ states of less energy, so that in correspondence with
$\epsilon= \mu$, a {\it Fermi Surface } is formed: all the states
inside this surface are filled, while those outside are empty. The
chemical potential is the energy needed to add a particle to the
system:
\begin{equation}\label{}
\mu= E_{0}(N+1)- E_{0}(N)= { \partial E_{0}\over \partial N},
\end{equation}
where the $E_{0}$ is the total energy:
\begin{equation}\label{}
E_{0}= \sum_{\epsilon< \mu} \, \epsilon \, f(\epsilon) .
\end{equation}
Exitations in the system are created by "exciting" a certain
numbers of particles across the Fermi Surface. In this way we
also create an equal number of {\it holes} in the space enclosed
by the Fermi Surface. For a non interacting system, the
exicitation energy $E$ of an excited state can be expressed with
respect to the energy $E_{0}$ of the ground state:
\begin{equation}\label{1.1bis}
E-E_{0}= \sum_{\epsilon} \, \epsilon \, \delta f(\epsilon) ,
\end{equation}
where $\delta f(\epsilon)$ is the departure from equilibrium of
the distribution function, and it is $+1$ if a particle is
created in the state $\epsilon$, or $-1$ if instead a particle is
destroyed (or equivalently a hole is created).

Sometimes it is more convenient to consider the chemical
potential $\mu$ conserved in the system, instead of the total
number of particles $N$. In such cases the relevant quantity is
the Free Energy: $F= E- \mu N$ at zero temperature. In this case
the Free Energy of an excited system is:
\begin{equation}\label{1.1a}
F-F_{0}= \sum_{\epsilon} \, ( \epsilon- \mu) \, \delta
f(\epsilon) .
\end{equation}

\subsection{The Quasiparticle Concept}
Let us now consider the case of an interacting Fermi Liquid. We
can imagine generating such a system starting from an eigenstate
of the ideal noninteracting gas and slowly turning on the
interaction between particles. Under such an adiabatic procedure,
the ideal eigenstate would progressively transform into certain
eigenstate of the real interacting system. {\it This is an
assumption }: there is no a priori justification to why in this
way all the eigenstates of the real system can be generated. The
case of a superconductor is in fact an evident example for which
this assumption is false. We will consider in the following only
Fermi liquids for which this adiabatic procedure is applicable.
This kind of Fermi liquids are called {\it Normal Fermi Liquids}.

To consider what happens in the real system when an excitation is
created, let us first add a particle with a certain momentum {\bf
p} in the ideal system, and then turn adiabatically on the
interaction between the particles, until the real excited state
is reached. If the interaction is turned on slowly enough, we can
assume the gas remains in equilibrium. But this time, because of
interaction, the particle's motion affects also the motion of its
neighbor particles, dragging them around. In field theory
language we say that the particle is now {\it dressed}, while in
the non-interacting case it was bare. The dressed particle is
considered as an independent entity, which is called {\it
quasiparticle}. The excited state created in this way corresponds
to the real ground state plus the added quasiparticle. If we wish
to created elementary exitations, we can imagine again to start
from the ideal non interacting case, moving a particle into a
state outside the Fermi surface and leaving a hole in a state
inside the Fermi surface, and then turning on the adiabatic
procedure until we reach the real state. In this way we establish
a one to one correspondence between ideal gas and real liquid.

What are the main limitations to the quasiparticle concept? The
limitations concern the finite lifetime of the excited states and
the idea of the adiabatic procedure, which requires an infinitely
long duration. Due to the particles interaction, the excited
states have, in fact, a lifetime that is shorter, the further the
excited particles are far from the Fermi Surface. This is due to
the uncertainty principle. It is not possible to generate an
excited state whose life is longer than the time required to
accomplish the adiabatic procedure. On the other hand, one cannot
be too fast in turning on the interaction, otherwise the procedure
would cease to be adiabatic. Accordingly, the quasiparticle idea
is well defined in the limit of quasiparticle excitations that are
very close to the Fermi Surface, for which the lifetime is
sufficiently long to consider the state stable (typically the
ground state, where the particles are all on the Fermi surface,
or for example in the zero temperature limit...).

\subsubsection{The Landau Fermi Liquid}

We wish now to consider how the energy of a real excited state
changes with respect to its ground state. To this porpouse, under
the FL viewpoint , we can move quasiparticles from states inside
to states outside the FS, as we would do for the noninteracting
Fermi gas. So, as in the latter case (\ref{1.1bis}), we can think
that the main contribution to the change in energy comes from the
change in the occupation of the energy levels. The free energy
$F$, which, as pointed out in the previous section, is the
quantity of real interest in studying physical systems, can be
written in the form (\ref{1.1a}). However, the apparently
first-order expansion of (\ref{1.1a}) is indeed a second order
one. Most of the properties we would like to study, in fact,
involve a thin region of states $\delta$ around the FS, where the
value of $\delta f_{p} $ is appreciably different from zero. For
such region the difference $\epsilon_{p}- \mu$ is also of order
$\delta$, so $F- F_{0}$ is in fact of order $\delta^{2}$. To be
consistent, we have to take into account also the interaction
between quasiparticles, and allow for a second order expansion of
the free energy:
\begin{equation}\label{1.1b}
F- F_{0}= \, \sum_{p} \, (\epsilon- \mu) \, \delta f_{p}+ {1
\over 2} \, \sum_{p,p'} \, f_{pp'} \, \delta f_{p}\ \delta
f_{p'}+ O(\delta f^{3}) .
\end{equation}
This expression is the main assumption in the phenomenological
theory of Landau (1956). The new coefficient $f_{pp'}$ describes
the interaction between quasiparticles, and it represents the
interaction energy of the excited particles $p$ and $p'$. It is
invariant under permutation of $p$ and $p'$, and it is supposed
continuous when crossing the FS. It is convenient to introduce the
single particle free energy in the following way:
\begin{equation}\label{1.2bis}
  \tilde \epsilon_{p} - \mu= (\epsilon_{p}- \mu)+ \sum_{p'} \, f_{pp'} \, \delta f_{p'} ,
\end{equation}
The quantity $\tilde \epsilon_{p}$ becomes crucial in taking into
account the effect on particle $p$ coming from the interactions
with all the other surroundings particles. It is mathematically
convenient also to introduce the distribution function:
\begin{equation}\label{1.2a}
\bar{f}_{p}^{0}= \, f^{0}(\tilde \epsilon- \mu)
\end{equation}
where $f^{0}$ is the usual Fermi-Dirac step function. Defining
\begin{equation}\label{1.2b}
\delta \overline{f}_{p}= f_{p}- \overline{f}_{p}^{0}
\end{equation}
the departure from equilibrium $\delta f_{p}$ is given by
\begin{equation}\label{1.3bis}
  \delta f_{p}= \delta \overline{f}_{p}+ { \partial f_{p}^{0}(\epsilon- \mu) \over \partial \epsilon_{p}} \, (\tilde
\epsilon_{p}- \epsilon_{p})
\end{equation}
or, using equation above (\ref{1.2bis}),
$$ \delta f_{p}= \delta \overline{f}_{p}+ { \partial f_{p}^{0}(\epsilon- \mu) \over
\partial \epsilon_{p}} \, \sum_{p'} \,
f_{p\,p'} \delta f_{p'}.$$

\section{The Transport Boltzmann Equation}
In order to study the transport properties of the Normal Fermi
Liquid, Landau considered the quasiparticles as independent
entities with velocity given by $\nabla_{p} \tilde \epsilon$ and
subjected to the force $- \nabla_{r} \tilde \epsilon$. Let us
suppose for the moment that there are no external forces acting
on the system. We can set up a kinematic equation, which
expresses the variation in time of the distribution function
$(df_{p}/ dt)$ due to quasiparticle collisions:
\begin{equation}\label{1.4bis}
  { \partial f_{p} \over \partial t}+ \nabla_{r}f_{p} \cdot \nabla_{p}\tilde
\epsilon_{p}- \nabla_{p}f_{p} \cdot \nabla_{r}\tilde
\epsilon_{p}= C_{p}.
\end{equation}
$C_{p}$ is the collision integral, which gives the balance in the
particles entering and going out of an element of phase space
$d^{3}p$ centered around $p$. If only binary collisions $$p+ q
\rightarrow p'+ q'$$ are considered relevant, $C_{p}$ is equal to:
\begin{eqnarray*}
 -\sum_{q,p',q} \, \delta^{4}(p,q,p',q') \mid
T(p,q,p',q') \mid^{2} \, \left[ f_{p} f_{q} (1-f_{p'})(1-f_{q'})-
(1-f_{p})(1-f_{q}) \, f_{p'} f_{q'}\right] ,
\end{eqnarray*}
where $T(p,q,p',q')$ is the transition matrix element between the
initial state $\mid p,q \rangle$ of the two particle before the
collision, and the final state $\langle p',q' \mid$ of the two
particles coming out of the collision:
$$T(p,q,p',q')= \langle p', q' \mid T(E) \mid p, q \rangle $$
$$ T(E)= V+ V (E-H_{0}+i\epsilon)^{-1} V+.....$$
with $H_{0}$ the unperturbed Hamiltonian, $V$ the potential, and
$\epsilon \rightarrow 0^{+}$.
$$\delta^{4}(p,q,p',q') \equiv \delta^{3}(p'+q'-p -q) \, \delta(\tilde \epsilon_{p'}+
\tilde \epsilon_{q'}- \tilde \epsilon_{p}- \tilde \epsilon_{q})$$
expresses the conservation of energy and momentum in the
collision.

Only excited quasiparticles close to the FS can contribute to the
flow in the phase space according to Landau's theory. Only the
departure from the equilibrium distribution $f^{0}_{p}$ in fact
enters in the description of all transport quantities. Let us
write $f_{p}= f_{p}^{0}+ g_{p}$, where $g_{p}$ represents the
departure from the equilibrium distribution. Plugging this
expression in (\ref{1.4bis}) and retaining only the first order
terms in $g_{p}$, we have:
\begin{equation}\label{1.5bis}
{ \partial g_{p} \over \partial t}+ \nabla_{r} g_{p} \cdot v_{p}-
\nabla_{p} f^{0}_{p} \cdot \sum_{p'} \, f_{pp'} \nabla_{r} g_{p'}
= C_{p} ,
\end{equation}
where we have taken into account the relation (\ref{1.2bis}), and
that:
$$ { \partial f_{p}^{0} \over \partial t}+ \nabla_{r}f_{p}^{0} \cdot \nabla_{p} \tilde
\epsilon_{p}- \nabla_{p}f_{p}^{0} \cdot \nabla_{r} \epsilon_{p}-
\nabla_{p} g_{p} \cdot \nabla_{r} \epsilon_{p}= 0 $$ as the
equilibrium distribution function $f^{0}_{p}$ and the particle's
energy $\epsilon$ do not depend on $t$ and $r$ in the non
interacting system. The first two terms in the expression
(\ref{1.5bis}) represent the flow of totally independent
quasiparticles. The last term comes from the interaction between
the quasiparticles, and it represents the flow of the ground
state quasiparticles dragged by interaction. If we use the
notation in (\ref{1.3bis}) to describe the departure from
equilibrium, and we retain only the linear terms in $g_{p}$, this
expression takes an easier form:
\begin{equation}\label{1.6bis}
  {\partial g_{p} \over \partial t}+ \nabla_{r} \bar{g_{p}} \cdot v_{p}=
C_{p} .
\end{equation}
The meaning of $\bar{g_{p}}$ is now more transparent: it describes
a diffusion term, and so the {\it local} departure from
equilibrium.

We wish now to linearize the collision integral $C_{p}$. It is
more convenient to start with the notation introduced in
(\ref{1.2a}) and (\ref{1.2b}), and write the distribution
function as $f_{p_{j}}= \bar{f}_{p_{j}^0}+ \bar{g}_{p_{j}}$.
Keeping again only the linear terms, $C_{p}$ is equal to:
\begin{eqnarray*}\label{}
  -\sum_{q,p',q} \, \delta(p,q,p',q')^{4} \mid
T(p,q,p',q') \mid^{2} \, [ \bar{f}_{p}^{0} \bar{f}_{q}^{0}
(1-\bar{f}_{p'}^{0})(1-\bar{f}_{q'}^{0})-
(1-\bar{f}_{p}^{0})(1-\bar{f}_{q}^{0}) \, \bar{f}_{p'}^{0}
\bar{f}_{q'}^{0}]
\end{eqnarray*}
plus the sum of the four terms
\begin{equation}\label{A}
 - \delta' \sum_{q,p',q'} \, \delta(p,q,p',q')^{4} \mid
T(p,q,p',q') \mid^{2} \, [ \bar{f}_{j}^{0} \, (1-
\bar{f}_{k}^{0})(1- \bar{f}_{l}^{0})+
  \bar{f}_{k}^{0} \bar{f}_{l}^{0} \, (1-\bar{f}_{j}^{0}) ] \,
\bar{g}_{i}
\end{equation}
that we can obtain starting with the set of indexes $(i,j,k,l)=
(p,q,p',q')$ and exchanging in the three remaining ways the
position of two indexes in each couple simultaneously. The
coefficient $\delta'$ is $+1$ if $i=p$ or $q$, $-1$ if $i=p'$ or
$q'$. Under the binary collision point of view, the first case
corresponds to the $i$ particle undergoing the collision, the
second to the $i$ particle coming out from the collision.

It is straightforward to check that the choice of the  Fermi
distribution at equilibrium $\bar{f}_{p_{j}}^{0}$ cancel the
first term (\ref{A}), because of the conservation of energy in the
binary collision $\tilde \epsilon_{p}+ \tilde \epsilon_{q}=
\tilde \epsilon_{p'}+ \tilde \epsilon_{q'}$ implied by the
$\delta^{4}$ factor. Neither spatial rotations, reflections, nor
time reversals change the physics of the binary collision. The
$T$ matrix is in fact invariant under all such transformations,
moreover the $\delta(p,q,p',q')^{4}$ factor appearing in the
integrand is invriant. We can in this way show that two of the 4
terms cancel out and the final expression for $C_{p}$ can be
written:
\begin{equation}\label{1.6a}
C_{p}= \sum_{p'} \, [ C_{pp'} \, \bar{g}_{p'}- C_{pp'} \,
\bar{g}_{p} ] ,
\end{equation}
where $C_{pp'}$ is the scattering matrix element defined as:
\begin{equation}\label{C}
C_{pp'}= \sum_{q,q'} \, \delta(p,q,p',q')^{4} \mid T(p,q,p',q')
\mid^{2} \, [ \bar{f}_{q}^{0} \, (1- \bar{f}_{p'}^{0})(1-
\bar{f}_{q'}^{0})+
  \bar{f}_{p'}^{0} \bar{f}_{q'}^{0} \, (1-\bar{f}_{q}^{0}) ] .
\end{equation}
It expresses the probability  that a particle with momentum $p$
scatters into the momentum $p'$ after the collision, and vice
versa.

To complete the transport equation we have only to add an
external force to the system:
$$ F_{p} \cdot \nabla_{p} f_{p} .$$
Usually $g_{p}$ is proportional to the external applied force
$F_{p}$, a typical example is an electric field force $F(r)= -e \,
E(r)$. As we are interested only in the first order terms we can
replace $f_{p}$ by the equilibrium ditribuition function
$f_{p}^{0}$. However, if we consider, for example, a magnetic
field force $F_{B}(r)= -e \, v_{p} \times B(r)$, this is clearly
orthogonal to $\nabla_{p} f_{p}^{0}= \partial f_{p}^{0}/ \partial
\tilde \epsilon_{p} \cdot v_{p}$, and it does not give any
contribution. So we have to consider as first non-zero term $F_{B}
\cdot \nabla_{p} \bar{g}_{p}$. We thus obtain at the end the
Linearized Boltzmann Transport Equation (BE):
\begin{equation}\label{1.7bis}
{ \partial g_{p} \over \partial t}+ \nabla_{r} \overline{g_{p}}
\cdot v_{p}+ F_{p} \cdot v_{p}{ \partial f^{0}_{p}\over \partial
\epsilon_{p}}- (e\, v_{p} \times B) {\partial \bar{g}_{p}\over
\partial p} = C_{p},
\end{equation}
with the collision integral $C_{p}$ given by (\ref{1.6a}). From
this equation we can derive all the transport properties of a
Fermi Liquid.

\subsection{Frequency Dependent External Force} We wish now to
consider the case in which the external force is time dependent,
so let us write
$$ F= F_{0} \, e^{-\imath \,  \omega t} .$$
Let us plug it in (\ref{1.7bis}) and, in the limit of small
$\omega$, let us seek for the the linear response solution
$g_{p}(t)= g_{p} \, \exp{(\imath \, \omega t)}$. We obtain:
$$ \imath \, \omega g_{p}- F_{0} \cdot v_{p}{ \partial f^{0}_{p} \over  \partial \epsilon_{p}}= C_{p},$$
where for simplicity we suppose an uniform medium for which $
\nabla_{r} g_{p}= 0$. Remembering we can write (\ref{1.6a})
$$ C_{p}= \sum_{q} \, [ C_{pq} \, \overline{g}_{q}- C_{pq} \, \overline{g}_{p} ] ,$$
and defining the relaxation time $\tau_{p}$
\begin{equation}
1/\tau_{p}= \sum_{q} \, C_{pq} ,\label{one_tau}
\end{equation}
we can write:
$$ \imath \, \omega g_{p}+ { 1 \over \tau_{p} } \overline{g}_{p}- \sum_{q} \, C_{pq} \overline{g}_{q}=
 F_{0} \cdot v_{p}{ \partial f^{0}_{p} \over  \partial \epsilon_{p}} .$$
In order to calculate the transport property we need to solve
this equation in $\overline{g}_{p}$ \cite{Pines}. We can use
relation (\ref{1.3bis}) to obtain
\begin{equation}\label{1.8bis}
[ \imath \, \omega + {1 \over \tau_{p}} ]  \ \overline{g}_{p}-
\sum_{q} \, C_{pq} \, \overline{g}_{q}+ \imath \, \omega ( {
\partial f^{0}_{p} \over \partial \epsilon_{p}} )\, \sum_{q} \,
f_{pq} g_{q}= F_{0} \cdot v_{p}{
\partial
f^{0}_{p} \over  \partial \epsilon_{p}}.
\end{equation}
We stress here that if we wish to recover the solutions for the
transport quantities in the time dependent case by simply
applying the substitution $1/\tau_{p} \rightarrow 1/ \tau_{p}+
\imath \, \omega$ to the solutions of the time independent case,
as is commonly done, we would neglect the third term in the
left-hand side of equation (\ref{1.8bis}).


\section{General Solution to the $N-$Patch Model}
\subsection{The linearized Boltzmann Equation in the $N-$Patch Model}
We briefly review the solution to the $N-$Patch Model proposed by
Perali, Sindel and Kotliar in \cite{PSK}. Starting from
Boltzmann's equation (\ref{1.8bis}), in the case of zero magnetic
field $B=0$, and for the steady case, we have

$$g_{k}/\tau_{k}- \sum_{k'} \, C_{k,k'}g_{k'}= [e v_{k} \vec{E}- \epsilon_{k}v_{k}
\bigtriangledown T/T ]\, (-\partial_{\epsilon_{k}} f_{k}^{0}) $$

where $g_{k}$ is the departure from the equilibrium distribution
function $f_{k}^{0}$, $C_{k,k'}$ is the scattering matrix
element, $\tau_{k}$ is the relaxation time for the state $k$
defined as $1/\tau_{k}\equiv \sum_{k'}  C_{k,k'} \label{tau}$. In
order to include a non trivial momentum dependence of the
scattering process and the division of the Brillouin Zone (BZ) in
{\it hot} and {\it cold} regions, as discussed in the
introduction to this chapter, one possibility is to expand the
scattering matrix $C_{k,k'}$ via a set of functions $\phi_{j}(k)$,
which weight differently the various regions of the BZ
\begin{equation}
C_{k,k'}= \sum_{\alpha ,\beta}^{N} \, \phi_{\alpha }(k) c_{\alpha
\beta } \phi_{\beta}(k'). \label{C_kk1_Npatch}
\end{equation}
$\phi_{j}(k)$ localizes a precise area in the
momentum space, the $j$th {\it patch}, which we want to have
 scattering properties different from the other regions. $N$ is the total number of patches in which the BZ
is divided. $c_{\alpha \beta }$ is the scattering amplitude
between patches $\alpha$th and $\beta$th, and it is a symmetric,
temperature dependent $N \times N$ matrix. In this way, the
Boltzmann's equation takes a form which allows an exact analytical
solution. In fact, we can write
\begin{equation}\label{}
  g_{k}= \tau_{k} \, \sum_{\alpha ,\beta } \, c_{\alpha \beta} \phi_{\alpha }(k) \,
\sum_{k'} \, \phi_{\beta }(k')g_{k'}+ \tau_{k} [e v_{k} \,
\vec{E}- \epsilon_{k}v_{k} \bigtriangledown T/T ]\,
(-\partial_{\epsilon_{k}} f_{k}^{0})
\end{equation}

that, for convenience, we rewrite in a more compact notation:
\begin{equation}\label{2x}
g_{k}= \tau_{k} \ \underline{\phi}_{k} \cdot \overline{c} \cdot
\sum_{k'}\, \underline{\phi}_{k'} g_{k'}+ \vec{\Theta _{k}} \cdot
\vec{v_{k}}
\end{equation}
where $\underline{\phi}_{k}$ is the $N$ dimensional vector
$(\phi_{1 }(k).....\phi_{N }(k))$, the $N\times N$ matrix
$(\overline{c})_{\alpha \beta }= c_{\alpha \beta }$ and
$$\vec{\Theta _{k}}= \tau_{k} [e v_{k} \, \vec{E}- \epsilon_{k}v_{k}\bigtriangledown T/T ]\,
(-\partial_{\epsilon_{k}} f_{k}^{0}) .$$ If in the quasimomentum
space we define the vector
$$ (\vec{g})_{k}= g_{k}$$
the matrix
$$ \Pi_{kk'}= \tau_{k} \, \underline{\phi}_{k} \cdot \overline{c} \cdot \underline{\phi}_{k'} $$
and the antisimmetric vector
$$ (\vec{u})_{k}= \vec{\Theta _{k}} \cdot \vec{v_{k}} $$
The BE  is a simple linear equation:
$$ \vec{g}= \Pi \cdot \vec{g}+ \vec{u},$$
or, defined $\chi = \vec{1}- \Pi $
\begin{equation}\label{3}
\chi \cdot \vec{g}= \vec{u} .
\end{equation}
To seek for the solution to the BE we have then to invert the
matrix $\chi$
\begin{equation}\label{4}
  \vec{g}= \chi^{-1} \cdot \vec{u} .
\end{equation}

However, before performing this operation, we should first
recognize that $\chi$ is not invertible in a strictly mathematical
sense. It has in fact the form
$$\chi_{kk'}= \delta_{kk'}- \tau_{k} C_{kk'}$$
and, bearing in mind $1/\tau_{k}\equiv \sum_{k'} C_{k,k'}$, it is
straightforward to observe that the sum over the columns labeled
$k'$ is identically zero for all $k$, i.e means that the column
vectors are not linearly independent. We will show later that
there is a superfluous unphysical solution, corresponding to
$\vec{g}=\,$ containing a constant term, which ought be
disregarded. In the solution we are going to present we shall
assume to act in the physical subspace of the total Hilbert space
where the unphysical solutions have been cut away, so that $\chi$
becomes invertible.

\subsection{$N$-Patch solution to the BE}

Let's consider in (\ref{2x}) the $k$-independent vector in the $N$
dimensional space
$$\underline{b}= \sum_{k} \, \underline{\phi}_{k'} g_{k'},$$
so that
\begin{equation}\label{5}
  g_{k}= \tau_{k} \ \underline{\phi}_{k} \cdot \overline{c} \cdot \underline{b}+ \vec{\Theta _{k}}
\cdot \vec{v_{k}} .
\end{equation}
Multiplying both side of the above equation for the vector
$\underline{\phi}_{k}$, and then summing all over $k$, we obtain
an equation for $\underline{b}$:
$$\overline{\Gamma} \cdot \underline{b}= \sum_{k} \, \underline{\phi}_{k}\,(\vec{\Theta _{k}} \cdot \vec{v_{k}})$$
where we introduce the $N\times N$ matrix
\begin{equation}\label{5.1}
\overline{\Gamma} = [\vec{1}- \sum_{k}\, \underline{\phi}_{k}
\tau_{k} (\underline{\phi}_{k} \cdot \overline{c} \cdot ...)].
\end{equation}
As $\overline{\Gamma}$ acts on a $N$ dimensional space, it is
more easily inverted than the $\chi$ matrix, which is instead
acting on the quasimomentum space. So the vector
$$\underline{b}=  \sum_{k} \, \overline{\Gamma}^{-1} \underline{\phi}_{k}\,(\vec{\Theta _{k}} \cdot \vec{v_{k}}),$$
and it can be plugged in eq. (\ref{5}), giving the solution
$$ g_{k}= \sum_{k'}\, \left[ \delta_{kk'}+ \tau_{k} \ (\underline{\phi}_{k} \cdot \overline{c} \cdot
\overline{\Gamma}^{-1} \underline{\phi}_{k'}) \right] \,
(\vec{\Theta _{k'}} \cdot \vec{v_{k'}}),$$ i.e.
$$ \vec{g}= \chi^{-1} \cdot \vec{u} ,$$
so that we finally obtain $\chi^{-1}$
\begin{equation}\label{6}
  \chi^{-1}_{kk'}=  \delta_{kk'}+ \tau_{k} \ (\underline{\phi}_{k} \cdot \overline{c} \cdot
\overline{\Gamma}^{-1} \underline{\phi}_{k'}) .
\end{equation}

\subsection{Invertibility and Inversion of the $\overline{\Gamma}$ Matrix}

As $\chi$ is not invertible, from eq. (\ref{6}) it follows that
$\overline{\Gamma}$ is not invertible. We show in fact that there
is at least a vector, $\underline{b}^{0}$, different from
$\underline{0}$, which $\in \hbox{ Ker }(\overline{\Gamma})$, and
$\overline{\Gamma} \cdot \underline{b}^{0}= 0$. Taken
$$ \underline{b}^{0}= \sum_{k}\, \underline{\phi}_{k},$$
we have
$$ \overline{\Gamma} \cdot \underline{b}^{0}= [\vec{1}- \sum_{k}\, \underline{\phi}_{k} \tau_{k}
(\underline{\phi}_{k} \cdot \overline{c} \cdot ...)] \cdot
(\sum_{k}\, \underline{\phi}_{k}) =$$
$$\sum_{k'} \, \underline{\phi}_{k'}- \sum_{kk'} \, \tau_{k} \underline{\phi}_{k}(\underline{\phi}_{k}\cdot
\overline{c} \cdot \underline{\phi}_{k'})$$ which is evidently
zero from the definition (\ref{one_tau}), $\tau_{k}= 1/(
\sum_{k'}\, \underline{\phi}_{k}\cdot \overline{c} \cdot
\underline{\phi}_{k'} )$. As already pointed out,
$\underline{b}^{0}$ corresponds to the "zero harmonic" solution
$\vec{g}_{0}=\,$ containing a constant term. From eq. (\ref{5}):
$$ g_{k}^{0}= \tau_{k} (\underline{\phi}_{k}\cdot \overline{c} \cdot \underline{b}^{0})+ \vec{\Theta _{k}}
\cdot \vec{v_{k}}= 1+ \vec{\Theta _{k}} \cdot \vec{v_{k}} .$$
This solution is not acceptable, because it does not conserve the
number of particles $n$ in the system. It is required that $\sum
\, f_{k}{}= \sum \, f_{k}^{0}+ g_{k}= n$, a condition evidently
not satisfied by $g_{k}^{0}$. Moreover, because $\sum \,
f_{k}^{0}= n$ in the equilibrium condition, we also observe that
for every physically acceptable solution $g_{k}$, it has to be
\begin{equation}\label{7}
\sum \, g_{k}= 0 .
\end{equation}

So, in order to invert $\overline{\Gamma}$, we have first to
exclude from the total Hilbert space the $\hbox{ Ker
}(\overline{\Gamma})$. This was done through the Singular Value
Decomposition technique (SVD) which allows one to separate the
total Hilbert space into all the eigenspaces corresponding to the
eigenvalues of $\overline{\Gamma}$, isolating the ones which have
zero eigenvalues. We could check that
dim[\,Ker($\overline{\Gamma}$)\,] is in fact 1, corresponding to
the eigenvector $\underline{b}^{0}$, invert $\overline{\Gamma}$ in
the physical subspace, and then we finally checked that the
condition Eq. (\ref{7}) is satisfied in the solution found.

\subsection{Transport Properties}

\subsubsection{Magnetic Field Dependence}

In order to study the magneto-transport properties in the system,
a magnetic field is added in the BE. Eq. (\ref{3}) and (\ref{4})
become:
\begin{equation}\label{8}
  \Xi  \cdot \vec{g}= \vec{u} .
\end{equation}
\begin{equation}\label{9}
\vec{g}= \Xi ^{-1} \cdot \vec{u} .
\end{equation}
with
$$\Xi = [1+ {e \over \hbar c}(\vec{v_{k}}\times B)\cdot \nabla_{k} ] \delta_{kk'}- \tau_{k} C_{kk'}.$$
Following \cite{PSK}, we consider weak magnetic fields, so that
the term containing $B$ in the equation above can be considered
as a small perturbaton. $\Xi$ is split in two parts, $\Xi= \chi+
\Omega_{B}$, so that the magnetic field dependence remains in
$(\Omega_{B})_{k,k'}= [{e \over \hbar c}(\vec{v_{k}}\times
B)\cdot \nabla_{k} ] \delta_{kk'}$. $\chi$ is the previously
considered zero magnetic field matrix. $\Xi$ can be then
perturbatively expanded in powers of $B$, and its inverse takes
the form:
\begin{equation}\label{10}
\Xi^{-1}= \chi^{-1}- \chi^{-1} \cdot \Omega_{B} \cdot \chi^{-1}+
\chi^{-1} \cdot \Omega_{B} \cdot \chi^{-1} \cdot \Omega_{B} \cdot
\chi^{-1}+ O(B^{3}).
\end{equation}
According to the quantity we want to calculate, we will consider
different terms of this expansion.

\subsubsection{Conductivity}
We shall mainly consider the electrical conductivity $\sigma$ in
the absence of a temperature gradient. Upon the application of an
electrical field $\vec{E}$, the electrical current $\vec{J}= e
\sum_{k} \, \vec{v_{k}} g_{k}$ is $\vec{J}= \overline{\sigma
}\vec{E}$, with the conductivity tensor $\overline{\sigma }$
given by:
$$ \sigma^{\mu \nu }= 2 e^{2} \, \sum_{k,k'} \, v_{k}^{\mu } \, \Xi^{-1} v_{k'}^{\nu} (-\partial_{k'}f^{0}_{k'}).$$

Considering the solution of $\chi$ (\ref{6}) and truncating to
first order in the expansion (\ref{10}), we get the direct
conductivity $\sigma_{xx}$
\begin{equation}\label{11*}
\sigma_{xx}= 2e^{2} \, \sum_{k} \, \tau_{k} (v_{k}^{x})^{2}+
2e^{2} \, \sum_{ijl}\, c_{ij}\, \overline{\Gamma }^{-1} \,
[\sum_{k} \, v_{k}^{x} \tau_{k} \phi_{i}(k)]\,[\sum_{k'} \,
v_{k'}^{x} \tau_{k'} \phi_{l}(k') (-\partial_{k'}f_{k'}^{0})]
\end{equation}
Notice how the first term in the right-hand side of this
expression coincides with the one used in \cite{PSK}. In fact,
because of the topological symmetry in the BZ of the two-patch
model used in \cite{PSK}, the second term in
$\overline{\Gamma}^{-1}$ is cancelled.

For the Hall conductivty $\sigma_{xy}$, we must consider instead
the second order in the expansion (\ref{10}), as the first gives
identically zero. It is:
$$ \sigma_{xy}= -2e^{2} \, \sum{k,k'} \, v_{k}^{x} \tau_{k} \, \Upsilon_{k,k'} \,  v_{k'}^{y} \tau_{k'}
(-\partial_{k'}f^{0}_{k'})$$ where the matrix $\Upsilon$ is
$$ \Upsilon_{k,k'}= ( \delta_{k,k'}+  \tau_{k'} \mu_{k,k'}) \, \vec{F}_{k'}\cdot \nabla_{k'}+ \sum_{l} \,
( \delta_{k,l}+  \tau_{l} \mu_{k,l}) \, \vec{F}_{l}\cdot
\nabla_{l} \, ( \tau_{l} \mu_{l,k'}) \, ,$$ with $\vec{F}_{k}= {e
\over \hbar c}(\vec{v_{k}}\times \vec{B})$ and where we have
defined $\mu_{k,k'}= (\underline{\phi}(k) \cdot \overline{c}
\cdot \overline{\Gamma}^{-1} \underline{\phi}(k')) $. Developing
this expression, we can also elucidate the dependence of
$\sigma_{xy}$ on the $\overline{\Gamma }^{-1}$ matrix:
\begin{equation}\label{11**}
\sigma_{xy}= -2e^{2} \, (X_{1}+ X_{2}+ X_{3}+ X_{4})
\end{equation}
where
$$ X_{1}= \sum_{k}\, v_{k}^{x} \tau_{k} \, \vec{F}_{k}\cdot \nabla_{k}   \,  (v_{k}^{y} \tau_{k})
(-\partial_{k}f^{0}_{k}) \hfil$$
$$ X_{2}= \, \sum_{ijl} \, c_{ij} \overline{\Gamma }^{-1}_{jl} \, [ \, \sum_{k} \, v_{k}^{x} \tau_{k} \phi_{i}(k)\, ]
\ [\, \sum_{k'} \, \tau_{k'} \phi_{l}(k') \, \vec{F}_{k'}\cdot
\nabla_{k'}   \,  (v_{k'}^{y} \tau_{k'})
(-\partial_{k'}f^{0}_{k'}) \, ]$$
$$ X_{3}= \,  \sum_{ijl} \, c_{ij} \overline{\Gamma }^{-1}_{jl} \, [ \, \sum_{k} \, v_{k}^{x} \tau_{k} \, \vec{F}_{k}\cdot
\nabla_{k} \, (\tau_{k} \phi_{i}(k)) \, ] \ [\, \sum_{k'} \,
v_{k'}^{x} \tau_{k'} \phi_{i}(k') \, (-\partial_{k'}f^{0}_{k'})
\, ]$$
$$ X_{4}= \,  \sum_{ijm} \, \sum_{pqr} \, c_{ij} c_{pq} \overline{\Gamma }^{-1}_{jm} \, \overline{\Gamma }^{-1}_{qr}
\, [ \, \sum_{k} \, v_{k}^{x} \tau_{k} \phi_{i}(k)\, ]$$ $$ \,
[\, \sum_{k'} \, v_{k'}^{x} \tau_{k'} \phi_{i}(k') \,
(-\partial_{k'}f^{0}_{k'}) \, ] \,  [ \, \sum_{l} \, \phi_{m}(l)
\tau_{l} \, \vec{F}_{l}\cdot \nabla_{l} \, (\tau_{l} \phi_{p}(l))
\, ] .$$ Again, the $\overline{\Gamma }^{-1}$ dependent terms
$X_{2}$, $X_{3}$, $X_{4}$ are identically zero if we consider the
two-patch model in \cite{PSK}.

In order to have a direct comparison with experimental results
\cite{Konsta} \cite{Ando}, it will be useful to consider the
cotangent of the Hall angle, defined as the ratio between the
direct conductivity and the Hall conductivity: \beq \cot
\theta_{H}= {\sigma_{xx} \over \sigma_{xy}} \label{12} .\eeq For
this quantity, from \cite{Konsta} \cite{Ando}, we expect a
behavior $\cot \theta_{H}\propto T^{\alpha}$, with $\alpha\simeq
2$ close and beyond optimal doping.
\section{The 5-Patch Model}

\begin{center}
\begin{figure}[!htbp]
\begin{center}
\includegraphics[width=12cm,height=9cm] {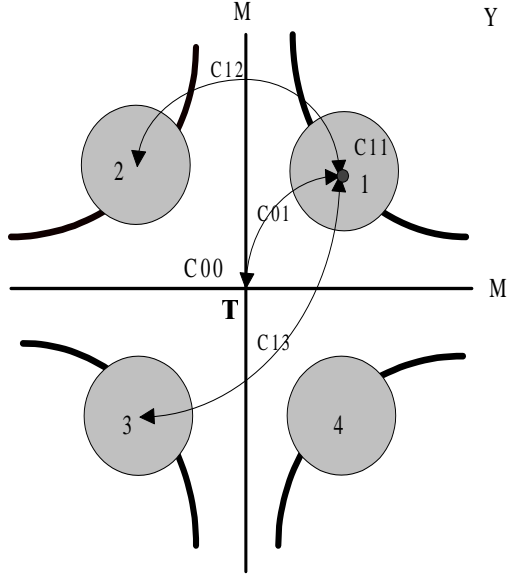}
\caption[]{5-Patch division of the BZ} \label{fig1}
\end{center}
\end{figure}
\end{center}

In order to extend the results of \cite{PSK}, we now introduce a
more general patch-division of the BZ, which could allow for
example to take into account the forward scattering terms. This
could verify the conjectures of \cite{Varma} on the different
temperature power-law dependence of the parallel and transverse
conductivity. The property is experimentally observed in the
linear temperature-dependence of the resistivity and the
T$^{\gamma}$-dependence of the cotangent of the Hall angle. The
weakness in all such models, which try to describe the transport
properties of cuprates using the assumption of a strong
momentum-dependent scattering rate, is the very different
power-law dependence of these two quantities. In a regular FL the
linear $T$-dependence in the resistivity completely determines
the $T$-dependence of the transverse conductivity too, which
therefore follows the same law. So let us describe the {\it cold}
regions as "disks" centered in the 4 points $\vec{k}_{j}$, with
$j= 1,2,3,4$ (Fig. \ref{fig1})
$$ \vec{k}_{j}= (k_{x},k_{y})= ( \pm {\pi \over 2}, \pm {\pi \over 2} ) ,$$
and, for the patch functions, we first chose a convenient
gaussian form:
\begin{equation}\label{13.a}
\phi_{j}(\vec{k}) = A^{2} \, e^{-2p(\vec{k}- \vec{k}_{j})/ \Delta
k^{2}} ,
\end{equation}
where $p(\vec{k}- \vec{k}_{j})$ is a function having the
periodicity of the Bravais lattice. We chose:
$$ p(\vec{k}- \vec{k}_{j})= 2[2-\cos(k_{x}-k_{jx})- \cos(k_{x}-k_{jx})] ,$$
so that in the limit $\vec{k}\rightarrow  \vec{k}_{j}$,
$p(\vec{k}- \vec{k}_{j})\sim (\vec{k}- \vec{k}_{j})^{2}$, and the
pure gaussian form of $\phi_{j}$ is restored. $\Delta k$ measures
the width of the cold patch function. For $\vec{k}- \vec{k}_{j} >>
\Delta  k$, $\phi_{j}$ falls to zero much faster than a simple
gaussian. The normalization constant $A$ has to be determined in
such a way that the sum of all the patch functions (including the
{\it hot} patch function $\phi_{0}$ ) in all the points $k$ of
the BZ is 1. We define the hot patch function:
$$\phi_{0}(\vec{k})
= 1- \sum_{j=1}^{4} \, \phi_{j}(\vec{k}) $$ and require that
$\phi_{0}$ totally disappears in the centers of the cold regions,
where we wish a complete Fermi liquid behavior (and where the
maximum of the sum of the 4 cold patch functions is). We obtain
in this way the constraint that gives the constant $A$:
\begin{equation}\label{13.b}
  A^{2}= (\sum_{j=1}^{4} \, e^{-2p({\pi \over 2},  {\pi \over 2} )/ \Delta k^{2}})^{-1} .
\end{equation}
We now consider the scattering matrix $\overline{c}$. As already
pointed out, it has to be symmetric and with all the elements
positive. Actually, the off-diagonal elements describe the
scattering amplitude between two electrons coming from different
patches, the diagonal elements, instead, describe the scattering
inside the same patch. Because of the symmetry of the patches on
the reciprocal lattice (see Fig. \ref{fig1}), it is: \beqa
c_{00}\neq c_{11}= c_{22}= c_{33}= c_{44} \cr
 c_{12}= c_{23}= c_{34}= c_{41}  \cr c_{01}= c_{02}= c_{03}= c_{04} \cr c_{13}=
c_{24} 
\eeqa At first approximation, we also choose for the diagonal
scattering amplitudes between cold patches:
$$ c_{13}= c_{12}.$$
The main difference with the two patches model considered in
\cite{PSK} is the introduction of different amplitudes for the
{\it in}-cold patch scattering and {\it between }cold patches
scattering:
\begin{equation}\label{13.c}
c_{11} \neq c_{12}
\end{equation}
This should allow us to characterize differently the small angles
scattering in the cold region and in this way to introduce also
the effect of the forward scattering on the transport properties.
On the contrary, if $c_{11}$ were equal to $c_{12}$, it is
straightforward to verify that the solution to the Boltzmann
equation coincides, except for the choice of the patch functions,
to the one used in \cite{PSK}. In fact, taken the Boltzmann
equation (\ref{3})
$$(\vec{1}- \Pi)  \cdot \vec{g}= \vec{u} ,$$
let's consider the solution in \cite{PSK}
$$ (\vec{g})_{k} \propto \vec{\Theta _{k}} \cdot \vec{v_{k}} . $$
If $c_{11}= c_{12}$ it is straightforward to verify that
$$\Pi \cdot  \vec{g}=  0 ,$$
because of the even symmetry of the cold region on the BZ and the
odd symmetry of the solution proportional to $v_{k}$. Therefore
also in our case $\vec{g}$ satisfies the Boltzmann equation.

The temperature dependence of the scattering matrix $\overline{c}$
is similar to the one chosen in \cite{PSK}:
\begin{equation}\label{14}
c_{00}= b \ ; \ c_{11}= a_{0}+ a_{1}\, T^{2}\ ; \ c_{12}= a_{1}\,
T^{2}\ ; \ c_{01}= c \, T
\end{equation}
In the cold region the scattering rate is low, and a Fermi Liquid
behavior with a $T^{2}-$dependence is a reasonable assumption. On
the other hand, in the hot region, where the scattering rate is
high, the temperature independence is suggested by ARPES
experiments. The coupling between hot and cold region is the key
assumption in \cite{PSK}: the inter-patch scattering is
considered to be linear in the temperature, and this is important
to obtain a linear temperature dependence in the resistivity. As
mentioned above, the new assumption introduced here is the
difference in the {\it in}-cold and {\it between}-cold patches
scattering, trough the temperature independent constant $a_{0}$.
Under a physical point of view, it is introduced with the aim to
represent the forward scattering in the cold region, which is
mainly temperature independent and involves only small angles.
Its effect is therefore confined inside each single cold patch
only.

\section{Results}
\subsection{Forward scattering: a second order effect}
\begin{center}
\begin{figure}[!htbp]
\begin{center}
\includegraphics[width=10cm,height=8cm] {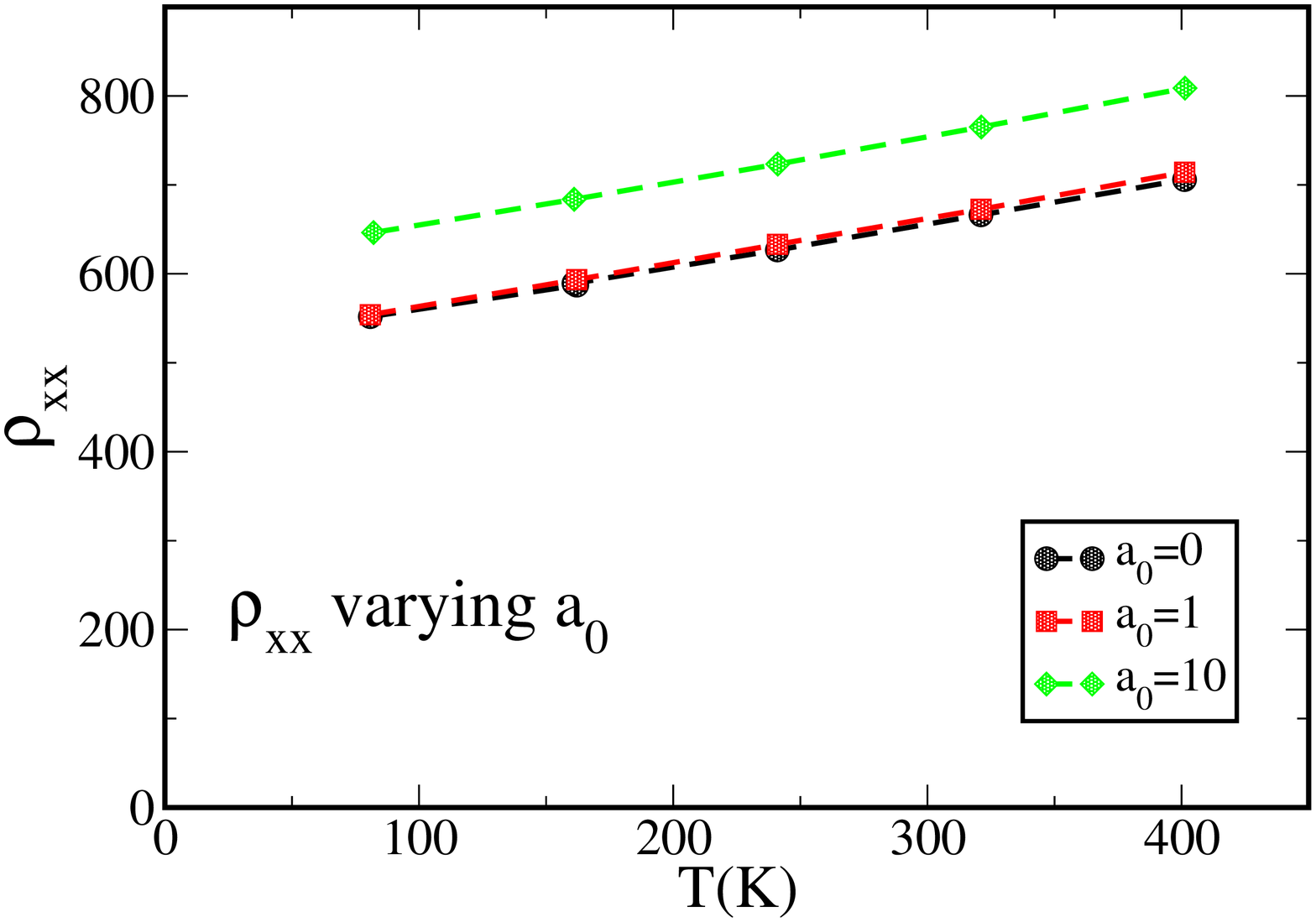}
\caption[]{$\rho_{xx}$ as a function of T varying $a_{0}$, $\Delta
k= 2\pi/5$, $a_1= 50$, $b=1$, $c=7$} \label{fig2}
\end{center}
\end{figure}
\end{center}

\begin{center}
\begin{figure}[!htbp]
\begin{center}
\includegraphics[width=10cm,height=8cm] {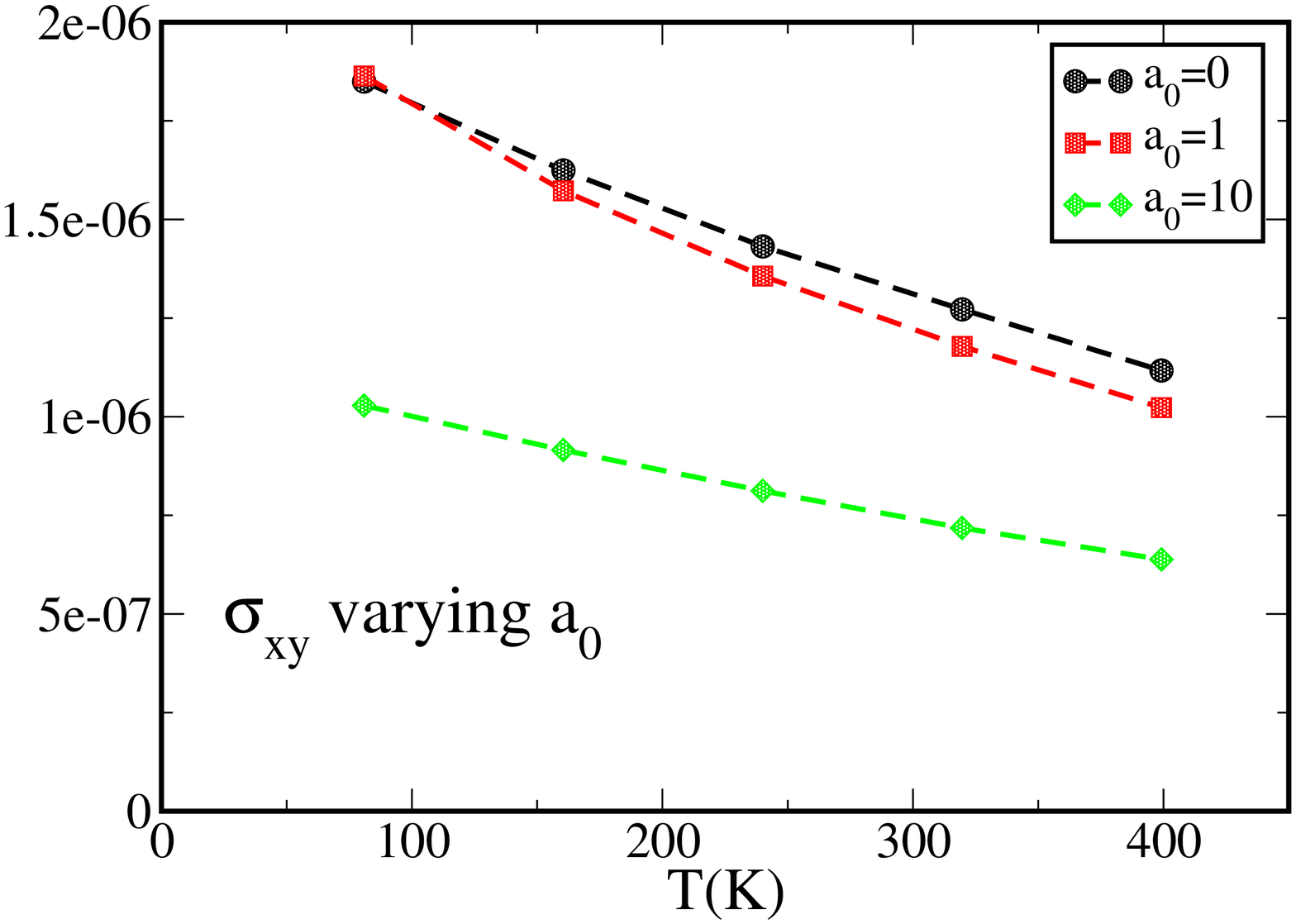}
\caption[]{$\sigma_{xy}$ as a function of T varying $a_{0}$,
$\Delta k= 2\pi/5$, $a_1= 50$, $b=1$, $c=7$} \label{fig3}
\end{center}
\end{figure}
\end{center}

\begin{center}
\begin{figure}[!htbp]
\begin{center}
\includegraphics[width=10cm,height=8cm] {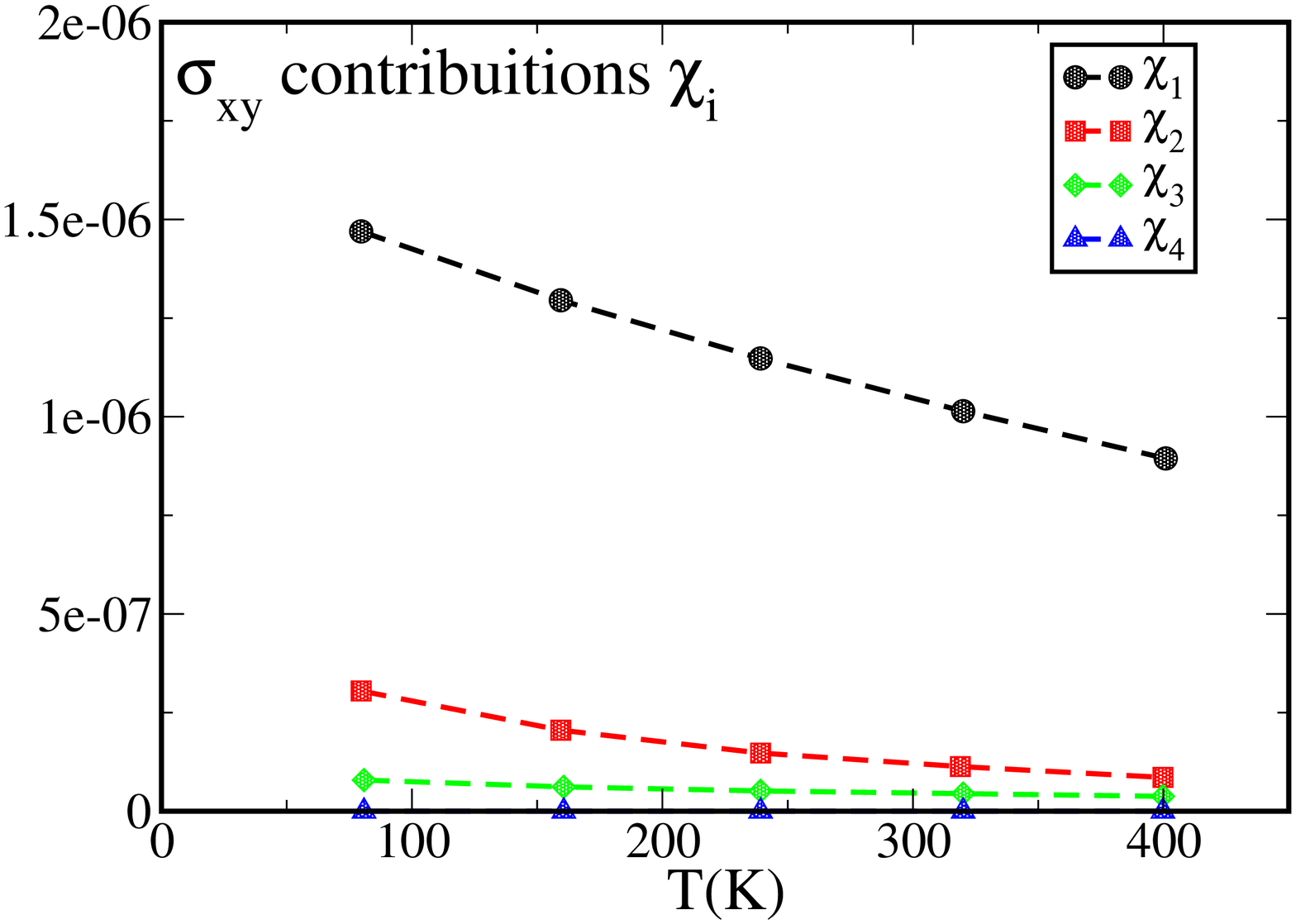}
\caption[]{Contributions to $\sigma_{xy}$, $a_{0}$, $\Delta k=
2\pi/5$, $a_1= 50$, $b=1$, $c=7$} \label{fig4}
\end{center}
\end{figure}
\end{center}

As mentioned in the previous section, in order to consider the
possible effect of the forward scattering due to impurity in the
cold region, we have introduced different scattering amplitudes
for the {\it in}-cold patch scattering $c_{11}$ and the {\it
between}-cold patches scattering $c_{12}$, choosing an $a_{0}
\neq 0$. The condition (\ref{13.c}) was in fact necessary to
obtain the $\overline\Gamma^{-1}$ dependent terms in the
expressions for the conductivity (\ref{11*}),(\ref{11**}), which
otherwise would be identically zero as in the 2-patch model. In
the following we show the results on the resistivity $\rho_{xx}$
(Fig. \ref{fig2}) and on the hall conductivity $\sigma_{xy}$
(Fig. \ref{fig3}). To simulate a small angle impurity scattering,
we chose a quite narrow {\it cold patch}, with $\Delta k= 2\pi/5$
(formula \ref{13.a}).
We will show in this section that the forward scattering is
indeed a second order perturbation effect which cannot explain
the different $T$-dependence of the in-plane parallel and
transverse conductivity. In both graphs, we display three possible
values of $a_{0}=0, 1 \hbox{ and } 10$. The introduction of the
small forward scattering effect is likely better simulated
passing from $a_{0}=0$ to $a_{0}=1$. But in this case only a
small change appears in either $\rho_{xx}$ and $\sigma_{xy}$. So
the impurity forward scattering should have only a small
influence on the conductivity. On the other hand, a big variation
from $a_{0}=0$ (for example the displayed $a_{0}=10$) causes a
vertical translation either for $\rho_{xx}$ and $\sigma_{xy}$. In
Fig. \ref{fig4} we show the four contributions  to $\sigma_{xy}$
$X_{i}$, $i=1,...4$ of eq. (14). The $\overline{\Gamma}^{-1}$
dependent corrections $X_{2}$ and $X_{3}$ are two and one order
of magnitude smaller than the leading term $X_{1}$ (the only
present when $a_{0}=0$), while the $(\overline{\Gamma}^{-1})^{2}$
dependent correction $X_{4}$ is even four orders of magnitude
smaller. This already shows that, although the forward scattering
term does indeed contributes to the bending of the cotangent of
the Hall angle as a function of temperature, the effect is a
second order one and it cannot account for the experimentally
observed temperature-exponent of the cotg of the Hall angle,
which is close to 2.
\subsection{Data Fitting Troubles with gaussian patches}
\begin{figure}[!htbp]
\begin{center}
\includegraphics[width=10cm,height=7.5cm] {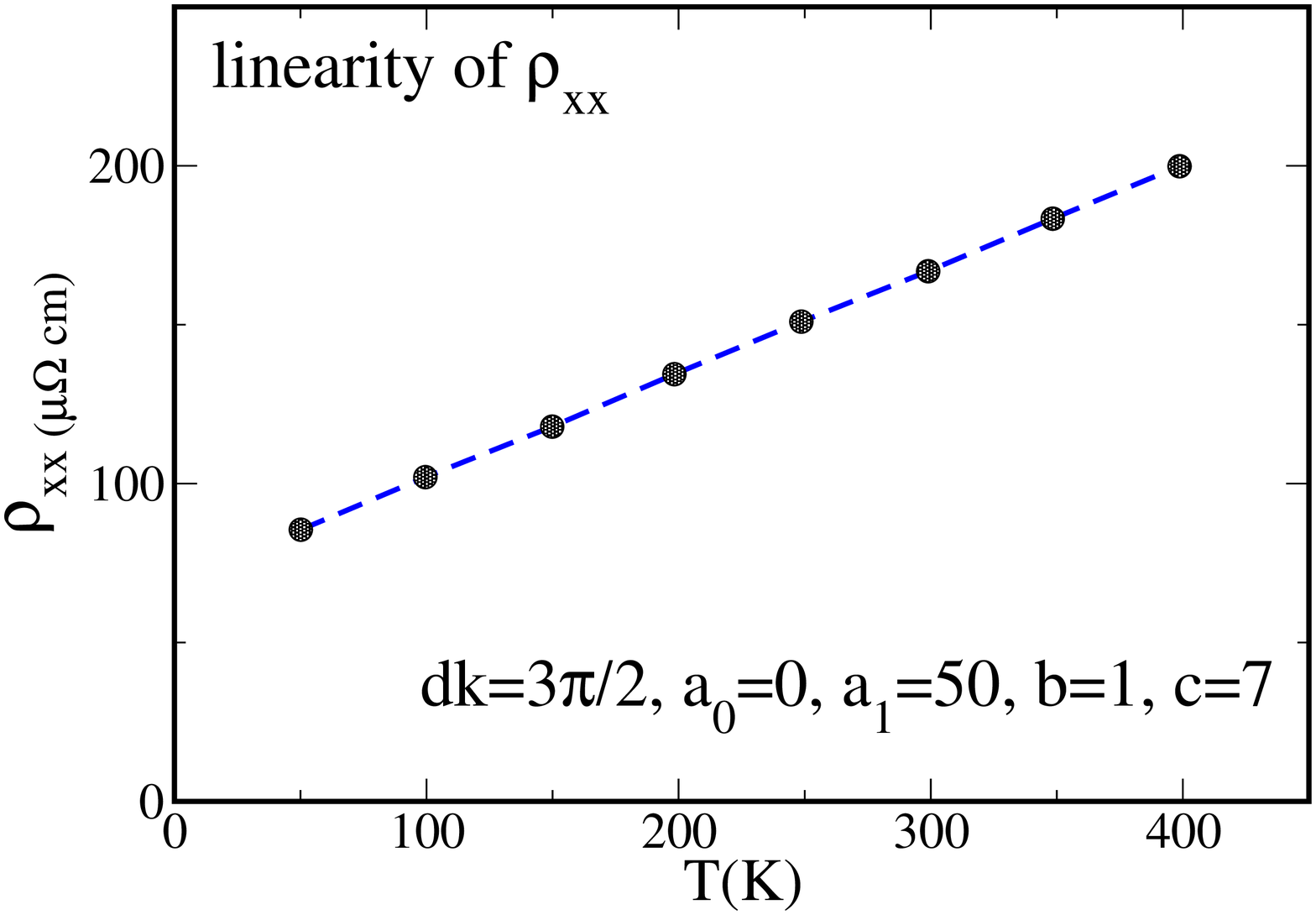} \\
\hspace{0.4cm}
\includegraphics[width=10.25cm,height=7.5cm] {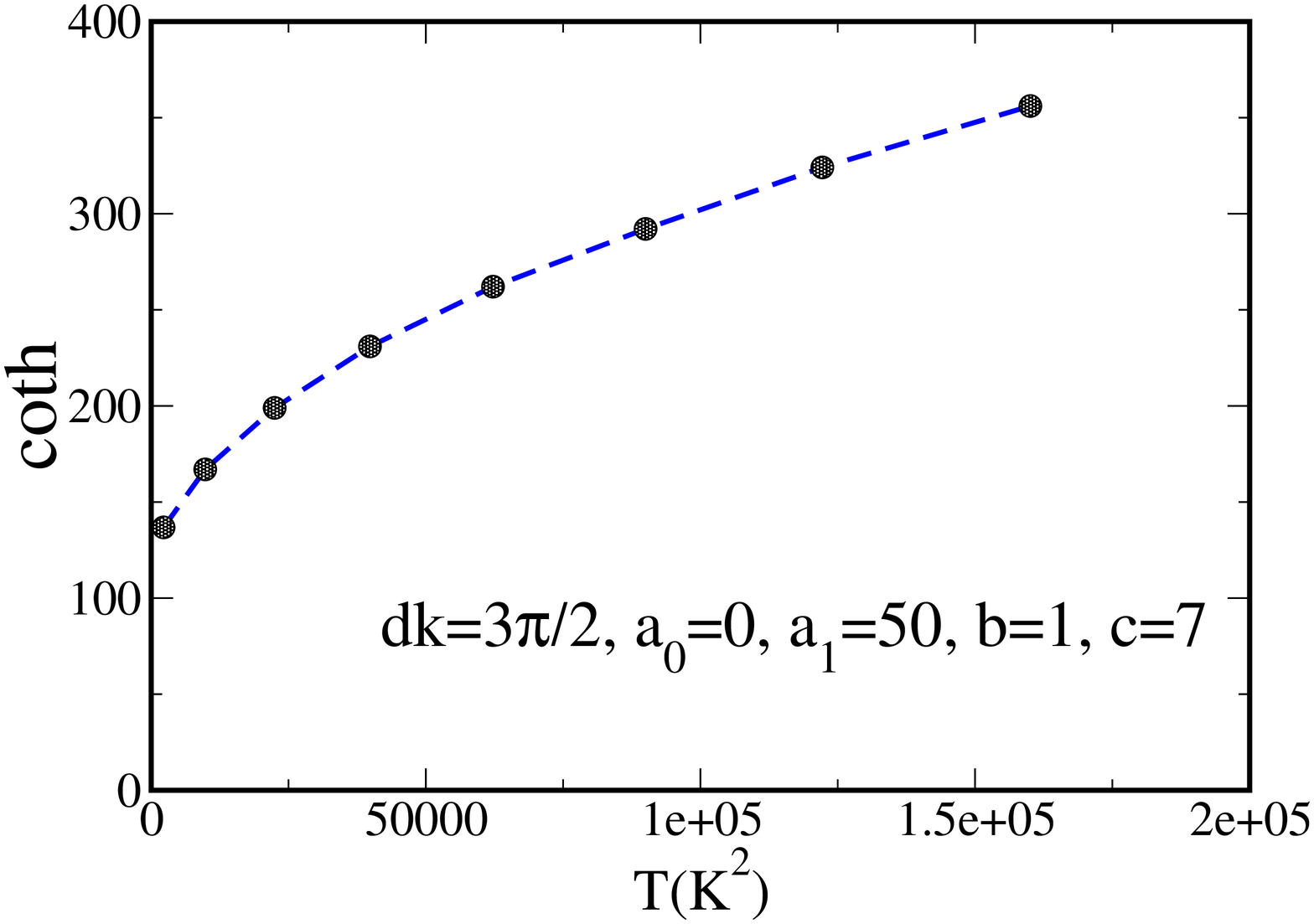}
\caption[]{$\rho_{xx}$ and $\cot_{\theta}$ vs. $T $ $\Delta k=
\pi/1.5$, $a_{0}= 0$, $a_1= 50$, $b=1$, $c=7$} \label{fig5}
\end{center}
\end{figure}

\begin{figure}[!htbp]
\begin{center}
\includegraphics[width=10cm,height=7.5cm] {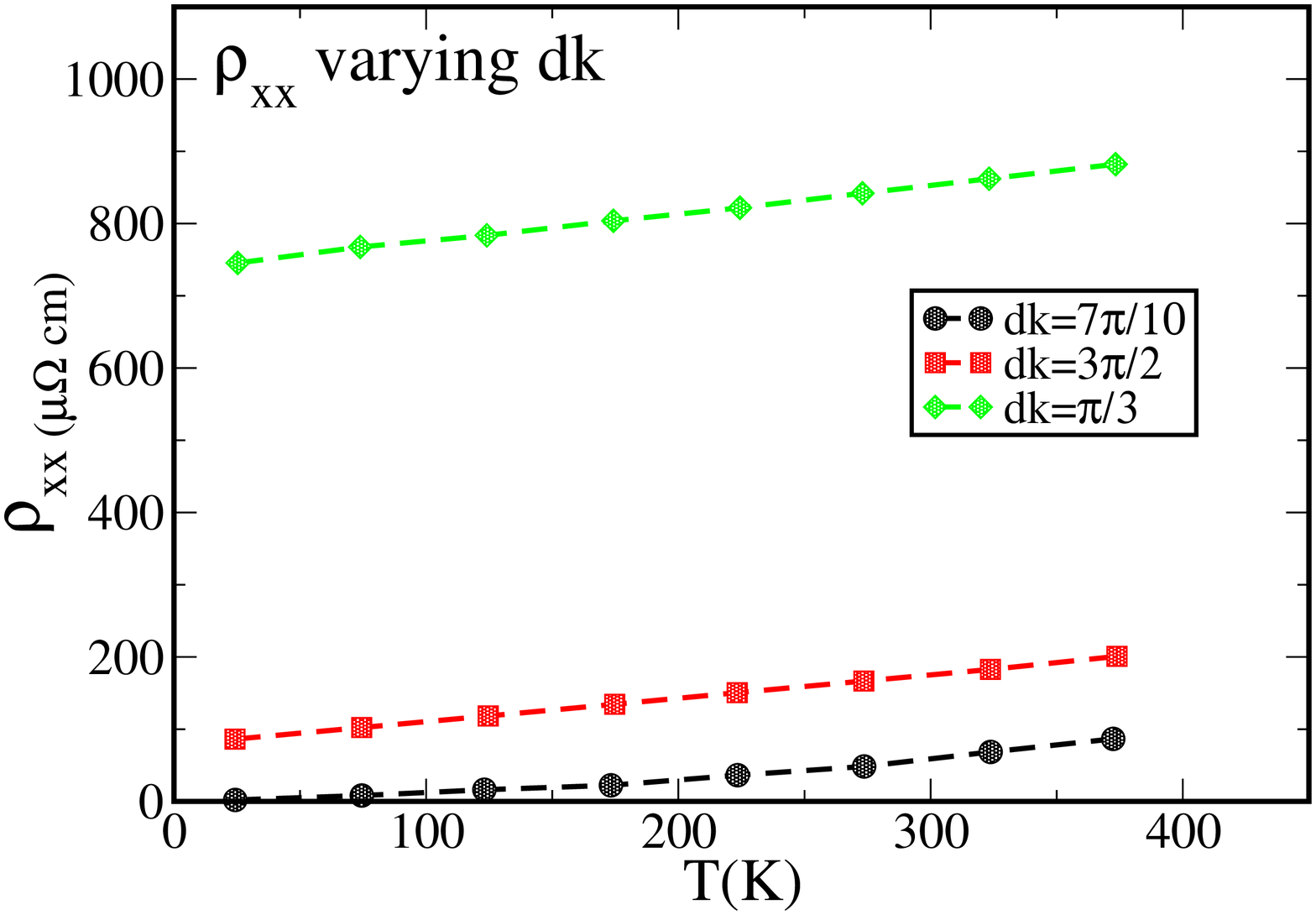}
\caption[]{$\rho_{xx}$ and $\cot_{\theta}$ vs. $T $ $\Delta k=
\pi/1.5$, $a_{0}= 0$, $a1= 50$, $b=1$, $c=7$} \label{fig6}
\end{center}
\end{figure}

Difficulties in finding different exponents in the $\rho_{xx}$
and the $\cot (\theta_H)$ are also encountered in the $5-$patch
model study if we attempt a fit of the experimental results, as
systematically done in \cite{PSK}. As first step, we could
analyze the behavior of the resistivity $\rho_{xx}$ and the Hall
angle $\cot(\theta_{H})$. We chose starting parameter values
close to the ones used in \cite{PSK}. The resistivity is inversely
increasing with the size of the cold region (Fig. \ref{fig6}),
i.e. it is inversely proportional to the doping, as observed for
the 2-patch model \cite{PSK}, and in agreement with the
experimental observation \cite{Ando}, \cite{Konsta}. It shows also
the expected linear behavior (Fig. \ref{fig5} and \ref{fig6}).
Deviation from linearity is, in fact, observed for very large size
of the cold patch, when the system becomes a FL: fitting with the
law $\rho_{xx}= A+ B\,T^{\alpha}$ for the case shown in the graph
$\Delta k= \pi/0.7$, $\alpha$ turns out to be 2.
\begin{figure}[!htbp]
\begin{center}
\includegraphics[width=10cm,height=7.5cm] {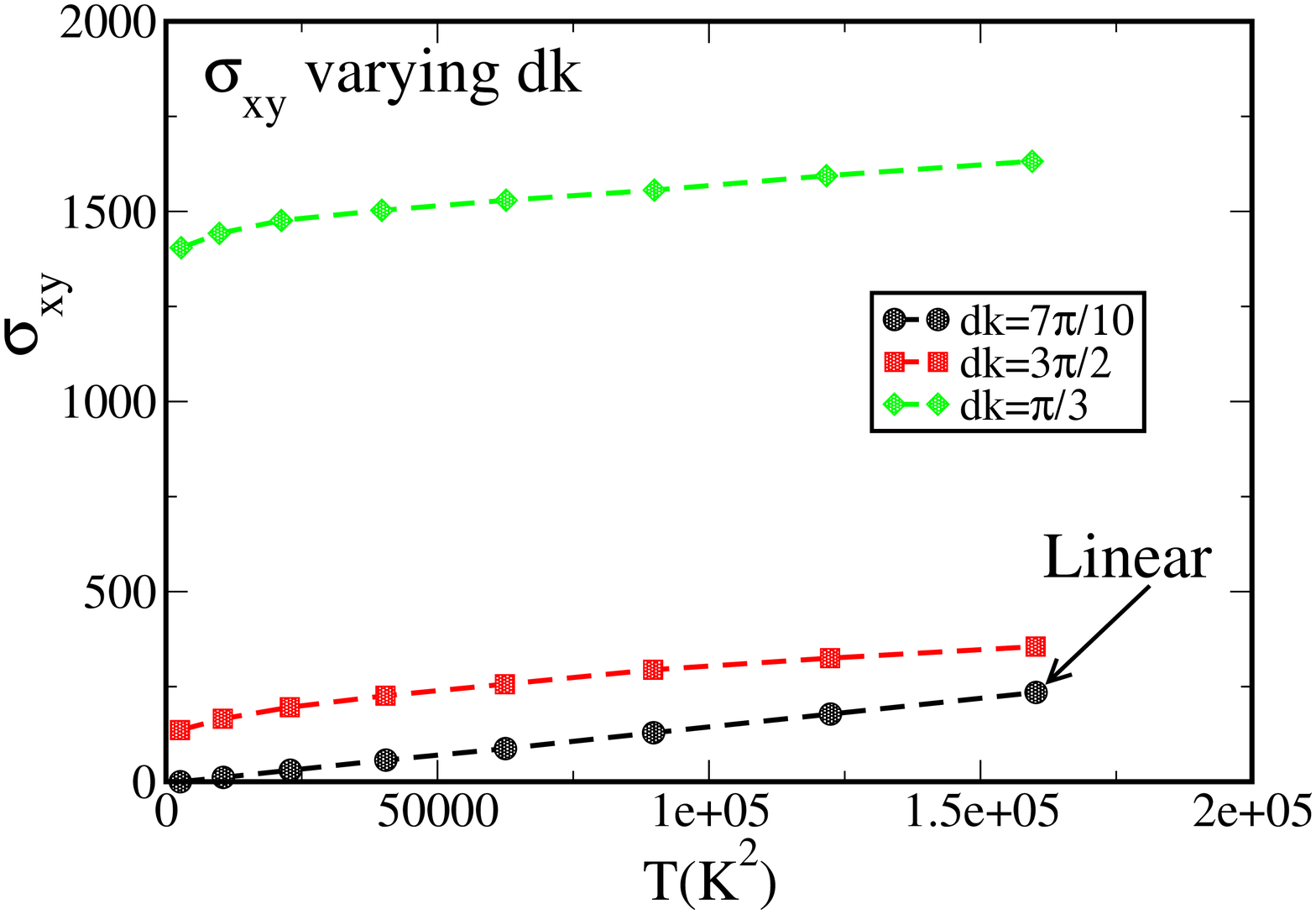}
\caption[]{$\cot_{\theta}$ vs. $T^2 $ varying doping, i.e.
$\Delta k$} \label{fig7}
\end{center}
\end{figure}
The $\cot(\theta_{H})$ behavior is shown in Fig. \ref{fig7}. Like
$\rho_{xx}$, it is decreasing with increasing doping, i.e. with
increasing cold patch size. This agrees with the experimental
data \cite{Ando}\cite{Konsta}. However, strong differences are
seen in our result. First of all, according to
\cite{Ando}\cite{Konsta}, $\cot(\theta_{H})$ should be smaller
than $\rho_{xx}$ (almost half) for the same set of parameters,
but in our case it is exactly the opposite, with a
$\cot(\theta_{H})$ that is almost twice $\rho_{xx}$, no matter the
set of parameters chosen. Second, $\cot(\theta_{H})$ should show a
$T^{\alpha}$ behavior, with $\alpha$ going from 2 for the
underdoped region to 1.5 in the overdoped ($\alpha\simeq 1.7$ for
optimal doping). So in a $\cot(\theta_{H})$ vs. $T^{2}$ plot, we
should observe straight lines (i.e. $\cot(\theta_{H}) \sim T^{2}
{}$) that get smaller and smaller in absolute value and start to
curve (\cite{Ando}\cite{Konsta}) with increasing doping, as
$\alpha$ becomes smaller than 2. In fact in Fig. \ref{fig7} we
observe exactly the opposite behavior, with straight lines that
curve when the doping is increased. The $\cot(\theta_{H})$ seem
to have a opposite behavior from the one expected given
experimental results and, as expected from the observations of
the previous section, its power-law behavior follows closely the
one of the resistivity.

\section{The "panettone"-patch model}

\begin{figure}[!htbp]
\begin{center}
\includegraphics[width=6cm,height=6cm] {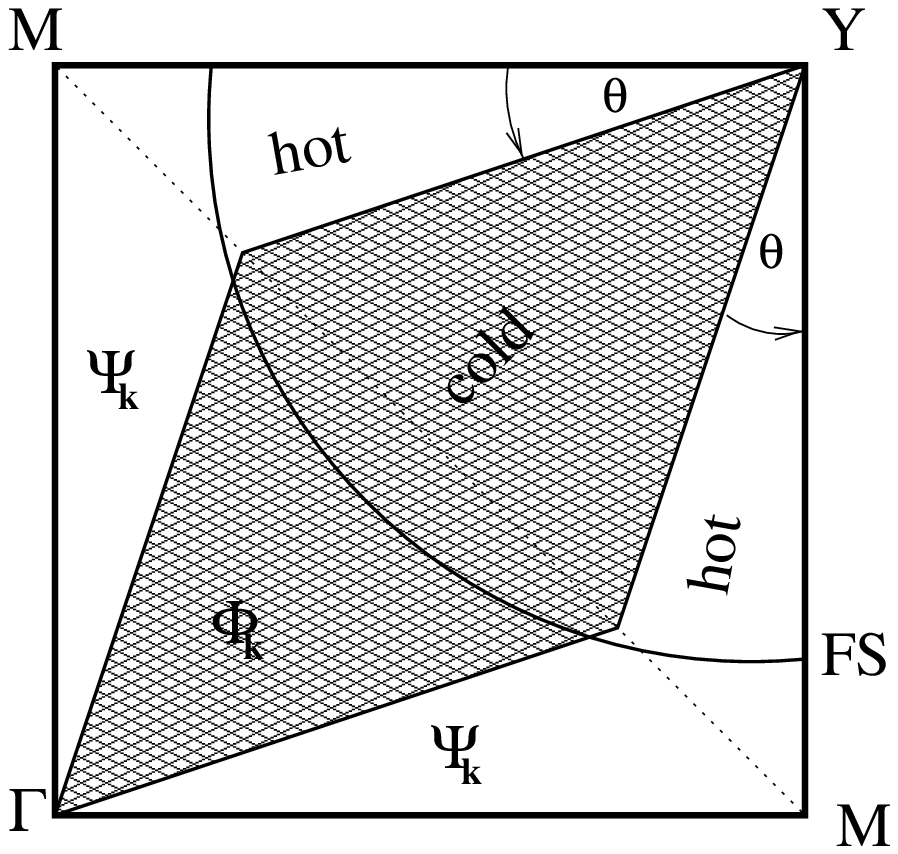}
\qquad \qquad \qquad
\includegraphics[width=5cm,height=5cm] {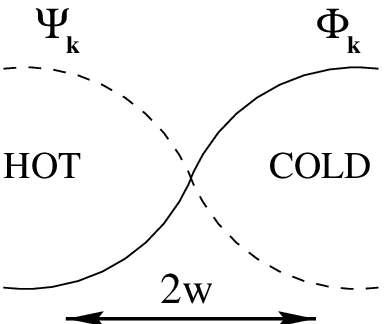}
\caption[]{Left hand side: Patch used in \cite{PSK} in the first
quadrant of the BZ. Right hand side: intermediate region of
overlapping hot/cold patch.} \label{PSK-Fig}
\end{center}
\end{figure}

\begin{figure}[!htbp]
\begin{center}
\includegraphics[width=12cm,height=9cm] {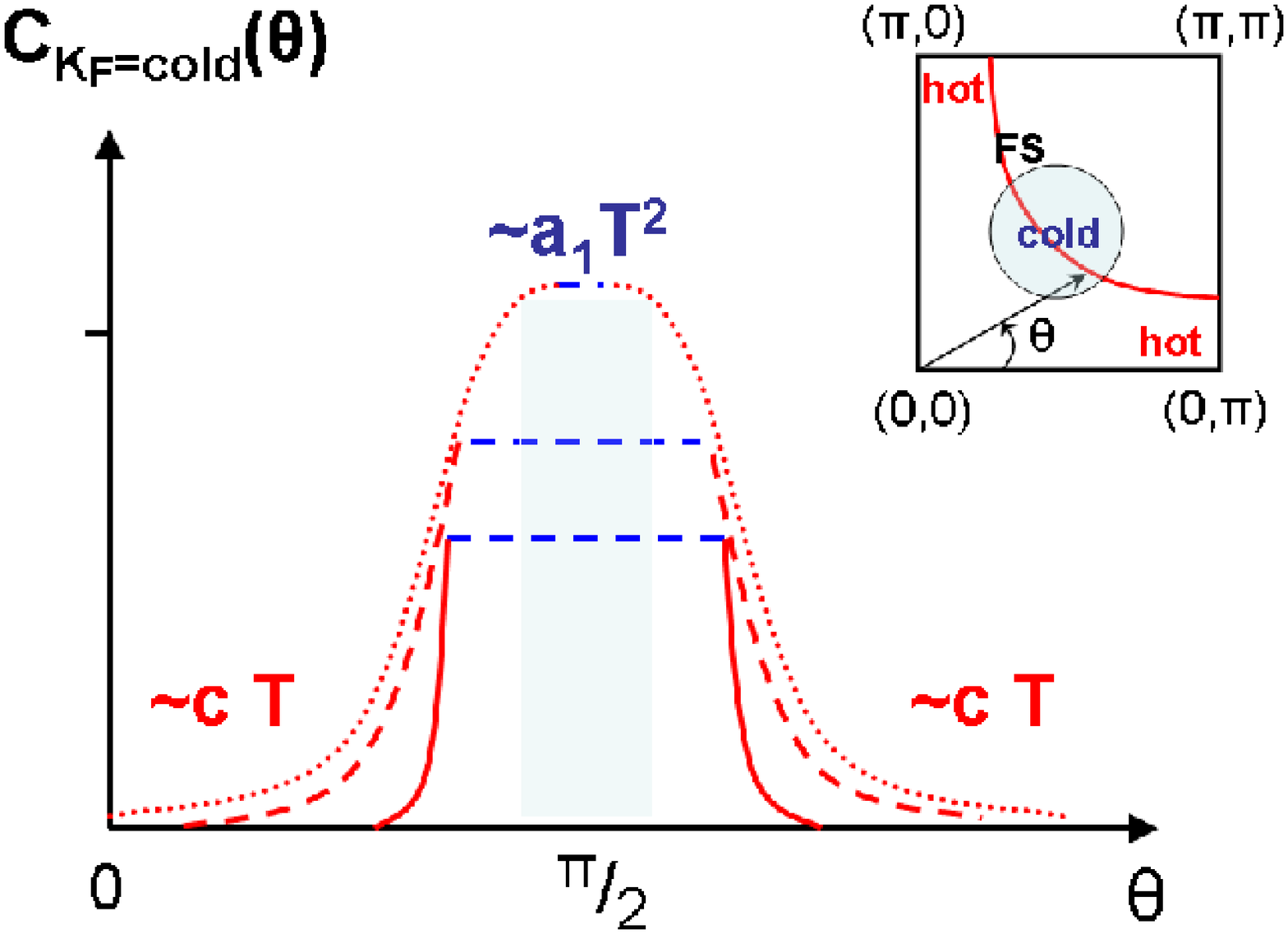}
\includegraphics[width=12cm,height=9cm] {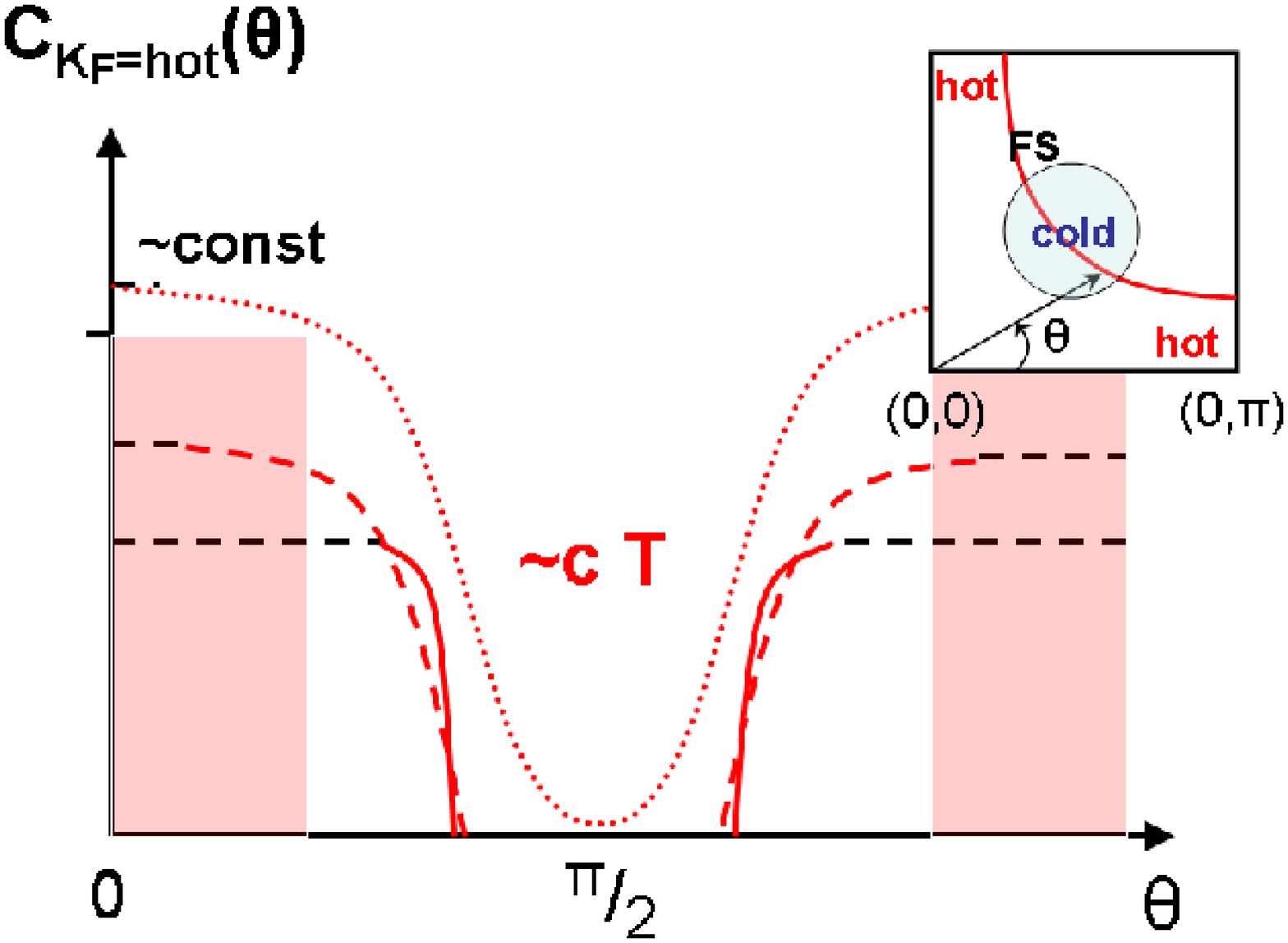}
\caption[]{Scattering matrix $C_{kk'}$ (formula
\ref{C_kk1_Npatch}) parametrized with a 5-patch model. In the top
panel we display $k \in$ cold-patch and $k'$ varying on the FS
(measured with the angle $\theta$ from the $x$-axis). In the
bottom panel we display $k \in$ hot-patch and $k'$ varying on the
FS (measured with the angle $\theta$ from the $x$-axis). We
elucidate the temperature $T$-dependence in the different regions.
Notice that in the 5-patch "panettone" model a temperature
Fermi-function-like dependent (three different temperature are
shown in the picture) patch is introduced in the smoothly varying
intermediate cold/hot region (according to formulas
\ref{panettone-patch} and \ref{w_o}), besides the phenomenological
linear-$T$ dependence of the scattering amplitude $c_{\alpha
\beta}$ used in the 2-patch model.} \label{Ckk1}
\end{center}
\end{figure}

The results derived in the previous sections show some problems in
the developing of a 5-patch model using gaussian patch-functions:
\begin{enumerate}
  \item The introduction of a $T$-independent forward scattering,
  made possible the a $5$-patch structure, appears as a second
  order perturbation effect which does not substantially modify
  the $\gamma$ exponent in the $\cot \theta_{H}$.
  \item the values we can obtain for the resistivity and the Hall
  conductivity do not match what is expected given experimental observation
  of cuprate systems. The $\cot \theta_{H}$ is in fact
   always bigger
   than the plane resistivity $\rho_{xx}$.
  \item while it is possible to obtain a linear in temperature resistivity
  $\rho_{xx}$,
   the $\cot \theta_{H}$ displays a $T^{\gamma}$-dependence with
   an exponent $1 < \gamma < 1.5$.
   Albeit $\gamma > 1$, its dependence from the model parameters tightly
   follows the resistivity behavior like in a regular FL.
\end{enumerate}
Accordingly, the fundamental question that remains is how does one
obtain two independent temperature exponents for the resistivity
$\rho_{xx}$ and for the $\cot \theta_{H}$? In particular $\cot
\theta_{H}\simeq T^{\gamma}$ with $\gamma \leq 2$. The results
presented here for the $5$-patch model with gaussian
patch-function seem even to worsen the results of the $2$-patch,
where a different patch was used \cite{PSK}. In the latter case
point (1) above was essentially solved and a $\gamma
> 1.5$ was found. It was in fact possible to obtain a lower
Hall conductivity $\sigma_{xy}$ (and therefore a lower $\cot
\theta_{ H }$ ) by introducing a steep interlacing region trough
hyperbolic patch-functions (see left hand side Fig.
\ref{PSK-Fig}):
\begin{equation}\label{PSK-patch}
  \phi(k)=\, \prod_{i=1}^{4} \, \phi_{i}(k)
\end{equation}
\begin{equation}\label{PSK-patch_eq}
\phi_{i}(k)= \,\frac{1}{2}\, \left[ \,1+(-1)^{i}\, \tanh\left(
\frac{k_{y}-m_{i}k_{x}-t_{i}}{w} \right) \, \right]
\end{equation}
where $m_{i}= m_{1}$ for $i=1,4$, $m_{i}= m_{2}$ for $i=2,3$,
$t_{i}= t_{1}$ for $i=3$, $t_{i}= t_{2}$ for $i=4$ and $t_{i}$ is
zero for $i=1,2$. For more details refer to the work in
\cite{PSK}. This creates a cold region like the one shadowed in
Fig. \ref{PSK-Fig}.
The hot patch $\psi_{k}$ is defined has:
$\psi_{k}=\, 1- \phi_{k}$
Introducing smoothly varying patch-functions is necessary (ref.
\cite{PSK}) to avoid un-physical ${\bf k}$-derivative-diverging
terms in the leading order of the magnetoresistance. The smooth
change between hot and cold regions (measured by a width
parameter $w$) is, in this case, described by hypebolic functions
(shown at the right hand side of Fig. \ref{PSK-Fig}). This
however looks to be important also in matching the right
quantitative order of the resistivity $\rho_{xx}$ and the $\cot
\theta_{H}$, as expressed in the point (2) above. The shape of
the patch enters in fact in the parametrization of the collision
integral $C_{kk'}$ (formula \ref{C_kk1_Npatch}), which determines
the collision integral (\ref{1.6a}) (or equivalently the lifetime
$\tau_k$ (\ref{one_tau}) of the quasiparticles). In turn, this
quantity completely determines the flow of quasiparticles in the
phase-space as determined by the BE (\ref{1.7bis}). It is
therefore important not only to well parametrize the cold patch
(with a Fermi-like $T^{2}$-dependence of the scattering amplitude)
and the hot patch (where the scattering amplitude is assumed
temperature-independent), but also the intermediate region, which
have to vary smoothly but fast enough. Hyperbolic-like tail seems
to work better than a gaussian tail, in spite they select a
similar region of the $k$-space. We display a schematic diagram of
the scattering matrix $C_{kk'}$ on the FS in Fig. \ref{Ckk1}.
Quasiparticle are present only on small stripe of width $\sim k_B
T$ around the FS (the only active states in a Fermi liquid close
to accessible quasiparticle states), therefore only in that
region the effect of $C_{kk'}$ is important in determining the
solution to the BE. In the top panel we fix $k=k_F$, $k \in$ the
cold region and we vary $k'$ on the FS in the first quadrant of
the BZ, measuring its position trough the angle $\theta$ with
respect to the $x$-axis. This describes a particle originally in
the cold region undergoing a collision and ending in the state
$k'$. In the bottom panel $k=k_F$ but $k \in$ the hot region. As
we said, in order to fit the right values of $\rho_{xx}$ and
$\cot \theta_H$, in this diagrams the tail of the cold/hot
intermediate region has to be steep, at least more than in a
gaussian function. This is the reason why the choice of gaussian
patches, albeit practical, does not work efficiently.
\begin{center}
\begin{figure}[!htbp]
\begin{center}
\includegraphics[width=6cm,height=6cm] {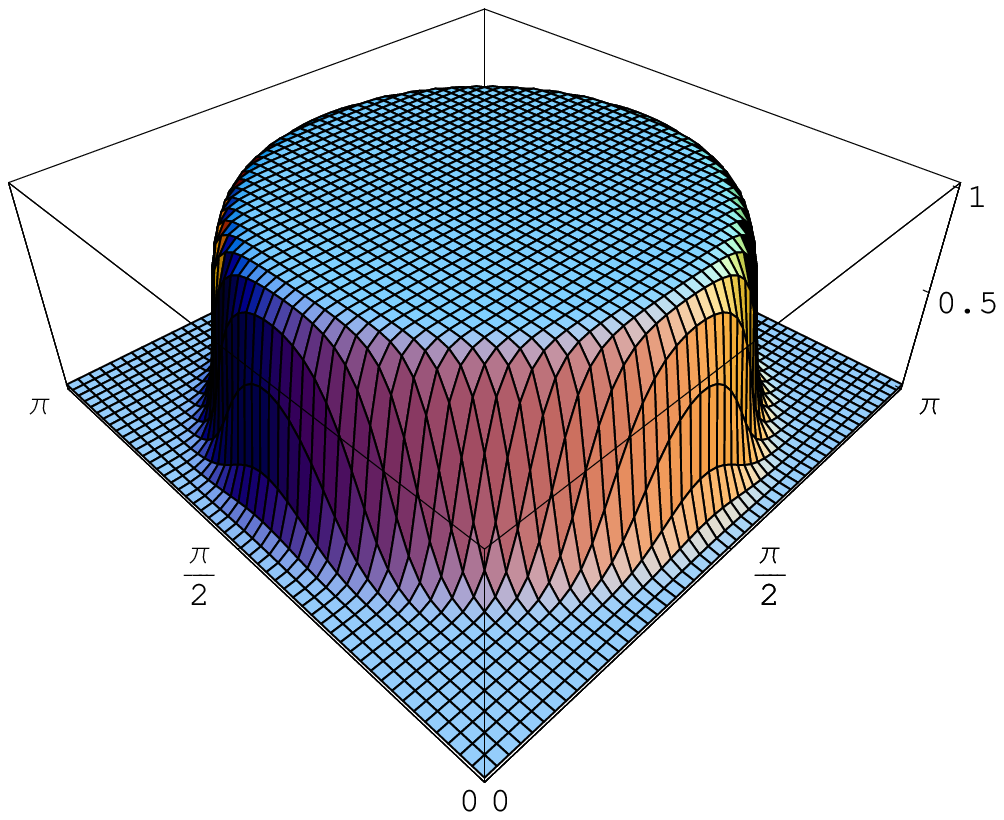}
\caption[]{"Panettone"-patch in the first-quadrant of the BZ.}
\label{panettone}
\end{center}
\end{figure}
\end{center}
We can restore the quality of results found in \cite{PSK} by
introducing a $5$-patch structure with a more steep tanh-like
varying {\it cold}-patch:
\begin{equation}\label{panettone-patch}
\phi_{i}(k)= \, \frac{1}{2}\, \left[ 1- \tanh\left( w^{-1}\,
(r_{\mathbf{k}}-r_{o})\right) \right] \hspace{2cm} i=1...4
\end{equation}
where $r_{\mathbf{k}}$ measures the distance from the center of
the patch $(c_{x},c_{y})=\,(\pm \frac{\pi}{2},\pm
\frac{\pi}{2})$, in the four center-quadrant of the Brillouin
Zone:
\begin{equation}\label{r_k}
r_{\mathbf{k}}=\, \sqrt{(k_{x}-c_{x})^{2}+(k_{y}-c_{y})^{2}}
\end{equation}
and $w$ is again the width parameter of the cold/hot smoothing
region. This is akin to substituting the gaussian patches
introduced in the previous section by "panettone"-like shaped
patches (see Fig.\ref{panettone}). The qualitative change with
respect to the topological division in hot/cold patches used in
\cite{PSK} (Fig. \ref{PSK-Fig}) is minimal, being the parts of
the $k$-space effectively involved in the calculation of
transport quantities small stripes around the Fermi surface
$-(\partial f_{\mathbf{k}} / \partial \mathbf{k})$.
\begin{figure}[!htbp]
\begin{center}
\begin{tabular}{|c|c|c|c|c|c|} \hline
$r_o$ (\ref{panettone-patch})  & $\overrightarrow{c}$ (\ref{r_k})
& $w_{o}$ (\ref{w_o}) & $a_0$ & $b$ &
  $c$\\ \hline
0.5       & (1.15,1.15)  & 16  & 70 & 0.4  & 2.8 \\ \hline
\end{tabular} \vspace{1cm}\\
\includegraphics[width=7cm,height=6cm] {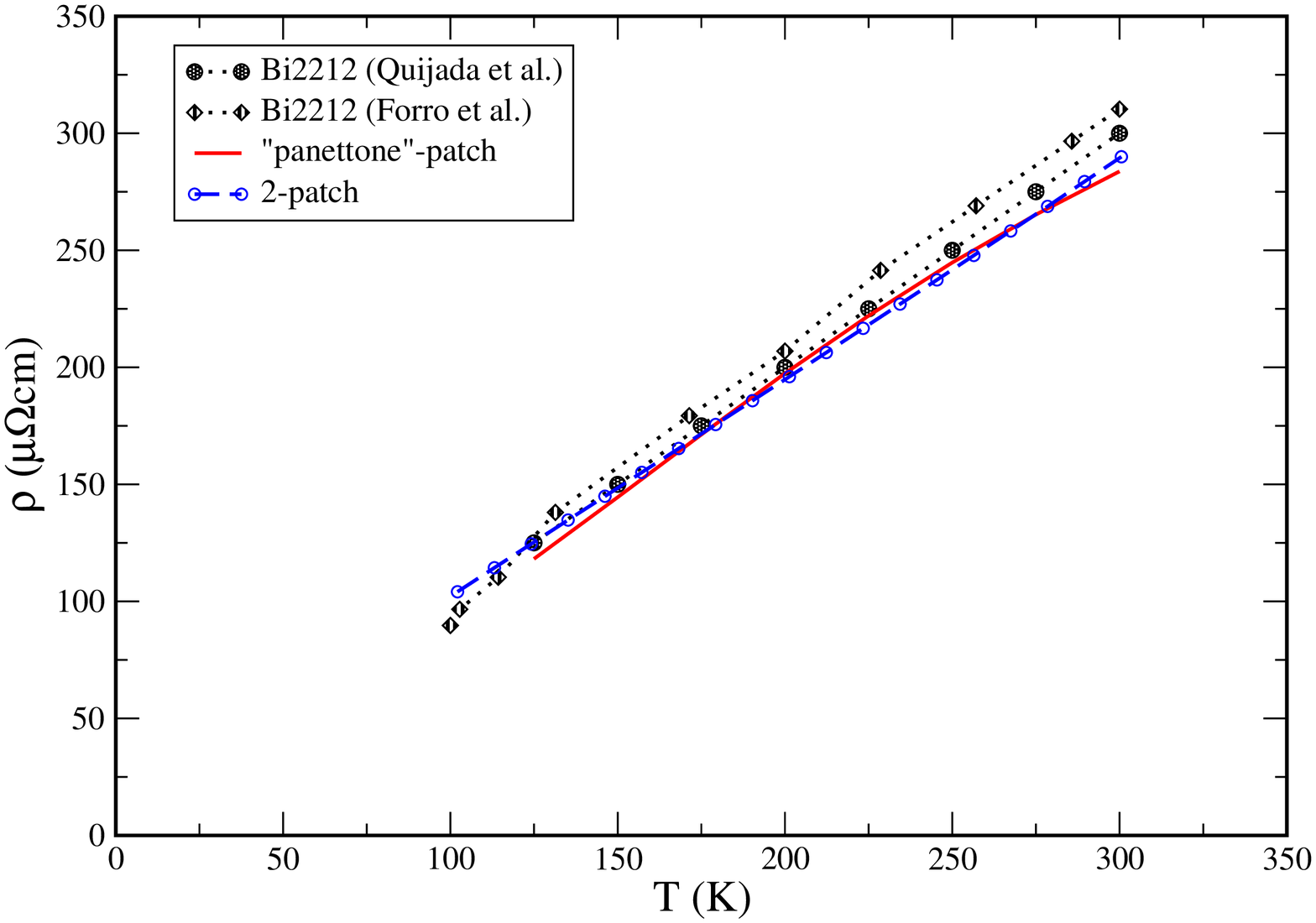}
\includegraphics[width=7cm,height=6cm] {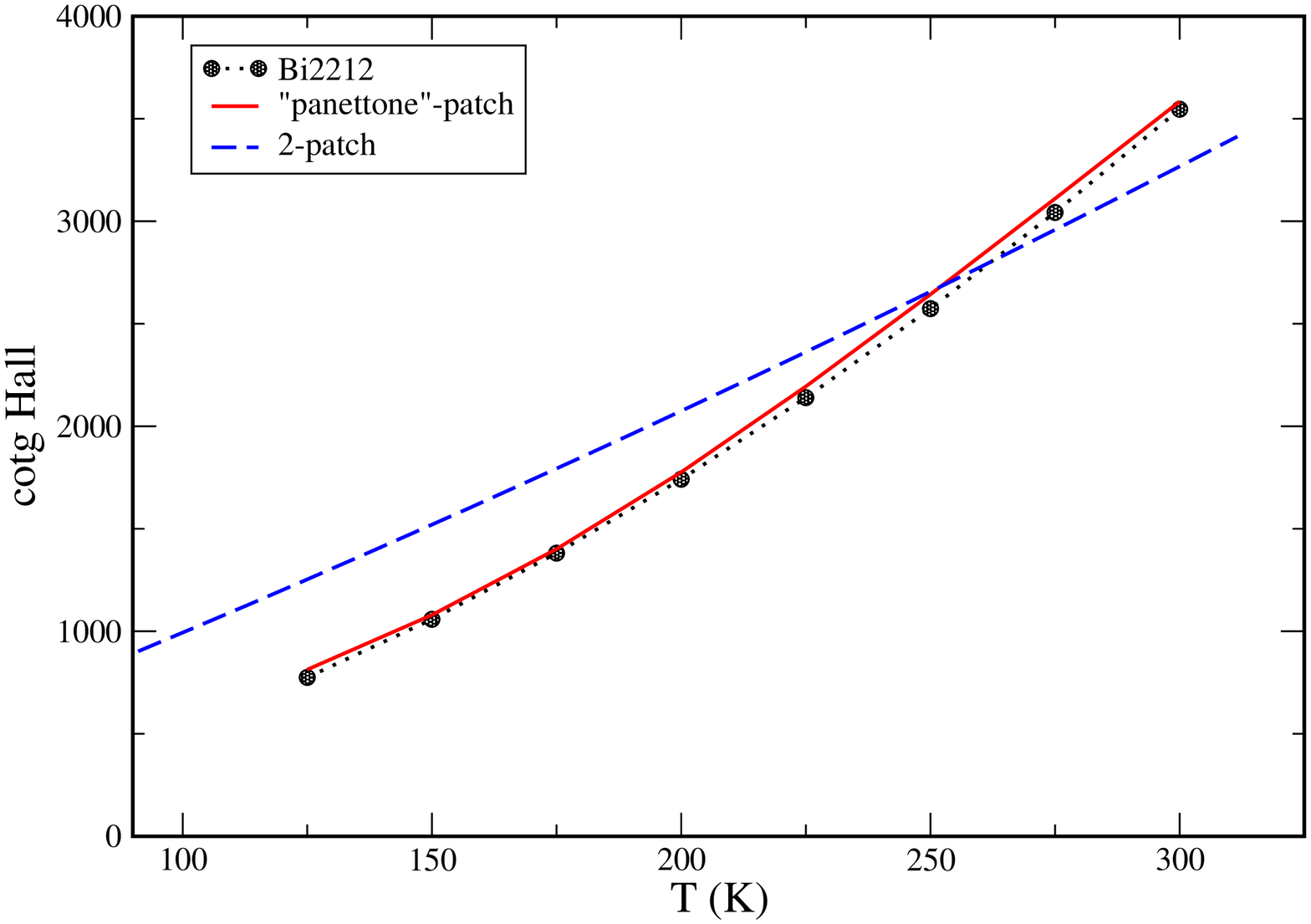}
\includegraphics[width=7cm,height=6cm] {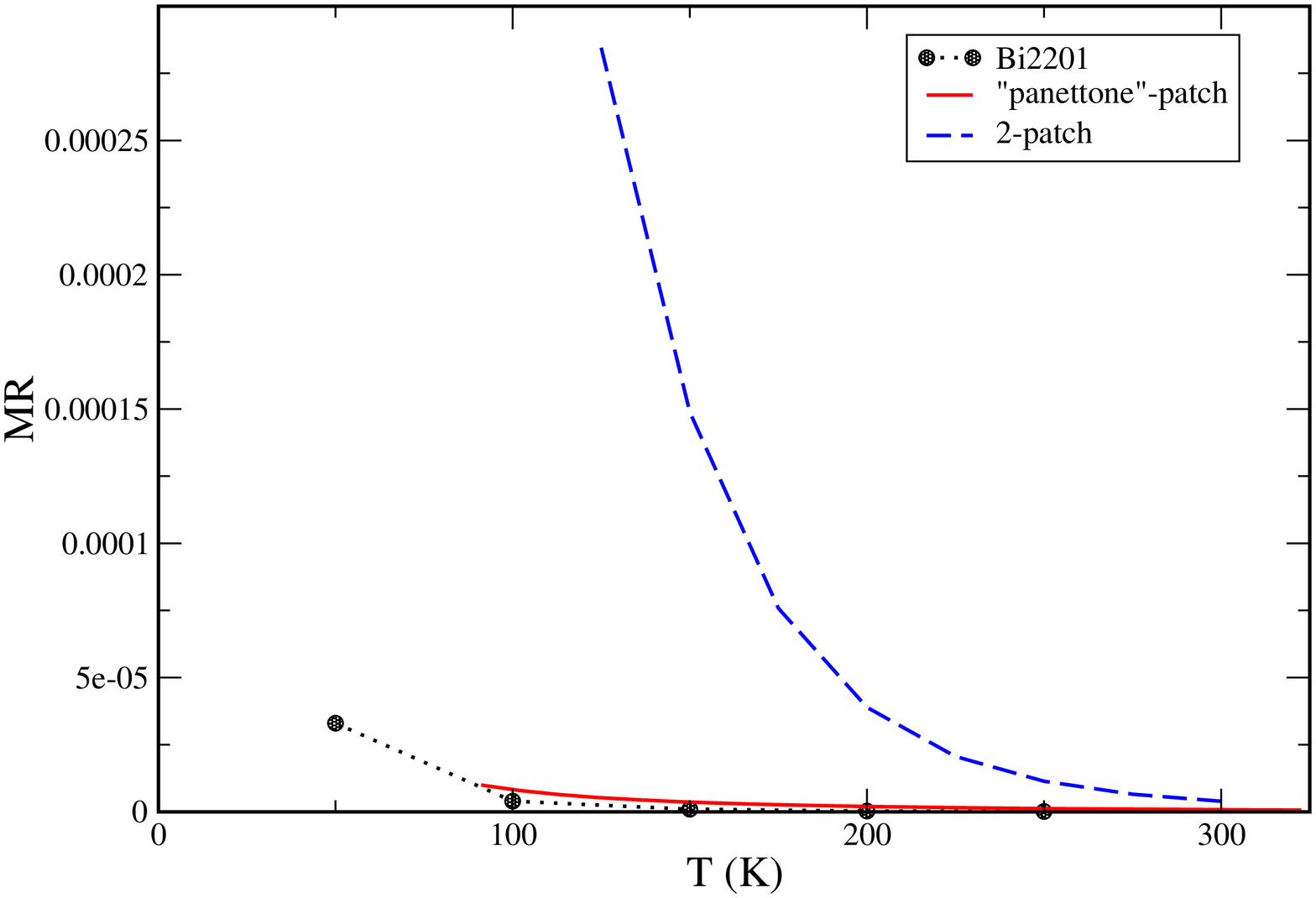}
\includegraphics[width=7cm,height=6cm] {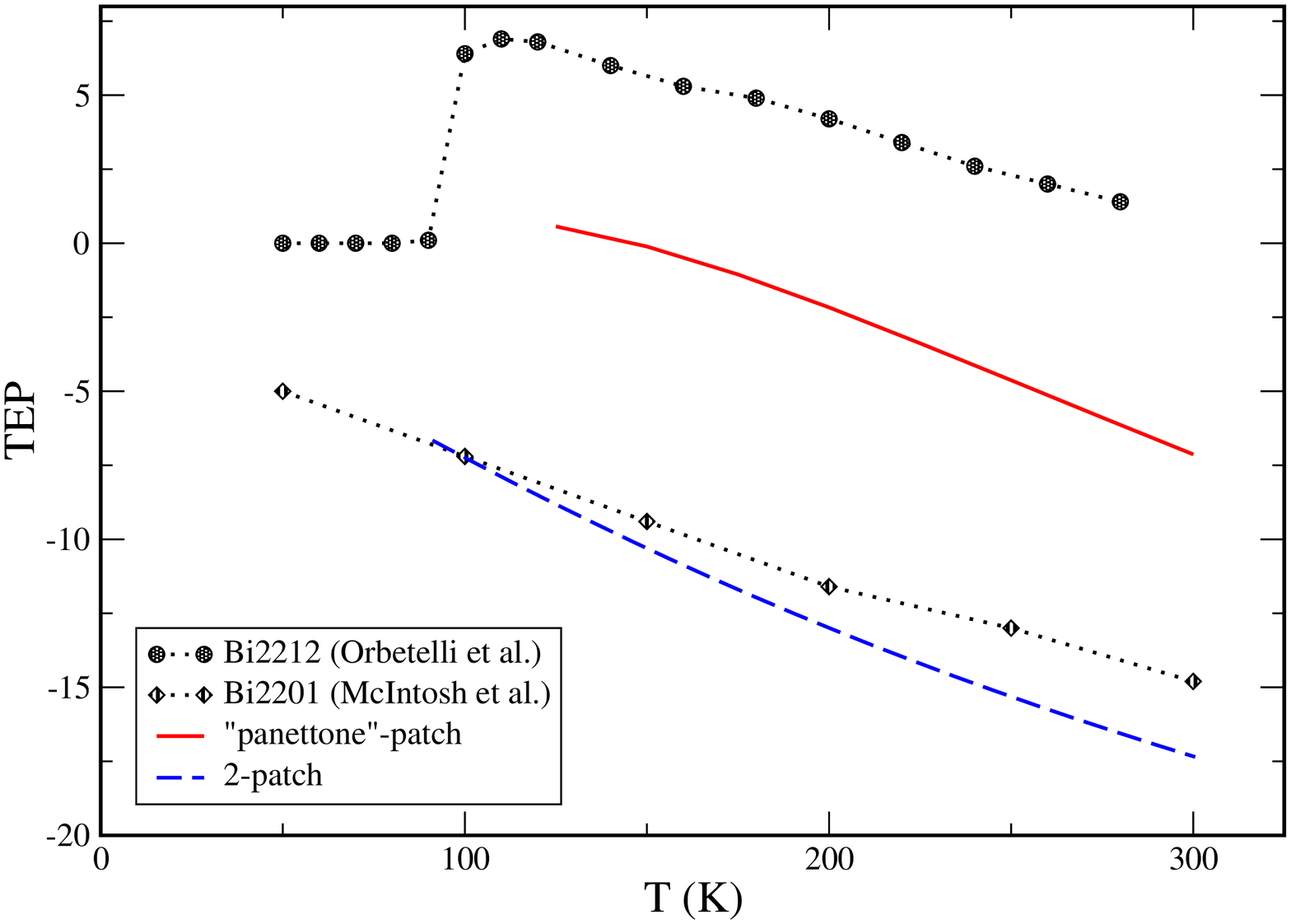}
\caption[]{Top left: the experimental linear dependence in
temperature in the resistivity $\rho_{xx}$ of Bi2212
\cite{Quijada}\cite{Forro} is confronted with the 2-patch model
of \cite{PSK} and the "panettone"-patch. Top right: The
improvement of the "panettone"-patch on the 2-patch model in
describing the $T^{\gamma}$-dependence of the $\coth \theta_{H}$
is evident. Bottom left: with the same fitting parameters used to
fit $\rho_{xx}$ and $\coth \theta_{H}$, the "panettone"-patch
magneto-resistance MR  is confronted with the experimental
results on of Bi2201 and the 2-patch model. Bottom right: the
experimental termo-electric power TEP of Bi2212 \cite{Obertelli}
and of Bi2201 \cite{McIntosh} are confronted with the
"panettone"-patch model and the 2-patch model. }
\label{transport-panettone}
\end{center}
\end{figure}

Here, however, we wish to add a further hypothesis: {\it we
introduce a temperature dependent "panettone", which presently
finds solely a "mathematical" justification in the way it is able
to reproduce the experimentally observed data}. We  therefore
naively assume that the parameter determining the spreading of
hot/cold intermediate region $w$ behaves like an effective
temperature:
\begin{equation}\label{w_o}
w(T)=\, w_{o} \, T
\end{equation}
We just indicate some observations:
\begin{itemize}
  \item the size of the cold patch has been heuristically linked to
  the doping in the system \cite{PSK}. The cold patch behaves in
  this sense as a pocket of fermionic particles
  (whose size and smoothness is dependent on the temperature).
  \item According to what observed above, the edge of a cold patch
  would be like a fermionic FS. This is right the shape we guessed
  for the panettone-patch edge, which decays at his border as fast
  as a Fermi function (which is another way to write $\tanh$).
  Such behavior in the intermediate hot/cold region is important
  not only to avoid a ill defined MR, but also to match
  quantitatively the order of magnitude of $\cot \theta_{H}$
  as compare to the resistivity $\rho_{xx}$.
\end{itemize}
The smaller the temperature $T$, the steeper the edge of the
panettone-patch, as with the edges of a Fermi distribution
function. $w_{o}$ enters like a Boltzmann's constant, and we treat
it as a fitting parameter. We can now re-do the same fitting
procedure on experimental data, as in \cite{PSK}. We do not aim
for a systematic comparison, nor look for the complete
explanation of the transport properties of cuprate materials. We
want only to verify if the introduction of this particular
T-dependent panrettone-patch, and therefore of a different
scattering matrix $C_{k k'}$ (formula \ref{C_kk1_Npatch}), may
improve the $2$-patch model of \cite{PSK}. In particular, we want
to see if it is possible to obtain a $T$-linear resistivity and,
at the same time, a pronounced curved $\coth \theta_{H}$ --the
main
problem in the $2$-patch model.  

The result is shown in \ref{transport-panettone}. The resistivity
$\rho_{xx}$ and the $\cot \theta_{H}$ vs temperature are shown in
the upper panels. As in \cite{PSK}, these 2 quantities have been
chosen to fit the parameters. We expect that the terms
$\nabla_{k} \phi_{k}$, which introduce new $T$-dependent
contributions, enter in the $\rho$ expression as second order
terms (formula \ref{11*}), while they should be more relevant in
determining the behaviour of $\cot \theta_{H}$, because the first
order terms are in this case zero. This is explicitly seen in
formula (\ref{11**}), where the terms $\nabla_{k} \phi_{k}$
appear in $\nabla_{k} \phi_{k}$. And in fact we succeed in
obtaining a result which reproduces incredibly well the $T^{1.7}$
curvature (red-dotted line) as compared to the $2$-patch model
(blue-dashed line) and follows very well experimental data, while
the resistivity keeps a good linearity. Having fit the parameters
with these two quantities, let's have a look to what happens to
other quantities, like for example the magneto-resistance (MR)
and the thermo-electric power (TEP) in the lower panels. We
observe that the MR shows a noticeable improvement too with the
panettone-patch (again red line) compared to the $2$-patch model
(blue-dashed line), either quantitatively and qualitatively,
being better portrayed the $\alpha\simeq -4$ exponent of the MR
$\sim T^{-\alpha}$ law. The TEP gives a quantitative result
comparable to \cite{PSK} and between experimental values. Notice
the correspondence with the TEP $\sim -T$ law. How much constant
lattice contribution to the thermal transport quantities may be
relevant is still an open debate. The {\it ad hoc} modifications
introduced seem therefore to generate the correct qualitative
behaviour for the transport properties of these materials. The
hypothesis was however introduced to fit experimental
observations and cannot be justified at this phenomenological
level: they can only be taken as steps in the right direction. To
make steps forward it is necessary to start our study from a
microscopic theory capable of deriving the cuprate properties,
e.g. the Hubbard model presented in the previous chapter, which
explains at least some of the fundamental properties of cuprate
material in the insulating state. For this, a method for
performing reliable calculations on strongly correlated many-body
systems is most required.

\chapter{Cellular Dynamical Mean Field Theory}
In the previous chapter we studied the normal state (NS)
properties of cuprate superconductors at optimal doping using a
macroscopic phenomenological Boltzmann theory. Given a few {\it ad
hoc} assumptions regarding the anisotropic scattering properties
of these systems on experimental observation, we constructed and
extended the $N$-patch solution to the Boltzman transport equation
found in \cite{PSK} and determined the temperature-dependence of
transport quantities. Both the qualitative and quantitative
agreement with experimental data was surprisingly satisfactory,
in spite of the simplicity of the Boltzmann quasiparticle
approach enriched by an anisotropic scattering rate and
parametrized by a only $N=2-5$ -patch model. However, to get
insight into the mechanism that arises in anomalous properties
from a simple FL already in its NS (as in going from high doping
to the Mott insulating state of cuprate systems), and to justify
the hypothesis introduced in the phenomenological Boltzmann
approach, we need a microscopic theory able to support and
explain the strong scattering anisotropy starting from a model
Hamiltonian, like the aforesaid Hubbard Model. For this a method
able to contend with strongly correlated many body electron system
is required. In recent years a great success was achieved trough
the use of Dynamical Mean Field Theory (DMFT) \cite{bibble}. This
method reduces the full lattice many-body problem to a local
impurity embedded in a self-consistent effective bath of free
electrons, which mimics the effect of the full lattice on the
local site. While in this way spatial correlations are
disregarded, time-dependent correlations are fully taken into
account by the dynamical bath. A self consistency condition links
the effective impurity model to the original lattice problem.
DMFT already proved its power giving the first unified scenario
of the Mott transition, where the high and low energy physics has
to be treated on equal footing trough a dynamical approach
\cite{bibble}. Despite DMFT being an exact theory only in the
limit of infinite dimension, it has proved an excellent
approximation to deal with real finite dimensional systems.
Combined with electronic structure method it was able for example
to explain transition metal oxides physics \cite{kotliar_P2day}.
The assumption of locality however, being the foundation on which
the mean field theory rests, may be too restrictive if applied to
systems where local correlations are fundamental in determining
their properties. This is the case for cuprate systems: if strong
momentum dependent properties are relevant, as seems to be the
case, they arise from non-local spatial correlations. Moreover
with a pure local theory we cannot describe
broken-symmetry-states where some spatial arrangement of the
order parameter is required, like for instance AF of $d$-wace SC.
In this chapter we review and develop an extension to DMFT which
is able to partially cure its spatial limitations, the Cellular
Dynamical Mean Field Theory (CDMFT). We replace the site-impurity
by a cluster of impurities embedded in a self-consistent bath.
Short-ranged spatial correlation are in this way treated exactly
inside the cluster, and a first momentum-dependence of the
properties of the system is recovered. In the next chapter we will
benchmark CDMFT with the exactly solvable one dimensional case,
which is the worst case for a mean field approach, and we will
discuss its possible implementation according to the shape and
size of the cluster. In chapters 5 and 6 we will apply it to the
experimentally relevant two dimensional case. A key-technique
used in our procedure the Exact Diagonalization (ED) algorithm
used to solve the impurity problem. ED allows access to zero
temperature physics and real frequency information, as well as a
wide range of model-parameters, as compared to other high
temperature methods (like for example Quantum Monte Carlo (QMC) ).

\section{Dynamical Mean Field Theory DMFT}

We briefly review Dynamical Mean Field Theory, as to discuss more
clearly in the next section the innovations of {\it Cluster}
Dynamical Mean Field Theory. For a full reference on this method
and a complete bibliography of technical implementations and
results we refer to \cite{bibble}. Here we wish to convey the
simple ideas underlying DMFT.
\begin{figure}[!htbp]
\begin{center}
\includegraphics[width=15cm,height=7cm] {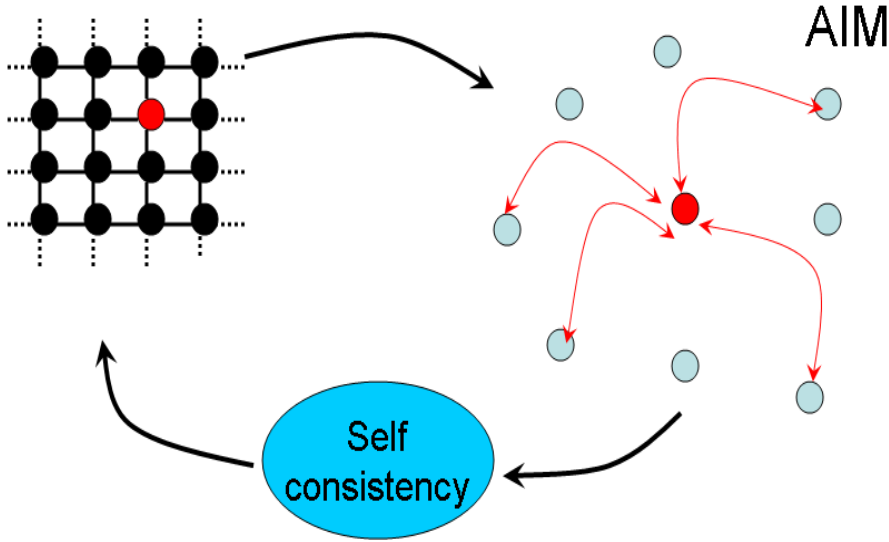}
\caption[]{DMFT loop} \label{fig3_DMFT}
\end{center}
\end{figure}
Accordingly, let us consider a general Hubbard problem on a
lattice of dimension $d$, described by the Hamiltonian:
\begin{equation}\label{Hubbard}
  H = -\sum_{ i,j,\sigma} t_{ij}\, (c^{\dagger}_{i\sigma} c_{j\sigma} +
h.c.) + U \sum_i n_{i\uparrow}n_{i\downarrow} -\mu\sum_i n_i
\end{equation}
where we indicate with $c_{j\sigma}$ the destruction operator of
a fermion with spin $\sigma= \pm \frac{1}{2}$ on site $j$ of the
$d$ dimensional lattice. The first term in the Eq. \ref{Hubbard}
describes a kinetic term where a particle can jump from site $i$
to site $j$ and viceversa with a probability measured by the
orbital overlapping integral parameter $t$. The second term acts
when two fermions of opposite spin occupy the same lattice
orbital $i$, and expresses mutual Coulomb repulsion of intensity
$U$. The two terms compete to determine the physics of the
system: the first would like to move electrons in order to lower
the energy, the second instead strongly un-favors the double
occupancy. As we explained in the first chapter, in spite of the
simplicity of this model it is extremely hard to extract reliable
information at finite dimension. Only in one dimension there
exists the exact {\it Bethe Ansatz} solution \cite{Lieb-Wu}. Now
let's see how DMFT approaches the Hubbard problem to give a
solution, which becomes exact in the opposite limit of
$d=\infty$). A schematic diagram is shown in Fig. \ref{fig3_DMFT}:
\begin{enumerate}
\item We do not attempt a solution of the full lattice problem and
focus instead on a single site (marked in red on the left-hand
side of Fig. (\ref{fig3_DMFT}).
\item We then
simulate the effect of all the other sites in the lattice by an
effective bath of free fermions (right hand side of Fig.
\ref{fig3_DMFT}). Differently from a normal static mean field
theory, the effective impurity model (Anderson impurity model AIM)
is able to mimic dynamical processes wherein particles jumping out
of a site and returning to it after some time.
\item the effective-bath construction has to be completed by imposing
a self-consistency condition which links the local Green's
function of the effective AIM to the one of the original lattice
Hubbard model.
\end{enumerate}
In mathematical language, this is well explained by the "cavity
construction" (see for example \cite{bibble}).
\begin{eqnarray} \label{Slatt}
\frac{ 1 }{ Z_{ef} }  \,
e^{-S_{eff}\,[c^{\dagger}_{o\sigma},c_{o\sigma}]} \equiv \,
\frac{ 1 }{ Z } \, \int  \prod_{(i,j,\sigma) \ne (o,o) } \, D
c^{\dagger}_{i\sigma} \, D c_{j\sigma} & e^{-S}
\end{eqnarray}
Given the imaginary-time the action $S[c^{\dagger},c]$, we
construct an effective action $S_{ef}[c_{o}^{\dagger},c_{o}]$
integrating out all the sites $(i,j) \neq o=(i_{o},j_{o})$, where
$o$ is the site chosen as impurity (Eq. \ref{Slatt}).
$S_{ef}[c_{o}^{\dagger},c_{o}]$ can be written in the quadratic
form:
\begin{eqnarray} \label{Seff}
S_{ef}= \, \int_{0}^{\beta} \, d\tau   \, \sum_{\sigma} \, c_{o
\sigma}^{\dagger}(\tau)\, {\cal G}_{\sigma}^{-1}(\tau)\,
{c_{o\sigma}}+ \, U \int_{0}^{\beta} \,d\tau  \,
n_{o\uparrow}(\tau) n_{o\downarrow}(\tau)
\end{eqnarray}
where ${\cal G}_{o}(\tau)$ plays the role of an effective Weiss
field. It expresses the probability of destroying a particle on
the impurity site $o$ at time $\tau=0$ and recreating it on the
same impurity site $o$ after an imaginary time $\tau$ elapsed.
The primary difference with a static mean field is that now ${\cal
G}_{o}(\tau)$ is a time-dependent Green's function which fully
takes into account the local temporal dynamics despite the
truncation of spatial degrees of freedom. It represents the bare
Green's function of the effective action, but it should not be
confused with the bare Green's function of the original lattice
model. Having the effective action at hand $S_{ef}$, we can now
calculate the local Green's function of the effective impurity
model (in a following section we will go over this point in more
detail):
\begin{eqnarray} \label{Gimp}
G_{imp}(\tau)=\, -T_{\tau} \, \langle \,c^{\dagger}_{o}(\tau) \,
c_{o} \, \rangle_{S_{ef}}
\end{eqnarray}
and through Dyson's equation:
\begin{eqnarray} \label{Dyson}
\Sigma_{\mu\nu}(\imath\omega_{n})= {\cal
G}_{\sigma}^{-1}(\imath\omega)- G_{\sigma}^{^{imp}-1}(\imath
\omega)
\end{eqnarray}
which allows to extract the local self-energy $\Sigma$, evaluate
the original lattice local Green's function:
\begin{eqnarray} \label{G_k}
G_{}(k,\imath \omega)= \,  \frac{1}{\imath \omega + \mu -
\varepsilon_{k} - \Sigma}
\end{eqnarray}
and obtain once again the local Green's function summing all over
the BZ:
\begin{eqnarray} \label{Gloc}
G_{loc}(\imath \omega)= \, \sum_{k} \frac{1}{\imath \omega + \mu
- \varepsilon_{k} - \Sigma}
\end{eqnarray}
Here we have expressed the green's function on the Matsubara
axis\\
$G(\imath \omega)= \frac{1}{2}\,\int_{0}^{\beta}\, e^{\imath
\omega \tau} \, G(\tau) $. The self-consistency conditions
requires the two local Green's functions (\ref{Gimp}) and
(\ref{Gloc}) to be the same:
\begin{eqnarray} \label{self-consistency}
G_{imp} \equiv G_{loc}
\end{eqnarray}
or equivalently
\begin{eqnarray} \label{self-consistency0}
\Sigma(\imath\omega)= {\cal G}_{\sigma}^{-1}(\imath\omega_{n})-
G_{\sigma}^{^{loc}-1}(\imath \omega, \Sigma )
\end{eqnarray}
Behind equations \ref{Seff} and \ref{Gloc} lies the foundational
assumption of DMFT. It is clear that we assume
\begin{eqnarray} \label{self_loc}
\Sigma(\imath \omega) \neq \, \Sigma(\mathbf{k},\imath \omega)
\end{eqnarray}
the self-energy is a purely local quantity. This is also the
assumption required for the gaussian form of $S_{ef}$ in Eq.
\ref{Seff} to be exact and it is true in the limit
$\mathbf{d\rightarrow \infty}$, where the DMFT solution is exact.
One can get an idea of this through the cavity construction.
Writing the quantum action:
\begin{eqnarray*}
S=\, S_{o}+ \Delta S+ S^{o}
\end{eqnarray*}
where $S^{o}$ contains the lattice sites except site $o$, $S_{o}$
only site $o$, $\Delta S$ contains terms connecting site $o$ with
the rest of lattice
\begin{eqnarray*}
S_{o}= \, \int_{0}^{\beta} \, d\tau   \, \sum_{\sigma} \, c_{o
\sigma}^{\dagger}(\tau)\, \left(\partial_{\tau} - \mu  \right)\,
{c_{o\sigma}}+ \, U \int_{0}^{\beta} \,d\tau  \,
n_{o\uparrow}(\tau) n_{o\downarrow}(\tau) \nonumber \nonumber
\end{eqnarray*}
\begin{eqnarray*}
\Delta S_{}= \, -\int_{0}^{\beta} \, d\tau   \, \sum_{i\sigma} \,
c_{o \sigma}^{\dagger}(\tau)\,{c_{i\sigma}}+
c_{i\sigma}^{\dagger}(\tau)\,{c_{o\sigma}}  \nonumber
\end{eqnarray*}
We have
\begin{eqnarray*} \label{dummy1}
\frac{ 1 }{ Z_{ef} }  \,
e^{-S_{eff}\,[c^{\dagger}_{o\mu\sigma},c_{o\nu\sigma}]} \sim \,
\, \int  \prod_{(i,j,\sigma) \ne (o,o) } \, D
c^{\dagger}_{i\sigma} \, D c_{j\sigma} & e^{-\Delta S} \,
e^{S^{o}} \ \nonumber
\end{eqnarray*}
i.e. the expectation value $\langle\, e^{-\Delta S}\,
\rangle_{S^{o}}$ calculated in the system with a "cavity" in
place of site $o$. The effective action can therefore be written:
\begin{eqnarray*} \label{dummy2}
S_{ef}\sim\, \sum_{n=1}^{\infty}\, \sum_{i_{1}...j_{n}}\, t^{2
n}\, \int\,
c^{\dagger}_{i_{1}}(\tau_{i_{1}})...c^{\dagger}_{i_{n}}(\tau_{i_{n}})
\,
G^{o}_{i_{1}...j_{n}}(\tau_{i_{1}}...\tau_{i_{n}},\tau_{j_{1}}...\tau_{j_{n}})
\, c^{}_{j_{1}}(\tau_{j_{1}})...c^{}_{j_{n}}(\tau_{j_{n}})
\end{eqnarray*}
Now in growing the dimension of the system, conservation of
energy implies that the hopping terms scale like $t_{ij}\sim
\sqrt{1/d^{|i-j|}}$ (the space traveled by the particle grows
like a random walk), so that in the $d\rightarrow \infty$ limit
only the nearest neighbor $n=1$ term survives. This implies a
local self-energy (Eq. \ref{Dyson}).


\section{The Cluster Dynamical Mean Field Theory CDMFT}

While DMFT has proved remarkably successful in describing problems
whose physics is mainly local \cite{bibble}, having provided for
example the first unified scenario of the Mott metal-insulator
transition, it is not very suitable when the hypothesis of pure
locality (Eq. \ref{self_loc}) does not describe the physical
properties of the system well. A natural extension of DMFT, able
to introduce short range spatial correlation, is the Cluster
Dynamical Mean Field Theory (CDMFT). The idea is very simple and
preserves the spirit of the DMFT.
\begin{figure}[!htb]
\begin{center}
\includegraphics[width=13cm,height=10cm] {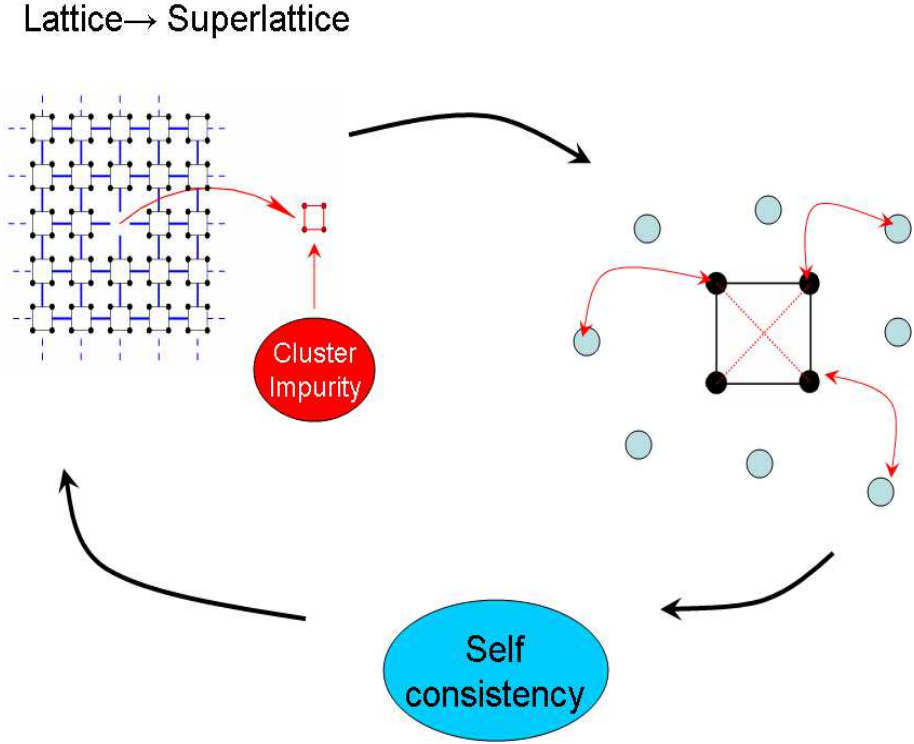}
\caption[]{CDMFT loop} \label{fig_CDMFT}
\end{center}
\end{figure}
Instead of considering a single site as impurity, we consider a
cluster of sites and, as in DMFT, we construct an effective
action by integrating out all the other degrees of freedom on the
lattice (see Fig. \ref{fig_CDMFT}):
\begin{eqnarray} \label{Slatt_cluster}
\frac{ 1 }{ Z_{ef} }  \,
e^{-S_{eff}\,[c^{\dagger}_{\mu\sigma},c_{\nu\sigma}]} \equiv \,
\frac{ 1 }{ Z } \, \int  \prod_{(i,j,\sigma) \ne (\mu,\nu) } \, D
c^{\dagger}_{i\sigma} \, D c_{j\sigma} & e^{-S} \ \nonumber
\end{eqnarray}
where with $(\mu,\nu)$ we indicate the coordinates inside the
cluster-impurity. We obtain cluster-impurity problem embedded in
a bath of free fermions, which again we write in a quadtratic
form:
\begin{eqnarray} \label{Seff_cluster}
S_{ef}= \, \int_{0}^{\beta} \, d\tau   \, \sum_{\mu\nu\sigma} \,
c_{\mu \sigma}^{\dagger}(\tau)\, {\cal
G}_{\mu\nu\sigma}^{-1}(\tau)\, {c_{\nu\sigma}}+ \, U
\int_{0}^{\beta} \,d\tau  \sum_{\mu}\,\, n_{\mu\uparrow}(\tau)
n_{o\mu\downarrow}(\tau) \nonumber \nonumber
\end{eqnarray}
The difference with respect to DMFT is that now the Weiss field
${\cal G}_{\mu\nu\sigma}(\tau)$ is a NxN matrix expressing all the
possible site-relations inside an impurity cluster of N$^{2}$
sites. At this point we once again impose the self-consistency
conditions requiring that the local cluster Green's functions is
equal to the cluster Green's functions of the original model.
"Local" this time means "inside the cluster", so that again we
have NxN Green's function relations:
\begin{eqnarray} \label{Gimp_cluster}
G^{}_{\mu\nu\sigma}(\tau)=\, -T_{\tau} \, \langle
\,c^{\dagger}_{\mu\sigma}(\tau) \, c_{\nu\sigma} \,
\rangle_{S_{ef}}
\end{eqnarray}
But how do we evaluate the lattice Greens's function having now
not only the local self-energy $\Sigma$ but a set of
cluster-self-energies $\Sigma_{\mu\nu}$? Here lies the central
idea in CDMFT. We apply the same concept of DMFT to the
cluster-impurity, considered as a site of a "super-lattice" whose
sites are in their turn cluster of sites of the original lattice
(top left hand-side of Fig.\ref {fig_CDMFT}). We construct a
super-lattice Green's function assuming the cluster-self energy
$\Sigma_{\mu\nu}$ local as in DMFT (but this time the term local
covers a range of the size of the cluster-impurity):
\begin{eqnarray} \label{Gk_superlatt}
\hat{G}(K,\imath\omega)= \, \frac{1}{(\imath\omega+\mu)\,\hat{1}+
\hat{t}_{K}- \hat{\Sigma} }
\end{eqnarray}
The Fourier space of the superlattice $K$ is defined in a
Brillouin Zone (RBZ), reduced by the partition of the original
lattice into N-site-clusters. Here the NxN relations inside the
cluster-impurity are expressed in a matrix notation, $\hat{1}$ is
the NxN identity matrix, $\hat{t}_{K}$ the intra-cluster hopping
matrix, and it expresses the band-dispersion in the superlattice,
like $\varepsilon_{k}$ was the band dispersion in the momentum
space of the original lattice in DMFT (expression \ref{Gloc}). As
usual, the local cluster-Green's function is the sum in the RBZ
of the super-lattice Green's function:
\begin{eqnarray} \label{Gloc_super}
 \hat{\mathbf{G}}_{loc}(\imath\omega_{n})=\, \sum_{K}\, \mathbf{\hat{G}}(K,\imath\omega)
 =\,\int_{RBZ} \,
{1 \over (\imath\omega_{n}+\,\mu) \, \mathbf{1}-\,
\hat{\mathbf{t}}_{K} -\, \hat{\mathbf{\Sigma}}(\imath\omega_{n})}
\ {dK^{d} \over (2\pi)^{d}/N_{c}}
\end{eqnarray}
and the we obtain a set of closed equation imposing this local
cluster-Green's function $ \hat{\mathbf{G}}_{loc}$ equal to the
one obtained from the effective quantum action:
\begin{eqnarray} \label{self-consistency-super}
\mathbf{\hat{G}}_{} \equiv \, \mathbf{\hat{G}}_{loc}
\end{eqnarray}
or equivalently
\begin{eqnarray} \label{self-consistency0-super}
\Sigma_{\mu\nu}(\imath\omega)= {\cal
G}_{\mu\nu\sigma}^{-1}(\imath\omega_{n})-
G_{\mu\nu\sigma}^{^{loc}-1}(\imath \omega, \hat{\Sigma} )
\end{eqnarray}
CDMFT allows for consideration of the correlation between
particles up to a distance of the order of the cluster-size. The
advantage is clear, as we can now cope with problems whose
relevant physical properties are not purely local. These are
typically encountered in real finite size systems. Taking the
limit $d\rightarrow \infty$ would be in this case not very
efficient, as $\Sigma_{ij}\rightarrow 0$ for every $i\neq j$,
i.e. CDMFT $\rightarrow$ DMFT. But in using CDMFT on finite size
systems, it is also clear that the gaussian representation of the
effective action $S_{ef}$ is no longer exact, unlike DMFT, and we
are using an approximation. The quality of the approximation is
dependent on the cluster-size as compared to how much local is the
physics of the problem specifically considered. We will make
clear examples and applications in the following chapter.

\section{Exact Diagonalization Method}
We now face the problem of solving the effective associated
impurity problem. Fortunately, as the Anderson impurity problem
has been studied for over 35 years, there exist various methods
well-established in literature, either analytical (perturbation
theory methods, projective methods \cite{George-Kotliar92}) and
numerical (Quantum Monte Carlo \cite{Hyrsch-Fye86} and Exact
Diagonalization \cite{Caffarel94}\cite{Rozenberg94}\cite{Si94}).
In this work we use the Exact Diagonalizarion (ED) method on
small (up to 12 sites) cluster. The advantage of ED is the high
precision in determining  the zero-temperature
groundstate-properties and the access to real frequency
information, as well as the ability to treat the large-U regime,
which is hardly accessible by Quantum Monte Carlo (QMC).
\begin{figure}[!hp]
\begin{center}
\includegraphics[angle=90,width=13cm,height=18cm] {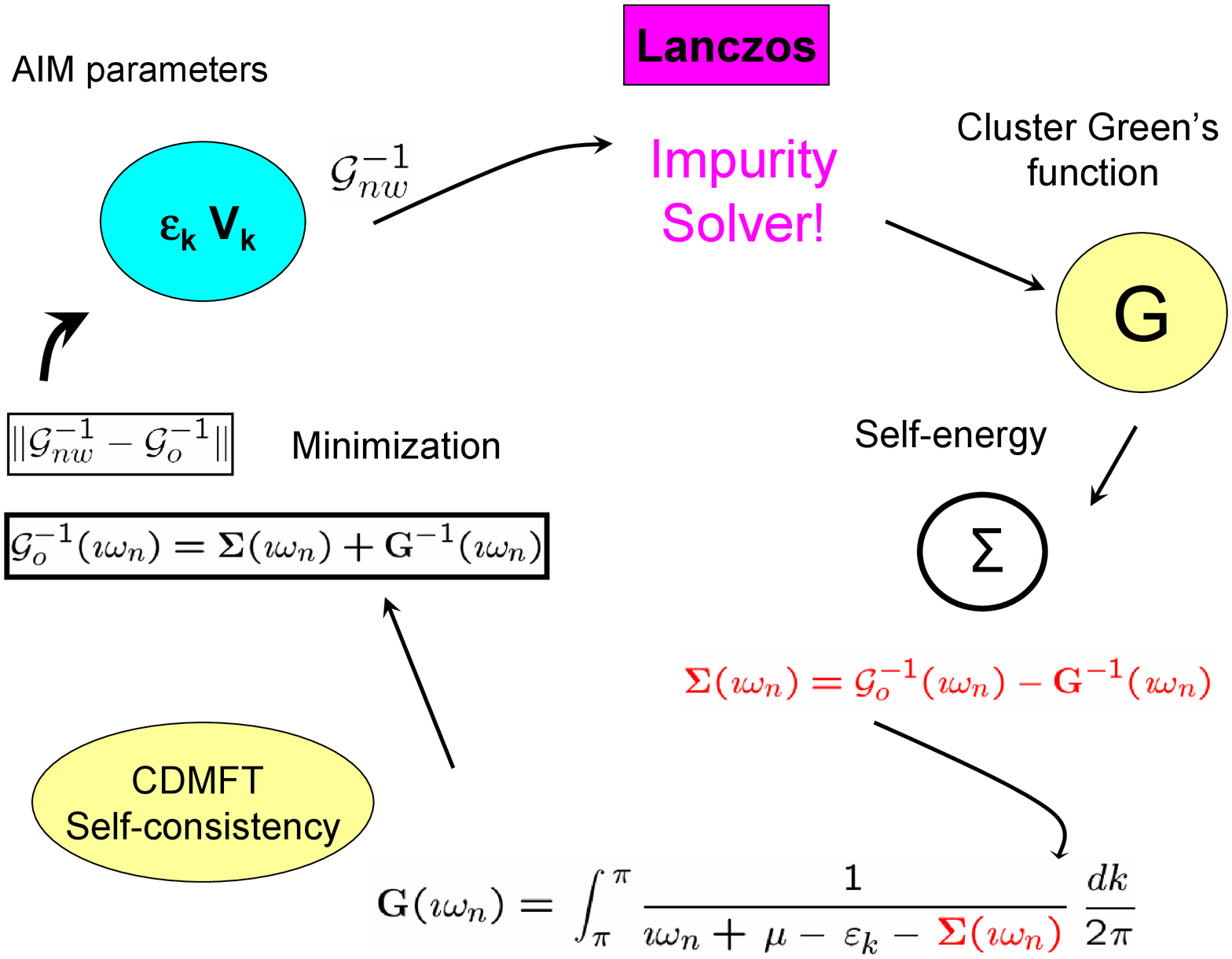}
\caption[]{ED loop} \label{ED_loop}
\end{center}
\end{figure}
\subsection{Truncation of the problem size} In practical
implementations the quadratic effective action $S_{ef}$ is
derived from an Anderson Impurity Hamiltonian ${\cal H_{AI}}$. In
CDMFT the site-impurity is replaced by a cluster-impurity:
\begin{eqnarray} \label{HAIM}
&&\mathbf{\cal{H}_{AI}}={\sum_{\mu \nu}^{n_{c}} \sum_{\sigma }\
{E_{\mu \nu \sigma }} \ {c_{\mu \sigma }^{+} \, c_{\nu \sigma } }
\, + {U} }\, {\sum_{\mu }^{n_{c}}\ \ {c_{\mu \uparrow }^{+} \,
c_{\mu \uparrow } \,
c_{\mu \downarrow }^{+} \, c_{\mu \downarrow }   } } \\
&&{\sum_{k}^{n_{b}}\sum_{\sigma }\ \varepsilon \,_{k\sigma }\
a_{k\sigma }^{+}a_{k\sigma }+\sum_{k}^{n_{b}}\sum_{\mu \sigma }\
V_{k\mu \sigma }\ a_{k\sigma }^{+}  {c_{\mu \sigma } +V_{k\mu
\sigma }^{\ast }\ {c_{\mu \sigma
}^{+}} a_{k\sigma } \, } }\\
\end{eqnarray}
Here we have introduced the fermionic operators $c_{\mu\sigma}$
which destroy an electron with spin $\sigma$ on site $\mu$ of the
cluster-impurity, and $a_{k\sigma}$ which destroys an electron on
site $k$ of the free electron bath. The matrix $E_{\mu\nu\sigma}$
contains the original model hopping parameters (for example for
the Hamiltonian \ref{Hubbard} it the chemical potential $\mu$ on
the diagonals and the hopping parameter $-t$ in nearest neighbor
sites $|\mu-\nu|=1$). The Hamiltonian describes the
cluster-impurity (labels $\mu,\nu$) embedded in a self-consistent
free electron bath with dispersion $\varepsilon_{k}$ which is
hybridized through the hopping parameters $V_{k\mu \sigma }$. The
$(\varepsilon_{k},V_{k\mu \sigma })$ are the variables to be
determined by the self-consistency condition. The primary
simplification of the ED-method lies in considering a finite-site
($n_{s}$) AIM, where the bath is truncated to a finite number of
orbitals $n_{b}$ (in practice the total number of sites handled
is $n_{s} \simeq 12$). The problem can be thereby solved using
standard {\it Lanczos} algorithms (see APPENDIX B for more
detail). We stress here that a truncation in the effective
Anderson impurity problem does not imply a truncation in the
original lattice problem, which fully retains its thermodynamical
limit. The free-electron bath determines the cluster-impurity
Weiss field,
\begin{eqnarray*}
{\cal G}^{\mu\nu}_{o}(\imath \omega_{n})^{-1}=\, \imath \omega_{n}
\, \delta_{\mu\nu}+ E_{\mu\nu}- \, \sum_{k}^{\infty}
\frac{V^{*}_{\mu k}V_{k \nu}}{\imath \omega_{n}- \varepsilon_{k}}
\end{eqnarray*}
which within this procedure is therefore truncated in a
finite-pole expansion:
\begin{eqnarray} \label{G0_ns}
{\cal G}^{\mu\nu}_{o_{nb}}(\imath \omega_{n})^{-1}=\, \imath
\omega_{n} \, \delta_{\mu\nu}+ E_{\mu\nu}- \, \sum_{k}^{n_{b}}
\frac{V^{*}_{\mu k}V_{k \nu}}{\imath \omega_{n}- \varepsilon_{k}}
\end{eqnarray} 
The AIM is then solved self-consistently, using the Dyson's
equation \ref{self-consistency0-super}.

\subsection{The ED-loop for CDMFT}
 The ED-loop consists of the following steps (see Fig. \ref{ED_loop}):
\begin{enumerate}
    \item The truncated impurity problem is solved with a {\it Lanczos} procedure,
    which accurately determines the ground state $ | gs \rangle$ of the associated
    AI Hamiltonian ${\cal H}_{AI}$. This is done in the usual way, i.e. by picking a random
    wave-function $| \varphi \rangle$ in an appropriate subspace of the full Hilbert space,
    according to the symmetries of ${\cal H}_{AI}(\varepsilon_{k}, V_{k})$, and
    diagonalizing ${\cal H}_{AI}$ in a
    linear hull of $ |\varphi \rangle$, $ {\cal H}_{AI}\, |\varphi
    \rangle$...$ {\cal H}^{n}_{AI}\, |\varphi \rangle$ , 
    The local zero-temperature cluster-impurity Green's $G_{\mu\nu}$ is
    obtained choosing as initial
    vector $| \varphi \rangle=\, c^{\dagger}\, | gs \rangle$ and implementing a
    second Lanczos procedure, which takes advantage of the continued-fraction representation
    of a Green's function describing the "particle" ($\omega >0$) and "hole" excitations \cite{bibble}:
\begin{eqnarray*}
G_{\mu\nu}(\omega)=\, G_{\mu\nu}^{>}(\omega)+
G_{\mu\nu}^{<}(\omega)
\end{eqnarray*}
with
\begin{eqnarray*}
G^{>}_{\mu\nu}(\omega)= \frac{\langle \,gs\, | \,c_{\mu\nu}\,
c_{\mu\nu}^{\dagger}\, | \,gs\, \rangle } {\omega- a_{0}^{>}-
\frac{b_{1}^{>2}} {\omega- a_{1}^{>}- \frac{b_{2}^{>2}} {\omega-
a_{2}^{>}- ...}}}
\end{eqnarray*}
\begin{eqnarray*}
G^{<}_{\mu\nu}(\omega)= \frac{\langle \,gs\, |
\,c_{\mu\nu}^{\dagger}\,c_{\mu\nu}\, | \,gs\, \rangle } {\omega-
a_{0}^{<}- \frac{b_{1}^{<2}} {\omega- a_{1}^{<}- \frac{b_{2}^{<2}}
{\omega- a_{2}^{<}- ...}}}
\end{eqnarray*}
The $a$ and $b$ parameters entering this expression are directly
determined in the diagonalization procedure of ${\cal H}_{AI}$. We
notice that the same equations can be used to extract the Green's
function on the real axis as well as on the Matsubara axis by
simply using analytical continuation $\omega_{n}\rightarrow
\omega+ \imath \delta$, where $\delta$ is a small number giving a
spreading around the real-axis pole. This is an advantage over
other techniques, in the QMC, for example, the analytic
continuation procedure is non-trivial.
    \item Using the self-consistency condition \ref{self-consistency0} to extract first
    $\Sigma$, we can
    close the DMFT equations determining the local cluster-Green's
    function:
\begin{eqnarray}
 \hat{G}(\omega,\hat{\Sigma})=\, \sum_{k\in\, RBZ} \, \frac{1}{\hat{1}\,
 (\omega+\mu)+\hat{t}_{k}-\hat{\Sigma}}
 \label{Gloc1} \nonumber
\end{eqnarray}
    \item We then use again the self consistency condition \ref{self-consistency0} to
    determine a new Weiss function:
\begin{eqnarray*}
{\cal G}^{\mu\nu}_{o_{new}}(\omega)^{-1}=\,
\Sigma_{\mu\nu}(\omega)+ \, G_{\mu\nu}(\omega)^{-1}
\end{eqnarray*}
     \item The final step, which determines the new bath
     parameters $\varepsilon_{k}$ and $V_{k}$ in the ${\cal
     H}_{AI}$, is the more subtle one. In the CDMFT-Lanczos-procedure the problem consists
     of
     finding the best set of $\varepsilon_{k}$ and $V_{k}$
     determining the new finite-pole expansion of the Weiss
     function
     ${\cal G}^{\mu\nu}_{o_{nb}}(\omega)$, which
     better describes the new ${\cal G}^{\mu\nu}_{o_{new}}(\omega)$
     coming from the CDMFT-self-consistency. This is equivalent to
     projecting ${\cal G}^{\mu\nu}_{o_{new}}(\omega)$ onto a space
     of functions $\{\, {\cal G}^{\mu\nu}_{o_{nb}}  \,\}$ built
     from a finite set of orbitals $n_{b}$. For this there is not one unique
     procedure, and different methods
     have been proposed (for a review \cite{bibble}). In this work we follow
     the method developed in the context of single site ED-DMFT by
     Caffarel and Krauth \cite{Caffarel94}, which defines a
     distance-function $f$
     $f=\, \| \, {\cal G}^{\mu\nu}_{o_{nb}}(\imath \omega_{n})^{-1}-
     {\cal G}^{\mu\nu}_{o_{new}}(\imath \omega_{n})^{-1}\, \| $
     and a {\it minimization} procedure to determine the set
     of bath-parameters which minimize this distance. We notice
     that:
     \begin{itemize}
        \item the definition of $f$ is arbitrary, being
        difficult to define a unique criterion of distance between two
        functions. We will follow physical intuition
        in defining a $f$, verifying {\it a posteriori} the
        results. We will adopt in this work mainly two
        distancefunctions $f$:
\begin{eqnarray} \label{distf1}
f=\, \sum_{n}^{n_{\tiny \hbox{max}}}\, \frac{1}{|\, \omega_{n}
\,|} \quad |\, {\cal G}^{\mu\nu}_{o_{nb}}(\imath \omega_{n})^{-1}-
     {\cal G}^{\mu\nu}_{o_{new}}(\imath \omega_{n})^{-1}   \,|
\end{eqnarray}
\begin{eqnarray} \label{distf2}
f=\, \sum_{n}^{n_{off}}\, \quad |\, {\cal
G}^{\mu\nu}_{o_{nb}}(\imath \omega_{n})^{-1}-
     {\cal G}^{\mu\nu}_{o_{new}}(\imath \omega_{n})^{-1}   \,|
\end{eqnarray}
    The distance $f$ is defined on the
    Matsubara axis where the
    functions have no pole and are smoothly behaved. In definition
    \ref{distf1} $ n_{\tiny
    \hbox{max}}$ is an upper energy-cutoff
    which includes the typical
    maximum scale of the problem (in a Hubbard problem, for example,
    some factor of the on-site interaction $U$). The $\frac{1}{|\,
    \omega_{n} \,|}$ is chosen with the task to weight more the
    low-energy scale, which is important in capturing the low-energy physics of
    the Mott metal-insulator transition. The success of this choice in describing the MT is
    showed in a one-dimensional study of the Hubbard Model (next chapter and \cite{marce}).
    In definition \ref{distf2}, $n_{off}$ is a low energy
    cut-off of the order of the bandwidth $\sim t$.
    This definition is more suitable to describe the low energy
    physics of the doped systems not very close to the MT, where
    the transfer of spectral weight from the low-energy peak to the
    Hubbard Bands is still a negligible effect.
        \item Introducing the distance $f$ on the Matstubara
        axis requires defining a fictitious temperature,
        which determines a low energy cutoff. This temperature acts
        like a virtual temperature in the system: despite the Green's
        function being evaluated on the ground
        state of the associated AIM, the self-consistency is implemented
        at a virtual temperature $1/\beta$ determined by the grid of points it
        selects on the Matsubara axis. In the following chapter we will
        show examples on how this effects the results in
        connection with QMC-impurity-problem studies at high temperature.
     \end{itemize}
\end{enumerate}

\chapter{Implementation of CDMFT}
\section{Benchmark on the 1D Hubbard Model}
In order to elucidate some of aspects of the ED-CDMFT procedure we
present in this section a comparison with the exact Bethe Ansatz
(BA) solution for the one dimensional Hubbard Model
\cite{Lieb-Wu}. This is the worst case scenario for a mean field
theory: the doped system is a Luttinger Liquid which presents a
non-analytic behavior in the low frequency part of the Green's
functions . Nevertheless we can study the metal-insulator
transition focusing the attention on thermodynamical quantities,
such as for example the density in the system or the charge
compressibility. We first present the result leaving for the
following subsections practical examples on the CDMFT
implementation.
\subsection{Improvement of CDMFT on DMFT}
\begin{figure}[!htbp]
\begin{center}
\includegraphics[width=10cm,height=8cm,angle=-0]
{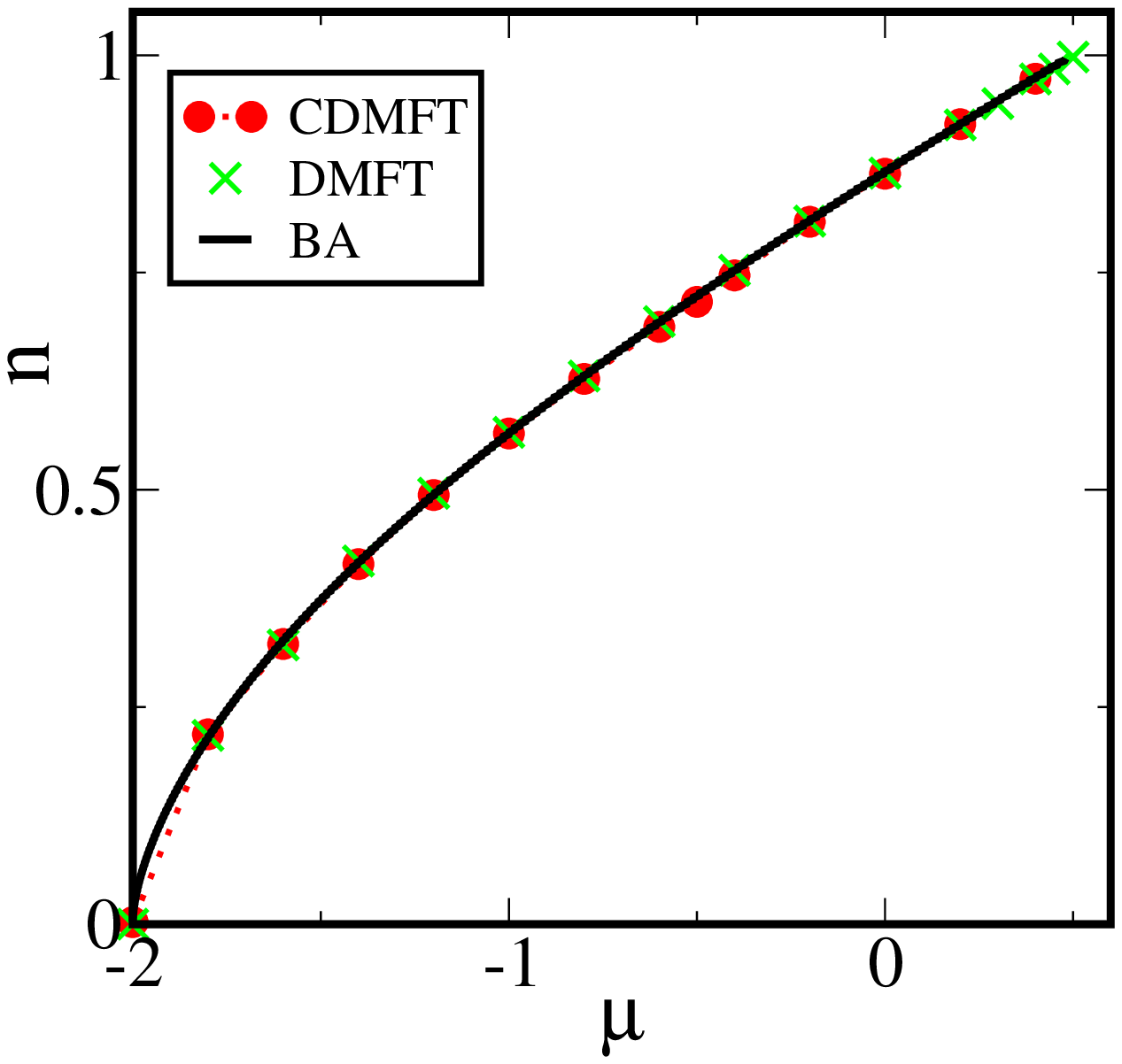}\vspace{1cm}\\
\includegraphics[width=10cm,height=8cm,angle=-0] {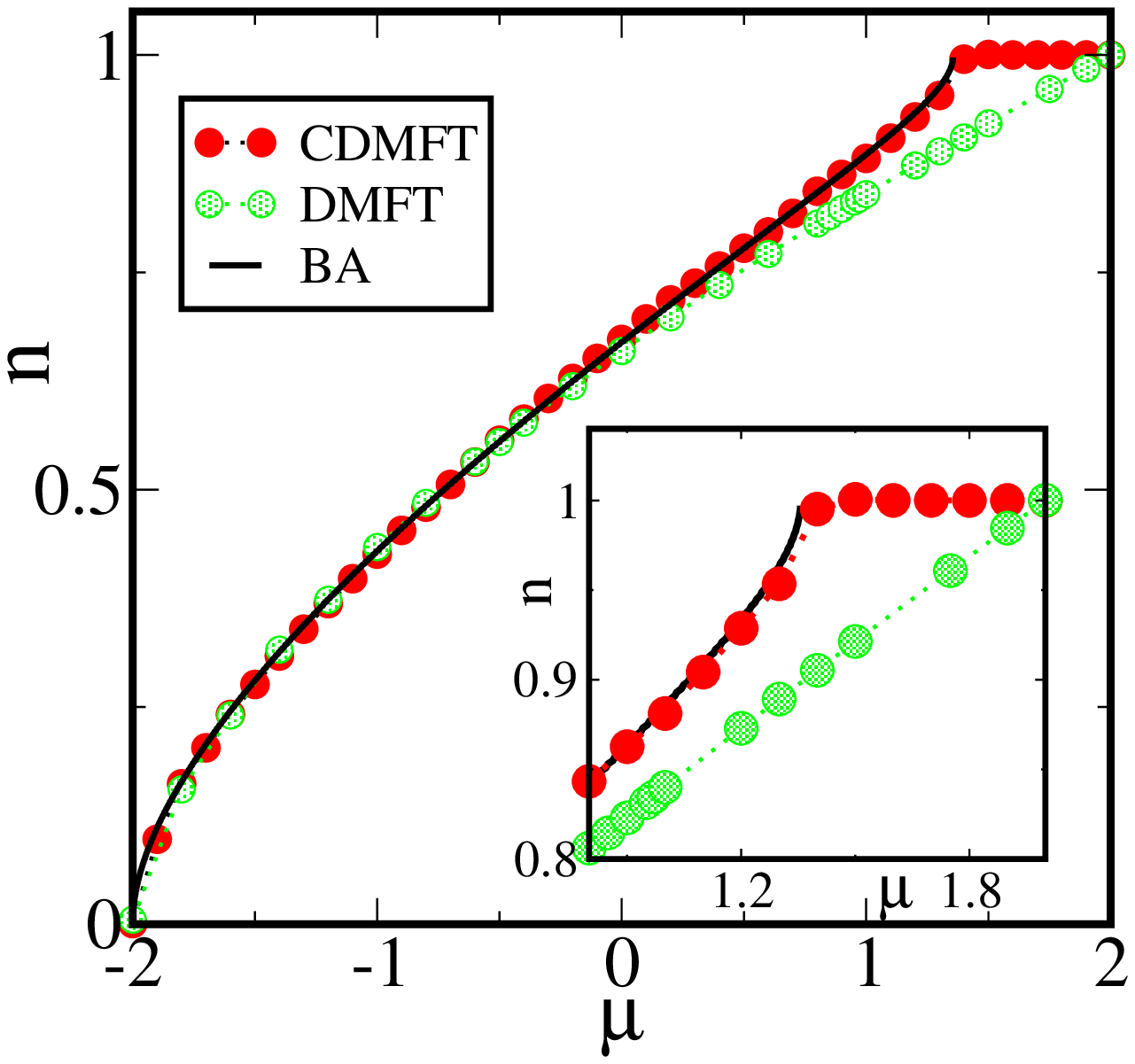}
\end{center}
\caption{Density $n$ vs. chemical potential $\mu$ in the 1D
Hubbard Model. The CDMFT method is compared with the single-site
DMFT and the exact Bethe Ansatz (BA) result. Upper panel: $U/t=
1.0$, $N_{c}=2$, $N_{b}=8$. Lower panel: $U/t= 4.0$, $N_{c}=2$,
$N_{b}=8$. The inverse virtual temperature is
$\beta=300$.}\label{mudens1D}
\end{figure}
\begin{figure}[!bth]
\begin{center}
\includegraphics[width=8cm,height=10cm,angle=-0] {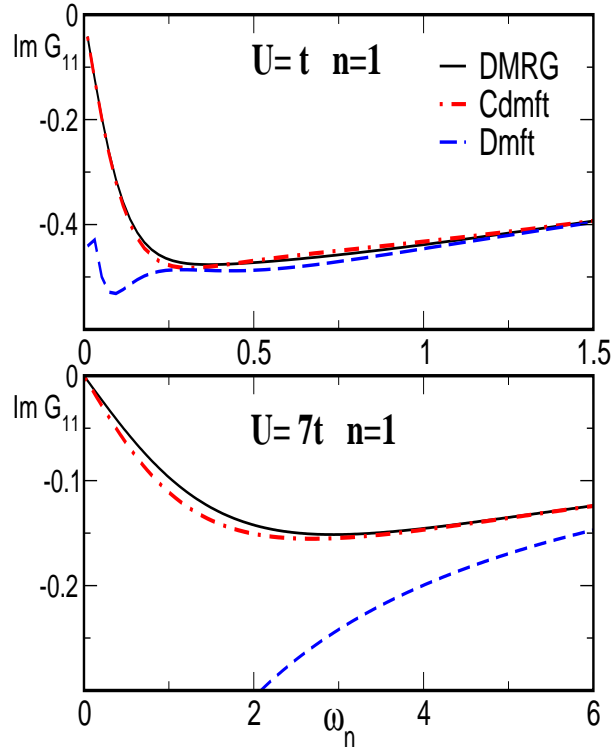}
\end{center}
\caption{Imaginary part of the local cluster Green's function at
half-filling. The CDMFT result is compared with the previous
single-site DMFT. DMRG calculation, provided in \cite{venky}, is
used as bench-mark. The upper panel shows the weakly interacting
case $U/t=1$, the lower panel the strongly interacting $U/t=7$.
$N_{c}=2$, $N_{b}=8$ and $\beta=300$. } \label{ImG11_1D}
\end{figure}

In Fig. \ref{mudens1D}, we display a plot of the density of the
system $n$ as a function of the chemical potential $\mu$ for the
Hubbard Model with only two sites in the cluster $N_{c}=2$ and
$N_{b}=8$ sites in the bath. Two case are shown: the weakly
interacting $U= t $ (upper panel) and the intermediate $U= 4.0t$
(lower panel). We compare the result of CDMFT with the
single-site DMFT and benchmark it with the exact Bethe Ansatz
(BA). For $U= t$ we observe that both CDMFT and DMFT are able to
well portray the exact density. All the CDMFT-circles and the
DMFT-crosses lay on the continuous BA-curve, up to the
half-filled state at density $n=1$. We will see later (Fig.
\ref{ImG11_1D}), however, that, contrary to CDMFT, DMFT is giving
a wrong insulating state even if the density is correctly
described. The differences between CDMFT and DMFT are seen
sharply in the intermediate coupling case $U= 4t$, where the
value of interaction is equal to the bandwidth, and the
interacting and kinetic parts of the Hubbard Hamiltonian are in
competition. At high doping (up to $n\sim 0.75$) CDMFT and DMFT
once again perform equally well in describing the correct density
of the system. The situation is clearly different near the
half-filled insulating state. The insulating phase, in fact,
persists around the particle-hole symmetric point $\mu= U/2 $,
where the $n$ vs. $\mu$ curve shows a plateau of values, until
sharply going into the metal phase at some critical $\mu$. The
charge compressibility is at this point divergent, as is shown by
the continuous black BA line. This behavior is difficult to
capture in a mean field approach. Single-site DMFT (green dotted
line) fails in fact in describing the MT. Cluster-DMFT instead,
with only two sites $N_{c}=2$ in the impurity-cluster, is able to
follow very successfully the BA line (as evidenced in the inset),
following then the half-filled insulating solution at constant
density $n=1.0$.

These considerations are also supported by the analysis of
dynamical quantities. As in Ref. \cite{venky}, we consider the
imaginary part of the on-site Green's function $G_{11}$ on the
Matsubara axis, which is plotted for $U/t=$1,7 and for $n=1$
(Fig. \ref{ImG11_1D}). We display CDMFT and single-site DMFT,
again for $N_c=2$ and $N_b=8$, compared with the results of
Density-Matrix Renormalization Group (DMRG), a numerical approach
which is known to provide essentially exact results for
one-dimensional systems (for more details on our calculation of
dynamical properties with DMRG, see Ref. \cite{venky}). The
agreement of CDMFT with the virtually exact DMRG results is
extremely good for $U/t=1$. Interestingly, the single-site DMFT
completely fails in the description of dynamical  properties even
if the $n$-$\mu$ curve shown in Fig. \ref{mudens1D} is close to
the exact solution. In the strong coupling case $U/t= 7$, CDMFT
closely follows the DMRG result while single-site DMFT totally
fails.

This examples show:
\begin{enumerate}
    \item how important it is to consider short-ranged local interaction in finite dimensional systems
    to describe the peculiarity of the the MT.
    \item the importance of correctly weighting the low
    energy scale as compared with the high energy scale of the Hubbard bands. This result has been obtained
    in fact using the low-energy weighted distance introduced in the previous section which
    gives importance to the low energy physics at the critical point of the transition.
\end{enumerate}
But how precise is the result and how much depends on the number
of sites used to solve the associated AIM?
\subsection{Implementation of CDMFT}
We answer the question at the end of the previous subsection by
explicitly showing how to implement the Lanczos-CDMFT in an
efficient way. We offer a clever way of exploiting symmetries
(particularly in one dimension) in order to reduce the number of
the degrees of freedom in the bath-parametrization and, in
addition, a fast way to select the ground-state of the associated
Anderson impurity problem by first freezing the particle number
and choosing then the right particle-sector.
\subsubsection{Reducing the parametrization: double-bath-degeneracy in one
dimension.}
\begin{figure}[!htb]
\begin{center}
\includegraphics[width=10.0cm,height=3.5cm,angle=-0] {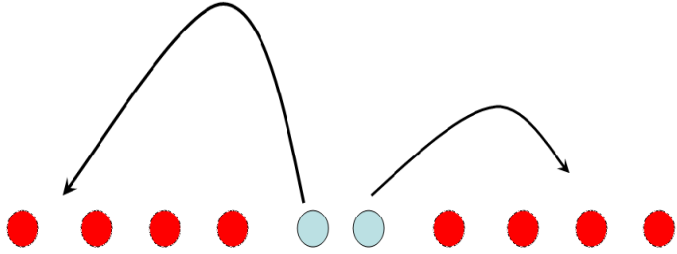}
\caption{2-site cluster-impurity in 1D. The particular topology
of this case splits the free-particle bath in two degenerate
sub-baths each one coupled to the correspondent site of the
cluster-impurity. } \label{reduced-chain}
\end{center}
\end{figure}

We discuss here a peculiarity of the topology of the 1D
CDMFT-implementation, which allows a reduction of the number of
free parameters used (a similar parametrization for the 2D case is
introduced at the end of this chapter). In 1D we can in fact
consider the full bath as divided into two sub-baths (for example
in the case Nb=8, we have 2 baths with Nb=4). Pictorially, one
sub-bath is placed at the left an the other on the right of the
Nc=2 cluster-dimer (Fig. \ref{reduced-chain}). The electrons in
the left site of the cluster can only jump to the left bath, while
electrons in the right site can only jump to the right.  We also
require for there to be a trivial mirror-symmetry, imposing the
bath energy $\varepsilon$ and the $V$s to be the same on the left
and on the right. The number of effectively used $\varepsilon$ is
reduced of $\frac{1}{2}$, and the number of $V$s of
$\frac{1}{4}$. This reduction in the number of bath parameters
derived from symmetry considerations in the one dimensional chain
has its roots in the CDMFT-self-consistency condition (formula
\ref{self-consistency}), which in 1D imposes that the nearest
neighbor term of the hybridazing function $\Delta_{12}$ (from
formula \ref{G0_ns})
\begin{eqnarray} \label{Delta_ns}
\Delta_{\mu\nu}(\imath \omega_{n})= \, \sum_{k}^{n_{b}}
\frac{V^{*}_{\mu k}V_{k \nu}}{\imath \omega_{n}- \varepsilon_{k}}
\end{eqnarray}
is identically zero (\cite{tudorpr}). This can be seen by
explicitly calculating $\Delta_{12}$ for the one-dimensional case
in formula \ref{self-consistency}. In other words, under the
viewpoint of the cluster-AIM, a particle can for example jump
from the right site of the 2-site-cluster impurity onto the right
free-interacting bath. There, it undergoes many dynamical
processes before jumping back into the cluster-impurity. However,
because of the 1D-topology of the chain, the particle in the
right bath is forbidden to enter directly into the left-side bath
or the left site in the cluster-impurity. In this way the nearest
neighbor hybridizing field $\Delta_{12}$ is zero, i.e. the nearest
neighbor hopping, is not renormalized by the bath.
\subsubsection{Regular implementation of CDMFT }
The CDMFT-solution depends on the initial guess of the AI bath
parameters $\varepsilon_{k}$ and $V_{k}$. In the ED method in
particular, given the truncation of the bath to a finite number of
orbitals $N_{b}$, the self-consistency condition formula
\ref{self-consistency} is only approximately fulfilled, being the
degree of precision given by the fitting distance $f$ between the
continuous and the discrete Weiss fields $\Delta_{\mu\nu}$ (see
section 3.4.1). It is in fact this minimization procedure,
ultimately dependent on the definition of distance function $f$,
that at each ED iteration determines the new AI bath parameters
$\varepsilon_{k}$ and $V_{k}$ to be fed back into the impurity
solver. The lowest-energy solution is then picked as the ground
state, searching into all the particle sectors ($n_{\uparrow},n_{
\downarrow}$) of the associate AIM. It is not clear {\it a
priori} in which particle sector of the AIM the solution has to
be looked for. It depends on the Hamiltonian parameters $U/t$ and
$\mu/t$, as well as on the starting guess of the AI bath
parameters $\varepsilon_{k}$ and $V_{k}$. Given a starting guess,
however, the algorithm moves into the region around the
($\varepsilon_{k}$, $V_{k}$)-starting-point, searching for the
best compromise between fulfillment of the self-consistency
condition (minimum in the distance-function $f$) and
determination of the ground state.
\subsubsection{Fast implementation of CDMFT }
\begin{figure}[!thb]
\begin{center}
\includegraphics[width=12cm,height=10cm,angle=-0] {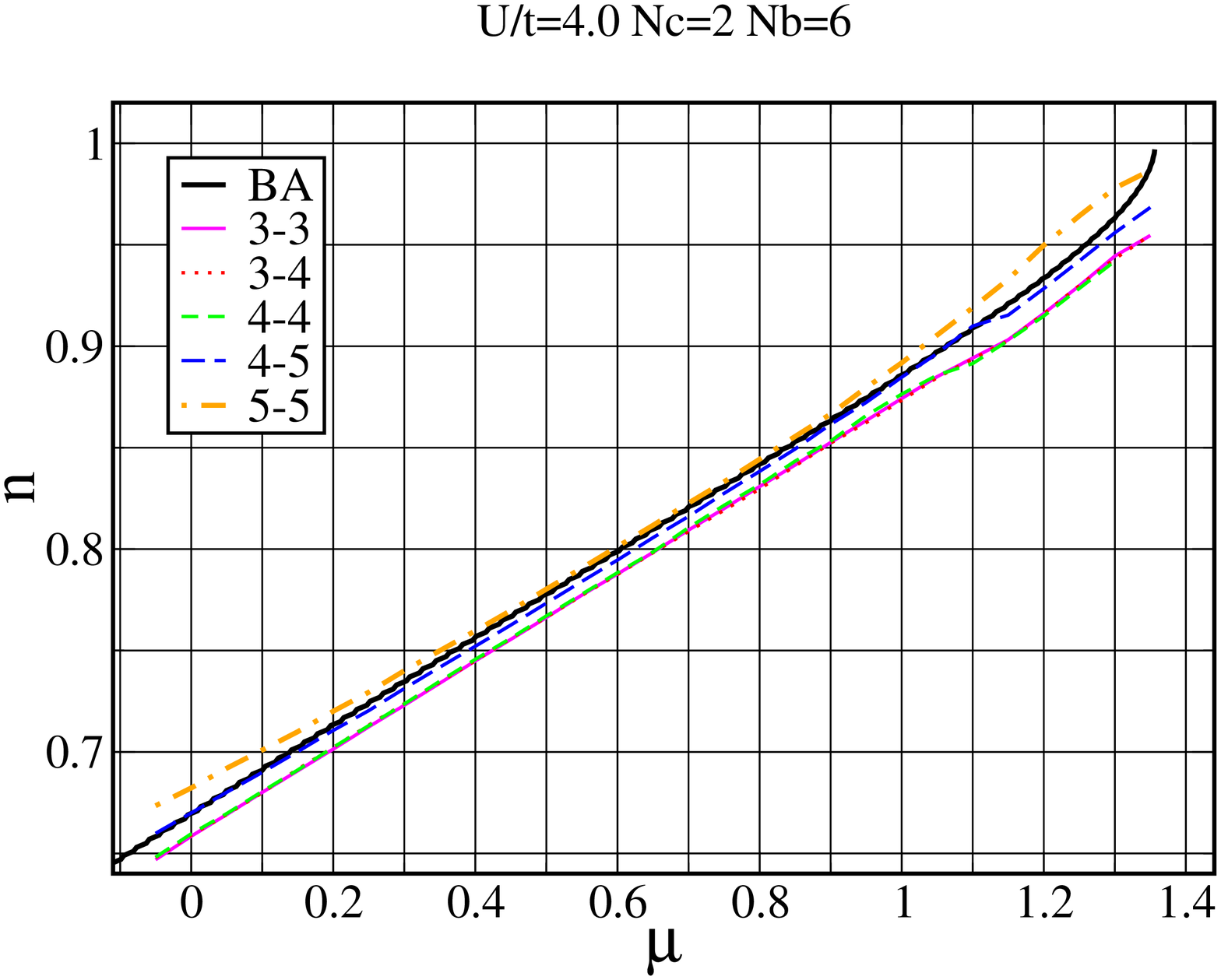}
\end{center}
\caption{Density vs. chemical potential, $n$ vs. $\mu$ in the 1D
Hubbard Model. Solutions are obtained starting from the same seed
and fixing the number of particle-sector in the AIM. Sector span
from $(n_{\uparrow},n_{\downarrow})=3-3$ to $5-5$. The black
continuous line is the BA solution. }\label{mu_densNb6}
\end{figure}

\begin{figure}[!thb]
\begin{center}
\includegraphics[width=8cm,height=5cm,angle=-0] {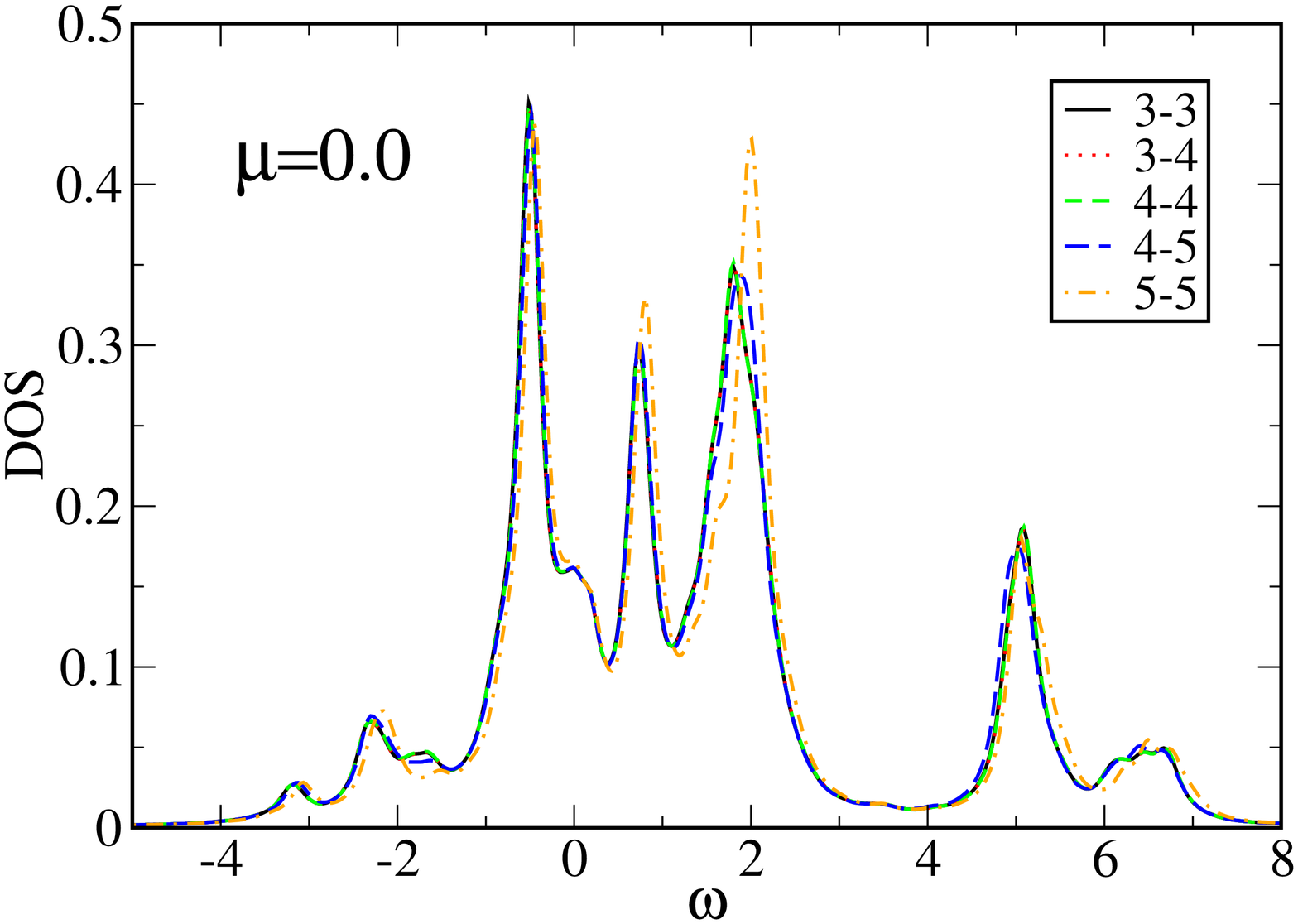}
\includegraphics[width=8cm,height=5cm,angle=-0] {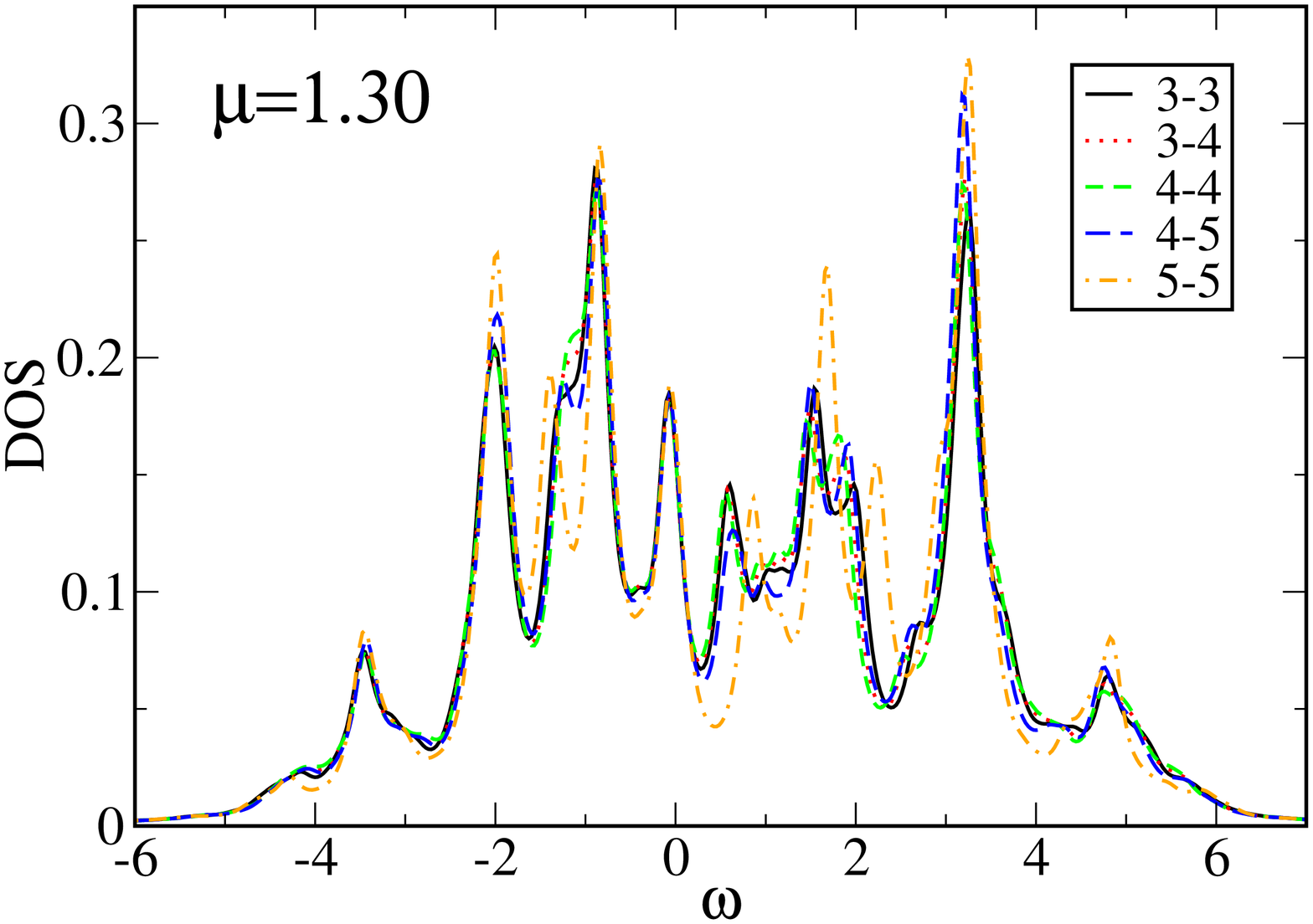}
\end{center}
\caption{Local DOS in two points, $\mu=0.0$ and $\mu=1.30$, of the
$n$ vs. $\mu$ curve displayed in Fig. \ref{mu_densNb6}. The curves
shown are for different particle-sectors, nevertheless the
physics described is the same.} \label{2pointsNb6}
\end{figure}


In order to clarify how the algorithm works in balancing between
the self-consistency constrain and the determination of
AIM-groundstate, we present in the following a procedure which
allows on one hand a faster determination of the AIM-groundstate,
on the other the resolution of converging problems that can arise
in frequent changes between particle-sectors at each
DMFT-iterational step. The latter situation may take place if the
ground-state energies of two different particle-sectors of the
AIM are very close. This can make it difficult to determine a
convergent DMFT-solution, as the algorithm gets trapped in a
resonant loop between the two different (yet close) solutions in
the two different particle-sectors. We therefore \begin{enumerate}
\item constrain the particle-sector
$(n_{\uparrow},n_{\downarrow})$, and there seek the
CDMFT-solution. Naturally more tasks can be attempted in parallel
for each particle sector: notably this saves computational time in
the determination of the impurity solver ground-state.
\item then check each particle sector to ensure that the solution
determined is a real ground-state of the AIM by allowing a loop
on all the other $(n_{\uparrow},n_{\downarrow})$ sectors: if the
bath [ i.e. the AI parameters $(\varepsilon_{k},V_{\mu k})$ ] do
not change the solution is accepted, otherwise it is disregarded.
\end{enumerate}
\subsubsection{A practical example of fast implementation}
We show now the advantages and equivalence of this fast
implementation with respect to the regular implementation. In
Fig. \ref{mu_densNb6} we freeze the particle sectors
($n_{\uparrow},n_{\downarrow}$) of the associated AIM and seek
for the solution in each of them separately. In the case
displayed $U/t=4.0$, the chain used is $N_{s}=8$ sites long with
$N_{c}=2$ sites in the impurity-cluster and $N_{b}=6$ sites in
the bath. We start with the same bath-parameter guess
$(\varepsilon_{k},V_{\mu k})$ and span the sectors from
$(n_{\uparrow},n_{\downarrow})= (3,3)$ to $(5,5)$, a reasonable
occupation for the densities expected in the range of chemical
potential displayed. Starting from the same seed we let the
solution develop in each particle sector. As evidenced from Fig.
\ref{mu_densNb6}, we obtain different curves which at their best
mimic the BA exact curve (continuous black line). Because of the
constrained $(n_{\uparrow},n_{\downarrow})$, the algorithm does
its best to adjust the bath parameters $\varepsilon_{k}$ and
$V_{k}$ in order to best fulfill self-consistency (looking for a
minimum in $f$). So, in each fixed particle sector, the paths in
the parameter space taken to the solution are all different, as
well as the final sets of $\varepsilon_{k}$ and $V_{k}$.
Nevertheless the curves are reasonably close, the difference
being of the order of 1\%. In fact from particle sector $(
n_{\uparrow},n_{\downarrow})= (3,3)$ to $(4,4)$ the solutions are
practically the same , more different are the ones in sector
$(4,5)$ and $(5,5)$. Notice that the $(4,5)$-sector solution is
the one that better reproduces the BA curve, except very close to
the MT point; this however cannot be considered as a criterion to
set the solution as the "best" one in the CDMFT procedure. We
show now that, in spite of the apparent quantitative difference
in the densities of the solutions in different sectors (which we
remark is nevertheless quite small, of the order of 5\% in the
worst case), the physics described is the same. In Fig.
\ref{2pointsNb6} we display the local density of states DOS=
$-\frac{1}{\pi}\, G_{11}(\omega)$ in two points of the $n$ vs.
$\mu$ diagram, one in the far doping region ($\mu=-0.10$ and
$n=0.68$, upper panel of Fig. \ref{2pointsNb6}) and close to the
MT point ($\mu=1.3$ and $n=0.96$, lower panel of Fig.
\ref{2pointsNb6}). It is once again evident that the solution
from sectors $(3,3)$ to $(4,4)$ are for all practical purpose the
same, despite the ground-state of the AIM is evaluated in
different particle sectors. The differences with sectors $(4,5)$
and $(5,5)$ are tiny even quantitatively in the far doped case
(upper panel), larger close to the MT (lower panel), overall they
describe the same solution. This is most comforting: even if we
constrain the impurity solver (in this case the particle sector),
CDMFT works to achieve the same physical results in the limits
imposed by the constrain. But among the many possible solutions
in different sectors, which one do we pick? As said above, we
choose the solution in the sector that is a real ground-state for
the associated AIM, i.e. the solution which does not change once
the fixed-$(n_{\uparrow},n_{\downarrow})$-constrain is relaxed.
In this case the winning sector is
$(n_{\uparrow},n_{\downarrow})=\, (4,4) $.
\begin{figure}[!htb]
\begin{center}
\includegraphics[width=14cm,height=12cm,angle=-0] {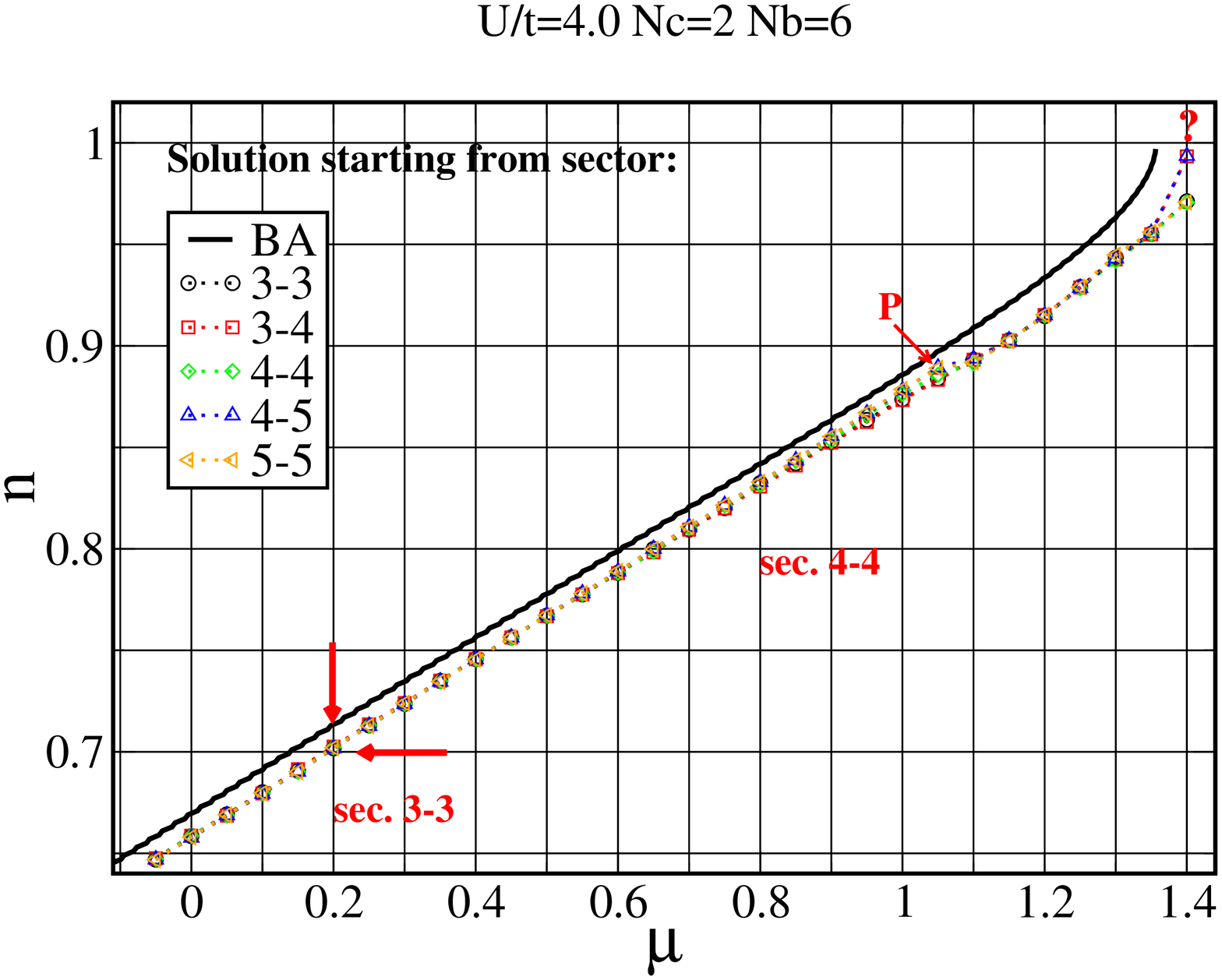}
\end{center}
\caption{We show the result $n$ vs. $\mu$ obtained starting in
each particle sector from the previously determined solution
(ref. Fig. \ref{mu_densNb6}) at fix number of particles in the AIM
and freeing the particle sectors. All the runs end onto the same
solution. } \label{mu_dens_confr_best3}
\end{figure}

We show that this fast procedure is in fact equivalent to the
regular procedure, taking advantage once again of the example we
used above. In Fig. \ref{mu_dens_confr_best3} we start from the
set of solutions determined in Fig. \ref{mu_densNb6} by
constraining the particle-sector, and we use them as starting seed
to implement a new calculation with free number of particles in
the impurity solver. All the solutions end on the same curve,
independent of the sector we started from.
In order to have a better understanding on how the result
develops, let us take a closer look to the curves of Fig.
\ref{mu_dens_confr_best3}. First of all, we observe that in all
the cases the solutions found share the same particle-sector
$(n_{\uparrow},n_{\downarrow})$, and that the solutions are the
same in the limit of computational precision. An estimate of the
latter is given close to the point $P$ indicated in Fig.
\ref{mu_dens_confr_best3}, where the solutions differ the most.
We stress that this tiny difference does not correspond to a
different physical solution. It indicates that, according to the
particle-sector we started from, in some cases the solution
chosen acquires a bigger distance function $f$ or a higher
G.S.E., contrary to the common intuition that these two quantities
should be as small or negative as possible. Secondly, we notice
that all the solutions start from far doping in the sector
$(n_{\uparrow},n_{\downarrow})=(3,3)$ and at some point (around
$\mu=0.20$ and $n=0.70$) they switch to the sector
$(n_{\uparrow},n_{ \downarrow})=(4,4)$ (two perpendicular arrows
in Fig. \ref{mu_densNb6}). This is natural: according to the
actual filling of the system which is monotonicaly increasing
with the chemical potential $\mu$, the associated AIM tends to
increase its particle density too. The point corresponding to
$\mu=0.20$ is highly degenerate, being the G.S.E of the sectors
$(n_{\uparrow},n_{\downarrow})= (3,3),(3,4),(4,4)$ very close, in
the limit of a continuous line of $n$ vs. $\mu$ equal. This
signals the passage from sector $(n_{\uparrow},n_{ \downarrow})=
(3,3)$ to $(4,4)$. The set of curves remains in the latter sector
until they reach the MT point in correspondence of $\mu=1.4$.
Here it is not possible to reach a converged solution very easily
(question mark in Fig. \ref{mu_densNb6}). The solutions
oscillates between two densities $n=0.97$ and $n=0.99$
corresponding to sectors $(n_{\uparrow},n_{ \downarrow})= (4,5)$
and $(n_{\uparrow},n_{ \downarrow})= (5,5)$.  Evidently the
solution attempts a switch of particle sector, this time from
$(n_{\uparrow},n_{ \downarrow})= (4,5)$ to $(n_{\uparrow},n_{
\downarrow})= (5,5)$, demonstrating in this way the diverging
compressibility of the 1D result. However the steep change in the
density for a small change in the chemical potential requires a
slow move from point to point on the $n$ vs $\mu$ line close to
the MT-region, in order to direct the solution to convergence (as
it was done on Fig. \ref{mudens1D}). Another efficient way to
determine this point is to use a finite-temperature
Lanczos-calculation (see APPENDIX B). We can start introducing a
temperature in the system (a virtual $\beta^{-1}$ was already
introduced in the zero-temperature-Lanczos-procedure to determine
a grid of points on the Matsubara axis) whose magnitude has the
same order of the difference in energy between the ground-states
in sector $(4,5)$ and $(5,5)$. A $\beta=64$ (in energy units of
$t$) satisfies the condition in this case. In principle we can
consider many excited states, either ground-states of different
particle sectors and excited states within the same sector,
according to the cut-off determined by the scale of $\beta^{-1}$.
The resulting density is a value between the two oscillating
$n=0.97$ and $n=0.9$ displayed in Fig. \ref{mu_dens_confr_best3},
resulting from the mixing of states weighted by the Boltzmann's
factors $e^{-\beta\, (E-E_{o})}$. As the temperature is
increasingly lowered, the state which has the minimum in energy
(in sector $(5,5)$ in this case) dominates the mixing and the
solutions approaches more and more this state.
\subsection{Results with increasing bath-size}
We wish now to study the effect of increasing bath size on the
CDMFT solution. We therefore consider the $N_{b}=6$ solution
determined in the previous section as starting point to build the
solution with a bigger bath.
\begin{figure}[!htb]
\begin{center}
\includegraphics[width=12cm,height=12cm,angle=-0] {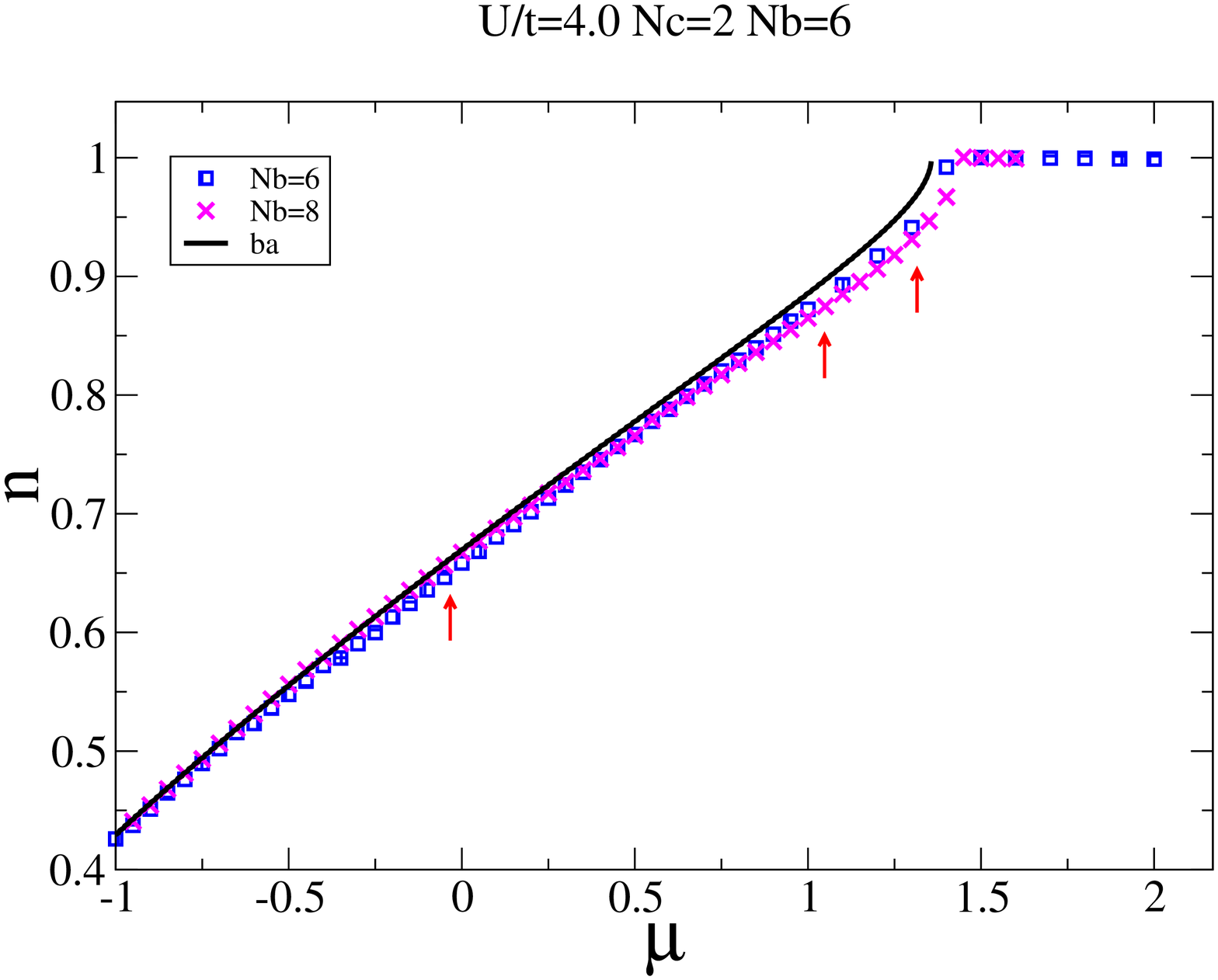}
\end{center}
\caption{$n$ vs. $\mu$ curve obtained with $N_{b}=8$ (magenta
crosses) from the $N_{b}=6$ (blue squares) of Fig.
\ref{mu_densNb6}. The black continuous line is the BA solution.
}\label{Nb6-2-Nb8b}
\end{figure}
\begin{figure}[!htbp]
\begin{center}
\includegraphics[width=14cm,height=14cm,angle=-0] {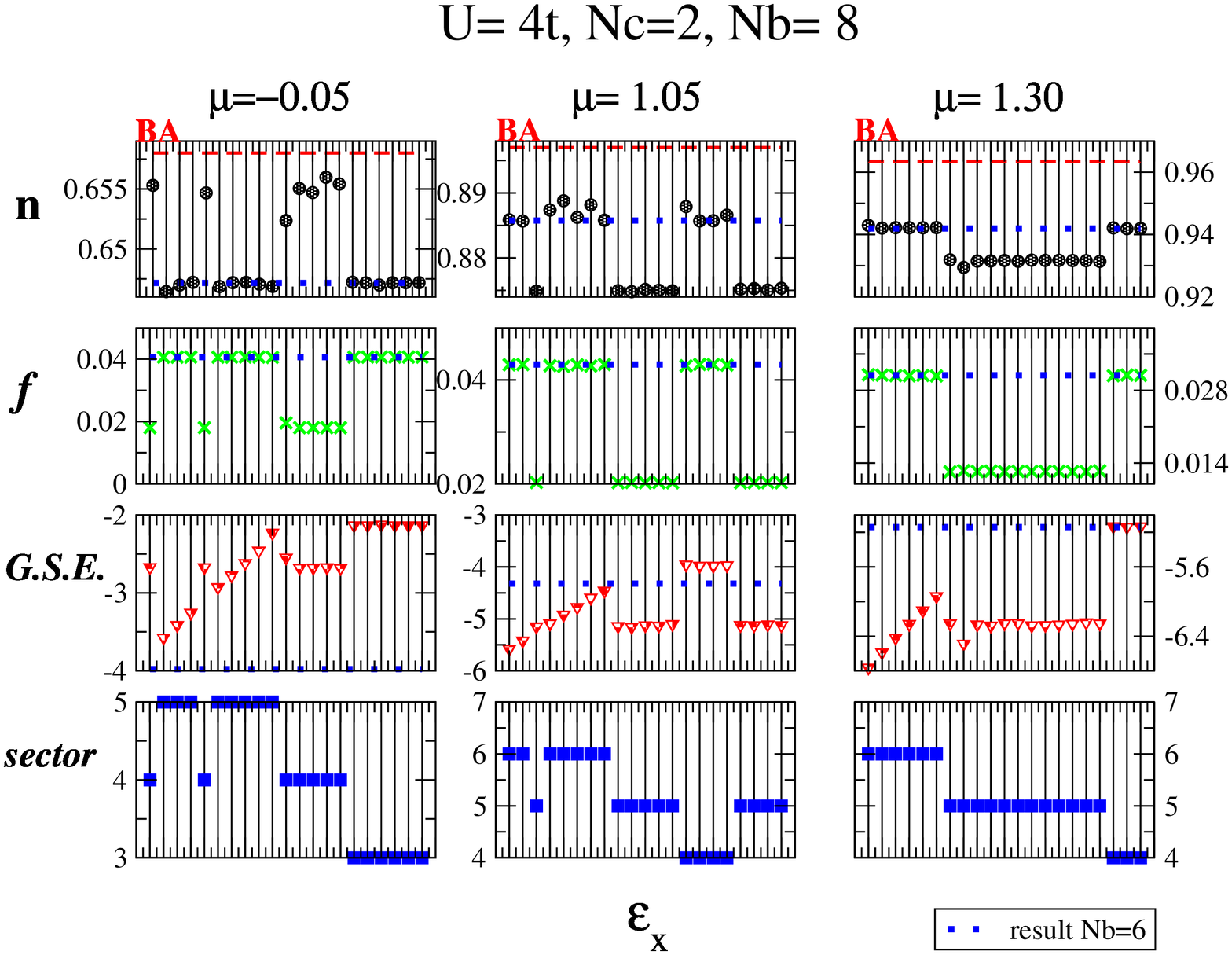}
\end{center}
\caption{Three points of the $N_{b}=8$ solution (red arrows of
Fig. \ref{Nb6-2-Nb8b}), corresponding to $\mu=-0.05$, $\mu=1.05$
and $\mu=1.30$ (each different column) are here analyzed in detail
for many possible starting guesses grown from the $Nb=6$ result
(labeled $\varepsilon_{x} \in [-0.4,0.4]$ is the value of the
bath-energy added to the $Nb=6$ solution). From the top to bottom
we display: the density $n$ (black dots) compared to the
$N_{b}=6$ density (blue dotted line) and the BA density (dashed
red line), the distance function $f$ (green crosses, blue dots
are always the $N_{b}=6$ result), the ground state energy of the
AIM and, in the last raw, the particle-sector of the AIM in which
the solution was found. Notice that the $Nb=8$ solutions which lie
on the blue dotted line do not improve the $Nb=6$ result, rather
they reproduce it with a higher computational effort.
}\label{Nb6-2-Nb8}
\end{figure}
We do not consider here a $N_{b}=7$-bath sites case. Because of
the breaking in the right-left double-chain symmetry
(Fig.\ref{reduced-chain}), this kind of system is frustrated and
yields worst result than $N_{b}=6$. We therefore study a $Nb=8$
system, i.e. we add a site in the right and left-hand bath. A
systematic way to proceed is to guess a value for the added pole
$\varepsilon_{X}$ and choose the hybridizing parameters $V_{kX}$
an order of magnitude smaller than the ones determined with the
$Nb=6$ solution, so that in the first iterational loop the result
will not change too much from the $N_{b}=6$ case. If the CDMFT
condition favors a bath-energy in correspondence of the guessed
$\varepsilon_{X}$, the $V_{kX}$ starts growing and a new solution
that better satisfies the CDMFT-condition develops. In this case
we expect the distant-function $f$, which measures the precision
in the self-consistency condition \ref{self-consistency}, to
decrease from its $N_{b}=6$ value. In Fig. \ref{Nb6-2-Nb8} we
show the result of this procedure for three selected points: a
far doping case at chemical potential $\mu=-0.05$ and BA density
$n=0.658$ (first column in Fig. \ref{Nb6-2-Nb8}), and close to
the MT region $\mu=1.05$ and BA density $n=0.898$ (second column
in Fig. \ref{Nb6-2-Nb8}) and $\mu=1.30$ and BA density $n=0.964$
(third column in Fig. \ref{Nb6-2-Nb8}). These three points are
also indicated by the arrows in Fig. \ref{Nb6-2-Nb8b}. We sample
the added-pole $\varepsilon_{x}$ in the range of values
$[-0.4,0.4]$ which includes all the bath-energy levels
$\varepsilon$ for the $N_{b}=6$ solutions (but we also checked
for values out of this range, not finding relevant modifications
for the discussion that follows). The graph  Fig.\ref{Nb6-2-Nb8}
displays four rows, the vertical grid lines elongating in each
column mark the same solution for a starting $\varepsilon_{x}$.
The graphs in the first row from the top display the density $n$
for each solution. The dashed red line is the BA value. The dotted
blue line is the value of the $N_{b}=6$ result. In the second raw
we show the distance function $f$ (again compared with the
$N_{b}=6$ value). The third raw displays the ground-state energy
G.S.E. of the associated AIM. In the last raw it is shown the
particle sector where the solution was found.

First we observe that in all cases some solutions give the same
result as in the $N_{b}=6$ case used as starting seed. These are
the points that overlap with the dotted blue lines representing
the value of the $N_{b}=6$ solution. This is true for the density
as well as for the distance function $f$ for all the three cases
displayed. This set of solutions does not improve the $N_{b}=6$
result, rather it reproduces it but with a higher computational
effort. Therefore these solutions are not to be considered good
$N_{b}=8$ results and should be disregarded. Another set of
solutions does change the $N_{b}=6$ result, lowering the distance
function $f$, as evidenced in the second raw (the set of green
crosses which departs from the blue-dotted line) and showing a
move toward the large size-limit solution. In the high-doped case
$\mu=-0.05$ (first column) this set of solutions corresponds not
only to a smaller $f$ (and so to a better satisfaction of the
self-consistency condition) but also to a net improvement in the
density with respect to the BA value. The other two solutions
close to the MT $\mu=1.05$ and $\mu=1.30$ (second and third
column respectively) do not correspond to an improvement in the
calculated density. The change is however small (less than 1\%),
and it is relevant that the divergent compressibility, a peculiar
and difficult to capture feature of the MT in 1D, is well
portrayed. So in the high doping region we observe a systematic
improvement of the density of the system with increasing
bath-size as compared to the exact BA value. At small doping,
close to the MT, though this improvement is not observed, the
self-consistency condition is better satisfied and the difficult
character of the MT is well described (Fig. \ref{Nb6-2-Nb8b}).
There is no evident connection with the G.S.E. of the associated
AIM, as evidenced in the third raw of Fig. \ref{Nb6-2-Nb8}.


\section{Comparison between QMC and CDMFT}
\begin{figure}[!hbt]
\begin{center}
\includegraphics[width=12cm,height=10cm,angle=-0] {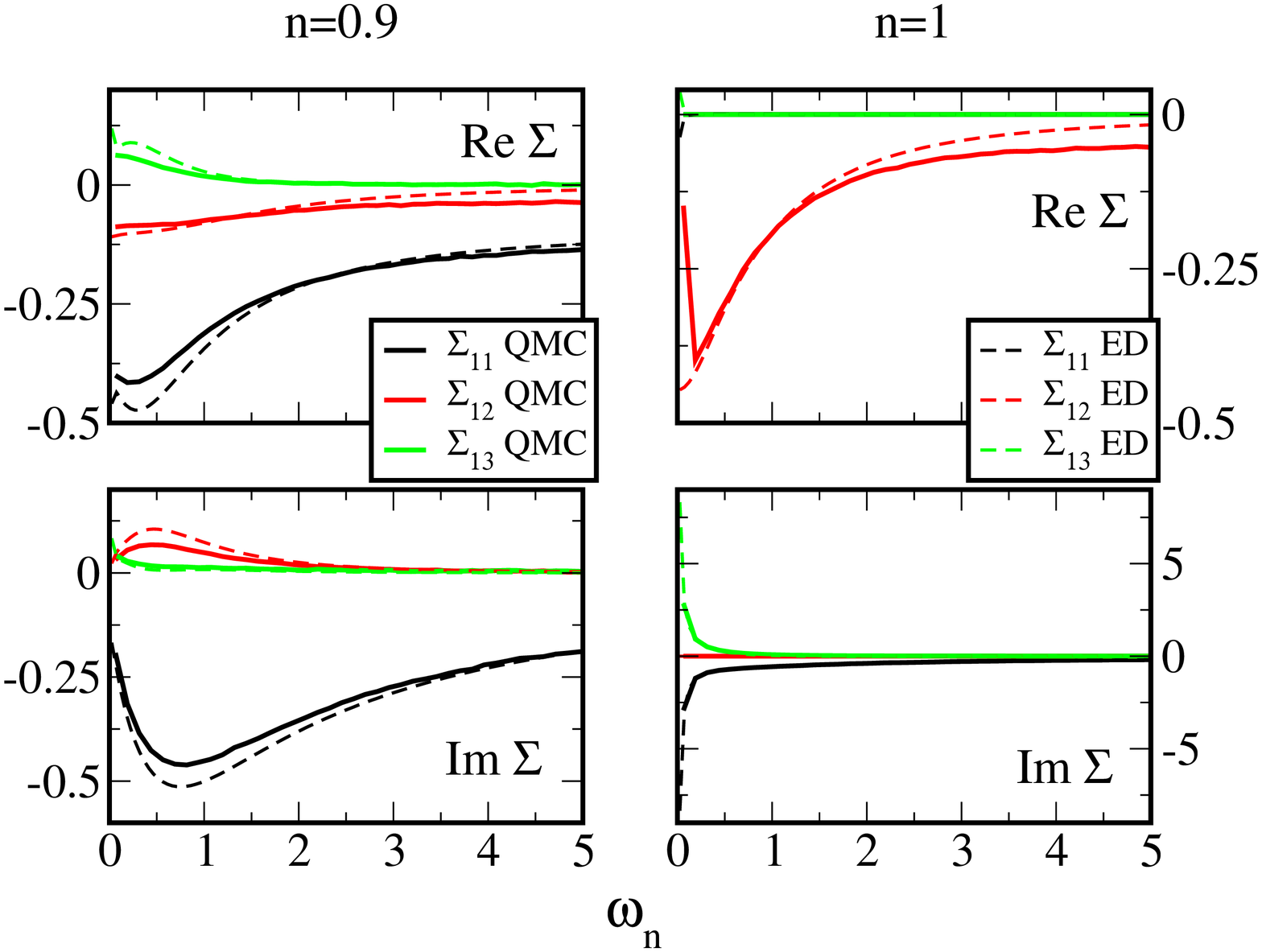}
\end{center}
\caption{Confront between ED (dashed lines) and QMC (continuous
lines), U/t=8.0 t'=0. QMC results are taken from \cite{bpk}, where
energy units on the axis are half-bandwidth $D= 4t$. The cluster
self-energies are shown for a doped system ($n=0.90$) on the
left-hand side, and in the insulating half-filled system($n=1.0$)
on the right-hand side. In the latter case the particle-hole
symmetry produces zero real parts in $\Sigma_{11}$ and
$\Sigma_{13}$ (top right hand side panel). The inverse QMC
temperature is $\beta=12.5$ (in the $t=1$ energy units used in
this thesis), the virtual temperature of the ED solution is
$\beta=32$. }\label{ED-QMC}
\end{figure}
\begin{figure}[!htbp]
\begin{center}
\includegraphics[width=7cm,height=5.5cm,angle=-0] {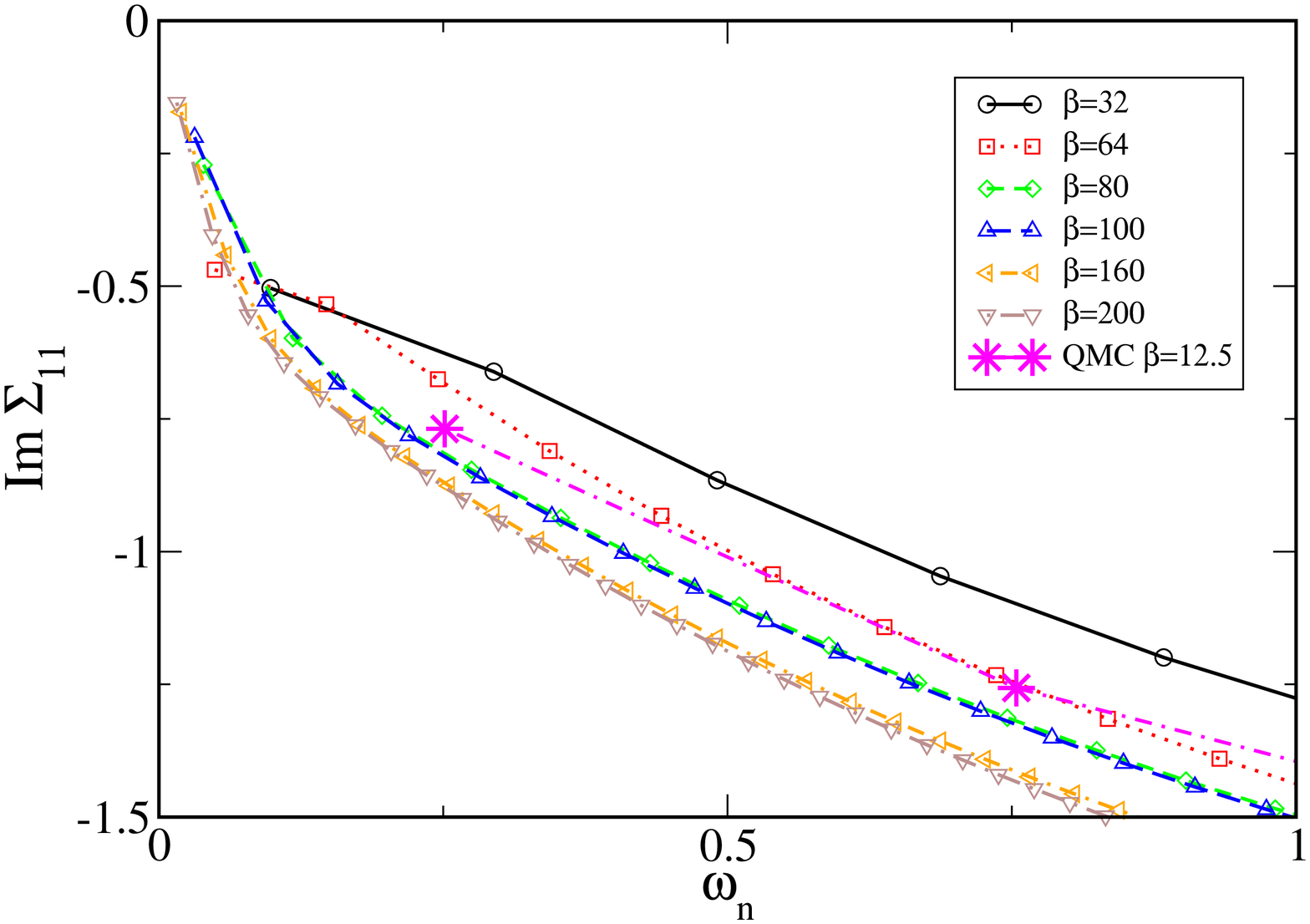}
\includegraphics[width=7cm,height=5.5cm,angle=-0] {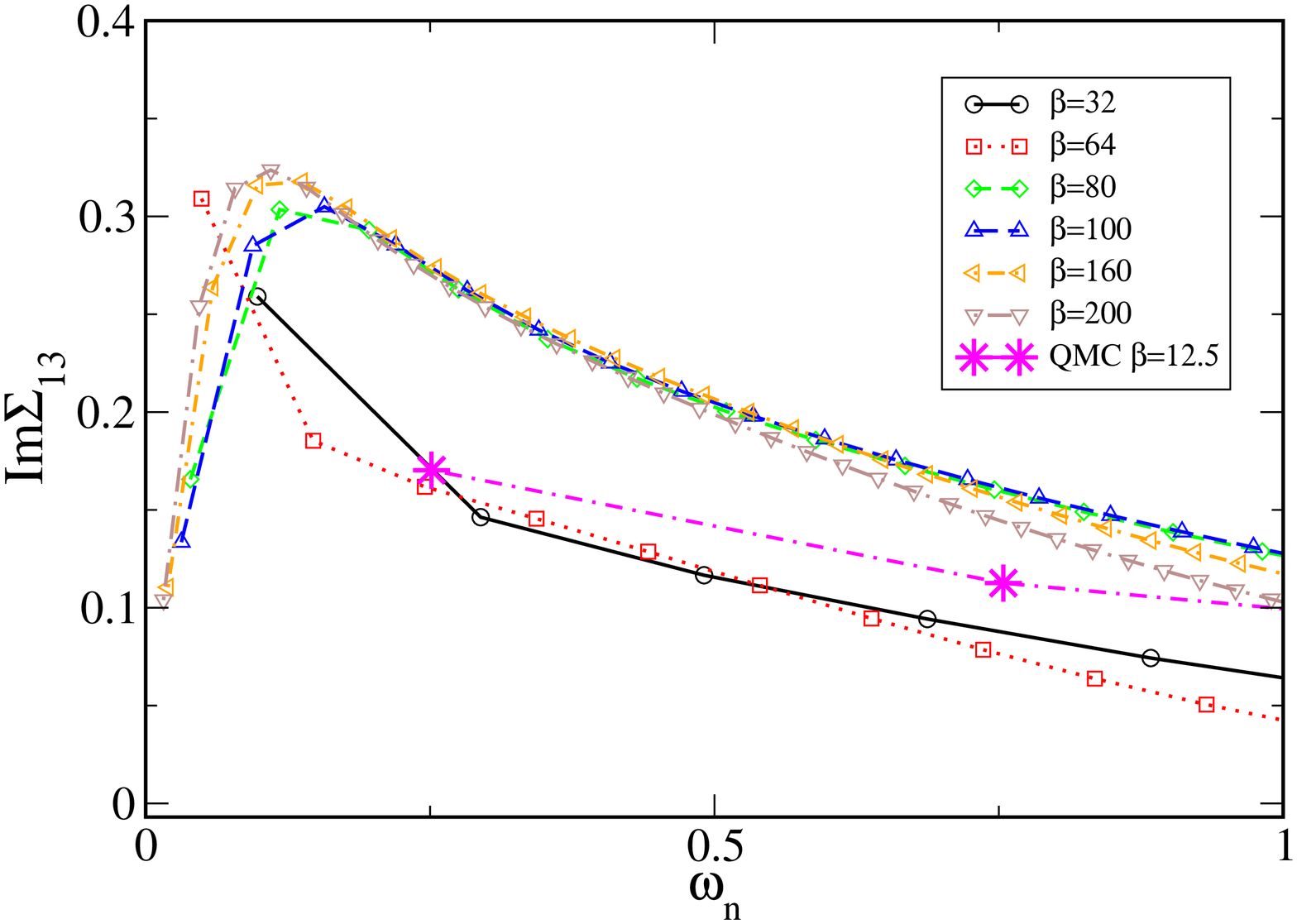}
\caption{Imaginary parts of the on-site cluster self-energy
$\Sigma_{11}$ and the next nearest-neighbor $\Sigma_{13}$ for
increasing values of the virtual temperature $\beta$ in the 2D
Hubbard Model. $U= 8.0 t$, the chemical potential of the system is
fixed at $\mu= 1.8$ and the demsity $n=0.9$. QMC results are
displayed as magenta stars with $\beta=12.5$. The energy unit is
set here $t=1.0$. }\label{Sigmas-beta}
\end{center}
\end{figure}
\begin{figure}[!htbp]
\begin{center}
\includegraphics[width=12cm,height=10cm,angle=-0] {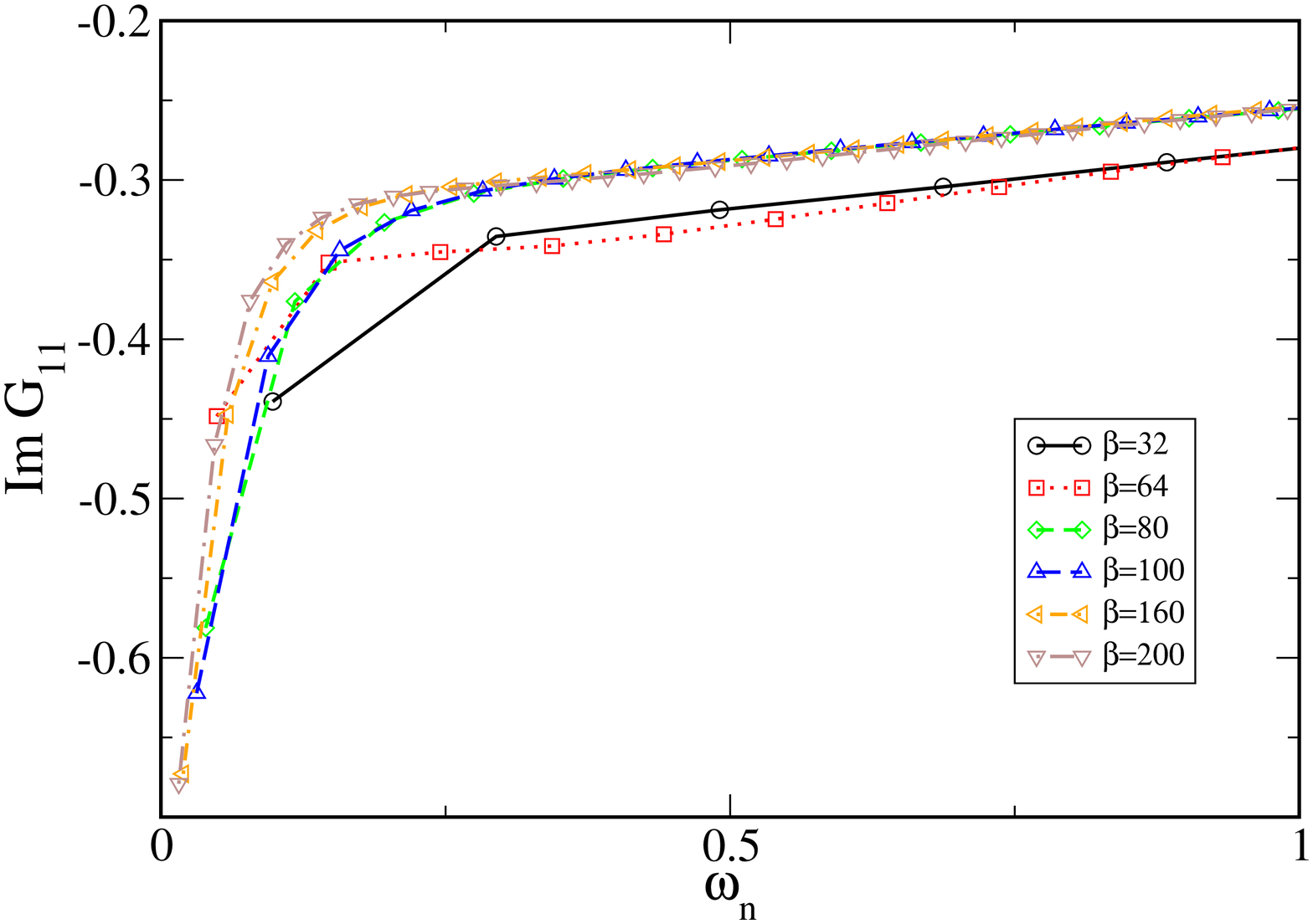}
\end{center}
\caption{Imaginary part of the on-site Green's function
$G_{11}(\omega_{n})$ for increasing values of the virtual
temperature $\beta$ in the 2D Hubbard Model. $U= 8.0 t$, the
chemical potential of the system is fixed at $\mu= 1.8$,
$n=0.90$. QMC results are displayed as magenta stars with
$\beta=12.5$. The energy unit is $t=1.0$.} \label{G11-beta}
\end{figure}
We want to stress here the differences as compared with the
previous high temperature Quantum Montecarlo study (QMC) of O.
Parcollet, G. Biroli and G. Kotliar \cite{parcollet}. In their
work they considered a strongly frustrated two-dimensional Hubbard
Model at half-filling varying the on-site repulsion $U$. The
cluster-impurity used was a 2X2 square of 4 sites strongly
anisotropic in the diagonal direction of the square. In their
result they found that the imaginary part of the on-site cluster
self-energy Im$\Sigma_{11}$ was going to zero for
$\omega\rightarrow 0$, while the diagonal cluster self-energy
Im$\Sigma_{13}$ was growing for small $\omega$ close to the MT.
This was the key-effect giving rise to the phenomenon of hot/cold
spot modulation in $k$-space spectral properties. It was argued
however that if Im$\Sigma_{11}\rightarrow 0$ for
$\omega\rightarrow0$, so Im$\Sigma_{13}$ had to do in order to
preserve casuality (i.e the cluster matrix $\Sigma_{\mu\nu}$ has
to be definite positive), an intrinsic property of the CDMFT
method. This effect could not be observed in the Montecarlo study,
which could not reach the required low energy scale because of the
well known sign problem. The ED-CDMFT method is alternative and
complementary to the QMC-CDMFT. It allows reaching a lower
temperature range and higher values of the local interaction $U$.
In Fig. \ref{ED-QMC} and \ref{Sigmas-beta} we show a comparison
between ED and QMC results for the two dimensional Hubbard Model
$U/t= 8.0$, $t'=0.0$ in the metal phase at density $n=0.90$ (left
panel of Fig. \ref{ED-QMC} and Fig. \ref{Sigmas-beta}) and at
half-filling ($n=1.0$, right panel Fig. \ref{ED-QMC}). Here the
cluster-impurity is a 4 site 2X2 isotropic plaquette, which we
will consider later in further two dimensional studies in the
following chapter (displayed in Fig. \ref{Fcluster}). The inverse
temperature of the QMC result is $\beta=12.5$ (in $t=1.0$ units),
the inverse virtual temperature of the ED varies from $\beta=32$
to $\beta=200$. Fig. \ref{ED-QMC} shows the real and imaginary
part of the cluster Green's function while Fig. \ref{Sigmas-beta}
the imaginary part of the cluster on-site self-energy
$\Sigma_{11}$ and diagonal $\Sigma_{13}$ on the Matsubara axis,
where the QMC result is guaranteed to converge quadratically to
the exact result [with an error $O(\Delta\tau^{2})]$ and ED does
not have the "artificial spreading" of the poles required in the
analytical continuation on the real axis. If we confront the
highest-temperature (lowest $\beta$) ED-run with the QMC result
the agreement is remarkably good in spite of the much lower
virtual temperature of ED. This provides a bridge connecting high
and low temperature physics and gives credibility to both methods.

In figures \ref{Sigmas-beta} and \ref{G11-beta} we also show an
example on how the virtual temperature defined on the Matsubara
axis enters into the ED-calculation with respect to the real
temperature used in QMC. The virtual temperature simulates the
effect of a real temperature in the system, even if the state
considered in the solution of the associated Anderson impurity
problem is the ground-state only. For low $\beta$ (high virtual
temperature), the curves behave similarly to the QMC results
(pink stars in \ref{Sigmas-beta}), and change quickly with
increasing $\beta$, until collapsing onto the same curve for
$\beta$ high enough ($\simeq 100$). We may therefore consider (at
least for U=8.0) this value of beta like a limiting case in order
to obtain genuine T=0 calculations.
\section{Evaluation of Lattice Quantities}
Once a converged solution has been determined, the final step in
the CDMFT procedure is to extract physically interpretable lattice
quantities. In the spirit of dealing with problems whose dominant
physics is "local", (like for example Hubbard-like problems), the
intuitive idea lies in using the cluster quantities like effective
Fourier coefficients in a truncated Fourier expansion of the
momentum-dependent quantities. The most natural way of determining
these effective Fourier coefficients from the original lattice
problem is by averaging the correspondent super-lattice
quantities. Given an impurity-cluster configuration in fact (for
example a 2-site dimer in a one dimensional chain or a 2X2
plaquette in a two dimensional lattice), there are more than one
possible super-lattice-partitions of the full lattice into such
clusters, corresponding to the many different symmetry operations
$S$ (translation, rotation...) with whom one possible
super-lattice-partition can be mapped into another. Given a
physically local quantity $W(r)$, that we assume going quickly to
zero for $r>$ the cluster size, and its corresponding
super-lattice quantity $W^{S}(r)$ calculated in the CDMDT-scheme
within the cluster, for a given lattice-partition $S$ of the
lattice into cluster-impurities, we evaluate:
\begin{eqnarray}
\label{Wij}
 W(r)\simeq \, \frac{1}{N_{s}}\, \sum_{S}\,W^{S}(r)
\end{eqnarray}
$N_{s}$ is the number of possible ways the the full lattice can
be parted, once a cluster-impurity configuration is chosen.
\begin{figure}[!htbp]
  \begin{center}
 \includegraphics[width=10.0cm,height=8cm,angle=-0] {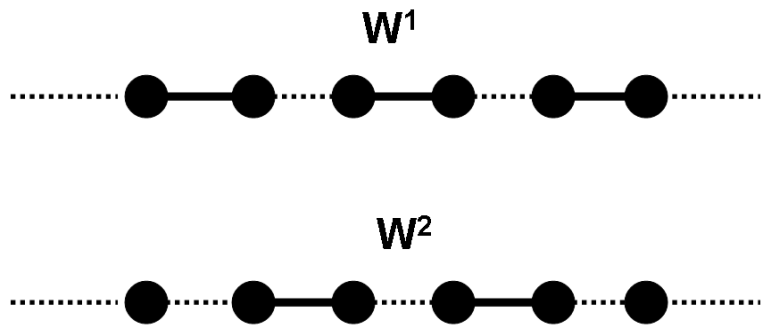}
  \caption{Possible partition in dimers for a 1D chain.}\label{Wchain}
  \end{center}
\end{figure}
We stress that Eq.(\ref{Wij}) represents a super-lattice average,
not a cluster average. Contributions $W^{S}(r)$, which connect two
points belonging to different clusters are systematically zero.
For example in Fig. \ref{Wchain} we present the simplest example
with a 1D chain parted in $N_{c}=2$-dimer clusters. There are two
possible partitions ($S=2$) of the 1D chain into dimers (drawn as
continuous links). Therefore we have two possible values for the
link-quantity $W^{S}$ (connecting two neighbor sites), $W^1$ and
$W^{2}=0$, i.e. :
$$
W= ( W^{1}+W^{2} ) / 2=\, W^{1}/2 \equiv W^{c}/2$$ where we
indicate with $W^{c}$ the in-cluster-value of $W$ evaluated for a
fixed partition of the lattice ($S=1$ in the present case).
Notice that typically for a square lattice the number of possible
partitions of the full lattice into clusters is $N_{s}= N_{c}$,
the number of cluster-sites. The momentum-dependent quantity is
then evaluated:
\begin{eqnarray}
\label{Wk} W(k,\omega)=\,\sum_{r}\, e^{\imath k r} \, W(r)=\,
\frac{1}{N_{c}}\, \sum_{\mu,\nu=\,1}^{N_{c}^{1/2}}\, e^{\imath
k\mu}\, W^{c}(| \mu-\nu |)\, e^{-\imath k\nu}
\end{eqnarray}
What is the right "local" quantity to periodize ? There is no a
trivial answer to this question. We present in this study three
possibilities:
\begin{enumerate}
    \item The $\Sigma-$scheme, where the cluster self-energy
    $\Sigma_{\mu\nu}$ is periodized \cite{Biroli:2005}:
\begin{eqnarray}
\label{Sk} \Sigma(k,\omega)=\, \frac{1}{N_{c}}\,
\sum_{\mu,\nu=\,1}^{N_{c}}\, e^{\imath k\mu}\, \Sigma(| \mu-\nu
|)\,  e^{-\imath k\nu}
\end{eqnarray}
    \item The $M-$scheme where the cluster-cumulant $M_{\mu\nu}$ is periodized \cite{tudor}:
\begin{eqnarray}
\label{Mk} M(k,\omega)=\, \frac{1}{N_{c}}\,
\sum_{\mu,\nu=\,1}^{N_{c}}\, e^{\imath k\mu}\, M(| \mu-\nu |)\,
e^{-\imath k\nu}
\end{eqnarray}
    \item The $G-$scheme where the Green's function on the super-lattice (a cluster
    $N_{c}$X$N_{c}$ matrix) is periodized \cite{senechal}:
\begin{eqnarray}
\label{Gk} G(k,\omega)=\, \frac{1}{N_{c}}\,
\sum_{\mu,\nu=\,1}^{N_{c}}\, e^{\imath k\mu}\,
 \left[
\mathbf{\hat{1}}\,\omega+ \hat{t}_{k}- \hat{\Sigma}
\right]^{-1}_{\mu \nu }\, e^{-\imath k\nu}
\end{eqnarray}
\end{enumerate}
Notice that the cluster-cumulant arising from the expansion of the
free energy around the atomic limit is related to the
cluster-self-energy (arising from the usual weak-coupling
expansion of the free-energy) by the expression:
\begin{eqnarray}
\label{S-M} \mathbf{M}(\omega)=\, \left[\,(\,\omega+\mu\,)
\mathbf{1}-\mathbf{\Sigma}(\omega) \, \right]^{-1}
\end{eqnarray}
and that the super-lattice Green's function entering in the
$G-$scheme is the same used in the CDMFT self-consistency
condition (\ref{Gloc}). These three methods are {\it a priori}
equally justified. We will in the following confront them, first
in the one dimensional case, where we can take advantage from the
comparison with the exact solution, then in the two dimensional
case, stressing the physical consequences and implications of the
different methods.
\subsection{Reconstruction of lattice-quantities and physical observables}
The problem of extracting momentum-dependent quantities is of
fundamental importance in identifying the right system observables
that are not generically the cluster ones. A DMFT approach is
based on the assumption that the important physics of the system
is local, thus the local quantities determined in the
cluster-impurity are a good representation of the
translational-invariant lattice ones. This is generally true for a
"pure local quantity", as for example the on-site density of the
system, which can be well determined directly from the
cluster-impurity density. With a small size-cluster however (which
is the only solvable case in practical applications with ED), the
boundary-sites are renormalized by the surrounding bath, and the
"non-pure local quantities" related to these kind of sites (as for
example, the Kinetic Energy) are also renormalized inside the
cluster. In the following section we will study this effect in the
bench-marking one dimensional case enlightening the dependence of
cluster-quantities from the cluster-size. In this section we show,
however, that in order to rightly evaluate physical quantities, it
is important to establish the correct link between CDMFT-cluster
quantities and lattice obsrvables. The three methods presented in
the previous section are three different ways to achieve this
goal. We present here an example applying again 2-site-cluster
CDMFT on the one-dimensional Hubbard Model, where the exact BA
solution can be used as a benchmark. As we said above, the number
of possible ways a one-dimensional chain can be parted into
2-sites dimers is $N_{s}=2$. So, if $W_{r}$ is a local quantity
that we assume small for $r\geq 2$ (for example the self-energy or
the cumulant..) calculated within the cluster (the $W^{S}$ in
formula \ref{Wij}), according to formula (\ref{Wk}) of the
previous section:
\begin{eqnarray}
\label{Wk1D} W(k,\omega)=\,W_{0}(\omega)+ W_{1}(\omega)\, \cos(k)
\end{eqnarray}
A truncated Fourier expansion of $W(k,\omega)$
\begin{eqnarray}
\label{Wk1D_latt} W(k,\omega)=\,W^{latt}_{0}(\omega)+ 2 \,
W^{latt}_{1}(\omega)\, \cos(k)
\end{eqnarray} implies that the next-nearest neighbor
cluster-coefficient
\begin{eqnarray}
W_{1}(\omega)= \,W^{latt}_{1}(\omega)/2
\end{eqnarray}
\begin{figure}[!thb]
\begin{center}
\includegraphics[width=8.0cm,height=6.0cm,angle=-0] {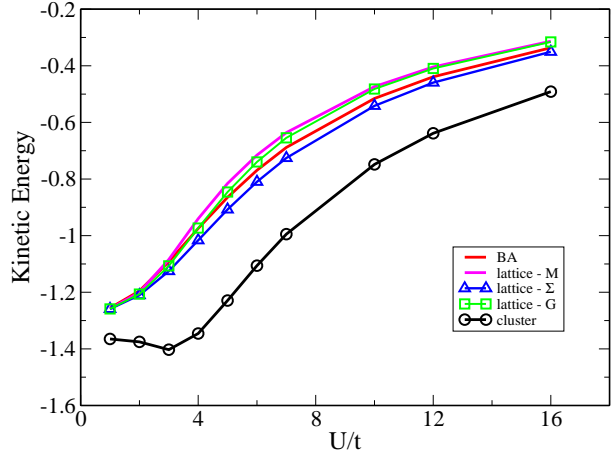}
\caption{Kinetic energy of the half-filled one-dimensional Hubbard
Model as a function of the on-site interaction U at zero
temperature. Results from the three periodization schemes $\Sigma$
(blue triangles), $M$ (magenta line) and $G$ (green squares) and
from the cluster Green's function (circles) are confronted with
the exact BA solution (red line). } \label{KineticEne1d}
\end{center}
\end{figure}
is renormalized by a geometrical factor 2 with respect to its
lattice correspondent, as expected from the discussion in formula
\ref{Wij}. To understand how this reflects on the determination of
non-pure local quantities, we can further calculate the kinetic
energy from the cluster and from the lattice. The former is the
energy integral of the nearest-neighbor cluster Green's function
$G_{12}(\omega)$ evaluated directly from the ground-state of the
AIM. The latter is evaluated by first periodizing the self energy
or the cumulant or the super-lattice Green's function with the
$\Sigma$-scheme, the $M$-scheme or the $G$-scheme respectively (as
explained above), in order to obtain the full lattice Green's
function $G^{lat}(k,\omega)$. $G^{lat}_{12}(\omega)$ is then
evaluated as second Fourier coefficient, and its energy integral
again gives the kinetic energy. These two values can be confronted
with the known BA result. This is shown in Fig.
\ref{KineticEne1d}, where the kinetic energy is plotted as a
function of the on-site interaction $U$. The values of the kinetic
energy given by the cluster Green's function are significantly
different from the exact BA result, while the values obtained from
the lattice Green's functions extracted with the three different
schemes are close to the BA line. This shows clearly that the
quantity which to be physically interpreted is the lattice and not
the cluster Green's function. We also notice that the lines
calculated with the $M$-scheme (magenta line) and the $G$-scheme
(green squares) are remarkably close, especially in the strong
coupling regime, and both underestimate the BA value, while the
line from the $\Sigma$-scheme (blue triangles) overestimates it.
In Fig. \ref{KineticEneInsets} we show the difference between the
BA kinetic energy (black line) and the cluster one (circles)
renormalized by a factor $1/\sqrt{2}:$ $\Delta E_{Kin}=
K_{BA}-K_{clu}/\sqrt{2}$ as a function of $U$. Notice that for a
value of the on-site interaction greater than the bandwidth $U>
4t$, this difference is very close to zero, indicating that that
the kinetic energy calculated from the cluster is underestimated
by the factor $\sqrt{2}$. This has a geometrical origin and
depends on the cluster size and the dimensionality of the problem.
On the left side of Fig. (\ref{KineticEneInsets}) we also plot the
difference between the BA kinetic energy and an averaged kinetic
energy from the $\Sigma$ and $M$ methods $\Delta E_{kin}=
K_{BA}-(K_{\Sigma}-K_{M})/2$ (squares), which appears to be a
reasonably good estimate of the true value.
\begin{figure}[!thb]
\begin{center}
\includegraphics[width=8.0cm,height=6.0cm,angle=-0] {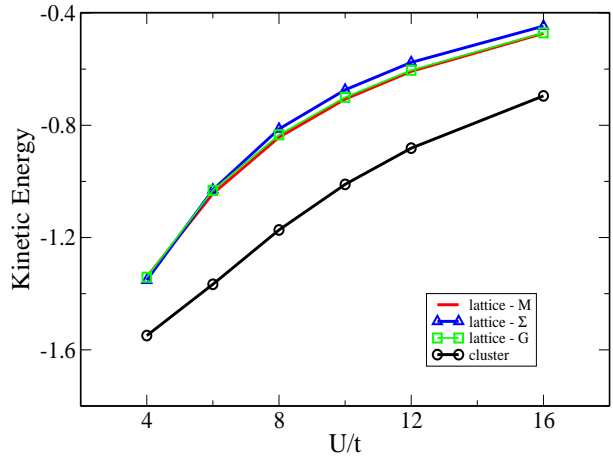}
\caption{Kinetic energy of the half-filled two-dimensional Hubbard
Model as a function of the on-site interaction U at zero
temperature. Results from the three periodizing schemes $\Sigma$
(blue triangles), $M$ (magenta line) and $G$ (green squares) are
confronted with the ones from the cluster Green's function
(circles) which are clearly different .} \label{KineticEne2d}
\end{center}
\end{figure}
\begin{figure}[!htb]
\begin{center}
\includegraphics[width=6.0cm,height=4.0cm,angle=-0] {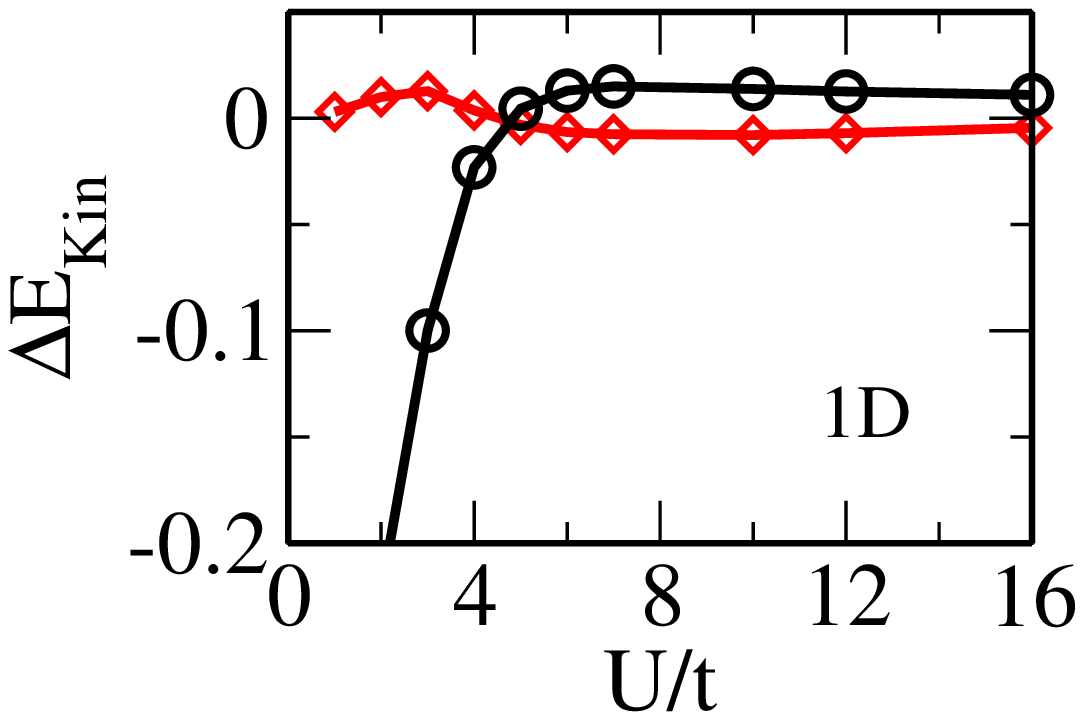}
\includegraphics[width=6.0cm,height=4.0cm,angle=-0] {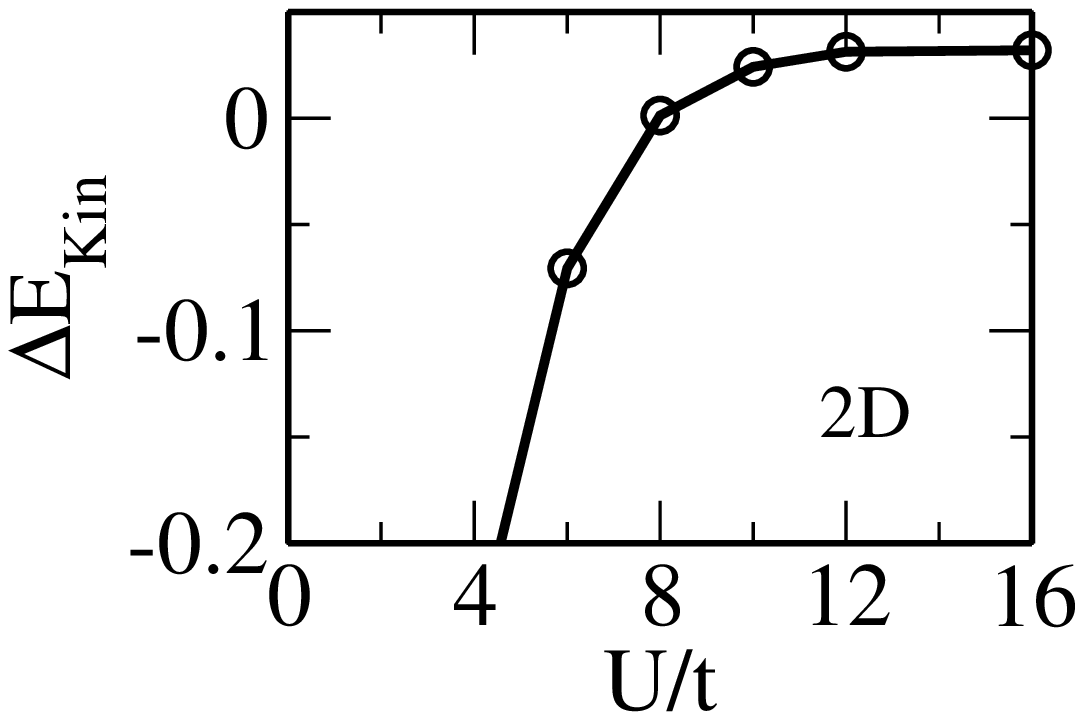}
\caption{On the left-hand side we display (black line with
circles) $\Delta E_{Kin}= K_{BA}-K_{clu}/\sqrt{2}$ as a function
of the interaction $U$ for the 1D Hubbard Model. Notice the
factor $\sqrt{2}$ which reflects the geometry of the
$N_{c}=2$-site cluster. We also show the difference between the
BA solution and an average of the results from the $\Sigma$ and
$M$-schemes $\Delta E_{kin}= K_{BA}-(K_{\Sigma}-K_{M})/2$ (red
line with squares). In the right-hand graph we display $\Delta
E_{Kin}= K_{BA}-K_{clu}/\sqrt{2}$ for the two-dimensional case.
The difference gets close to zero again for $U> 4t$, the
bandwidth of the model in 2D. } \label{KineticEneInsets}
\end{center}
\end{figure}
Let us now turn to the two-dimensional case, where we studied a
2X2 cluster-impurity. For this case lattice quantities $W(k)$,
which could be the self-energy $\Sigma_{k}(\omega)$ in the
$\Sigma$-method or the cumulant $M_{k}(\omega)$ in the $M$-method,
can be explicitly evaluated using formulas \ref{Sk} and \ref{Mk}
and exploiting the symmetry of the square-lattice:
\begin{eqnarray}
\label{Wk2D_latt} W(k,\omega)=\,W^{latt}_{0}(\omega)+ 2 \,
W^{latt}_{1}(\omega)\, \left( \, \cos k_{x}+ \cos k_{y} \right)+\,
W^{latt}_{2}(\omega)\, \, \cos k_{x}\, \cos k_{y}
\end{eqnarray}
For the $G$-method we once again refer to formula \ref{Gk}.
The expression for the self-energy $\Sigma_{k}(\omega)$ is:
\begin{equation}\label{selflat}
\Sigma_\latt(k,\omega)= \Sigma_{11}+ \Sigma_{12} \left( \, \cos kx
+ \cos ky \, \right)+ \Sigma_{13} \, \cos kx \, \cos ky
\end{equation}
and for the cumulant:
\begin{equation}\label{Mlat}
M_\latt(k,\omega)= M_{11}+ M_{12} \left( \, \cos kx + \cos ky \,
\right)+ M_{13} \, \cos kx \, \cos ky
\end{equation}
In Fig. \ref{KineticEne2d} we show the kinetic energy for the
half-filled Hubbard Model at zero temperature obtained by
calculating the lattice Green's function $G_{k}(\omega)$ using the
3 different periodizing methods. The qualitative picture is
strikingly similar to the 1D case. Although a benchmark exact
solution is missing in 2D, the difference between the results
coming from the lattice Green's functions calculated with the
three periodizing schemes and the one coming directly from the
cluster Green's function supports the same conclusions as in the
1D case. This time we plot in Fig. \ref{KineticEneInsets} the
difference between the averaged kinetic energy from the $M$ and
$\Sigma$ schemes and the kinetic energy from the cluster, the
latter renormalized by the factor $1/\sqrt{2}:$ $\Delta
E_{Kin}= K_{BA}-K_{clu}/\sqrt{2}$. Once again, for a local interaction  
greater than the bandwidth $U> 8t$, the difference tends to zero,
showing a dimensional-geometrical dependence as in 1D. This
discussion shows the importance of extracting the lattice
quantities, which have physical meaning, from the impurity-cluster
results. For this it is necessary to fully understand the
differences between the periodizing schemes proposed, and, in
turn, understand which of the schemes represents the best
procedure to obtain the correct physical interpretation.

\subsection{Lattice-quantities and local physics}
\begin{figure}[!htb]
\begin{center}
\includegraphics[width=8.0cm,height=6.0cm,angle=-0] {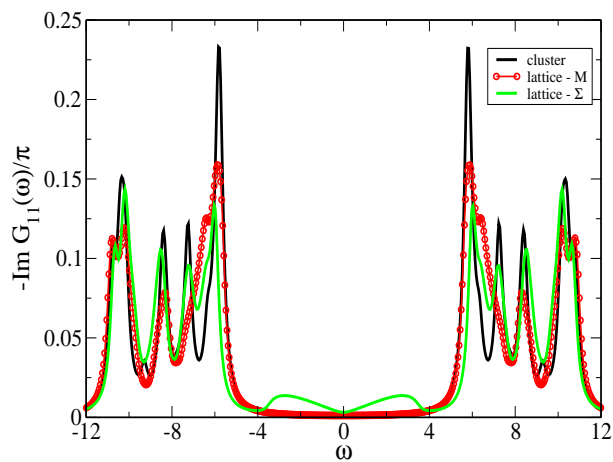}
\caption{Density of states as a function of frequency for the
half-filled 2D Hubbard Model with $U= 16t$ at zero temperature
(ED-CDMFT). The black line represents the local density of states
for the cluster, the red circle where obtained by periodizing the
cumulant, the green line by periodizing the self-energy.}
\label{DOSU162d}
\end{center}
\end{figure}
Are the three periodization schemes equivalent? We recall that a
cluster DMFT procedure is based on the assumption that the
physically relevant quantities are local, we therefore expect that
a good periodization scheme preserves them. If we, for example,
extract the lattice Green's function $G(\mathbf{k},\omega)$ using
the $\Sigma$, $M$ or $G$-schemes, we expect that the local Green's
function $G_{loc}(\mathbf{k},\omega)$ (the only "purely local"
quantity in the small cluster we will consider in practice)
obtained evaluating the energy-integral of the lattice Green's
function all over the BZ
$G_{loc}(\mathbf{k},\omega)=\,\sum_{k}\,G(\mathbf{k},\omega)$ well
describes the local cluster Green's function obtained directly
from the AIM. This is a tautology by construction in the
$G$-scheme, but it is not trivially true for the $\Sigma$ and
$M$-schemes. In Fig. \ref{DOSU162d} we present the DOS
($-(1/\pi)\,\Im G_{loc}(\omega)$) of the two-dimensional Hubbard
Model for $U/t=16$ at half-filling calculated within the cluster
with ED-CDMFT (black line), confronted with the results obtained
from the lattice Green's function trough the $\Sigma$ (green line)
and $M$-schemes (red line with circles). We observe right away
that the $M$-scheme well portrays the original DOS, preserving in
particular the Mott gap (in this case quite large due to the
choice of a high on-site interaction $U$). On the contrary, the
$\Sigma$-scheme creates states inside the Mott gap. These states
are obviously un-physical, and they signal a failure of the
$\Sigma$-scheme to describe the insulating state of the Hubbard
Model. It is expected that the $M$-scheme, based on cumulants
arising from the atomic limit of the original Hamiltonian,
describes better the Mott-insulating state with respect the
$\Sigma$-scheme, which is instead arising from the weak coupling
perturbative expansion of the original problem. Cumulants are
better quantities than the self-energy in describing a localized
insulating state. We expect the $\Sigma$-scheme to better work in
the metal phase, where electrons are itinerant and the Fermi
Liquid picture is valid.
\begin{figure}[!tb]
\begin{center}
\includegraphics[width=8.0cm,height=6.0cm,angle=-0] {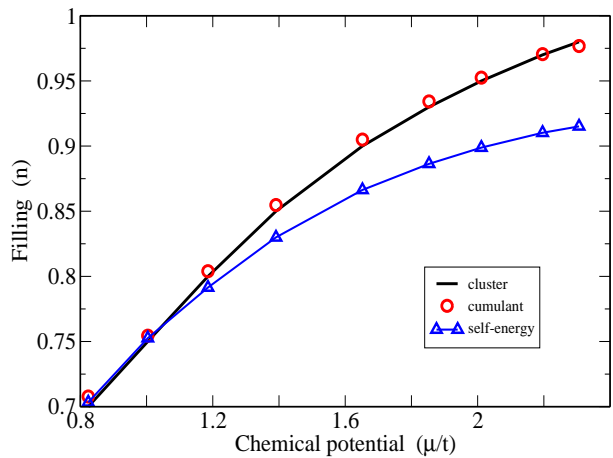}
\caption{Density as a function of $\mu$ for the 2D hubbard Model
with $U= 8t$ calculated with NCA solver at a temperature $T=0.1t$.
The black line is the filling calculated directly from the
cluster, the filling calculated from the periodized Green's
function are the circles for the $M$-scheme and the triangles for
the $\Sigma$-scheme.} \label{FillingU82d}
\end{center}
\end{figure}
In Fig. \ref{FillingU82d} we show the filling of the system
calculated from the DOS as a function of the chemical potential
$\mu$ in the doped state of the Hubbard Model in 2D for $U/t=8.0$.
Again the black line is the filling calculated from the
impurity-cluster, confronted with the $\Sigma$ (blue triangles)
and the $M$ schemes (red circles). The two schemes rapidly
approach the cluster-density at high doping, close to the M-I
transition the $\Sigma$-scheme creates spurious states in the gap
of the DOS, and it gives lower values of the density. The
$M$-scheme, instead, follows the cluster-line, until reaching the
insulator, appearing to be more appropriate for describing lattice
quantities in correspondence of the MT transition. The failure of
the $\Sigma$-scheme at the MT region is also the reason also for
the failure of the Periodizing CDMFT (PCDMFT) method introduced in
\cite{parcollet} and compared with CDMFT in \cite{marce}. In this
method, the lattice translational invariance of the system is
restored inside the DMFT loop itself by evaluating the lattice
self-energy with a $\Sigma$-scheme. An alternative to the PCDMFT
based on the periodization of the cumulant, the explicit cavity
construction (ECC)-DMFT, proposed by T, Stanescu and G. Kotliar
\cite{tudor}, cures the defects of PCDMFT.
\section{Cluster-size and cluster-lattice partitions dependencies}

\begin{figure}[!hb]
\begin{center}
\includegraphics[width=8.0cm,height=5.0cm,angle=-0] {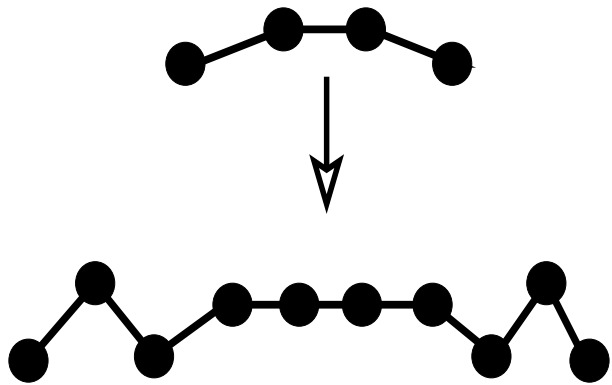}
\caption{1D chain-density profile as increasing the number of
cluster-sites $N_{c}$.} \label{1Dchain}
\end{center}
\end{figure}

\begin{figure}[!htb]
\begin{center}
\includegraphics[width=12.0cm,height=8.0cm,angle=-0] {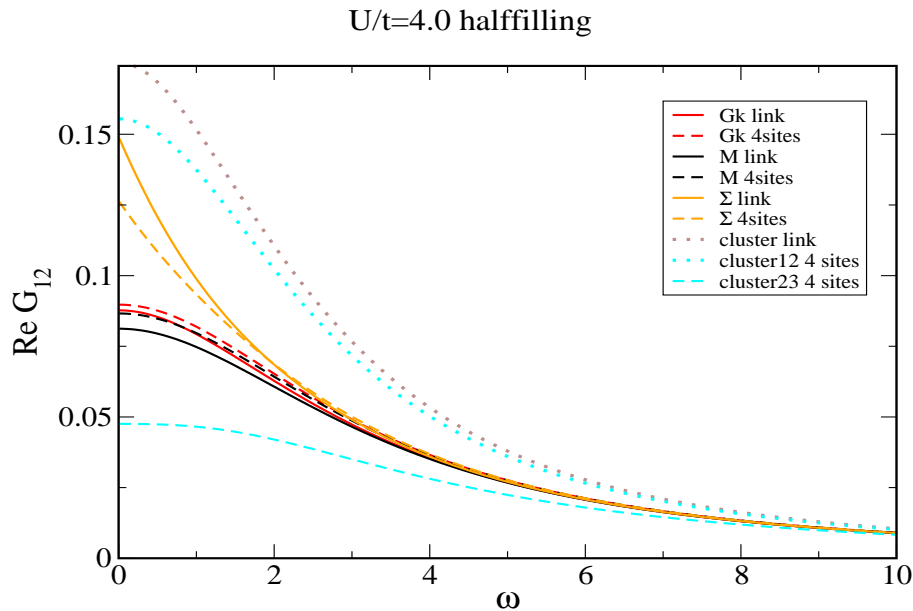}
\caption{Real part of the nearest neighbor propagator
$G_{12}(\omega)$ for a 1D case at half-filling calculated with the
different schemes and for increasing cluster size $N_{c}=2$ to
$N_{c}=4$. Also $G_{12}$ calculated directly within the cluster on
the ground-state of the AIM is displayed. The energy integral is
equal to the kinetic energy.} \label{ReG12_1D}
\end{center}
\end{figure}

\begin{figure}[!htb]
\begin{center}
\includegraphics[width=8.0cm,height=8.0cm,angle=-0] {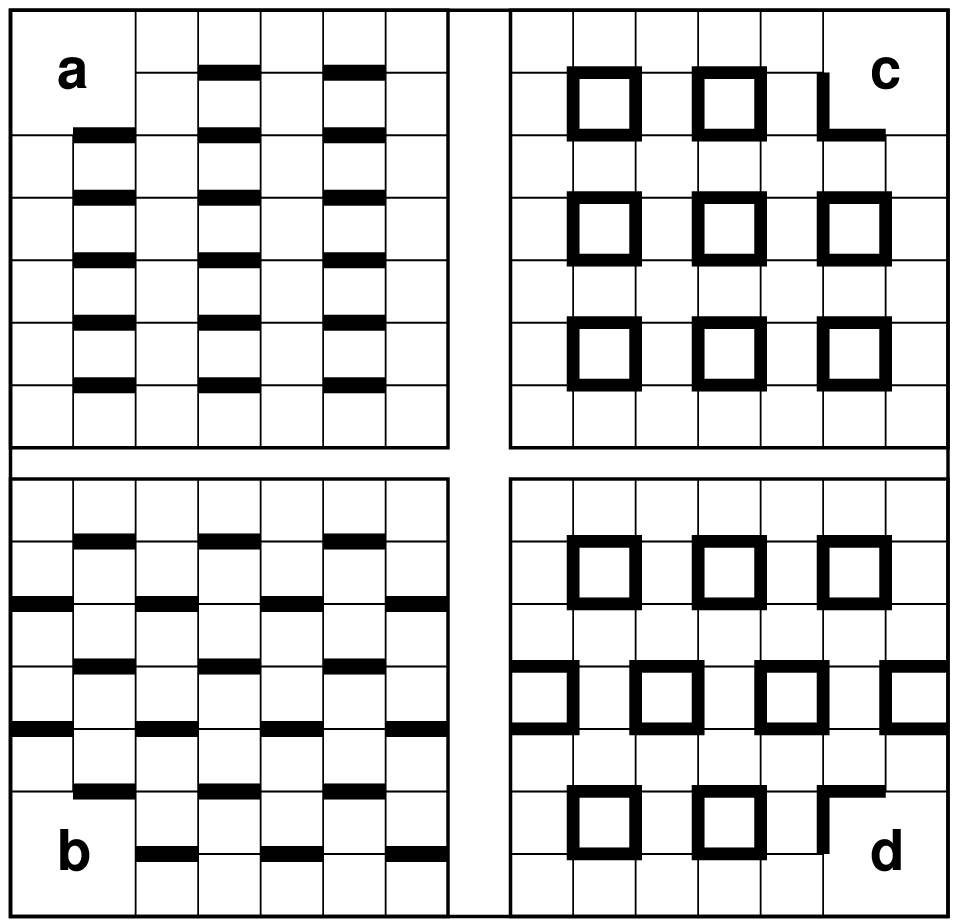}
\caption{Possible choices for a CDMFT super-lattice starting from
a two-dimensional square lattice and a two-site unit cell (a, b),
or a four-site unit cell (plaquette) (c, d). In the auxiliary
super-lattice problem the hopping is the same on all the links
while the non-local irreducible cumulants or the self-energies
are equal to the cluster values for the links marked with thick
lines and zero otherwise.} \label{cluster-size}
\end{center}
\end{figure}
CDMFT breaks the translational invariance of the system as soon as
a cluster shape and size have been chosen. The translational
invariance is restored only at the end, once the solution has been
determined by imposing the CDMFT self-consistency condition on the
cluster-impurity, and lattice quantities are extracted through one
of the periodizing schemes presented above. It is therefore
important to understand the extend to which the physical results
are dependent on the choice of the cluster and on the way lattice
is parted by these clusters. We start addressing the issue of the
dependence of the result on the number of sites in the cluster
$Nc$, and we start to consider the 1D case as we did in previous
sections. Even in this case, however, it is difficult to study in
a systematic way the dependence of static quantities, e.g. the
density versus the chemical potential, using ED. In the Lanczos
for example the total number of sites $Ns$ is limited by computer
power, and increasing the number of sites in the cluster $Nc$
reduces the number of sites available for the bath $Nb$. For
example, looking only at the density of the system $n$ for fixed
chemical potential $\mu$ we notice that a $Nc=4$ and $Nb=8$ run
reproduces the same values of a $Nc=2$ and $Nb=8$ run. In a $Nc=4$
and $Nb=8$ chain, the result is indeed an un-homogeneous system,
with the 2 central sites having different densities than the
external ones (see upper Fig.\ref{1Dchain}). If we look at the
values of the hybridizing hoppings $V_{\mu k}$, which connect the
sites of the chain with the bath, we notice that the central ones
are systematically sent to zero, i.e. the central chain chemical
potential and hopping are not renormalized by the bath.
Nevertheless, if confronted with the true BA value, the density of
the 2 central sites is not the best approximation: the best result
is obtained averaging over the cluster sites. We may think that
the boundary effect of the bath not only affects the most external
sites but propagates inside the cluster, and that only with a
larger cluster than the one we consider with ED ($Nc=4$) it is
possible to reach the bulk limit of the density (a qualitative
picture is shown in the lower line of Fig.\ref{1Dchain}, we refer
for a more systematic study to \cite{Biroli:2005}, and for a
similar picture obtained on larger cluster with QMC-CDMFT
\cite{bumsoo06}, in particular Fig. 6). In spite the local
(density) quantity does not seem to improve with a the $Nc=4$
cluster size in growing from $Nc=2$ to $Nc=4$, a real improvement
in the nearest neighbor quantities takes place, as shown in Fig.
\ref{ReG12_1D}. Here we display the real part of the nearest
neighbor Green's function as a function of the frequency (the
integral of this function is the lattice kinetic energy of the
Hubbard Model). This run has been obtained for strong coupling
$U/t=8$ at half-filling. The lattice Green's functions have been
obtained with the different periodizing schemes the $G$-method
(red line), the cumulant $M$-method (black line) and the
$\Sigma$-method (orange line). A comparison is displayed with the
link cluster ($Nc=2$) and the 4-sites ($Nc=4$), $Nb=8$ in both
cases. The nearest neighbor Green's function derived directly from
the cluster is also displayed for the link (brown dotted line) and
for the 4-sites (light blue lines), labelled the latter $G_{12}$
and $G_{23}$ in a 4 sites cluster 1-2-3-4. We can observe first of
that the cluster Greens' functions 1-2 and 2-3 are strongly
in-homogeneous, far from the bulk limit. The averages for any
scheme, however, are already better and lie between the 2 cluster
value. Also we see that all 3 periodizing methods move toward a
common curve for increasing cluster size $Nc$. As expected the
$M$-method and the $G$-method are very close with respect to the
$\Sigma$-method, and look to reach very quickly the same curve.
Also they display a similar qualitative behaviour in the shape of
the curve with respect to the $\Sigma$-method, indicating they
describe the same physics. So even if the results for $Nc=4$ are
still far from the their bulk values, the average represent a good
approximation which converges for increasing $Nc$ (as shown in
Fig. \ref{ReG12_1D} by the small change in the average-curve of
$\Re G_{12}$ in going from $Nc=2$ to $Nc=4$ in all the periodizing
schemes).

We now investigate how the solutions may be affected by the
partition into cluster of the original lattice. In fact, given a
cluster-impurity there are many possible ways the clusters can be
arranged on the lattice (see for example panels (a), (b) and (c),
(d) of Fig. \ref{cluster-size} for a two-dimensional square
lattice). A clear understanding of this feature may be helpful
determine the minimal dimension of the impurity-cluster that has
to be considered in order to properly describe the local physics.
Reasonably, if the cluster is chosen big enough the solution
should be weakly dependent on its shape and on the way the lattice
is parted. We investigate this issue with some practical example
in the two-dimensional square lattice of Fig. \ref{cluster-size},
where we use two different cluster sizes: a two-site link with two
different partitions of the lattice (panels (a) and (b)), and a
2X2 plaquette, which also results in two possible different
partitions of the lattice (panels (c) and (d)).
\begin{figure}[!htb]
\begin{center}
\includegraphics[width=10.0cm,height=8.0cm,angle=-0] {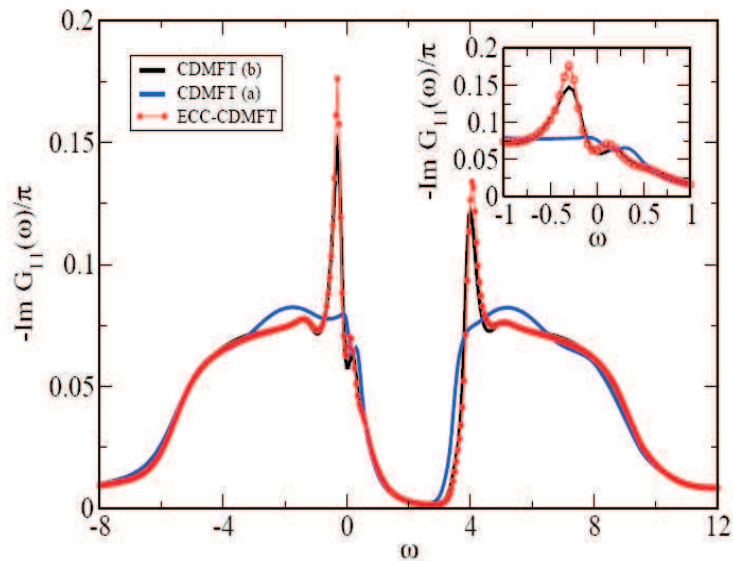}
\caption{Comparison between the on-site spectral functions for the
2D Hubbard model obtained using: 1) CDMFT on a super-lattice of
type ``a'' (blue curve); 2) CDMFT on a super-lattice of type ``b''
(black curve); 3) ECC-DMFT on a link (red circles), which
preserves translational invariance by construction. The
calculations were done for a  model with a hopping matrix element
t and on-site interaction U=8t at  a temperature T=0.15t and
filling n=0.95 using NCA for the impurity solver. A detail of the
low-energy behavior is shown in the inset.} \label{link}
\end{center}
\end{figure}
\begin{figure}[!htbp]
\begin{center}
\includegraphics[width=10.0cm,height=8.0cm,angle=-0] {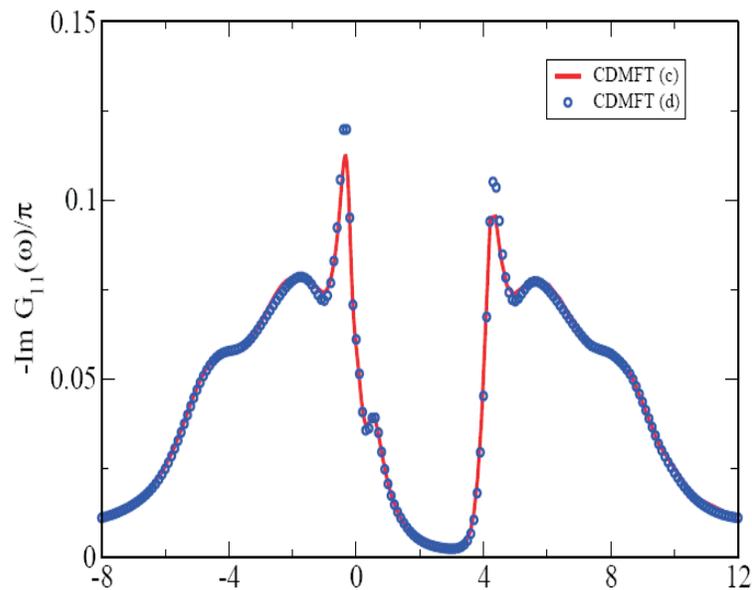}
\caption{Comparison between the on-site spectral functions for
the 2D Hubbard model obtained using  CDMFT on a super-lattice of
type ``c'' (red curve) and CDMFT on a super-lattice of type ``d''
(blue circles). The parameters for this calculation are U=8t,
T=0.15t,  n=0.95 and we used an NCA impurity solver.}
\label{plaquette}
\end{center}
\end{figure}
\begin{figure}[!p]
\begin{center}
\includegraphics[width=8.0cm,height=6.0cm,angle=-0] {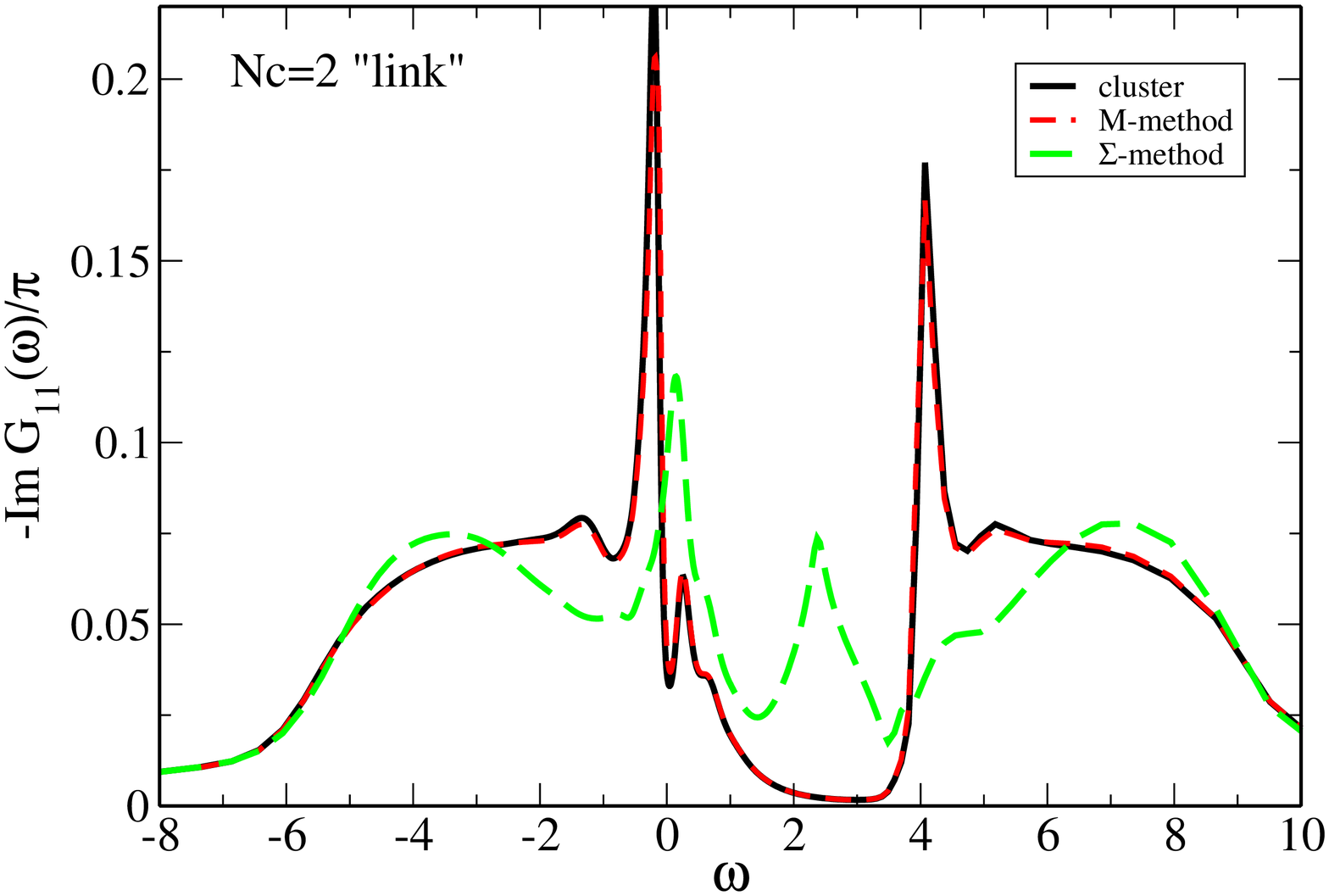}
\includegraphics[width=8.0cm,height=6.0cm,angle=-0] {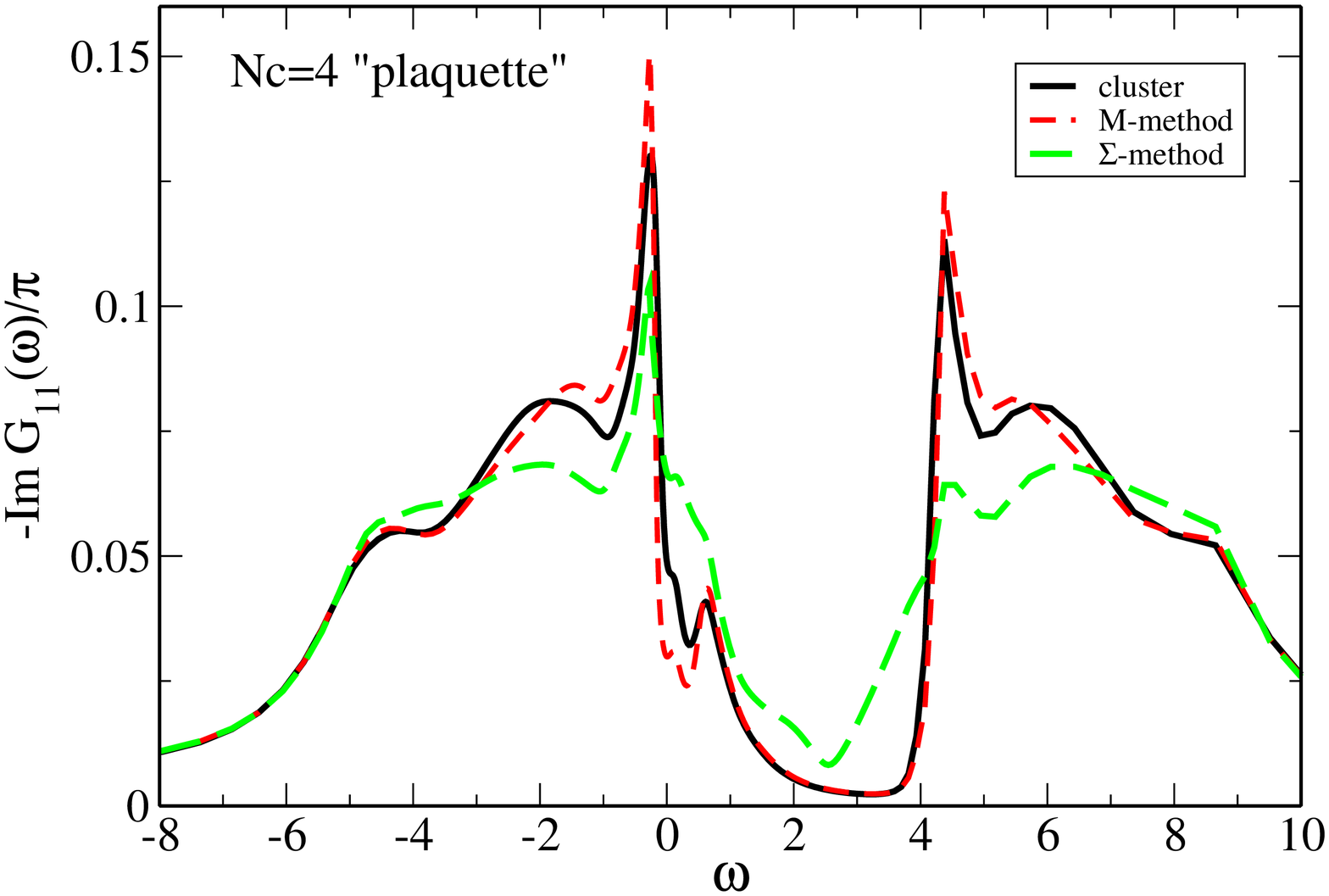}
\caption{The local DOS is compared between the $\Sigma$ and
$M$-schemes. The cluster DOS (black line) is also displayed. The
on-site interaction in the system is $U/t=8.0$, the doping is at
5\%. The impurity problem has been solved by NCA at a temperature
$T= 0.1t$ (courtesy of T. Stanescu and C. Haule). Notice the
Hubbard bands structure is preserved in all cases. The left hand
picture is a two-site link arrange as in configuration
\ref{cluster-size}b), the right hand picture is for a 2X2
plaquette arranged according to partition \ref{cluster-size}d).}
\label{Link-plaquette}
\end{center}
\end{figure}
\begin{figure}[!p]
\begin{center}
\includegraphics[width=8.0cm,height=6.0cm,angle=-0] {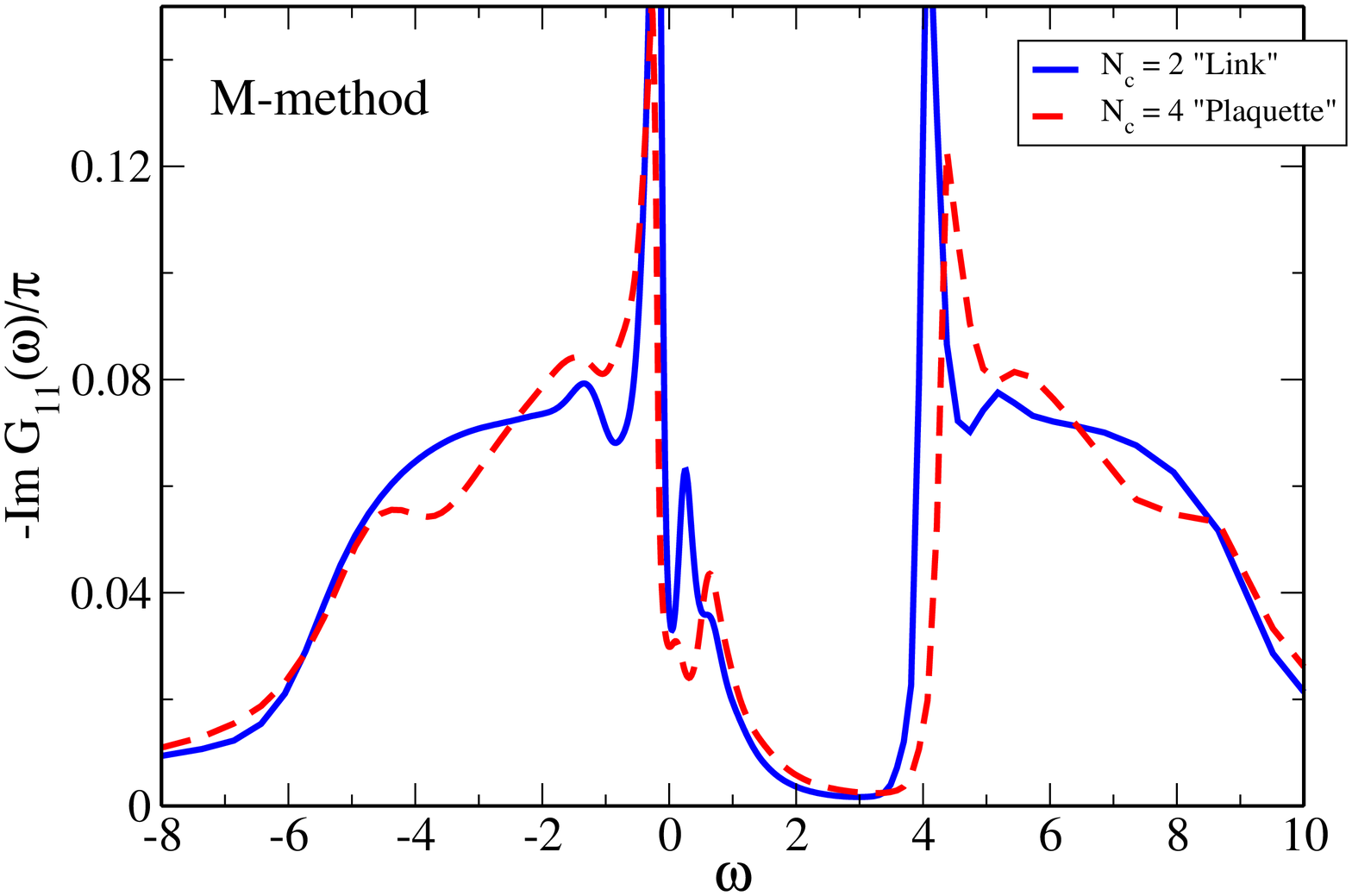}
\includegraphics[width=8.0cm,height=6.0cm,angle=-0] {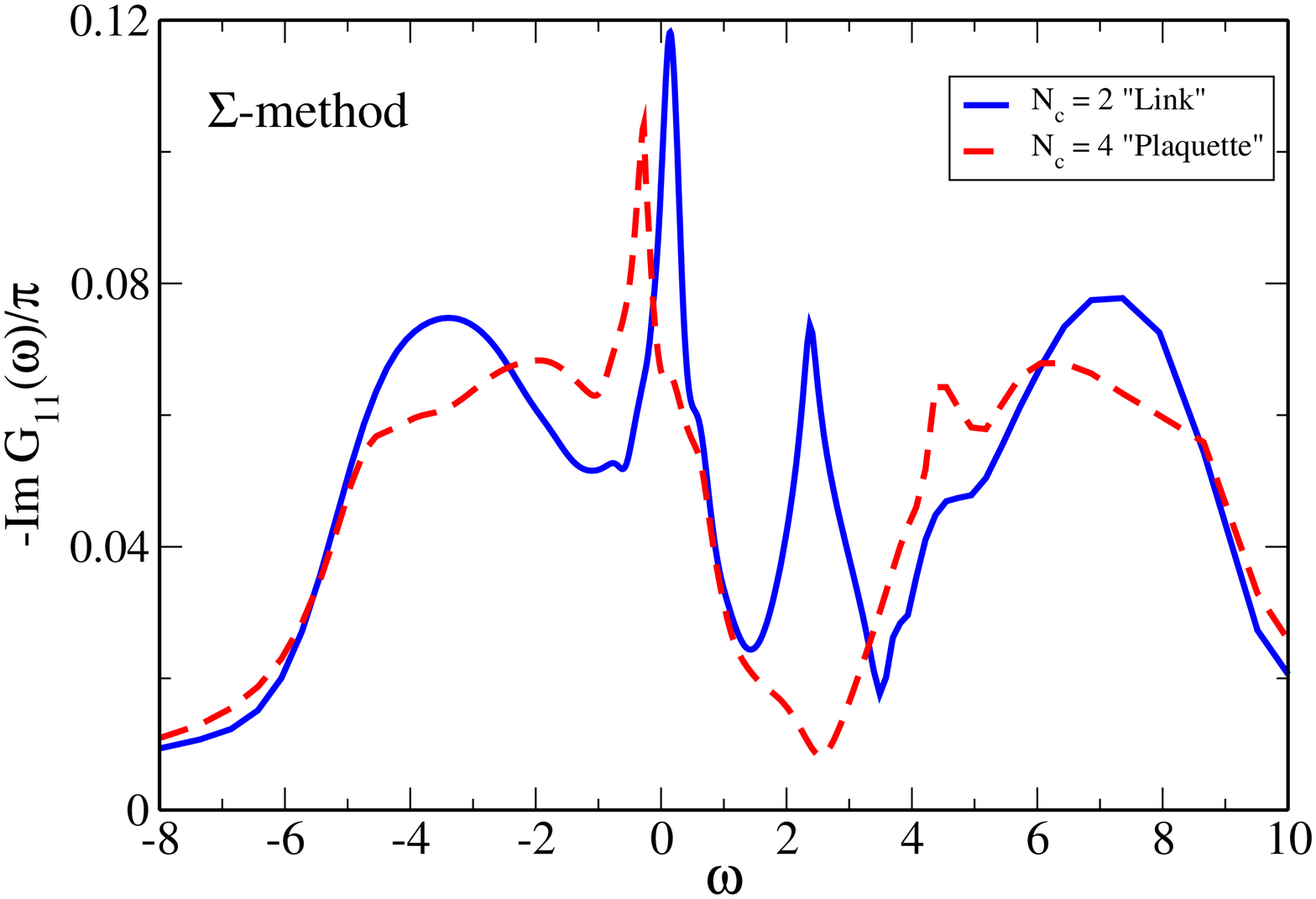}
\caption{Same as in Fig. \ref{Link-plaquette}, but the evolution
of the DOS from the two-site link to the four-site plaquette is
evidenced for the $M$-scheme (left-hand side) and the
$\Sigma$-scheme (right-hand side). Notice the more detailed
feature at the Fermi level $\omega=0$ of the plaquette calculation
in comparison with the link. For the $\Sigma$-scheme (right-hand
side) less sates accupy the Mott gap part of the spectrum.}
\label{Nc2-4MS}
\end{center}
\end{figure}
\begin{figure}[!tbhp]
\begin{center}
\includegraphics[width=8.0cm,height=6.0cm,angle=-0] {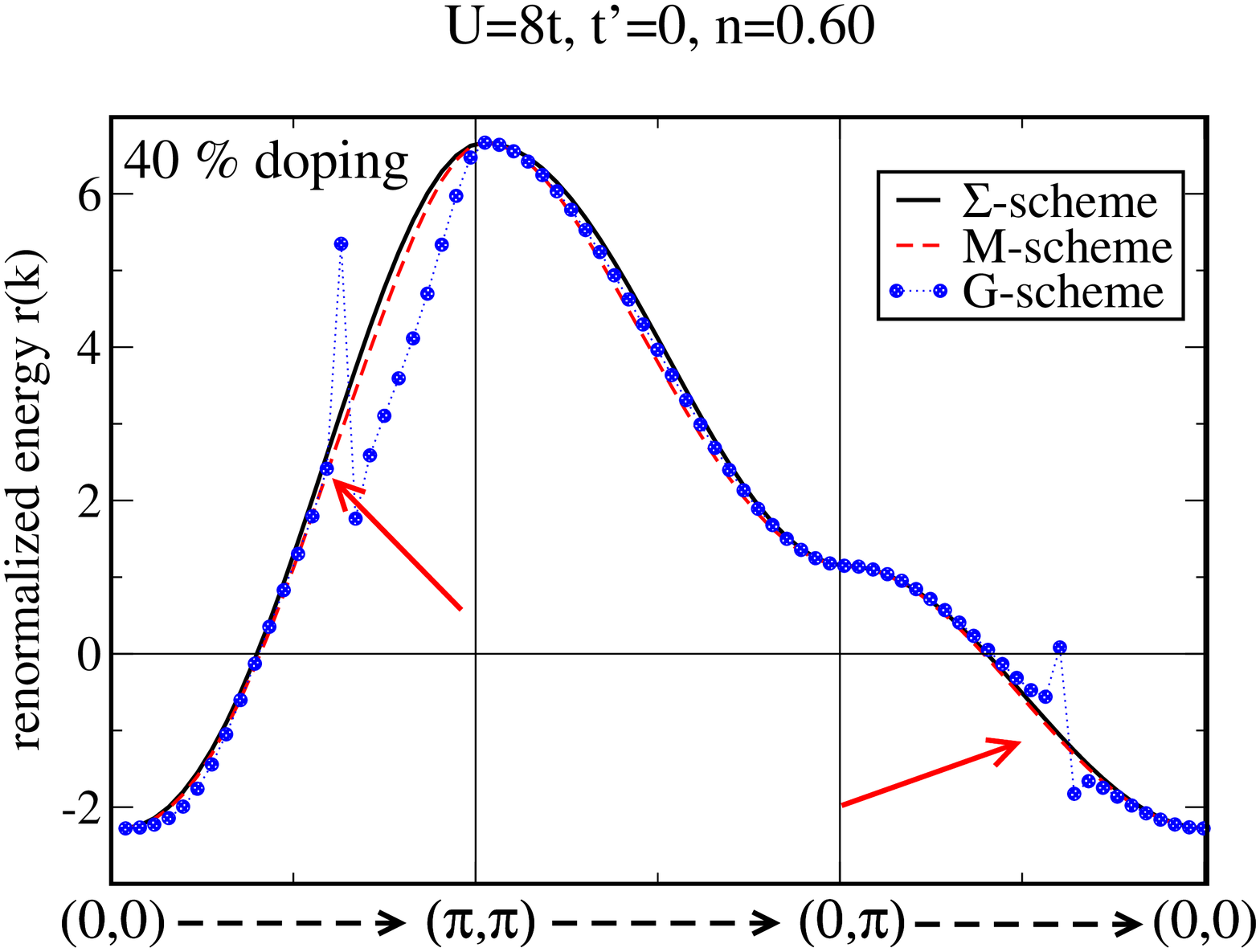}
\caption{Renormalized energy $r(k)= \varepsilon_k+\mu-\Re
\Sigma_k$ calculated with the three periodizing schemes:
$\Sigma$, $M$ and $G$. The quantity is displayed along the path
$(0,0)\rightarrow (\pi,\pi)\rightarrow (0,\pi)\rightarrow (0,0)$
of the first quadrant of the BZ. The system is highly doped
(40\%), and it is in a FL state. The point of zeroes in $r(k)$
correspond to the quasiparticle-poles in the lattice Green's
function. Notice that, in the case of the $G$-scheme, also
singular points appear in $r(k)$ (indicated by the red arrows)
corresponding to zeroes in the Green's function. }
\label{rk_Gk-zeroes}
\end{center}
\end{figure}
The CDMFT results for the local spectral functions in the two
link-geometries (a) and (b) are represented in Fig. (\ref{link})
by the blue line (for super-lattice ``a'') and the black line (for
super-lattice ``b''). We can clearly see that the two results are
qualitatively different, especially in the regions near the Mott
gap edges. A possible way to decide which of the two is closer to
the real solution is to confront them with ECC-DMFT \cite{tudor},
an improvement of PCDMFT based on the periodization of the
cumulant instead of the self-energy, which is independent of the
geometry of the super-lattice due to its built-in translational
invariance. The corresponding values for the local spectral
function are also shown in Fig. (\ref{link}) (red line with
circles). The solid agreement between the ECC-DMFT result and the
black curve (super-lattice ``b'') strongly suggests that the
partition (\ref{cluster-size}b) is the best choice for this case.
The qualitative difference between the solution based on the
super-lattice (\ref{cluster-size}a) and (\ref{cluster-size}b)
indicates that in CDMFT the physics of a two-dimensional Hubbard
model may not well captured  by a two-site cluster approximation.
This problem should in principle be cured by increasing the
cluster size. In Fig. (\ref{cluster-size} - c,d) we show two
possible super-lattices having a plaquette as the unit cell. The
corresponding spectral functions are represented in Fig.
(\ref{plaquette}) by two almost identical curves (the red line and
the blue circles correspond to the super-lattices
(\ref{cluster-size}c) and (\ref{cluster-size}d), respectively).
This example shows that the implementation of a CDMFT-type
approach depends on the partition of the super-lattice. If
different geometries give different results, the ambiguity can be
possibly eliminated by increasing the cluster size (in the case
where this is not possible due to technical limitations, a
possible way to discriminate is to use the ECC-DMFT scheme for the
same cluster size \cite{tudor}). Finally, but most importantly,
the fact that a plaquette-cluster in the 2D Hubbard Model is
independent of the choice of the super-lattice, suggests that a
four-site cluster approximation is able to capture the basic
physics of the model.

In Fig.s \ref{Link-plaquette} we show how the DOS calculated with
the $\Sigma$ and $M$-schemes evolves with increasing cluster size
$N_{c}$ from the link configuration  (\ref{cluster-size}b) to the
plaquette configuration \ref{cluster-size}d). The on-site
interaction is $U/t=8.0$ and the system is at 5\% doping. The
impurity problem has been solved again with NCA, at a temperature
$T=0.1 t$. For comparison the DOS directly calculated from the
impurity-cluster is also displayed. It is clear how the $M$-scheme
and the latter DOS are always closer in comparison to the
$\Sigma$-scheme DOS. Notice the more detailed feature appearing at
the Fermi energy $\omega=0$ in the plaquette calculation of these
two curves. The $\Sigma$-scheme presents always some spurious
states in the Mott-gap, and a less refined structure at the FS.
However the weight of the spurious states is less in the plaquette
calculation than in the link. We do expect the two methods
$\Sigma$ and $M$ to converge to the same result for increasing
$N_{c}$. But the $M$-scheme seems to be able to reproduce local
quantities better for small cluster (at least in the insulating
state and possibly in the metallic region that precedes the MT,
as explained in Fig.\ref{FillingU82d}).  

As we previously mentioned, the third scheme proposed, the
$G$-scheme, produces results similar to the $M$-scheme in the
low-doping region of the parameter space. In fact it is by
construction exact in re-building the local cluster-quantity $\Im
G_{11}(\omega_n)$. This scheme, however, may have some
inconvenient in describing $k$-dependent properties of the
high-doped region, where the system is a simple FL. In Fig.
\ref{rk_Gk-zeroes} we show for a highly doped case ($40 \%$
doping) the renormalized quasiparticle spectrum $r(k)=
\varepsilon_k+\mu-\Re \Sigma_k$ along the path $(0,0)\rightarrow
(\pi,\pi)\rightarrow (0,\pi)\rightarrow (0,0)$ of the first
quadrant of the BZ. The quasiparticle poles in the Green's
function $G(k,\omega)=\, \frac{1} {\omega +r(k)+ \imath\delta}$
(with $\Im\Sigma_k\rightarrow 0$) correspond to zeroes of $r(k)$,
as expected from a FL viewpoint. In Fig. \ref{rk_Gk-zeroes} we see
that for all the three periodizing schemes there are two points
where quasiparticles are formed, corresponding to the crossing of
the FS with the path in $k$-space considered. In the $G$-scheme,
however, we observe the appearance of points of {\it infinities}
in $r(k)$ (red arrows in Fig. \ref{rk_Gk-zeroes}), i.e. zeroes in
the Green's functions. In spite this is a real possible property
in a strongly interacting many-body system, as we will see in
detail in the next chapter studying the approach to the Mott
transition, in this case it is probably an artifact of the
$G$-scheme, as the system is expected to be a simple FL for this
value of doping. Therefore the $G$-scheme shows to introduce
spurious zeroes in the Green's function, not interpretable as
physical properties of the system.

\section{Reduced bath-parametrization for the Exact Diagonalization procedure}

We want in this section to present an operative method for search
out the solution with the ED-CDMDT method. In this procedure a
most delicate point is represented by the minimization step that
determines the best finite set of bath parameters
$(\varepsilon_{k},V_{\mu k})$ describing the continuous
hybridization function $\Delta_{\mu\nu}(\omega)$, born out of the
self-consistent CDMFT-loop. The minimization searches in the
$(\varepsilon_{k},V_{\mu k})$-parameter landscape for the minimum
of the distance function $f=
|\hat{\Delta(\omega)}-\hat{\Delta_{ns}(\omega)}|/\omega$, and
re-inputs the values it finds into the CDMFT-loop. The solution is
therefore determined by a compromise between optimal approximation
of $\Delta_{\mu\nu}(\omega)$, given by a minimum of $f$, and
fulfillment of the CDMFT-self-consistency. There is always the
risk of finding and remaining trapped in a minimum of $f$ which at
some numerical degree of precision satisfies the CDMFT-equations
but that is not physically meaningful. In order to drive the
solutions towards physically interesting regions of the
bath-parameters space, we have introduced a reduced
bath-parametrization which allows to exploit the symmetries in the
square lattice to gain a better physical insight of the Green's
function properties we look for. Moreover using fewer parameters
the work required by the minimization procedure is faster and the
result simpler to interpret.
\begin{figure}[!htbp]
\begin{center}
{\Large {\bf Free parametrization }   }
\includegraphics[width=11.0cm,height=7.0cm,angle=-0]{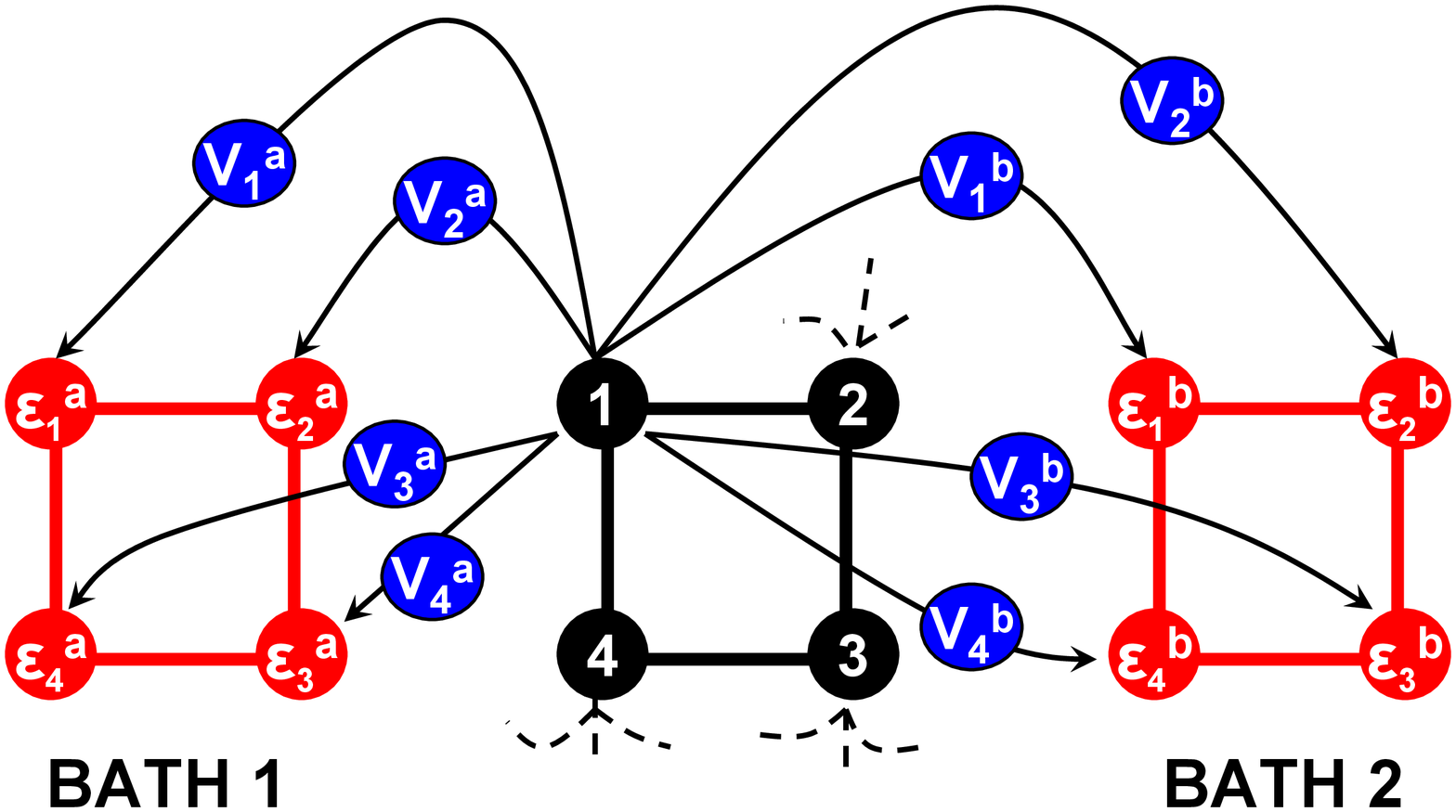}\vspace{1cm}\\
{\Large {\bf Constrained parametrization }  }
\includegraphics[width=11.0cm,height=7.0cm,angle=-0]{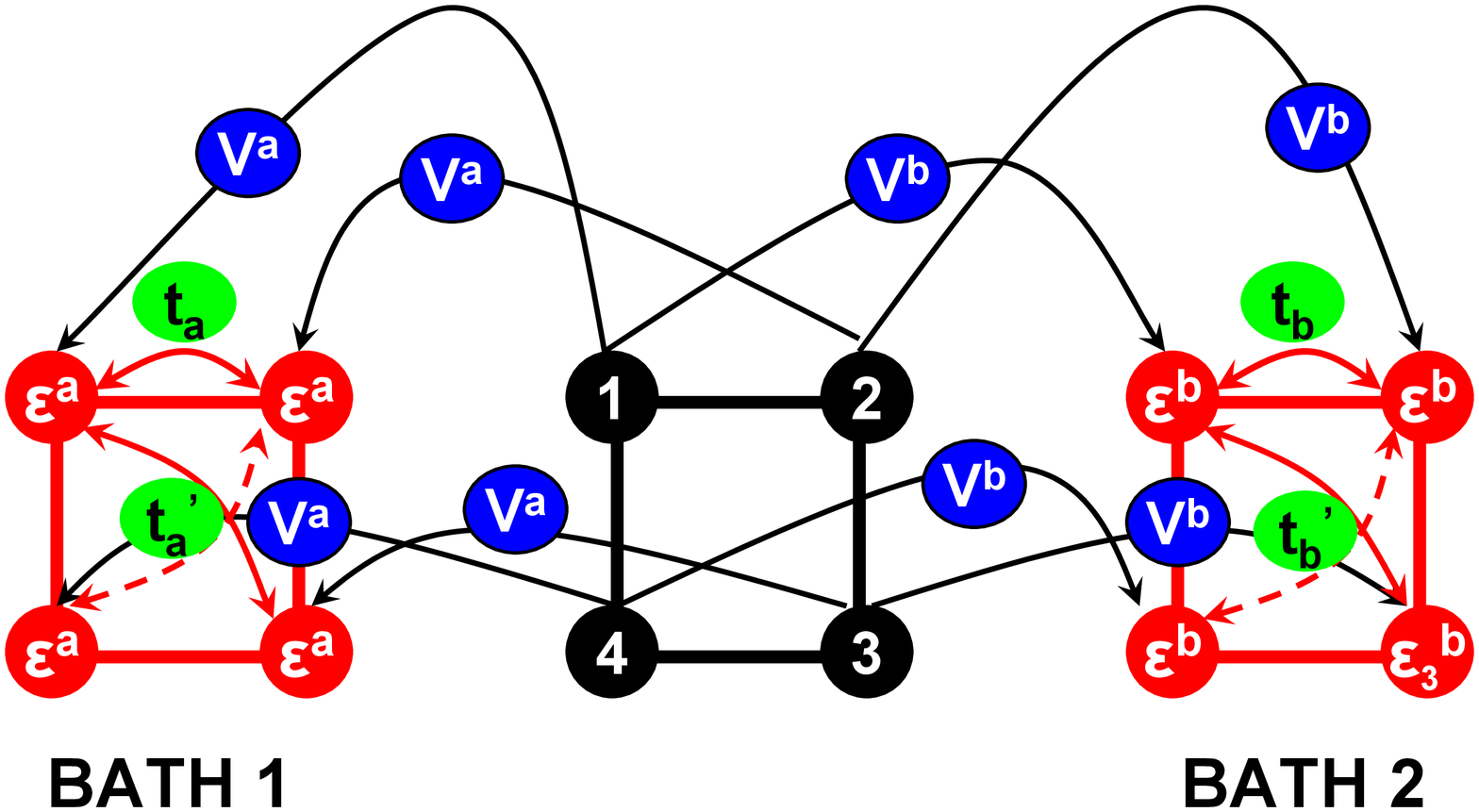}
\caption{Schematic drawing of a free bath-parametrization (upper
raw) and the constrained bath-paramerrization (lower row). }
\label{Multibath}
\end{center}
\end{figure}
The most general parametrization of the bath is given in the
Hamiltonian by
\begin{eqnarray}
\sum_{k}^{n_{b}}\sum_{\sigma }\ \varepsilon \,_{k\sigma }\
a_{k\sigma }^{+} a_{k\sigma }+\sum_{k}^{n_{b}} \sum_{\mu \sigma }\
V_{k\mu \sigma }\ a_{k\sigma }^{+}  c_{\mu \sigma}  +V_{k\mu
\sigma }^{\ast }\ c_{\mu \sigma }^{+} a_{k\sigma }
\label{BathHamil}
\end{eqnarray}
here as usual $c_{\mu \sigma }$ is the destruction operator of an
electron with spin $\sigma$ on site $\mu=1...4$ of the
cluster-impurity, $a_{k\sigma }$ is the destruction operator of an
electron in the site $k=1,..,N_{b}$ of the bath, $\varepsilon
\,_{k\sigma }$ are the level energies of the bath and $V_{k\mu
\sigma }$ the hybridazing couplings between cluster and bath. A
sketch of this bath is shown in the top of Fig. (\ref{Multibath}),
where, for convenience's sake, only the hybridization of the
cluster-site $\mu=1$ with 8 bath sites (also arranged in two 2X2
plaquette for convenience's sake) is shown. An electron can jump
from site $\mu=1$ to every site of the bath, which in general has
an energy $\varepsilon_{k}$ different from the other bath-sites,
with a hopping parameter $V_{\mu k}$ which is also different for
different bath-site. In the lower row of Fig. (\ref{Multibath}) we
show the constrained bath-parametrization. The bath-sites are
grouped in 4-site plaquettes to mimic the original lattice
structure parted by the plaquette-cluster (in the picture we show
the two plaquette-bath on the left-hand side and right-hand side
respectively ). For each of this bath-plaquette we impose the same
bath-energy $\varepsilon_j$ ($j=a,b$ for different
sub-bath-plaquette), to respect a translational invariance inside
the bath-plaquette, and we allow for hopping of electrons inside
the bath-plaquette introducing a nearest neighbor hopping $t_{j}$
($j=a,b$) and next-nearest neighbor hopping $t'_{j}$ ($j=a,b$). An
electron in a cluster-site can hop to the corresponding
bath-plaquette-site with hopping parameter $V_j$ ($j=a,b$), which
is the same for all the cluster-site and it is only
plaquette-dependent. The reduction in the number of parameters is
clear. Considering the simple example displayed in Fig.
(\ref{Multibath}), with $N_{c}=4$ and $N_{b}=8$, a free
bath-parametrization requires a search for 2X$N_{c}$X$N_{b}= 64$
hybridazation parameters $V_{k\mu}$ plus 2X$N_{b}= 16$ bath
energies $\varepsilon \,_{k\sigma }$, giving a total of $80$
parameters to optimize (the factor 2 is for the spin). In the case
of the constrained parametrization, instead, we have 2 hybridazing
$V_j$, one for each sub-plaquette-bath, 2 bath energies
$\varepsilon_j$, 2 bath hopping terms $t_{j}$ and 2 $t'_{j}$
($j=a,b$), in total 8 parameters only! It is now much easier to
associate some physical property of the system to some specific
set of parameters which can be turned on or off, according to the
kind of solution sought. But this is not the only advantage: the
bath-plaquette constructed to be "copies" of the cluster-impurity
do respect the square symmetry of the original lattice model,
producing Weiss fields ${\cal G}_{o}$ which automatically respect
these symmetries. This property garantess obtaining results
physically meaningful and also helps in determining good starting
seeds for a full free-parameters DMFT calculation. The risk with
too many parameters is that one ends up with a solution which,
although of a small distance and respecting (in the limits of
numerical precision) the CDMFT-self-consistency condition, breaks
the physical symmetries expected for the solution.

It is easily shown how the constrained parametrization is included
in the most general expression (\ref{BathHamil}). To this purpose
it is convenient to introduce spinor notation in the example
presented above with $N_{c}=4$ and $N_{b}=8$:
\begin{eqnarray}
\psi_{j\sigma}= \, (c^j_{1\sigma},...c^j_{4\sigma})\\ \nonumber
 \phi_{j\sigma}= \, (a^j_{1\sigma},...a^j_{4\sigma}) \label{spinor}
\end{eqnarray}
where $j=a,b$ indicates the 2 sub-bath-plaquettes, and write the
bath (formula \ref{BathHamil}):
\begin{eqnarray}
\sum_{j=a,b} \phi^{\dagger}_{j\sigma} \, \mathbf{E}_{j\sigma} \,
\phi_{j\sigma}+\, \phi^{\dagger}_{j\sigma} \, \mathbf{V}_{j\sigma}
\, \psi_{j\sigma}+\,\psi^{\dagger}_{j\sigma} \,
\mathbf{V}^{\dagger}_{j\sigma} \, \phi_{j\sigma}
\label{BathHaml-matrix}
\end{eqnarray}
where $\mathbf{E}_{j}$ is a matrix containing the bath energies in
the free parametrization (for which is diagonal) and also the
bath-hopping in the constrained parametrization:

\begin{eqnarray*}\label{Ebath}
\mathbf{E}_{j\sigma}= &\begin{pmatrix}
\varepsilon_{1j\sigma}& 0                     & 0                     & 0 \\
                    0& \varepsilon_{2j\sigma} & 0                     & 0 \\
                    0& 0                     & \varepsilon_{3j\sigma} & 0 \\
                    0& 0                     & 0                     & \varepsilon_{4j\sigma}
\end{pmatrix} & \qquad \hbox{ free paramerization }\\
                      &                & \\
                      &                & \\
\mathbf{E}^{'}_{j\sigma}= &\begin{pmatrix}
\varepsilon_{j\sigma}    & t_{j}                     & t'_{j}                & t_{j} \\
t_{j}                    & \varepsilon_{j\sigma}     & t_{j}                 & t'_{j} \\
t'_{j}                   & t_{j}                     & \varepsilon_{j\sigma} & t_{j} \\
t_{j}                   & t'_{j}                     &t_{j}
&\varepsilon_{j\sigma}
\end{pmatrix} & \qquad \hbox{ cons. paramerization }\\
\end{eqnarray*}
while in general for the free parametrization the hybridazing
hopping $\mathbf{V}_{j}$ is

\begin{equation*}\label{Vbath} \mathbf{V}_{j\sigma}= \begin{pmatrix}
  V^j_{11} & V^j_{12} & V^j_{13} & V^j_{14} \\
  V^j_{21} & V^j_{22} & V^j_{23} & V^j_{24} \\
  V^j_{31} & V^j_{32} & V^j_{33} & V^j_{34} \\
  V^j_{41} & V^j_{42} & V^j_{43} & V^j_{44}
\end{pmatrix}
\end{equation*}
in the constrained it is simply diagonal

\begin{equation*}\label{Vbath_constr}
\mathbf{V}'_{j\sigma}= \begin{pmatrix}
  V_{j} & 0    & 0    & 0 \\
  0    & V_{j} & 0    & 0 \\
  0    & 0    & V_{j} & 0 \\
  0    & 0    & 0    & V_{j}
\end{pmatrix}
\end{equation*}
So to express the constrained parametrization in terms of the
free one it is enough to look for the congruence transformation
$\mathbf{S}$ ($\mathbf{S}\mathbf{S}^{T}= \mathbf{1}$) which
diagonalizes $\mathbf{E}'_{j\sigma}$:

\begin{equation}\label{SEbSt}
\mathbf{E}_{j\sigma}= \mathbf{S}\, \mathbf{E}'_{j\sigma}\,
\mathbf{S}^{T}=\,\begin{pmatrix}
\varepsilon_{j\sigma}-t'_{j}& 0                     & 0                     & 0 \\
                    0& \varepsilon_{j\sigma}+t'_{j} & 0                     & 0 \\
                    0& 0                     & \varepsilon_{j\sigma}-\sqrt{t_{j}^{\prime 2}+4 t_{j}^{2}} & 0 \\
                    0& 0                     & 0                     &
                    \varepsilon_{j\sigma}+\sqrt{t_{j}^{\prime 2}+4 t_{j}^{2}}
\end{pmatrix}  \vspace{0.5cm}\\
\end{equation}
and clearly $\mathbf{V}_{j\sigma}$ is not diagonal anymore:

\begin{equation}\label{VbSt}
\mathbf{V}_{j\sigma}= \, \mathbf{V}'_{j\sigma}\,
\mathbf{S}^{T}=\,\begin{pmatrix}
0                 & -V_j/\sqrt{2}         & 0                     & V_j/\sqrt{2}  \\
-V_j/\sqrt{2}       & 0                   & V_j/\sqrt{2}            & 0             \\
-\Theta^{+} V_j/2\Lambda& t_{b}V_j/\Theta^{+}       & -\Theta^{+}
V_j/2\Lambda &
t_{j}V_j/\Theta^{+} \\
-\Theta^{-} V_j/2\Lambda& t_{j}V_j/\Theta^{-}      & -\Theta^{-}
V_j/2\Lambda & t_{j}V_j/\Theta^{-}
\end{pmatrix} \vspace{0.5cm}\\
\end{equation}
where we indicate $\Lambda= \sqrt{t_{j}^{\prime 2}+4 t_{j}^{2}}$,
$\, \Theta^{+(-)}= \,\sqrt{4t_{j}^{2}+ t'_{j}\,
(t'_{j}+(-)\Lambda)}$. Notice that in the case $t^{\prime}_{j}=0$
the expressions \ref{SEbSt} and \ref{VbSt} further simplify:
\begin{equation}\label{SEbSt2}
\mathbf{E}_{j\sigma}= \mathbf{S}\, \mathbf{E}'_{j\sigma}\,
\mathbf{S}^{T}=\,\begin{pmatrix}
\varepsilon_{j\sigma}& 0                     & 0                     & 0 \\
                    0& \varepsilon_{j\sigma} & 0                     & 0 \\
                    0& 0                     & \varepsilon_{j\sigma}- 2t_{j} & 0 \\
                    0& 0                     & 0                     &
                    \varepsilon_{j\sigma}+ 2t_{j}
\end{pmatrix}  \vspace{0.5cm}\\
\end{equation}
and
\begin{equation}\label{VbSt2}
\mathbf{V}_{j\sigma}= \, \mathbf{V}'_{j\sigma}\,
\mathbf{S}^{T}=\,\begin{pmatrix}
0                 & -V_j/\sqrt{2}         & 0                     & V_j/\sqrt{2}  \\
-V_j/\sqrt{2}       & 0                   & V_j/\sqrt{2}            & 0             \\
-V_j/2              & V_j/2               & -V_j/2                   &V_j/2\\
-V_j/2               & V_j/2               &  -V_j/2 &                V_j/2       \\
\end{pmatrix} \vspace{0.5cm}\\
\end{equation}
a degeneracy appears in the poles of each cluster bath, which are
symmetrically coupled to sites 1-3 and 2-4 of the
cluster-plaquette (first two row in the hybridizing matrix
\ref{VbSt2}).
This transformation also shows that the form (\ref{BathHamil}) is
indeed the most general we can adopt. A good operative procedure
for obtaining results is to start using the more controllable
constrained parametrization, and, starting from this solution and
exploiting the relations above (\ref{VbSt})(\ref{SEbSt}), relax
the parameters using the free parametrization. In this way we
first get close to a region of the parameter-space of physical
interest, and then we improve the result freeing all the degrees
of freedom we can.

\chapter{ Results in 2D }
\section{ Strong Coupling U/t=16 }
\begin{figure}[!htb]
\begin{center}
\includegraphics[width=8.0cm,height=3.0cm,angle=-0] {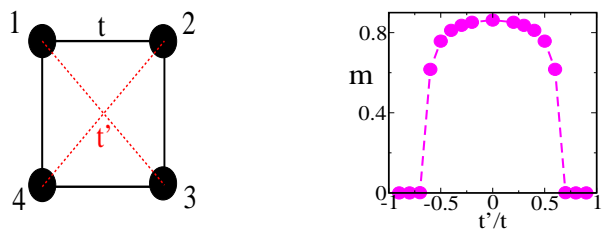}
\caption{Left side: the CDMFT plaquette. Right side: staggered
magnetization as a function of the next-nearest hopping
$t^{\prime}$ at half-filling. The parameter $t^{\prime}$ controls
the magnetic frustration in the system.} \label{Fcluster}
\end{center}
\end{figure}

In this chapter we apply CDMFT to the study of the MT in two
dimensions. Anticipating the cuprate material physics, we consider
once again the two-dimensional Hubbard model on the square
lattice:
\begin{eqnarray}
H = -\sum_{ i,j,\sigma} t_{ij}\, (c^{\dagger}_{i\sigma}
c_{j\sigma} + h.c.) + U \sum_i n_{i\uparrow}n_{i\downarrow}
-\mu\sum_i n_i \nonumber \label{hamiltonian}
\end{eqnarray}
where $c_{i\sigma}$ ($c^{\dagger}_{i\sigma}$) are destruction
(creation) operators for electrons of spin $\sigma$, $n_{i\sigma}
= c^{\dagger}_{i\sigma}c_{i\sigma}$ is the number operator, $U$ is
the on-site repulsion and $\mu$ the chemical potential which
determines the electron density $n=1/N\sum_{i,\sigma}\langle
n_{i\sigma}\rangle$ ($N$ being the number of sites). The hopping
amplitude $t_{ij}$ is limited to nearest-neighbors $t$ and to
next-nearest-neighbors $t^{\prime}$. We want to focus our
attention on the "Mottness". In the half-filled insulating case
and for intermediate values of the coupling $U$, a crossover takes
place from a AF-band insulator to a AF-Mott insulator for
increasing $U$. Our task is to study the doped AF-Mott state, so
we choose an on-site repulsion $U= 16t$ in order to be deep in the
strongly correlated regime. For the sake of convenience we also
confine the study to the hole-doped system ($n<1$) at different
levels of frustration controlled by the value of $t^{\prime}/t$.
This is equivalent to the electron-doped system ($n>1$) after a
particle-hole transformation which reverses the sign of
$t^{\prime}$. In particular we consider relatively small values of
$t^{\prime}/t = \pm 0.3$ suitable to describe the cuprates and a
large value $t^{\prime}/t =  0.9$, which, despite having no direct
correspondence with real materials, is able to completely destroy
any long-ranged AF ordering. (see the right panel of Fig.
\ref{Fcluster}, displaying the staggered magnetization $m=
1/N\sum_i (-1)^{i} (n_{i\uparrow}-n_{i\downarrow})$ at
half-filling as a function of $t^{\prime}/t$).
The quantum impurity model chosen for this case consists of a
$2\times 2$ plaquette (see Fig. \ref{Fcluster}), embedded in an
effective medium described by a self-consistently determined Weiss
function. We consider here a metallic phase which does not break
any symmetry and follow its evolution as a function of doping. The
choice of a plaquette-cluster is minimal in order to respect the
lattice square-symmetry and it allows describing various broken
symmetries, such as antiferromagnetism (AF) or $d$-wave
superconductivity. With CDMFT on a plaquette we can describe the
evolution of the electronic structure of the model in terms of
just a few (three is the present case) functions of frequency
which have a simple physical interpretation as parameterizing the
lattice self energy, and which show a systematic evolution towards
the Mott insulator (see Fig. \ref{fImS} and discussion below). It
can be considered as a dynamical generalization of the early slave
boson mean field theory \cite{grilli}\cite{liu} which is able to
treat both the coherent and the incoherent excitations
(quasiparticle peak and Hubbard bands) on the same footing,
capturing the short-range physics of singlet formation on bonds.

In order to perform an ED solution, the quantum impurity model is
truncated to a finite number (in this case 8) bath levels, whose
energies and hybridizations are self-consistently determined
through the minimization procedure. As we explained in chapter 3,
to implement the self-consistency condition, we need to introduce
Matsubara frequencies and hence an effective inverse temperature
$\beta$ which we set to $\beta=128$ in units of the half bandwidth
$4 t$. At low $\beta$ and relatively small $U$ our results are
qualitatively similar to those obtained solving the impurity by
QMC (chapter 4.2). Details on the implementation of ED within
CDMFT and a benchmark against the exact solution of the Hubbard
model in one dimension were presented in the previous chapter. As
already stressed, ED method's main limitations are the small
number of sites in the bath and the effective temperature, which
induces a limited energy resolution \cite{marce,bumsoo}. The small
size of the cluster induces a finite $k$ resolution : for the
$2\times 2$ plaquette with the square symmetry we only have two
independent directions in $k$ space (along the diagonal and along
the lattice axis).

\subsection{Cluster quantities}
\begin{figure}[!phtb]
\begin{center}
\includegraphics[width=10.0cm,height=10.0cm,angle=-0] {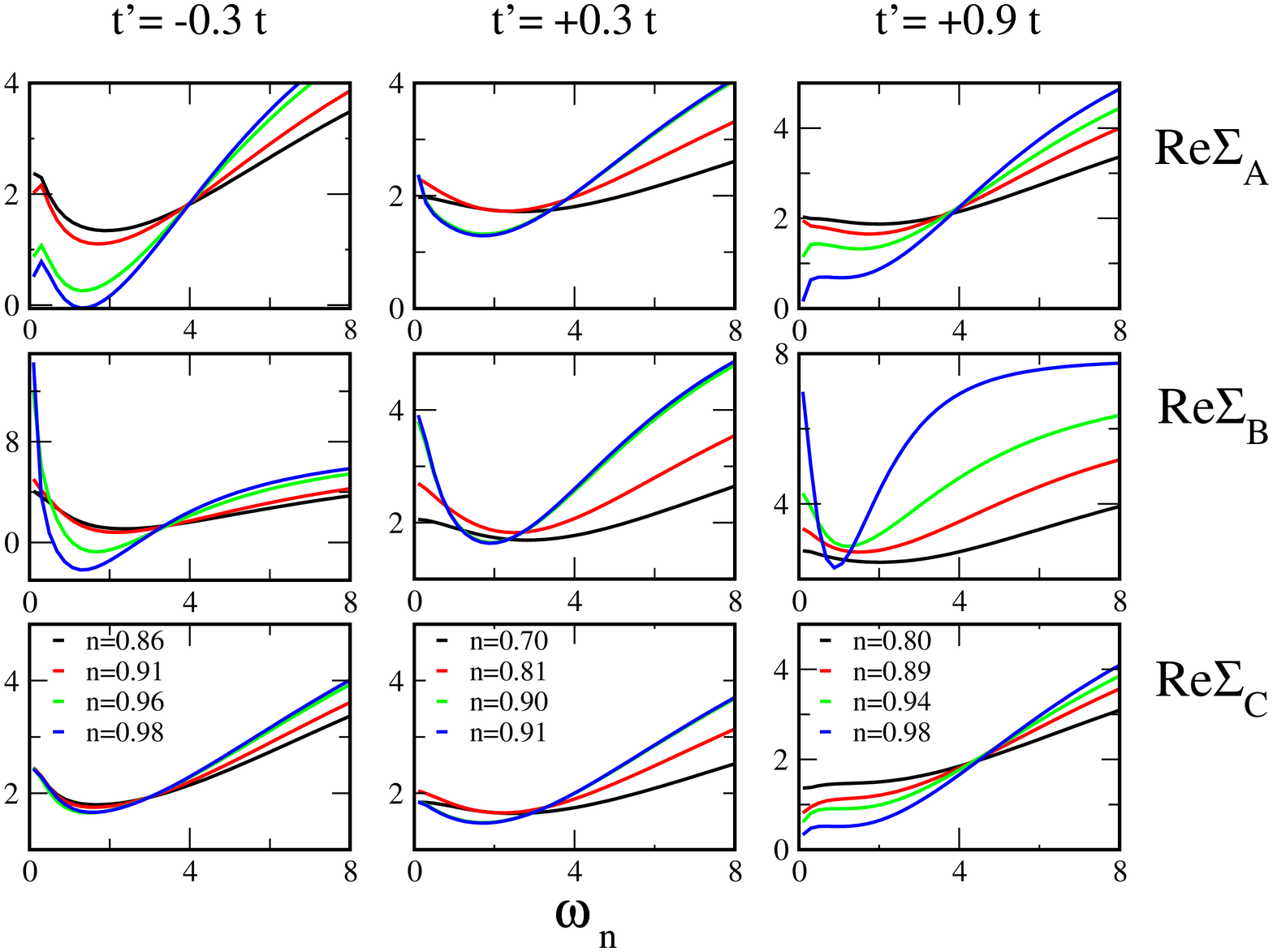}
\includegraphics[width=10cm,height=10cm,angle=-0] {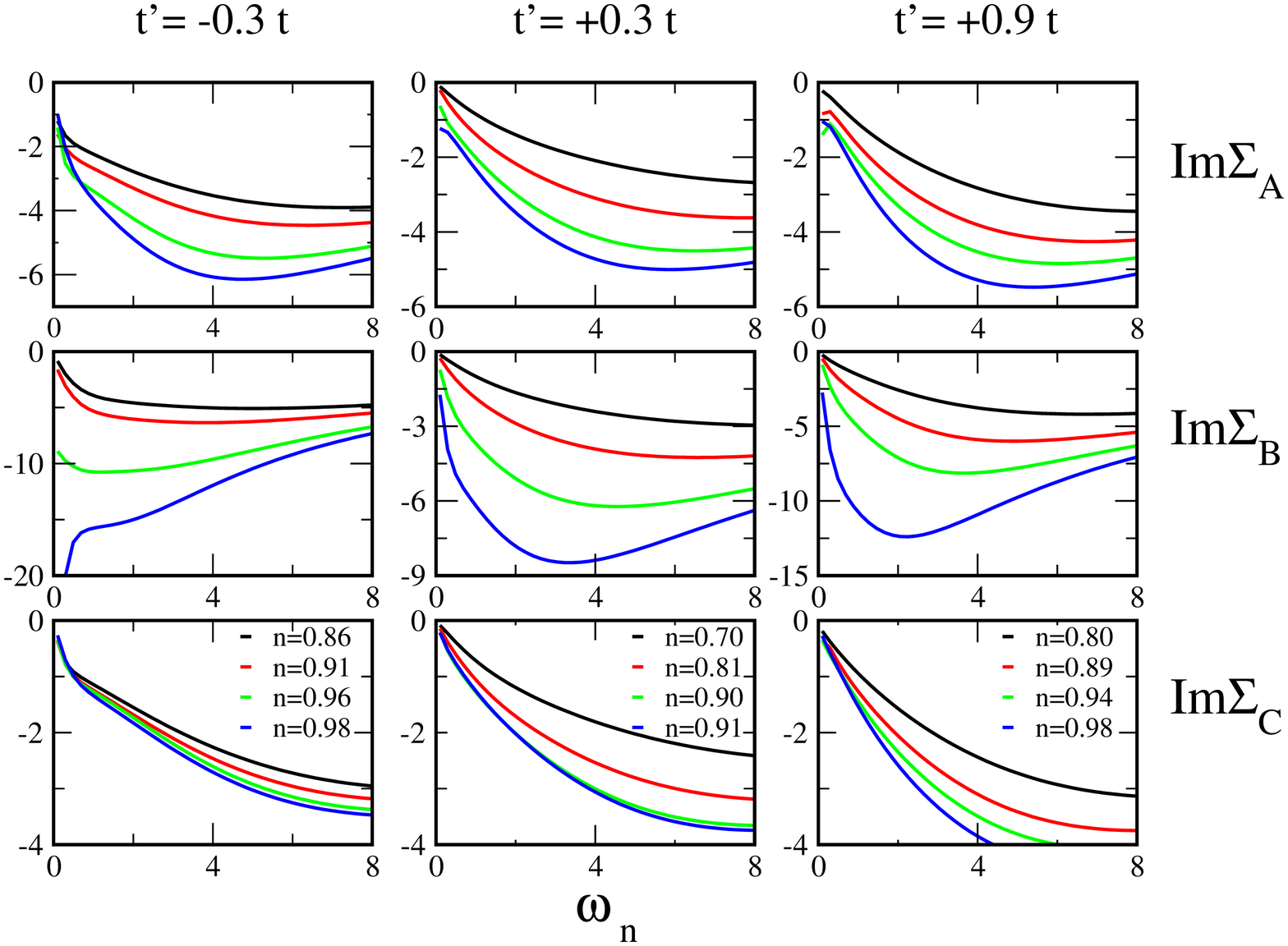}
\end{center}
\caption{Real (top) and imaginary (bottom) parts of the
eigenvalues of the cluster self-energies for
 $t^{\prime}/t= \pm 0.3, +0.9$. Interaction $U=16t$, the inverse virtual
 temperature $\beta=32/t$.}\label{fImS}
\end{figure}

\begin{figure}[!htb]
\begin{center}
\includegraphics[width=6.0cm,height=8.0cm,angle=-0] {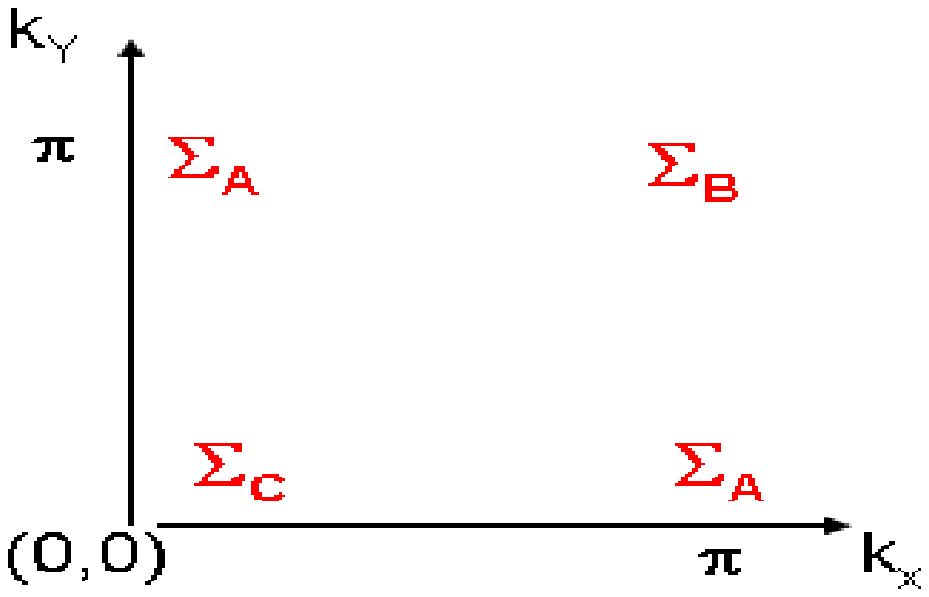}
\caption{Eigenvalues of the cluster self-energy
$\Sigma_{\mu\nu}$. They can be interpreted as the "corners" points
in the first quadrant of the BZ } \label{EigenSigma}
\end{center}
\end{figure}

We first analyze cluster quantities, postponing to the next
subsection results in the momentum space. In the latter case in
fact we will see that the periodizing method we use is important.
Using the square symmetry, the CDMFT results for the plaquette
are succinctly expressed in terms of three self-energies
$\Sigma_{11}$, $\Sigma_{12}$, $\Sigma_{13}$, or alternatively in
terms of the eigenvalues of the cluster self-energy matrix
$\Sigma_{ij}$, which can be thought as the lattice self-energies
in specific points of the momentum space, namely $\Sigma_{A}$ in
$(0,\pi)$ and $(\pi,0)$, $\Sigma_{B}$ in $(\pi,\pi)$ and
$\Sigma_{C}$ in $(0,0)$ (Fig.\ref{EigenSigma}).
\begin{eqnarray}
\left.
\begin{array}{l}
  \Sigma_{A}=\, \Sigma_{11}-\Sigma_{13} \\
  \Sigma_{B}=\, \Sigma_{11}- 2 \Sigma_{12}+ \Sigma_{13} \\
  \Sigma_{C}=\, \Sigma_{11}+ 2 \Sigma_{12}+ \Sigma_{13} \\
\end{array}
\right. \label{SelfABC}
\end{eqnarray}
CDMFT causality requires that the imaginary part of all the
self-energy eigenvalues is negative (i.e. $\Im \hat{\Sigma}$ is
definitive negative). As shown in Fig. \ref{fImS}, $\Sigma_A$,
$\Sigma_B$ and $\Sigma_C $ exhibit a clear systematic behavior as
the Mott transition is approached.
\begin{figure}[!htb]
\begin{center}
\includegraphics[width=7cm,height=5cm,angle=-0] {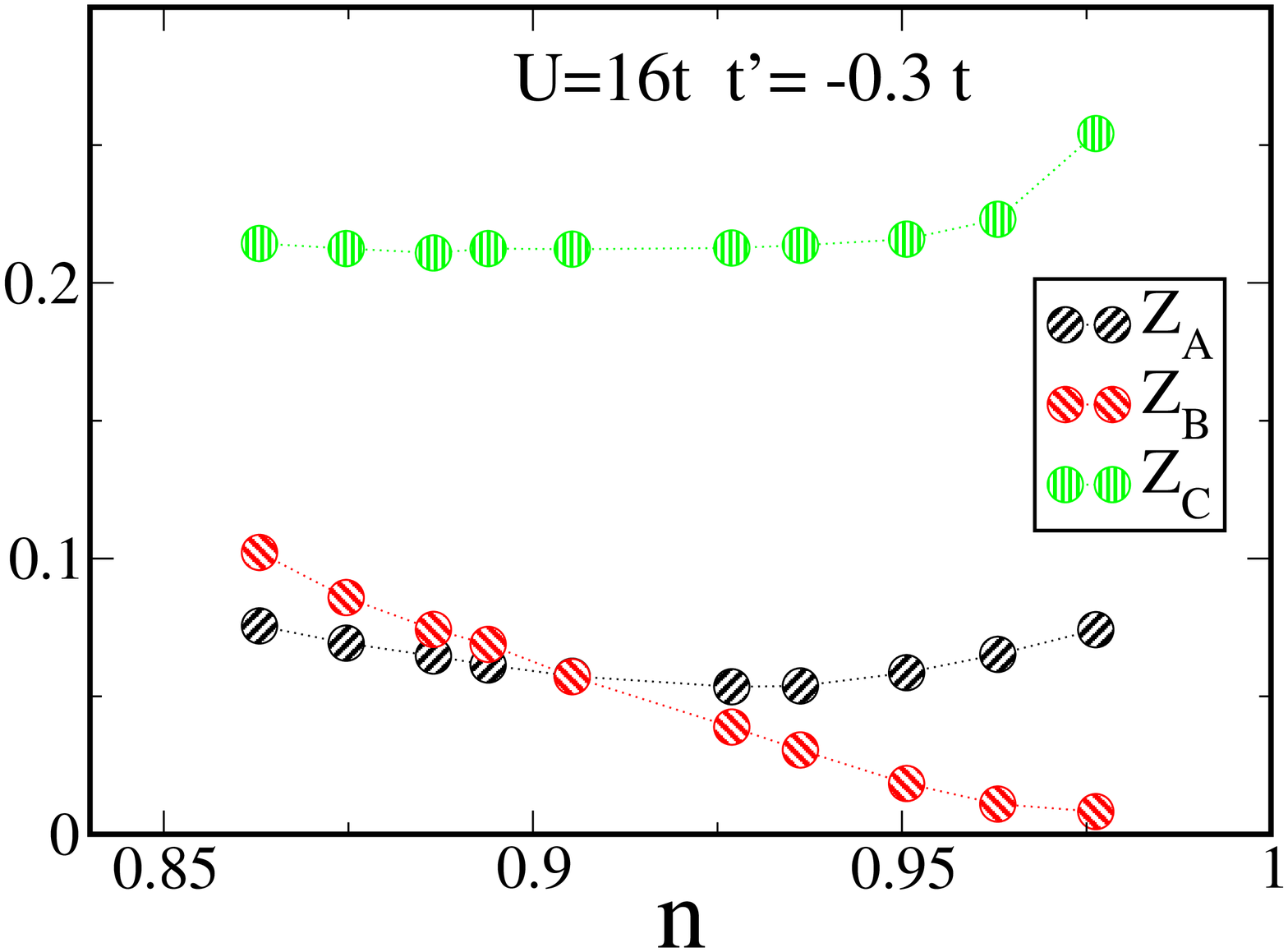}
\includegraphics[width=7cm,height=5cm,angle=-0] {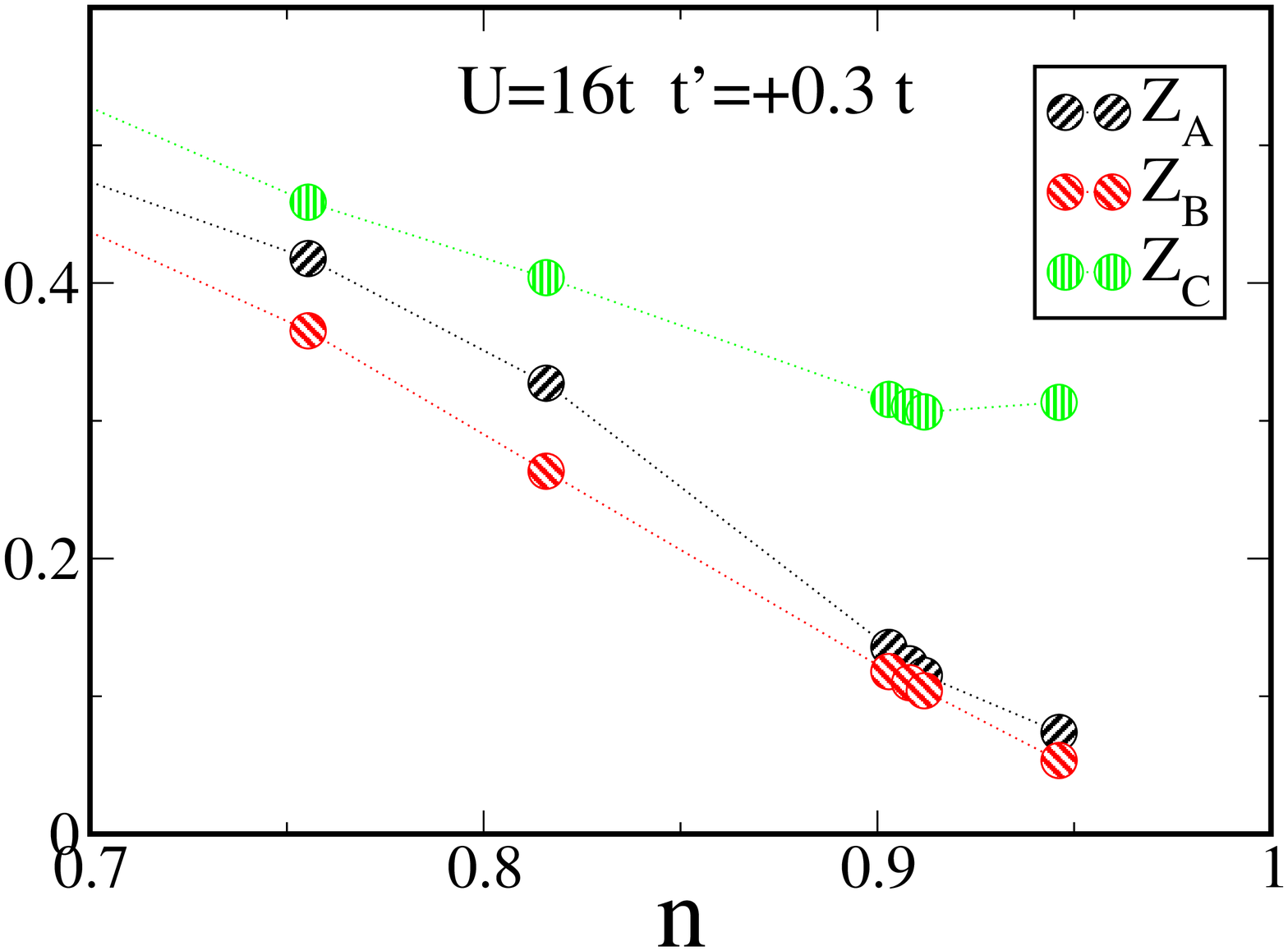}
\includegraphics[width=7cm,height=5cm,angle=-0] {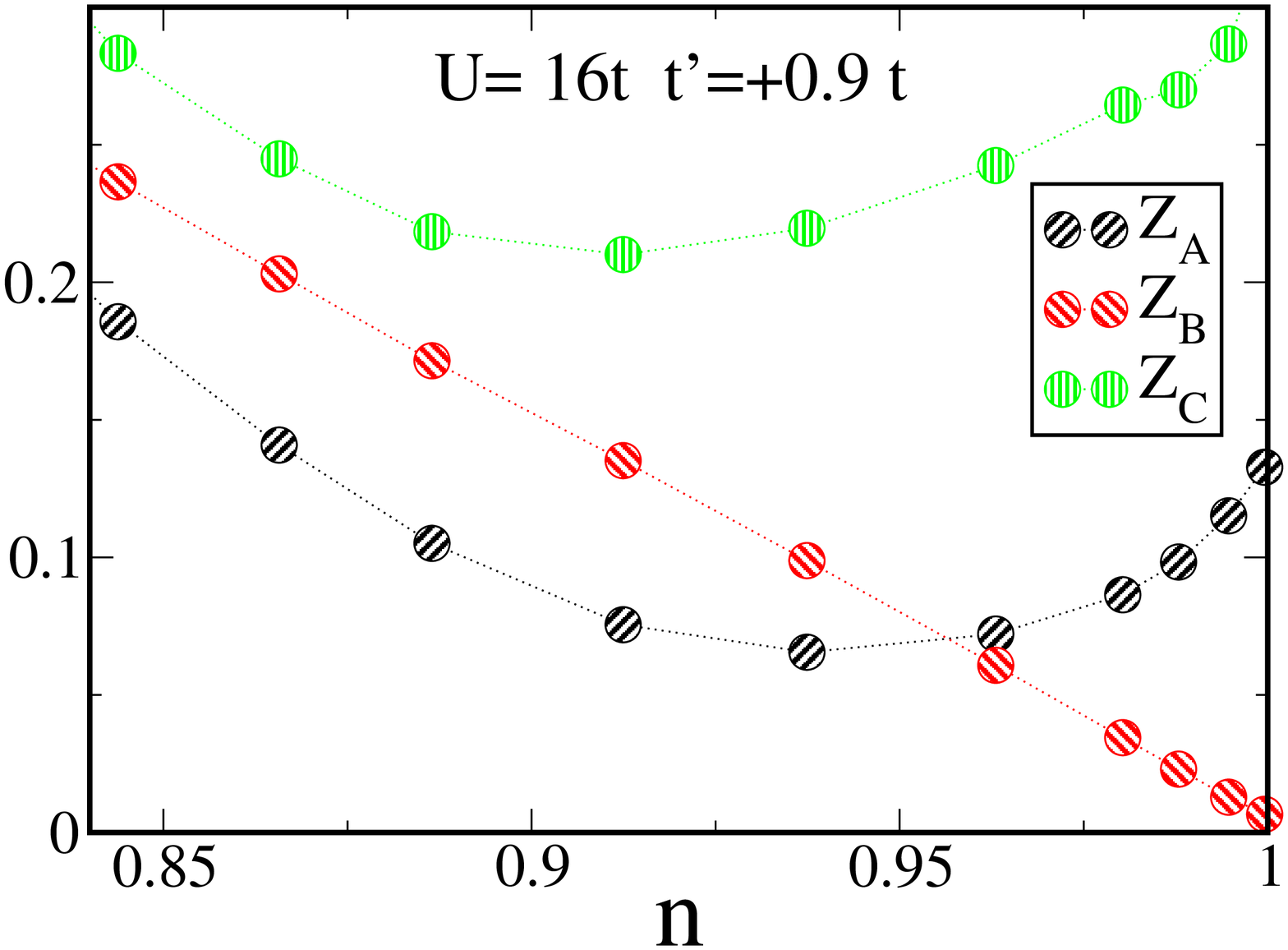}
\caption{Cluster quasiparticle residua $Z_X$ ($X= A,B,C$) as a
function of density $n$, for $U= 16t$ and $t^{\prime}= \pm 0.3$
and $t^{\prime}= + 0.9$. Virtual temperature is $\beta= 32/t$.}
\label{ZCluster}
\end{center}
\end{figure}
This systematic tendency describes the approaching of the system
to the Mott insulator. In particular we notice that, at the
virtual temperature used in this study $\beta^{-1}$, the $\Im
\Sigma_{X} \sim -\alpha \omega$ goes to zero for $\omega
\rightarrow 0$, as expected in a regular FL. We keep here a
conservative point of view, and assume at zeroth order
approximation that the system is indeed a FL. As the transition
is approached, we study how the quaiparticles residua $Z_{X}=( 1-
\frac{\partial \Im \Sigma_{X}}{\partial \omega} )^{-1}$ ($X=
A,B,C$) behave in the 4 corner points of the momentum space that
we have (Fig. \ref{EigenSigma}). The result is shown in Fig.
\ref{ZCluster}, where the 3 cases $t^{\prime}= \pm 0.3,$ and the
strongly frustrated one $t^{\prime}= +0.9$ are displayed.
In all of them the quasiparticle residuum $Z_{B}$ corresponding to
the point $(\pi,\pi)$ in $k$-space goes in linearly to zero as the
density $n\rightarrow 1$.
This is reminiscent of previous results with DMFT \cite{bibble},
where at the Mott transition point the quasiparticles disappear
because the residuum was going to zero. However in CDMFT, the
residua in the other points $A,C$ extrapolate to a finite value.
This clearly shows that the Mott transition is not approached in
the same way in the different regions of the BZ.

\begin{figure}[!htb]
\begin{center}
\includegraphics[width=7cm,height=5cm,angle=-0] {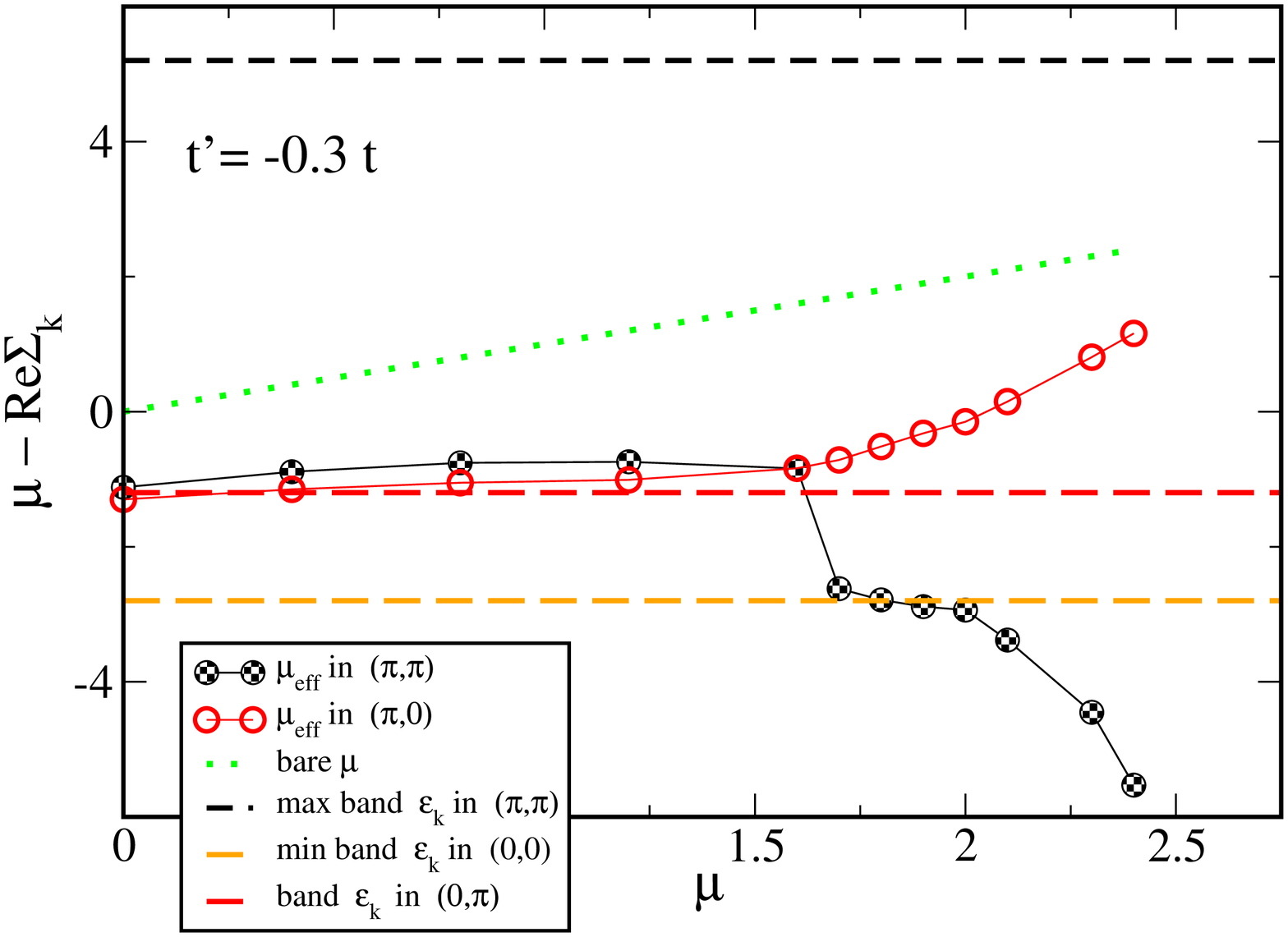}
\includegraphics[width=7cm,height=5cm,angle=-0] {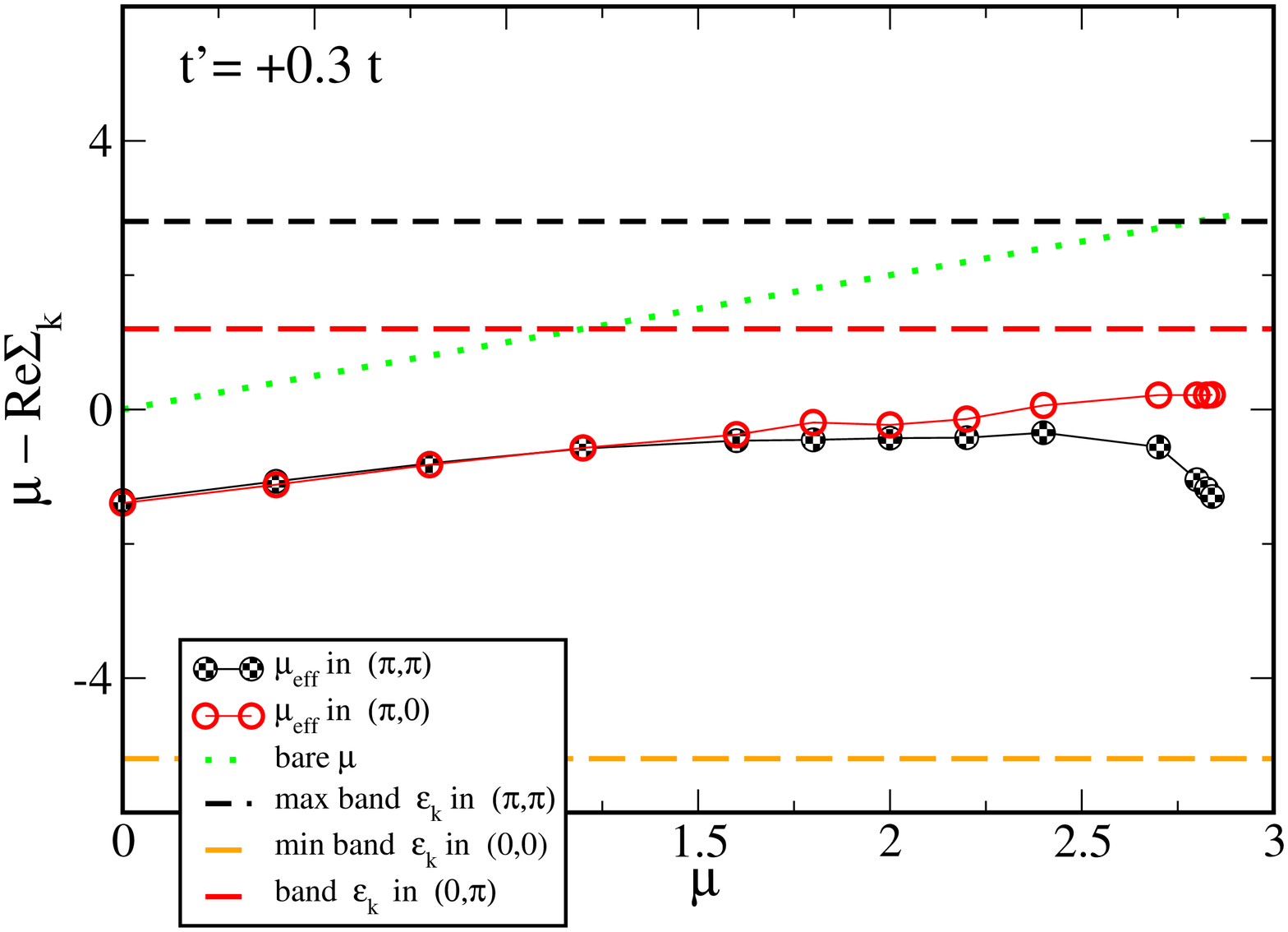}
\includegraphics[width=7cm,height=5cm,angle=-0] {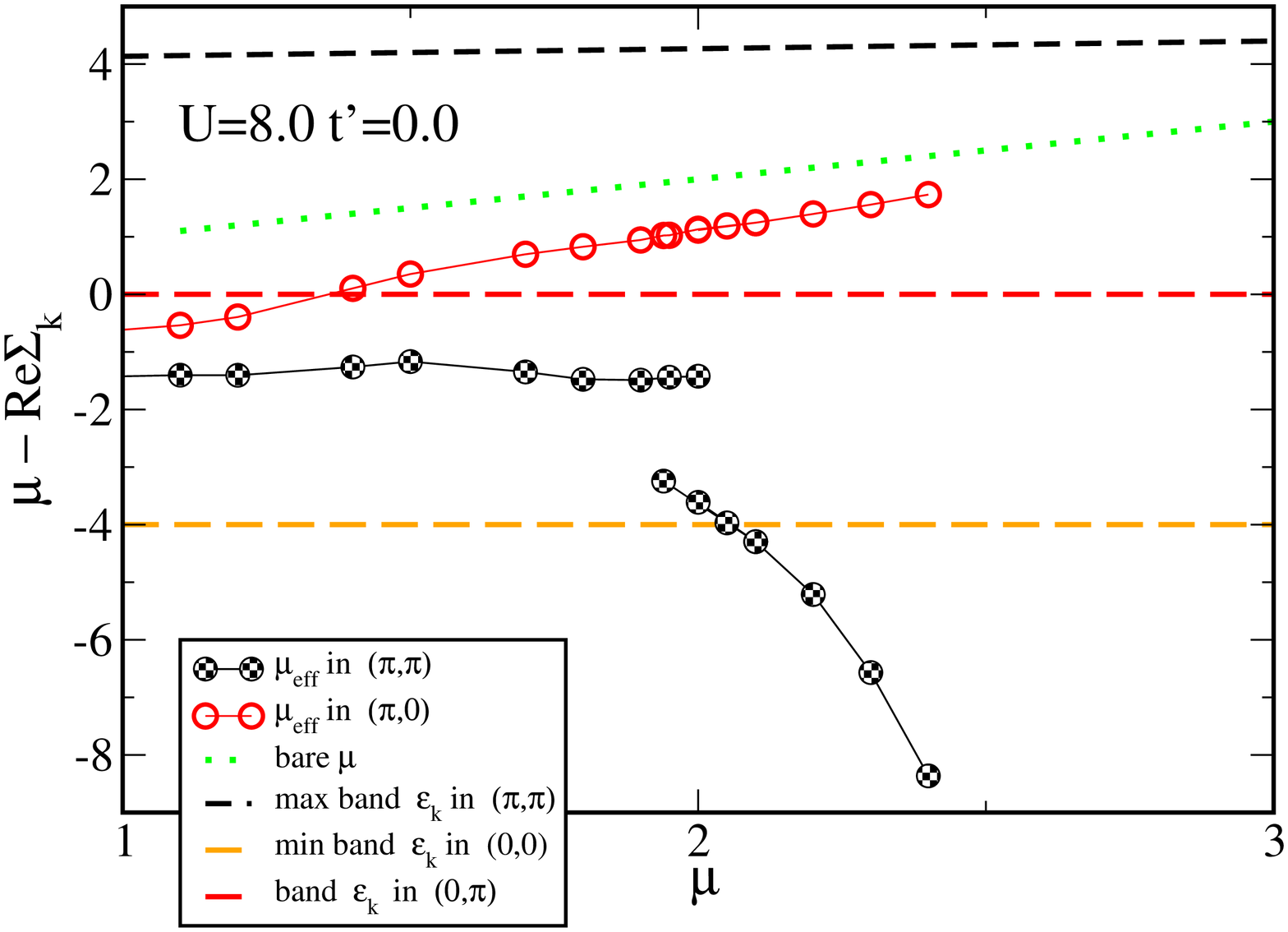}
\caption{Renormalized chemical potential $\mu_{eff}= \mu- \Re
\Sigma_k$ as a function of the bare chemical potential $\mu$. The
maximum of the one particle spectrum $\varepsilon_k$ [ $k=
(\pi,\pi)$], the minimum  [ $k= (0,0)$], and the its value in $k=
(0,\pi)$ are also displayed. A pole in the Green's function
appears when $\mu_{eff}\equiv \varepsilon_k$.} \label{mu_eff-fig}
\end{center}
\end{figure}

The real parts of the eigenvalues of the cluster self-energy
(Fig.\ref{fImS}) can be seen as renormalizing the bare chemical
potential in the special corner points of the BZ
(Fig.\ref{EigenSigma}:
\begin{eqnarray}
\mu_{eff}=\, \mu-\Re \Sigma_{X} \label{mu_eff}
\end{eqnarray}
where $X=A, B, C$. The way this occurs is shown in Fig.
\ref{mu_eff-fig}, where we compare the renormalized chemical
potential $\mu_{eff}$ in the corner points B [ $k=(\pi,\pi)$,
black filled circles ] and C [ $k=(0,\pi)$, open red circles ]
with the bare band energy $\varepsilon_{k}$ [ the minimum of the
band in $k=(0,0)$ and the maximum of the band in $k=(\pi,\pi)$ ].
For reference, we also plot the value of the band in $(0,\pi)$ and
the value of the bare chemical potential $\mu$. The graph displays
how $\mu_{eff}$ varies as a function of increasing bare chemical
potential $\mu$ (i.e. approaching he MT) for the two cases
$U=16t$, $t^{\prime}= \pm 0.3t$. In order for the FS to form, the
value of the chemical potential has to lie inside the band, so
that the equation $\varepsilon_{k}-\mu_{eff}=0$ can be satisfied
and a pole is formed in the one particle Green's function, i.e.
the chemical potential $\mu$ has to meet the value of the band
$\varepsilon_{k}$ for some point $k$ in the BZ. Now in all the
three cases displayed something dramatic happens in correspondence
of the B point $k=(\pi,\pi)$: the $\mu_{eff}$ diverges from the
value of the band in that point (black dashed line top of the
band): this region in $k$-space is becoming insulating-like before
all the others. In the cases $U=16t$, $t^{\prime}= -0.3t$ and
$U=8t$, $t^{\prime}= 0.0$ it even escapes out of the bare band.
When this happens in the $k$-point B there are no particle-states
available at any energy. Though the effect is less pronounced,
this is also true in the case $U=16t$, $t^{\prime}= +0.3t$. In
cases $U=16t$, $t^{\prime}= -0.3t$ and $U=8t$, $t^{\prime}= 0.0$,
even if less than in the B point, $\mu_{eff}$ in the C $k$-point
at $(0,\pi)$ is also escaping its band value, becoming
insulating-like. For $U=16t$, $t^{\prime}= +0.3$ instead we can
observe that $\mu_{eff}$ moves towards the band value (red dashed
line), meeting it in correspondence of the MT. From this picture
we can infer that the FS will pass in $(0,\pi)$ right at the MT.
We will verify this observation in the next section. This result
extends the previous DMFT conclusion that strong inelastic
scattering is enhanced close to the MT and that the break-up of
the metal is signed by the renormalized chemical potential $\mu$
going out of the bare band \cite{haule}. In DMFT, however, the
renormalized $\mu$ is an uniform quantity in $k$-space. CDMFT,
allowing for $k$-dependent self-energy, shows that the chemical
potential $\mu$ is renormalized at different rates in different
points of the BZ, and that the MT may take place in an anysotropic
fashion. In order to make some further steps, however, it is
necessary to understand what happens to quasiparticles in
correspondence of the FS, which lies between the 4 corner points
A,B and C. For this it is necessary to periodize the cluster
self-energy onto the $k$-space.

\section{$\Sigma$-scheme periodization: a conservative FL perspective}
We first study the MT in our system by applying the
$\Sigma$-scheme periodization, introduced in the previous chapter.
This interpretation of the result is the more
classical-FL-approach, as the self-energy is a simple linear
combination of the cluster self-energies \cite{bpk} and they
cannot give rise to any singularity in the lattice Green's
functions (if not already present in the cluster self-energies):
\begin{equation}\label{selflat1}
\Sigma_\latt(k,\omega)= \Sigma_{A} S_{A}(k)+ \Sigma_{B} S_{B}(k)+
                                 \Sigma_{C} S_{C}(k)
\end{equation}
where the functions
\begin{eqnarray} \label{S_X}
\left.
\begin{array}{l}
S_{A}(k)=\, ( 1- \cos k_{x}\, \cos k_{y}  )/2 \\
S_{B}(k)=\, ( 1- \cos k_{x} - \cos k_{y} + \cos k_{x} \,\cos k_{y} )/4\\
S_{C}(k)=\, ( 1+ \cos k_{x} + \cos k_{y} + \cos k_{x} \, \cos
k_{y} )/4
\end{array}
\right.
\end{eqnarray}
are positive ( and notice that $S_{A}(k)+ S_{B}(k)+ S_{C}(k)=1
\,)$. At large doping, only $\Sigma_{11}$ is appreciably different
from zero, while $\Sigma_{12},\Sigma_{13}\simeq 0$. The lattice
self-energy $\Sigma_\latt(k,\omega)$ [from Eq. (\ref{selflat1})]
is therefore $k$-independent and single-site DMFT results are
recovered. However the cluster self-energies increase sizably at
low doping, making $\Sigma_\latt$ strongly $k$-dependent. So let
us first look at the behavior of the FS as we get close to the MT.
The zero frequency limit of the real part of $\Sigma(k,\omega)$
determines the shape of the interacting FS, which we define as
$t_{\text{eff}}(k)=\mu$, where $t_{\text{eff}}(k)\equiv \,
t(k)-\Re \Sigma_{\text{latt}}(k,\omega=0^{+})$, $t(k)$ being the
Fourier transform of the hopping  $t_{ij}$ and the $\omega=0^{+}$
limit is extrapolated from the lowest Matsubara frequencies.
\begin{figure}[!htb]
\begin{center}
\includegraphics[width=8.5cm,height=9.5cm,angle=-0] {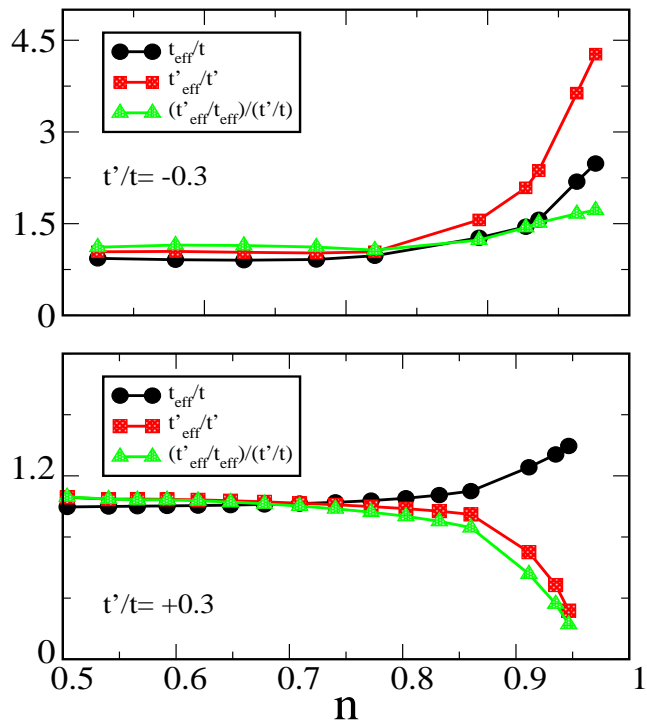}
\caption{Renormalization of the  hopping coefficients and of
their ratio as a function of density for $U= 16t$ and
$t^{\prime}/t=\pm 0.3$, $\beta=32/t$.} \label{ft10.9}
\end{center}
\end{figure}
The renormalization of the FS becomes appreciable close to the MT.
The self-energy itself depends weakly on the sign of $t^{\prime}$,
and in particular it has the same sign for both positive and
negative $t^{\prime}$. However, given its large magnitude, when
combined with $t^{\prime}$ of different signs, it produces
interacting FS's of very different shape in the electron-doped
and hole-doped case (dashed lines in Fig. \ref{fA}). This can be
understood in terms  of  the renormalized low energy hopping
coefficients $t_{\text{eff}}= t- \Re\Sigma_{12}(\omega=0^{+})/2$
and $t^{\prime}_{\text{eff}}= t^{\prime}-
\Re\Sigma_{13}(\omega=0^{+})/4$ presented in Fig. \ref{ft10.9}.
\begin{figure}[!htb]
\begin{center}
\includegraphics[width=8.5cm,height=9.0cm,angle=-0] {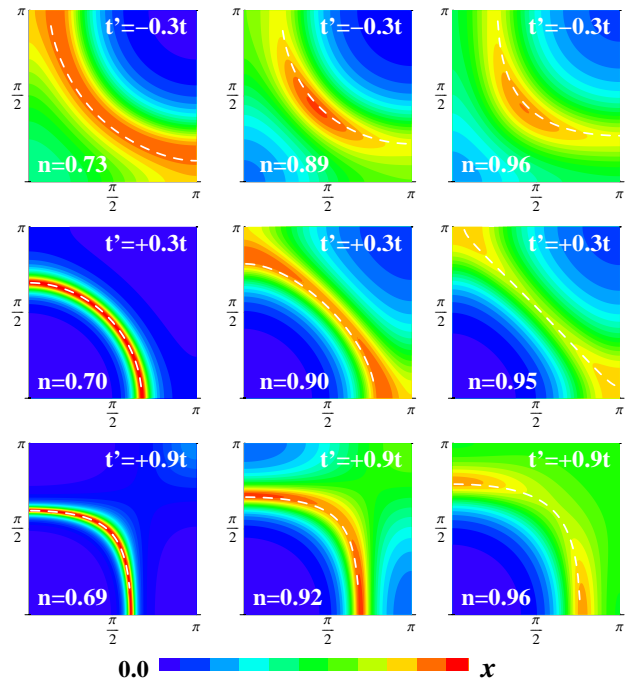}
\caption{$A(k,\omega=0^{+})$ in the first quadrant of the
Brillouin zone. In the first row from the top $t^{\prime}= -0.3
t$, densities $n = 0.73, 0.89, 0.96$, color scale $x= 0.28, 0.22,
0.12$; in the second row $t^{\prime}= +0.3 t$, $n = 0.70, 0.90,
0.95$, color scale $x= 0.82, 0.34, 0.27$; in the lowest row
$t^{\prime}= +0.9 t$, $n = 0.69, 0.92, 0.96$, color scale $x=
0.90, 0.32, 0.22$. The white dashed line is the FS given by
$t_{\text{eff}}(k)=\mu$.} \label{fA}
\end{center}
\end{figure}
Regardless the value of $t^{\prime}/t$, correlations act to {\it
increase } the value of $t_{\text{eff}}$. This  physical effect,
predicted by earlier slave-boson studies \cite{grilli}\cite{liu}
reduces the mass divergence  characteristic   of single-site
DMFT  where the effective mass scales inversely proportionally to
the quasiparticle residue. The renormalization of $t^{\prime}$
depends instead on the sign of $t^{\prime}/t$. This is  an effect
which is not present in slave-boson theories
\cite{grilli}\cite{liu}. For $t^{\prime}=-0.3 t$,
$|t^{\prime}_{\text{eff}}|$ increases in such a way that the ratio
$(t^{\prime}_{\text{eff}}/t_{\text{eff}})/ (t^{\prime}/t)$ weakly
increases approaching the Mott insulator, thereby enhancing the
hole-like curvature of the FS. On the other hand,
$t^{\prime}_{\text{eff}}$ decreases for  $t^{\prime}=+0.3 t$,
giving rise to an almost nested FS as half-filling is approached.
This is also clearly seen in Fig. \ref{fA}, where the FS is shown
on top of the spectral function (see below). In the hole-doped
case we observe also a horizontal flattening of the FS close to
$(0,\pi)$ or $(\pi,0)$ approaching the MT. The shape of the FS is
similar to what observed in ARPES \cite{KShen}, and thereby
interpreted as resulting from a doping independent nesting vector.
\begin{figure}[htb!]
\begin{center}
\includegraphics[width=12.0cm,height=3.75cm,angle=-0] {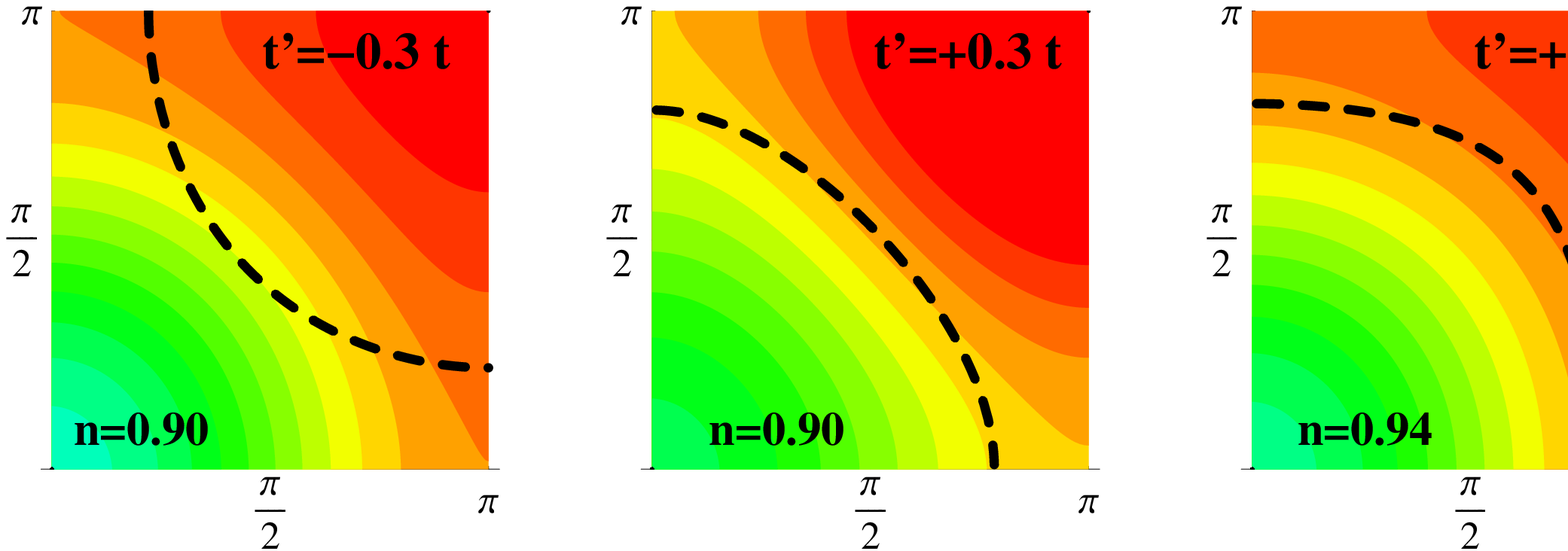}
\caption{$\tau^{-1}_{k}$ in the first quadrant of the BZ as
calculated from Eq. \ref{selflat1}. Red regions are high
scattering, green low scattering. Color scale is relative. The
dashed line is the renormalized FS given by
$t_{\text{eff}}(k)=\mu$.} \label{ImSigmak}
\end{center}
\end{figure}

Inelastic scattering strongly increases as the Mott insulator is
approached \cite{parcollet,haule,tudor-phillips}. In order to
investigate the anisotropy of this quantity we study the
imaginary part of the self-energy, plotted in Fig. \ref{fImS}.
Using Eq. (\ref{selflat1}), we can evaluate the lifetime, or
inverse scattering rate $\tau_{k}^{-1}=
-\Im\Sigma(k,\omega=0^{+})$ (again extrapolating to zero the
Matsubara values). Let us emphasize that our calculation is
performed at a finite effective temperature. In a Fermi liquid
this quantity would be small and vanish as $T^2$ as the
temperature goes to zero. Here we find a strong modulation of
$\tau_{k}^{-1}$ in the Brillouin zone that develops when the Mott
point is approached. Our results, as well as the QMC-CDMFT of
Ref. \cite{bpk},  may be interpreted in terms of a strongly
anisotropic coherence scale, which decreases at low doping. When
the scale becomes  smaller than the energy resolution of our
calculation we can not follow the decrease of $\Im \Sigma$ with
decreasing frequency (as evidenced by the line for $n=0.98$ for
$t^{\prime}/t = -0.3$ of Fig. \ref{fImS}). Therefore it not
possible to decide between a scenario where the Fermi liquid
picture breaks down or where the Fermi liquid coherence scale is
smaller than our energy resolution.

In Fig. \ref{ImSigmak} we show an intensity plot of $\tau_k^{-1}$
for $n=0.9-0.94$ and the values of $t^{\prime}$ previously
considered. We observe that approaching the MT the scattering rate
is enhanced in the region around the $(\pi,\pi)$ point of the
momentum space, independently of the value of $t^{\prime}$. This
is also the point where the cluster-quasiparticle-residuum $Z_{B}$
effectively goes to zero when approaching the MT. The FS (dashed
line in Fig. \ref{ImSigmak}) hits this region at different points,
according to its hole-like or electron-like curvature. Hence the
finite temperature lifetime is strongly modulated in momentum
space. Extrapolating to zero temperature, quasiparticles disappear
first in some regions of the momentum space and survive in others.
In the hole-like case (${t^{\prime}}= -0.3t$) the FS hits the
large scattering region around $(0,\pi)$ and $(\pi,0)$, while for
the electron-like cases (${t^{\prime}}= +0.3, 0.9$) the crossing
occurs close to $(\pi/2,\pi/2)$. Thus the combined effect of
different scattering properties in the momentum space and the
renormalization of the FS gives rise to the formation of a {\it
cold [hot]} spot  in $(\pi/2,\pi/2)$ $[(0,\pi),(\pi,0) ]$ in the
hole-like system. In the electron-doped system the position of hot
and cold spots is inverted. The presence of hot/cold regions is
reflected also by the spectral function $A(k,\omega=0^{+})= -1/\pi
\, \Im G(k,0^{+})$, shown in Fig. \ref{fA}, which agrees with the
qualitative behavior of experimental ARPES spectra
\cite{damascelli}\cite{campuzano}. These results have a direct
experimental interpretation: in particular photoemission data
should take into account the renormalization of the shape of the
FS in order to extract the model Hamiltonian parameters. They also
suggest a new viewpoint concerning the origin of the
experimentally observed asymmetry between electron and hole doped
cuprates. We will explain this in the next chapter in more detail,
where we will analyze the superconducting phase. Here we observe
that in the hole-doped case the quasiparticles survive in the
diagonal of the Brillouin zone, near $(\pi/2,\pi/2)$. This state,
which has a fermionic spectrum with point zeroes can be connected
to the quasiparticles of the d-wave superconducting state. The
electron-doped case is completely different. On a technical level,
it is harder to approach the MT closely, and a first-order phase
transition may preempt a continuous approach to the insulating
state. Furthermore, the FS is renormalized towards nesting, and
the quasiparticles survive in a small region around $(\pi,0)$ and
$(0,\pi)$. These quasiparticles cannot be easily deformed into the
superconducting state as compared to the quasiparticles which live
around $(\pi/2,\pi/2)$. These properties of the underlying normal
state of the Hubbard model have striking resemblance to what is
observed in the cuprates. The hole-doped materials have a
superconducting region  which appears almost immediately after
doping the Mott insulator. This superconducting state  evolves
continuously into the pseudogap state which in turn evolves
continuously from the Mott insulator. In the electron-doped case,
the pseudogap region is small, and a much larger doping is needed
to reach the superconducting phase.

Finally we emphasize that our results for large frustration
$t^{\prime}=+0.9 t$  (third panel in Fig. \ref{ImSigmak} and
third row in Fig. \ref{fA}, are always qualitatively similar to
the weakly frustrated system with the same sign of $t^{\prime}$.
This clearly shows that, at the $k-$resolution considered in this
study, the momentum-space differentiation does not depend on the
AF ordering of the parent insulator, since it occurs also for a
system in which AF is destroyed by frustration and the insulator
is likely to have a more exotic form of long-range order.
Moreover, since at high doping (or temperature) we find that
non-local self-energies are negligible and single-site DMFT is
not corrected by cluster DMFT, we expect similarly that there is
an intermediate doping (or temperature) region where the results
of the $2\times 2$ cluster will not be modified by increasing the
cluster size. Our conclusion is that in this region there is
$k$-space differentiation, which is independent of the value of
$t^{\prime}$ and is therefore due to short-range correlations,
captured within the plaquette, rather than to the specific order
of the parent Mott insulator.

So, analyzing our CDMFT results under a
$\Sigma$-scheme-periodization we can state that:
\begin{enumerate}
 \item the FS is strongly renormalized by the interaction,
 \item there is a breakdown of the FS and the appearance
 of {\it cold} and {\it hot} regions, whose precise location
 is the result of an interplay of the renormalization of the
 real and the imaginary parts of the self-energy,
 \item the emergence of these hot/cold regions
 is a consequence of the proximity of the Mott transition and
 long-ranged AF correlation is not a necessary condition for its
 existence.
 \end{enumerate}
These are zeroth order results, whose general statement is
independent of the periodizing scheme adopted to evaluate the
momentum-space quantities. The point of view is conservative, in
the sense that no particular anomaly in the FL is allowed by the
$\Sigma$-scheme which has the FL hypothesis at its foundation.
\section{Dependence of the Results on the periodizing-scheme}
\begin{figure}[!h]
\begin{center}
\includegraphics[width=12.5cm,height=10.0cm,angle=-0] {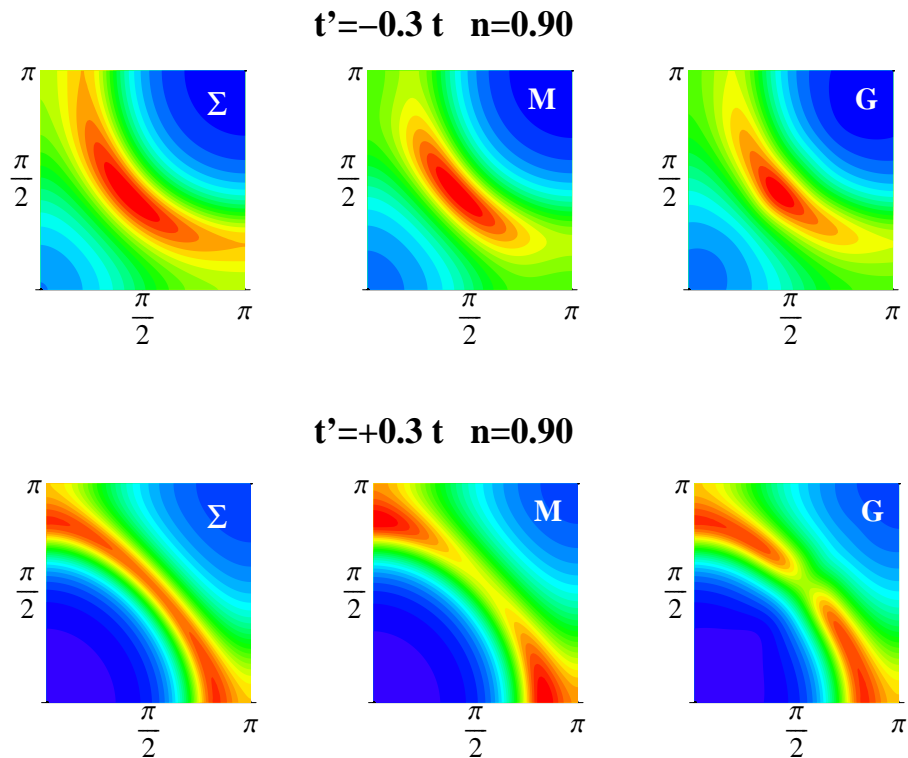}
\caption{Spectral function at the Fermi level in the $k-$space.
We present here the cases $t^{\prime}= -0.3 t$ (upper row) and
$t^{\prime}= +0.3 t$ (lower row)  at a density $n=0.90$. The
results from the three different periodizing schemes $\Sigma$,
$M$ and $G$ are presented.} \label{fAk_S_M}
\end{center}
\end{figure}
\begin{figure}[!htb]
\begin{center}
\includegraphics[width=14.0cm,height=6.0cm,angle=-0] {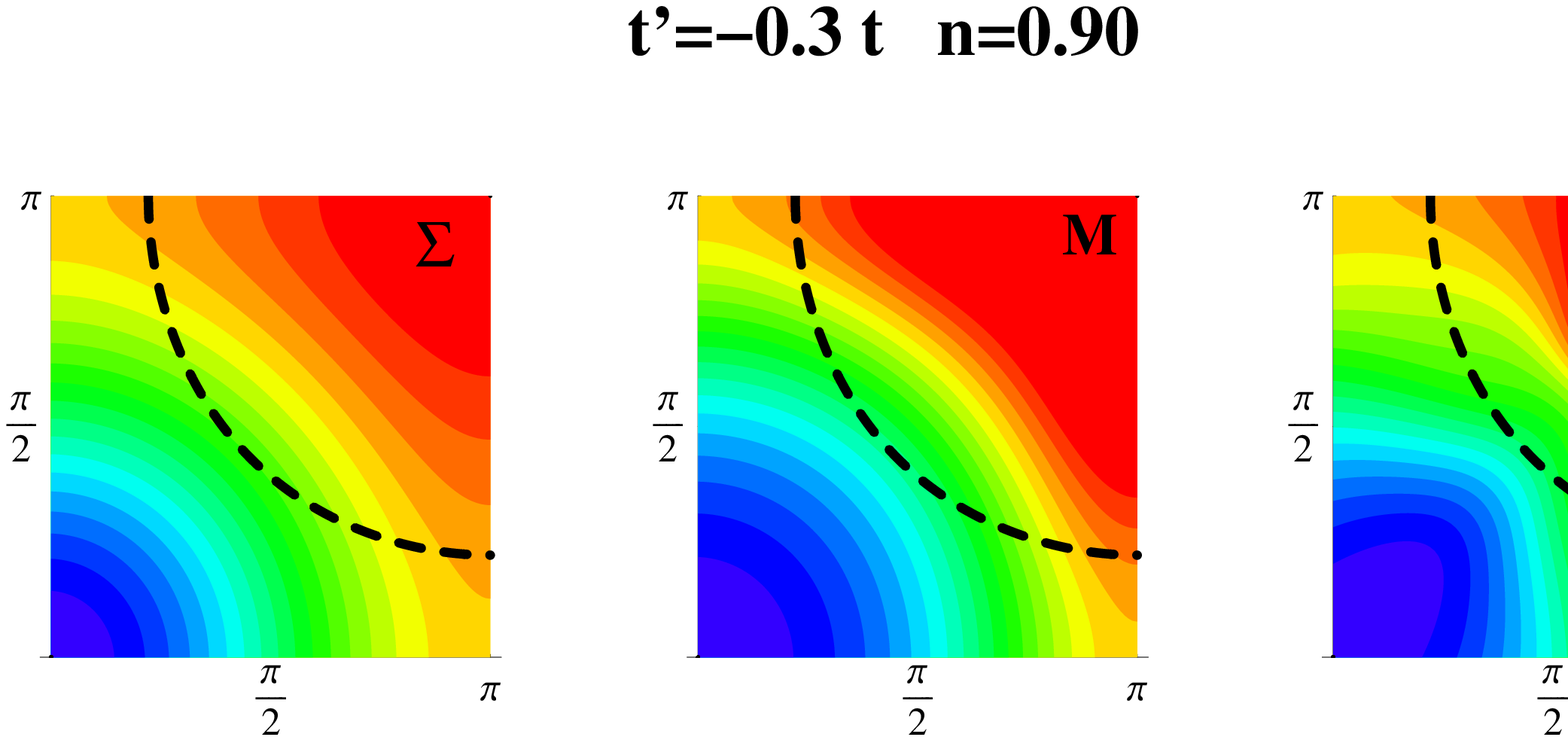}
\caption{$\tau_{k}^{-1}$ in the first quadrant of the BZ for the
case $t^{\prime}= -0.3t$ and density $n=0.90$. Results from the 3
periodizing-scheme $\Sigma$, $M$ and $G$ are confronted. The
color scale is relative. } \label{tauk_S_M_G}
\end{center}
\end{figure}
\begin{figure}[!h]
\begin{center}
\includegraphics[width=7.0cm,height=8.5cm,angle=-0] {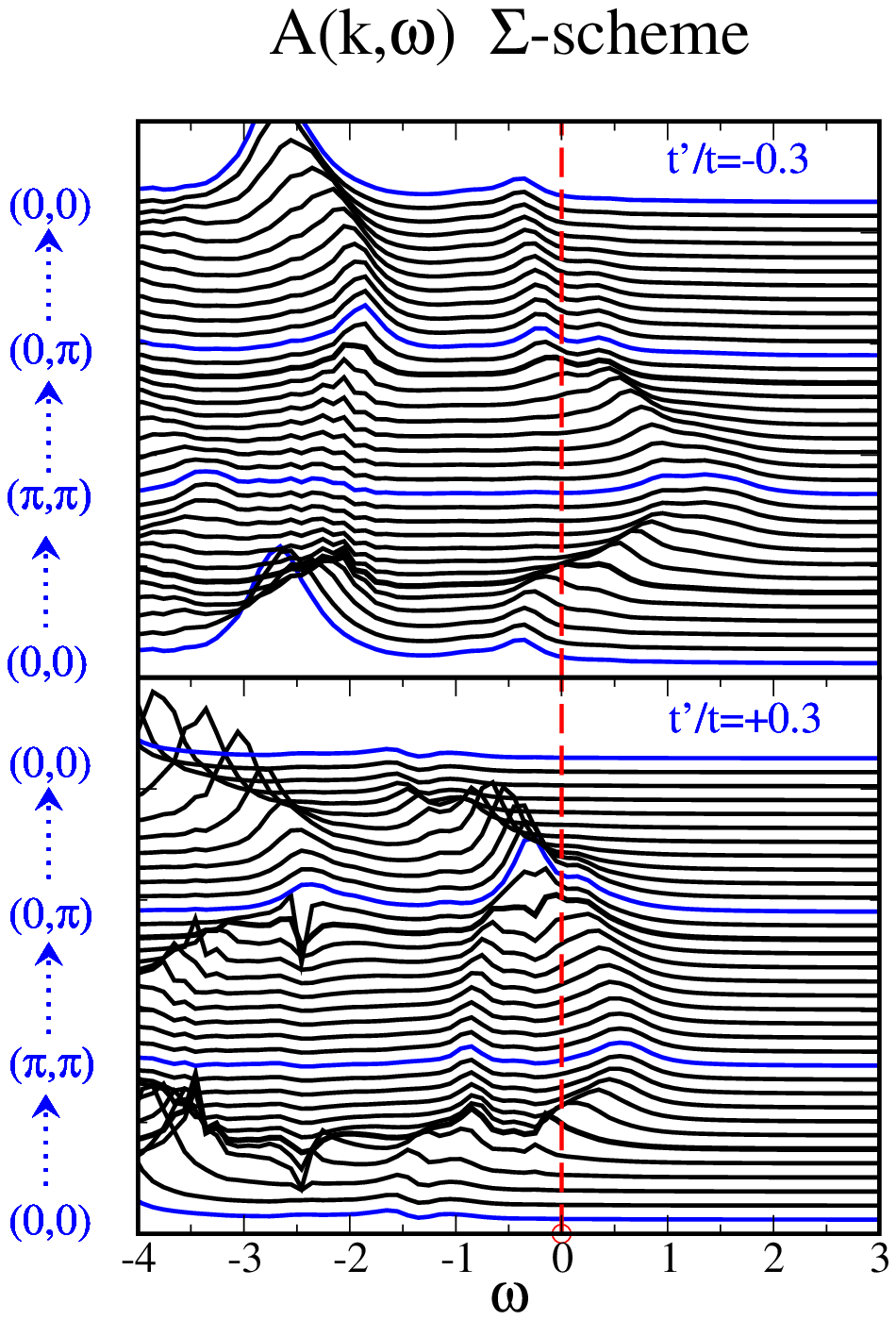}
\includegraphics[width=7.0cm,height=8.5cm,angle=-0] {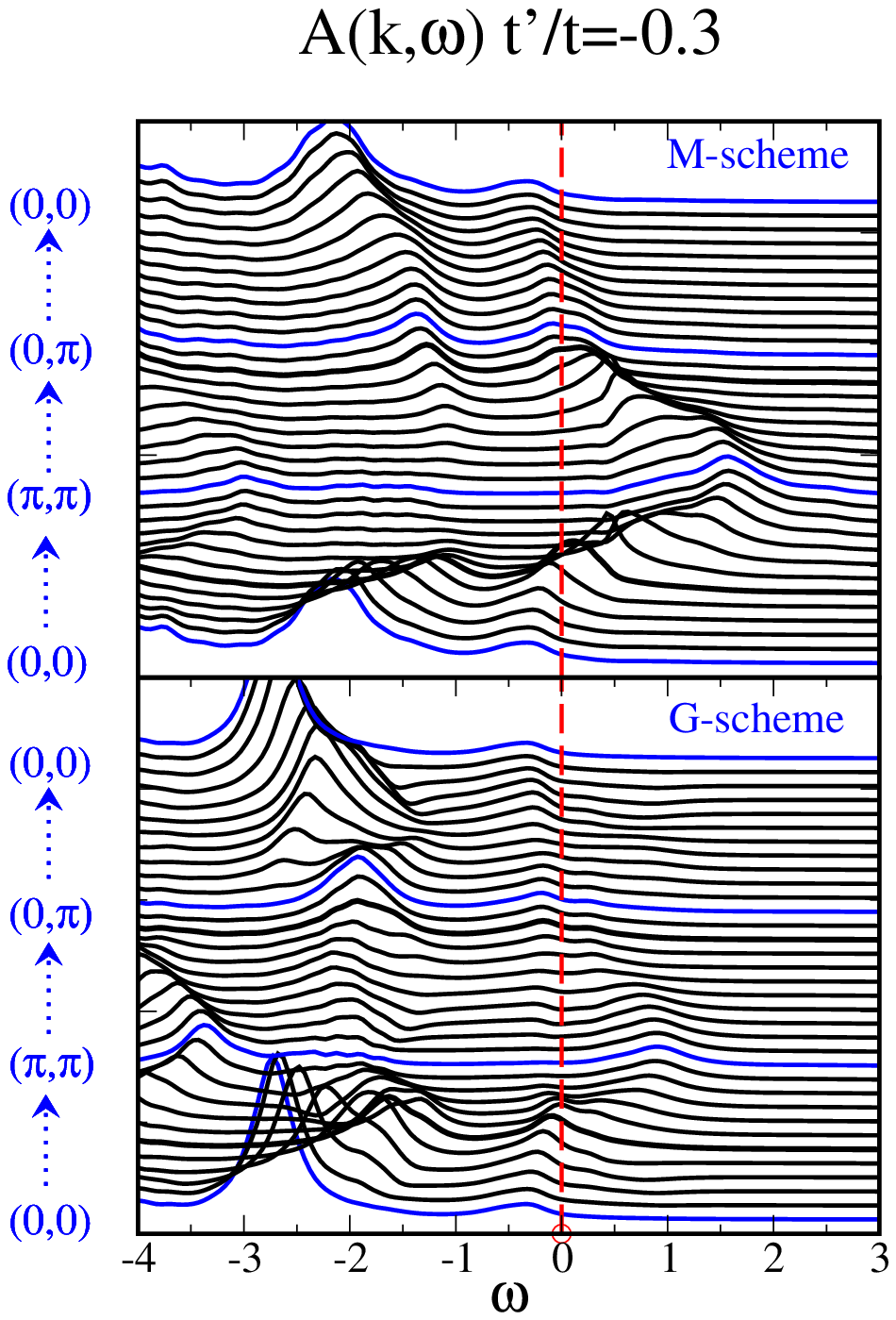}
\caption{$A(k,\omega)$ along the path $(0,0)\rightarrow (\pi,\pi)
\rightarrow (0,\pi)\rightarrow (0,0)$ in the first quadrant of
the BZ. The density is $n=0.9$. In the left hand side we present
results obtained with the $\Sigma$-scheme periodization. The
hole-like $t^{\prime}/t= -0.3$ case is on the upper graph, and
electron-like $t^{\prime}/t= -0.3$ on the lower graph. On the
right hand side we once again present the results for
$t^{\prime}/t= -0.3$ case but obtained with the two other
schemes: $M$-scheme periodization was used in the upper graph,
$G$-scheme in the lower graph.} \label{fAk}
\end{center}
\end{figure}

We want now to discuss how the results presented in the previous
section, obtained in the framework of the $\Sigma$-scheme, depend
on the periodizing scheme. Alternative periodizing schemes were
presented in chapter 4.3 (formulas \ref{Sk} and \ref{Gk}). We will
show that the gross features are not in fact periodizing-scheme
dependent, while the low-energy behavior indeed is. We focus here
our attention on how the MT is approached. In Fig.\ref{fAk_S_M},
we show that the phenomenon of momentum space differentiation
plotting the spectral function $A(k,\omega\rightarrow 0)$ in the
first quadrant of the BZ for the 2 cases already considered
$t^{\prime}= -0.3 t, +0.3 t$ (first, second rows), for the 3
different methods ($\Sigma$, $M$, $G$-schemes in the first, second
and third columns respectively), and for a small value of doping
where the effect is most evident (10\% doping). We see a strong
similarity in the shape of the {\it cold/hot} scattering regions.
Moreover the effect of scattering differentiation is even enhanced
in the case of the $M$-method and $G$-method, appearing to be a
robust feature. If we further investigate the mechanism of
interplay between the "bending" of the FS in the hole/e$^{-}$-like
system and the $k$-space-modulation of the scattering rate
$\tau_{k}= -\Im \Sigma_{k}$ (Fig.\ref{tauk_S_M_G}), we see that
the shape of the modulation is method dependent, but that the
$(\pi,\pi)$ point is the highest-scattering driving the system
into the MT remain un-altered. This gives, as was mentioned, a
naive zeroth order picture on how the MI is approached and why the
the hot/cold spots appear as effect of the proximity to the MT. We
show however that there are also fundamental differences, which
start to appear while considering the frequency dependence of the
Green's functions. For example in Fig. \ref{fAk} (left hand side)
\begin{figure}[!h]
\begin{center}
\includegraphics[width=8.75cm,height=8.0cm,angle=-0] {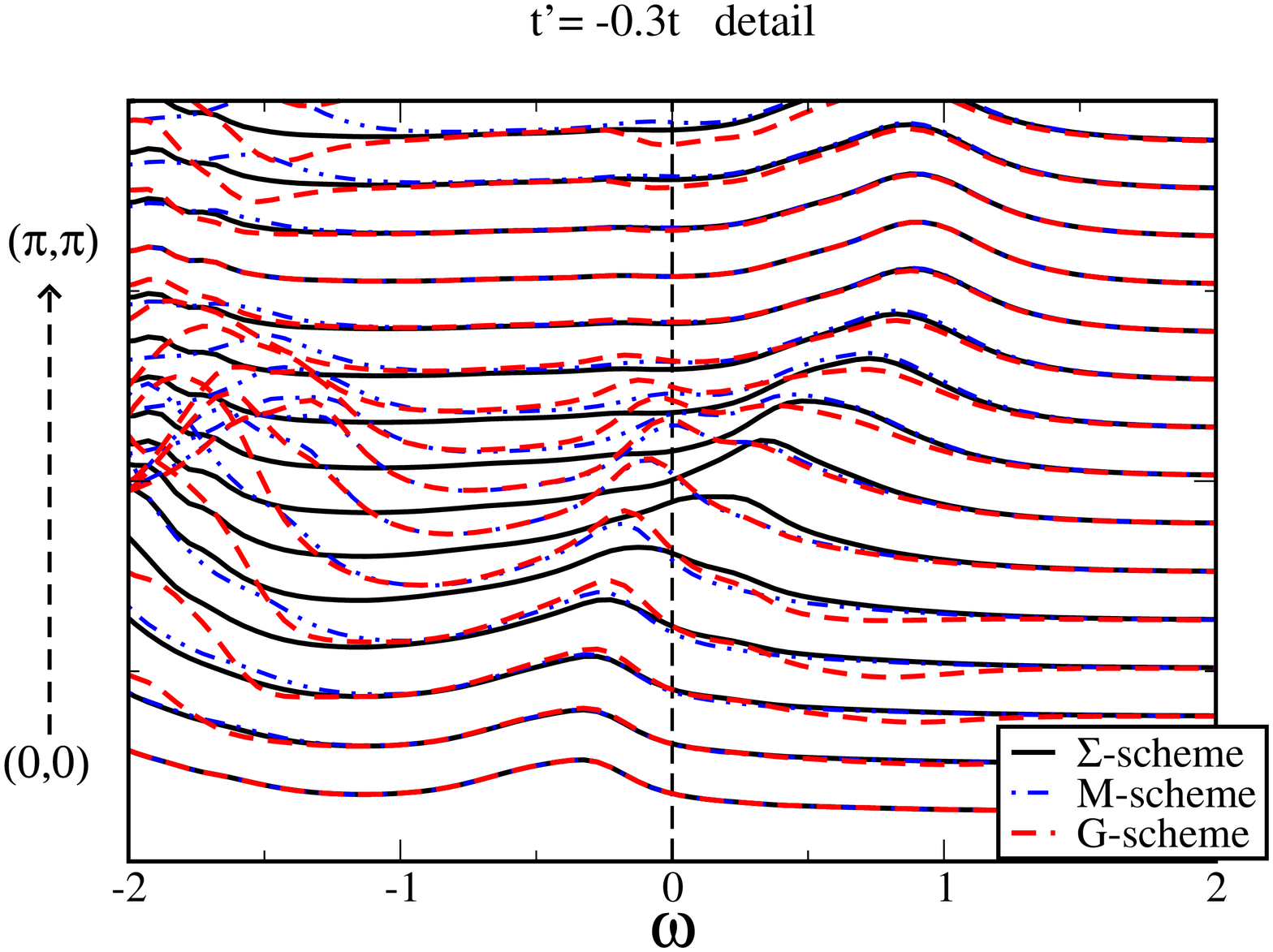}
\caption{A detail of the spectral function $A(k,\omega)$ as a
function of $\omega$ is presented along the path
$(0,0)\rightarrow (\pi,\pi)$ in the first quadrant of the BZ. The
3 periodizing schemes $\Sigma$, $M$ and $G$ are confronted in
correspondence of the Fermi level at $\omega=0$. } \label{fAkdet}
\end{center}
\end{figure}
we show the spectral function at $\sim 10\%$ doping $A(\omega,k)$
as a function of $\omega$ along a path in the BZ
$(0,0)\rightarrow (\pi,\pi) \rightarrow (\pi,\pi) \rightarrow
(0,0)$, calculated with the $\Sigma$-method and for the two
relevant cases $t{\prime}= \pm 0.3$. The $\Sigma$-scheme is
"fitting" the results to a FL, producing a momentum-dependent
self-energy which does not present any kind of singularity. In
the case $t^{\prime}/t= -0.3$ (the upper graph), a quasiparticle
peak  disperses along the $(0,0)\rightarrow (\pi,\pi)$, where a
cold spot is formed at the Fermi level; on the contrary the peak
is less dispersive passing through the region around $(0,\pi)$,
where there is the hot spot. On the other hand, in the case
$t^{\prime}/t=+0.3$ (lower graph in the left hand side) the
quasiparticle peak is dispersing more in the vicinity of the
region $(0,\pi)$ and, while dispersing less, evaporates around
$(\pi/2,\pi/2)$. The hot-cold spots are in this way switched. This
is the FL viewpoint for the hot/cold spot formation as observed
in ARPES experimental data \cite{damascelli}\cite{campuzano},
where a quasiparticle peak is detected in the spectral function
$A_{k}(\omega)$ in correspondence of the cold spot, while in the
hot spot region $A_{k}(\omega)$ present broad incoherent
features. The effect of momentum-differentiation is however in
this case quite weak. Looking at the same plots obtained using
the $M$-scheme and the $G$-scheme, we observe a qualitatively
similar behavior of the spectral function $A(k,\omega)$, as for
example is shown in the right hand side of Fig. \ref{fAk}, where
we display for the $M$-scheme (upper graph) and $G$-scheme (lower
graph) for the case $t^{\prime}= -0.3t$, to be confronted with the
$\Sigma$ scheme (upper graph, left hand side) . However, a closer
look in correspondence of the Fermi level (for example around the
point $(\pi/2,\pi/2)$, as shown in detail in Fig. \ref{fAkdet})
shows that:
\begin{itemize}
  \item the results of the $M$-scheme and the $G$-scheme are very
  close in spite the 2 methods are a priori different (this was
  explained by Tudor Stanescu \cite{tudorpr}).
  \item the low energy features, in particular the $\omega \simeq 0$
  corresponding to the Fermi level, are indeed quite different
  from the one obtained with the $\Sigma$-scheme (see detail Fig. \ref{fAkdet}).
\end{itemize}
In the $M$ and $G$-scheme, and indeed around the cold region, the
$\Sigma$-scheme quasiparticle peak appears now with a more
complicate substructure, split in two other peaks. At the crossing
of the FS in correspondence of the cold region the lower peak
starts evaporating while the upper one (which is above the FS)
increases. At the cold region, instead, we observe that the double
pick feature with a depression in the middle creates a bigger
difference in spectral weight in correspondence of the Fermi level
at $\omega=0$, and hence a more pronounced hot/cold spot
modulation. This indicates a departure from a simply FL viewpoint
given with the $\Sigma$-scheme.
\begin{figure}[!h]
\begin{center}
\includegraphics[width=7cm,height=5cm,angle=-0] {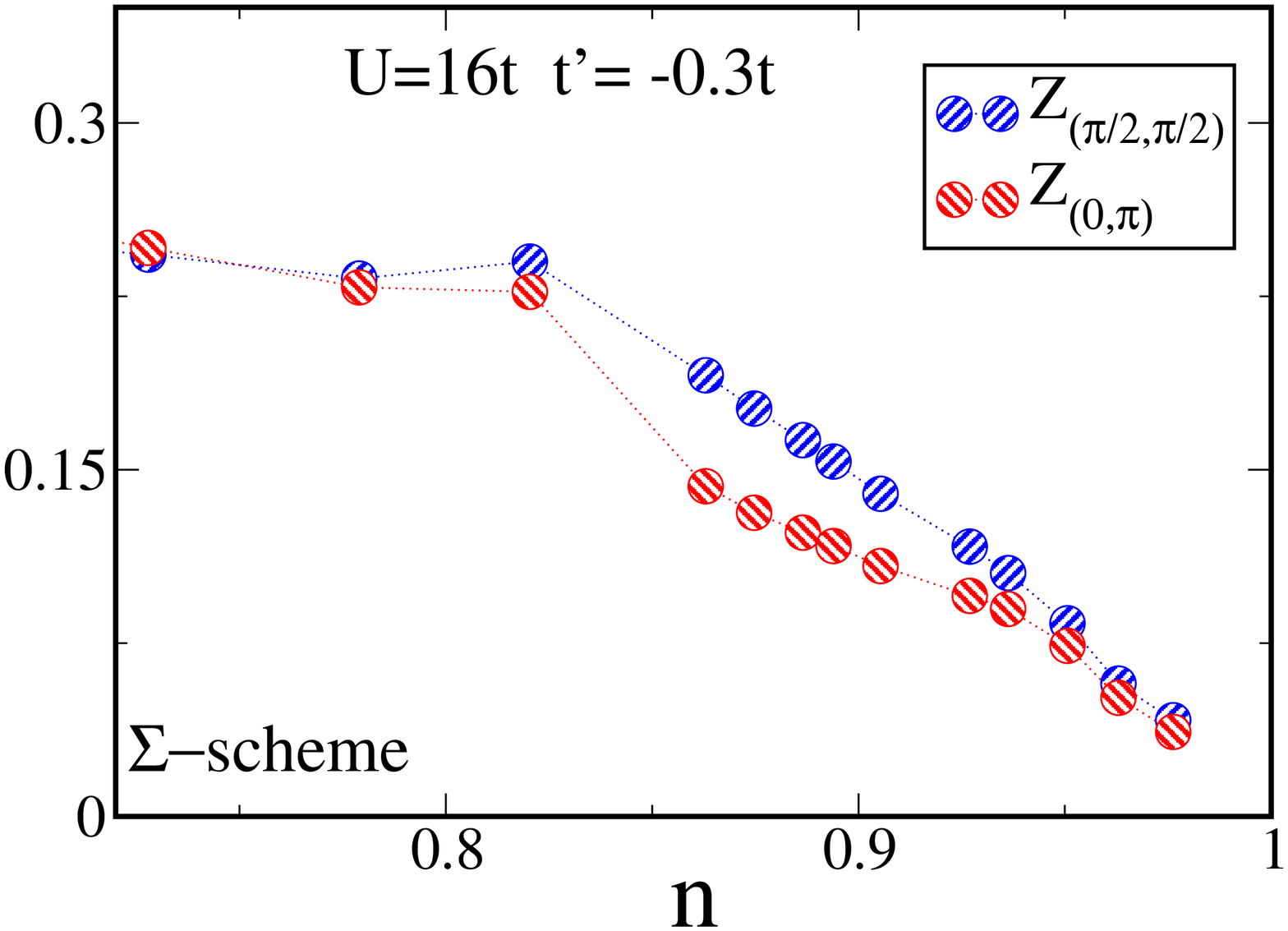}
\includegraphics[width=7cm,height=5cm,angle=-0] {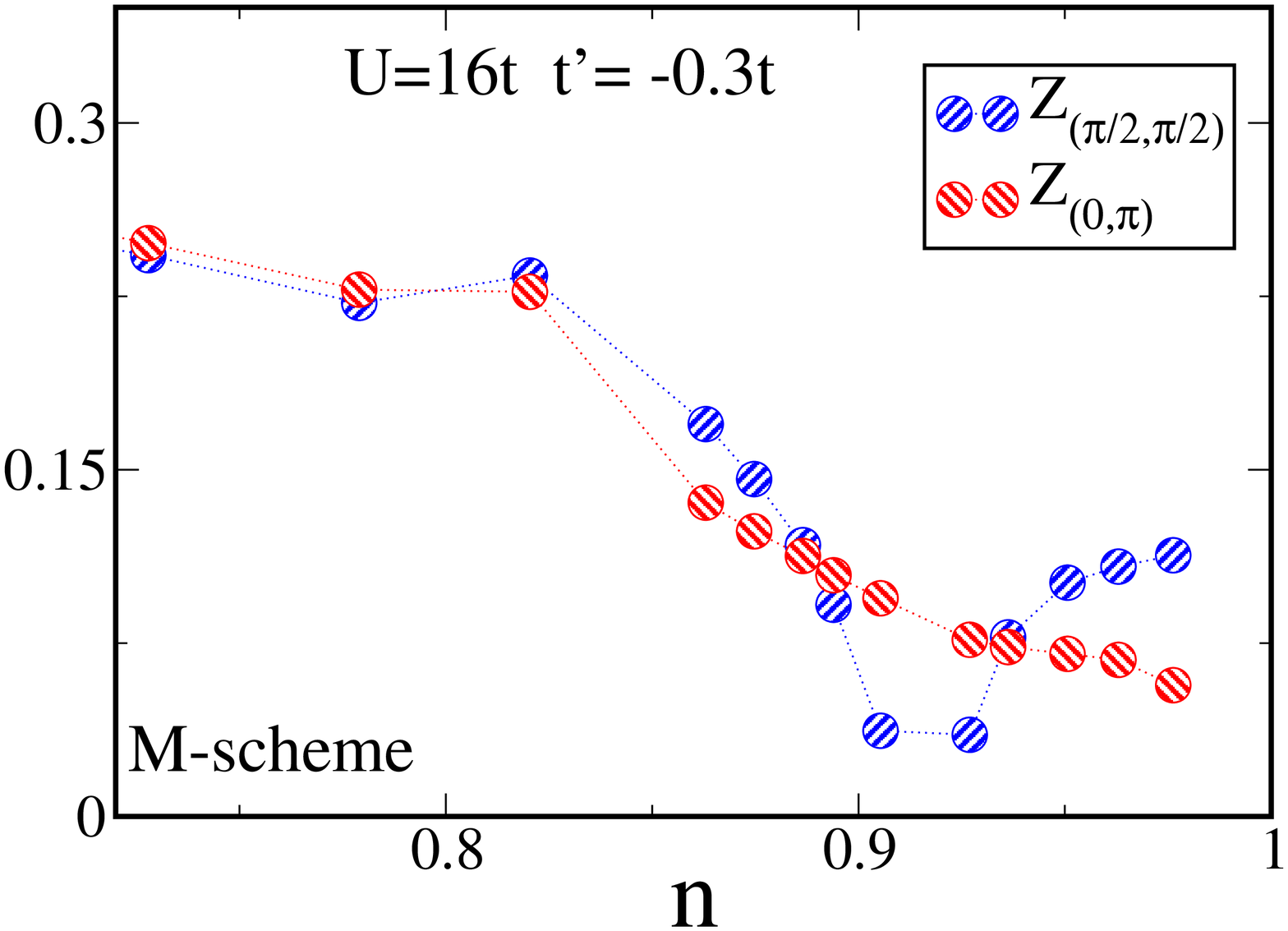}
\end{center}
\caption{Quasiparticle residua in the {\it hot/cold} regions of
the $k$-space. Left panel extracted from the cluster
self-energies with the $\Sigma$-scheme, right panel from the
cluster cumulant with the $M$-scheme. The virtual temperature
$\beta= 32/t$.} \label{Zc-h}
\end{figure}

In order to achieve a better understanding of these differences
between schemes, let's continue to assume a FL state, and imagine
that a good FL description of the system is valid in all methods.
We can see if this hypothesis works or eventually breaks down. In
the following we confront the $\Sigma$-method with the $M$-method,
which, as we have seen, provides results very close to the
$G$-method and it is more easy to hand from the theoretical
viewpoint, as the $k$-dependent quantities we want to study can be
directly linked to its cluster counterparts (as we show in the
following). An indicator of the quasiparticle behavior is their
residuum that can generally be extracted from the slope of the
imaginary part of the self-energy on the Matsubara axis:
\begin{equation}\label{Zk}
  Z(k)= \lim_{\omega\rightarrow 0} \left( 1- \frac{\partial \Im \Sigma_{k}}{\partial \omega}
  \right)^{-1}
\end{equation}
For convenience we link the $k$-dependent quantities to the
corresponding cluster self-energies already used in the previous
section:
\begin{equation}\label{ZX}
  Z_{X}= \lim_{\omega\rightarrow 0} \left( 1- \frac{\partial \Im \Sigma_{X}}{\partial \omega}
  \right)^{-1}
\end{equation}
where $X=A,B,C$ are the three possible cluster-eigenvalues of the
cluster-matrix. As we said, for the $\Sigma$-scheme and the
$M$-scheme we can directly connect the lattice-residua to the
cluster ones. In the $\Sigma$-method :
\begin{eqnarray}
\label{Zflat} Z_\latt(k,\omega)=
                              [ Z^{-1}_{A}(\omega)S_{A}(k)+ Z^{-1}_{B}(\omega) S_{B}(k)+
                                 Z^{-1}_{C}(\omega)S_{C}(k)]^{-1}
\end{eqnarray}
In the $M$-method we can evaluate the cluster cumulant
$\hat{\mathbf{M}}_{c}$:
\begin{eqnarray}
\hat{\mathbf{M}}_{c}=\, [ (\omega+\mu)\mathbf{1}-\,
\hat{\mathbf{\Sigma}}_{c} ]^{-1} \label{Mcluater}
\end{eqnarray}
and this is extracted onto the lattice:
\begin{eqnarray}
\label{Mlat1} M_\latt(k,\omega)=
                                M_{A}(\omega)S_{A}(k)+ M_{B}(\omega) S_{B}(k)+
                                M_{C}(\omega)S_{C}(k)
\end{eqnarray}
where, like with the $\hat{\Sigma}$ cluster matrix, we have
defined cluster cumulant eigenvalues. We can equally express
$M_\latt(k,\omega)$ directly in terms of the cluster
self-energies:
\begin{eqnarray}
\label{Mlat_S} M^{-1}_\latt(k,\omega)=
\left[ \frac{S_{A}(k)}{\omega +\mu - \Sigma_A} +
\frac{S_{B}(k)}{\omega +\mu - \Sigma_B} + \frac{S_{C}(k)}{\omega
+\mu - \Sigma_C}\right]^{-1}
\end{eqnarray}
It is then possible to extract the quasiparticle residua from the
lattice self-energy:
\begin{eqnarray}
\Sigma_{\latt}(k,\omega)=\,  (\omega+\mu)-\, M_{\latt}(k,\omega)
^{-1} \label{SigmaLattM}
\end{eqnarray}
using the formula \ref{Zk} and directly connecting it to the
cluster-resisua:
\begin{eqnarray}
\label{Zflat_M} Z_{\latt}(k,\omega)=
M_{k}^{2}(\omega=0)\,\,\left[ \frac{Z^{-1}_A S_{A}(k)}{(\mu -
\Sigma_A)^2} + \frac{Z^{-1}_B S_{B}(k)}{(\mu - \Sigma_B)^2} +
\frac{Z^{-1}_C S_{C}(k)}{(\mu- \Sigma_C)^2}\right]^{-1}
\end{eqnarray}
\begin{figure}[!!htb]
\begin{center}
\includegraphics[width=12cm,height=9cm,angle=-0] {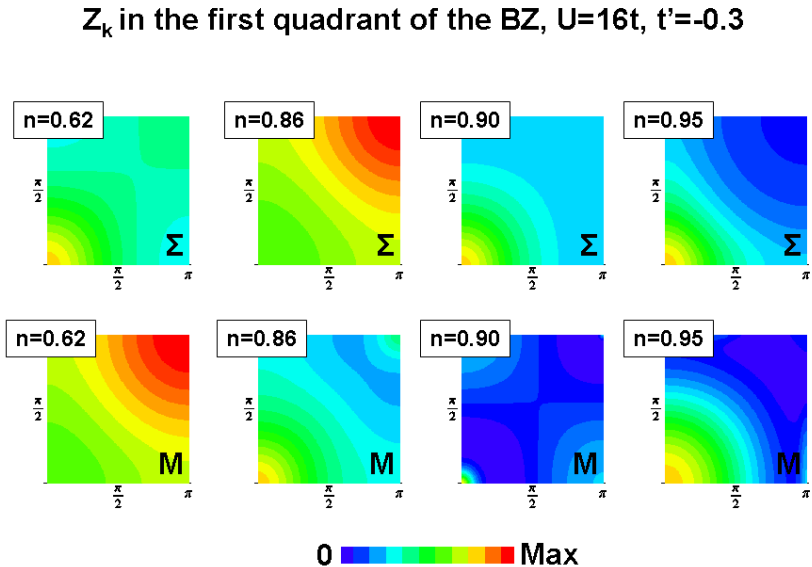}
\end{center}
\caption{Quasiparticle residua in the $k-$space. The first raw
pictures are extracted from the cluster self-energy $\Sigma$,
while the second raw from the cumulant method $M$. The density of
the system with $U=16t$ and $\mathbf{t'=-0.3}$ is: $n=0.62, 0.86,
0.90, 0.95$ from left to right, the inverse virtual temperature
$\beta=32/t$.} \label{fZ}
\end{figure}
\begin{figure}[!t]
\begin{center}
\includegraphics[width=7cm,height=6cm,angle=-0] {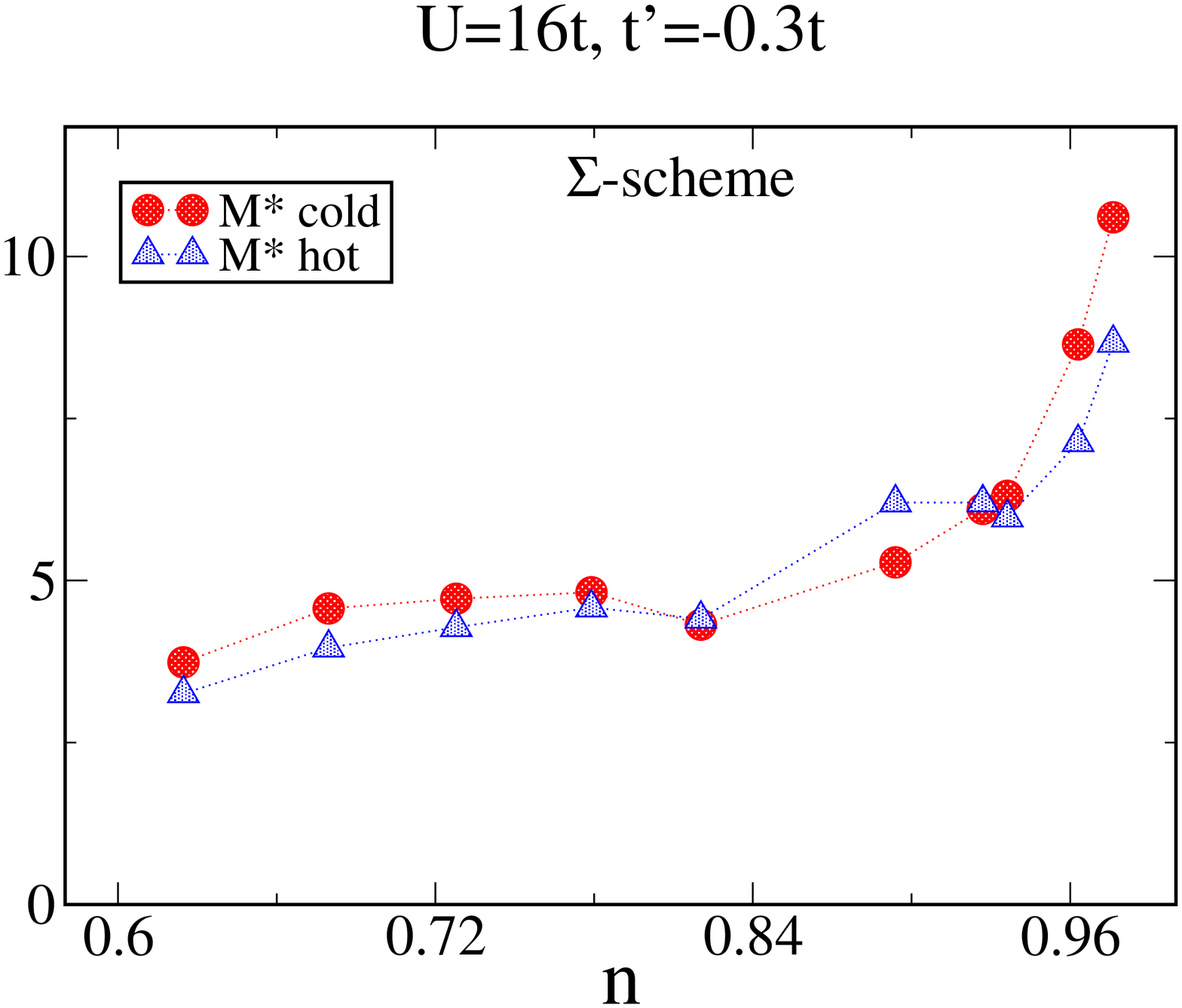}
\includegraphics[width=7cm,height=6cm,angle=-0] {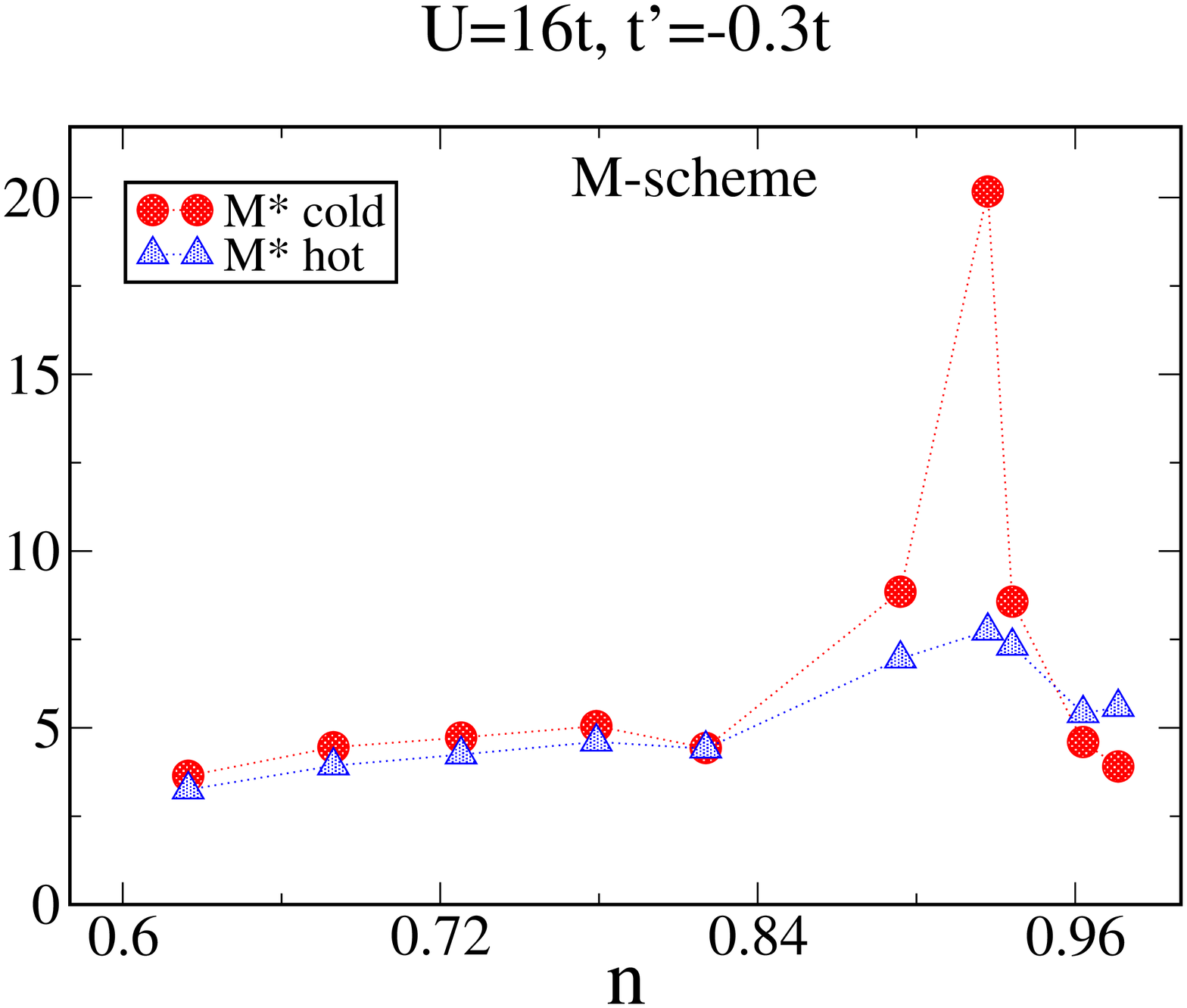}
\end{center}
\caption{Quasiparticle renormalized mass in the {\it hot/cold}
regions of the $k$-space. The $\Sigma$-scheme result is shown in
the left-hand panel , the $M$-scheme result in the right-hand
panel. The virtual temperature $\beta= 32/t$. } \label{MCluster}
\end{figure}
In Fig. \ref{Zc-h} we show the $Z$ extracted with the
$\Sigma$-method on the left and the $M$-method on the right,
evaluated in the $k$-points at the FS in correspondence of the
cold (red circles) and  hot (blue circles) regions. For
convenience's sake we plot only the hole-like case $U= 16t$ and
$t^{\prime}=-0.3$, but the qualitative result is the same for
other values of the parameters. In fact we see that the FL-picture
given by the $\Sigma$-scheme is well realized by the behavior of
the Z's, either in the hot and cold region, which can be seen to
be proceeding roughly linearly to zero as the transition point is
reached at density $n=1.0$. The hot/cold spot modulation is here
evident mainly around 15-10\% doping in the different values of
the Zs, higher for the low scattering cold spot. This is the
extension of the old DMFT results, where a $k$-space modulation
could not be described.
The panel on the right hand side of Fig. \ref{Zc-h} shows the same
Zs extracted with the $M$-method, formula \ref{Mlat1}.
Surprisingly, the behavior in this case is very different, despite
the very similar results we observe in the hot/cold spot
modulation (Fig.\ref{fAk_S_M}). We once again stress that at high
doping values ($> 20\%$ doping), where the modulation in $k$-space
is irrelevant, the methods give essentially the same values.
Approaching the MT however (doping $< 15\%$), in correspondence
with the cold spot (right-hand panel, red circles), where the
quasiparticle should be better defined and the spectral function
$A(k,\omega\rightarrow 0)$ presents a maximum, we see the $Z$
collapsing to zero at around $10\%$ doping, showing after this
point a roughly constant value.
The collapse of $Z_k$ at $10\%$ doping is driven by the lattice
cumulant $M_k$ in formula \ref{Zflat_M}. This is in fact the
quantity which goes to zero by tuning doping. According to
formula \ref{SigmaLattM}, this implies that the lattice
self-energy $\Sigma_k(\omega_n=0^+)$ is infinity exactly in that
point, i.e. the lattice Green's function $G_k(\omega_n=0^+)$ is
zero. The constant value of $Z_k$ approaching the MT point
results from the term $Z_B\,(\mu-\Re\Sigma_B)^2$ in formula
\ref{Zflat_M}, which remains finite despite $Z_B\rightarrow 0$
[which is the quasiparticle-residuum contribution from the
$k$-point $(\pi,\pi)$] for $n\rightarrow 1$, as seen in Fig.
\ref{ZCluster}. This is because the renormalized chemical
potential $\mu-\Re\Sigma_B$ is rapidly increasing in absolute
value, going out of the bare band as compared to the other points
in the $k$-space, as we discussed in Fig. \ref{mu_eff-fig}.
The aforementioned collapse of the $Zs$ in the $M$-scheme
interpretation is extraordinary evident in Fig.\ref{fZ}, where the
full $Z(k)$ are plotted in the first quadrant of the BZ for the
$\Sigma$-scheme (upper row) and the $M$-scheme (lower row), for
the case $t^{\prime}= -0.3 t$. As usual we indicate with blue
color the lower values of $Z$ and with red color the highest
values. Four values of doping (from left to right) $38, 24, 10, 5
\%$ are displayed. The highest doping panels (left hand side $38
\%$ and $24 \%$ doping ) do not display essential differences
between the two methods. When however the MT is approached (third
plot from the left) at $10 \%$ doping, we see that for the
$M$-method the $Z$ literally goes down to zero on a vast region of
the $k$-space (as evident by the wide blue area that appears).
In Fig. \ref{MCluster} we also display the renormalized
quasiparticle mass $m^{*}/m$, obtained from
\[
Z\frac{m^{*}}{m}  =\frac{1}{1+\frac{d \Sigma' (k,0)}{dk_{\perp}
}/\frac{dE_{k}}{dk_{\perp}}}
\]
where $m^{*}$ and $m$ are respectively the renormalized and the
bare effective mass, $k$ is along the Fermi surface and
$\frac{d}{dk_{\perp}}$ means the derivative perpendicular to the
Fermi surface. The $\Sigma$-scheme result (left side of Fig.
\ref{MCluster}) once again displays the behavior expected at in a
FL system which approaches a MT: the quasiparticle renormalized
mass increases close to the phase-transition point (possibly
diverging). This is in agreement with the previous single-site
DMFT results. In the $M$-scheme instead (right side of Fig.
\ref{MCluster}) the mass divergence takes place in the cold
region around $10\%$ doping, where the self-energy $\Sigma_k$
diverges and the Green's function is zero. After this critical
point, $m^{*}/m$ comes back to an approximately constant value
which maintains until the MT point. In the hot region, the
renormalized mass remains instead always approximately constant.
In the $M$-scheme framework, this behavior hints at the existence
of a transition around $10\%$, which is, instead, washed out by
the FL-perspective given by the $\Sigma$-scheme. Such a
phenomenon indicates that something dramatic may take place at
$10\%$. This is worth to be investigated in the light of the very
different physical interpretation that the $M$-scheme presents
with respect to a FL $\Sigma$-scheme viewpoint.
\section{The M-scheme perspective: an anomalous FL}
How is the $M$-scheme periodization able to break down the simple
FL description? As already mentioned, in the $\Sigma$-scheme the
lattice self energy is a simple linear extrapolation (formula
\ref{selflat1}):
\begin{equation}
\Sigma_\latt(k,\omega)= \Sigma_{A} S_{A}(k)+ \Sigma_{B} S_{B}(k)+
                                 \Sigma_{C} S_{C}(k) \nonumber
\end{equation}
where the function $S_{X}$ ($X= A,B,C$) was given in formula
\ref{S_X}. In the $M$-case instead the lattice self-energy is
given by a highly non-linear relation
\begin{equation}
\label{selflat_M2} \Sigma({\bf k}, \omega) = \,\imath \omega+ \mu-
\,\left[ \frac{S_{A}(k)}{\omega +\mu - \Sigma_A} +
\frac{S_{B}(k)}{\omega +\mu - \Sigma_B} + \frac{S_{C}(k)}{\omega
+\mu - \Sigma_C}\right]^{-1},
\end{equation}
We find that at zero temperature the imaginary parts of the
cluster self-energies go to zero at zero frequency (Fig.
\ref{fImS}). For the following discussion we can therefore write
the lattice Green's function as
\begin{equation}
G({\bf k},\omega) = \frac{1}{\omega - r({\bf k},\omega) -
i\eta({\bf k},\omega)},    \label{rk}
\end{equation}
where $\eta({\bf k},\omega),\, \lim_{\omega\rightarrow 0} \eta=0 $
represents the imaginary part of the self-energy and $r({\bf
k},\omega) = \epsilon({\bf k}) - \mu + \Re\Sigma({\bf k}, \omega)$
is the renormalized energy. Concerning the real part of the
cluster self-energy, we distinguish two regimes. At large doping
the diagonal cluster self-energies are dominated by the local
component and Eq. (\ref{selflat_M2}) reduces in the first
approximation to Eq. (\ref{selflat1}). In this regime the physics
is almost local with small corrections due to short-range
correlations. All the periodization schemes converge and the
single-site DMFT represents a good first order approximation. In
contrast, close to the Mott transition the short-range
correlations become important and the off-diagonal components of
the cluster self-energy become comparable with the local one. As a
consequence, at zero frequency the denominators in Eq.
(\ref{selflat_M2}) may acquire opposite signs generating a
divergence in the lattice self-energy. This divergence of
$\Sigma({\bf k}, \omega=0)$, or equivalently of $r({\bf k})$,
corresponds to a zero of the lattice Green's function. Here the
main differences between the two schemes stands. The $M$-method
allows for the appearances of lines of zeroes in the Green's
function, a possibility not admitted with the $\Sigma$-scheme,
and that is likely to be at the origin of the differences and
anomalous $Zs$ we found in the previous section. We now study
this $M$-effect in a specific case.
\subsection {FS topology transition and Pseudogap regime}

We present in this section results obtained with distance
\ref{distf2} between the continuous hybridization function from
the impurity solver $\Delta(\imath \omega)$ and
$\Delta_{nb}(\imath \omega)$, parametrized by the truncated bath:
\beqa \hbox{ dist }= \sum_{n}^{Nw_{off}} \, \| \Delta(\imath
\omega_{n}) - \Delta_{nb}(\imath \omega_{n}) \| \label{BumDist}
\eeqa where $Nw_{off}$ is a cut-off frequency chosen so that
$\omega_{n}\simeq 2\,t$. This distance is designed to best
describe the low-energy scale in the system, with the aim of
grasping the behavior at $10\%$ doping, more than the MT point.
\begin{figure}[!htb]
\begin{center}
\includegraphics[width=10.0cm,height=8.0cm,angle=-0] {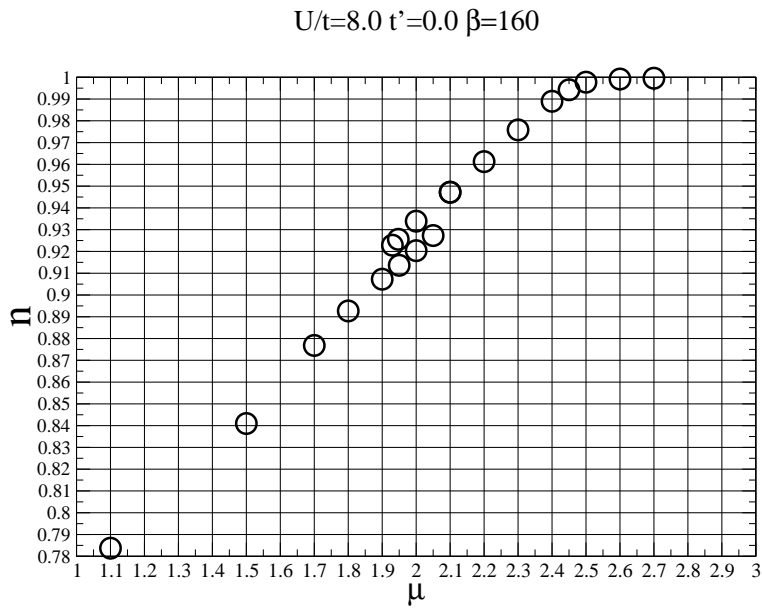}
\caption{Density $n$ vs. chemical potential $\mu$} \label{mu_dens}
\end{center}
\end{figure}
\begin{figure}[!htbp]
\begin{center}
\includegraphics[width=10.0cm,height=8.0cm,angle=-0] {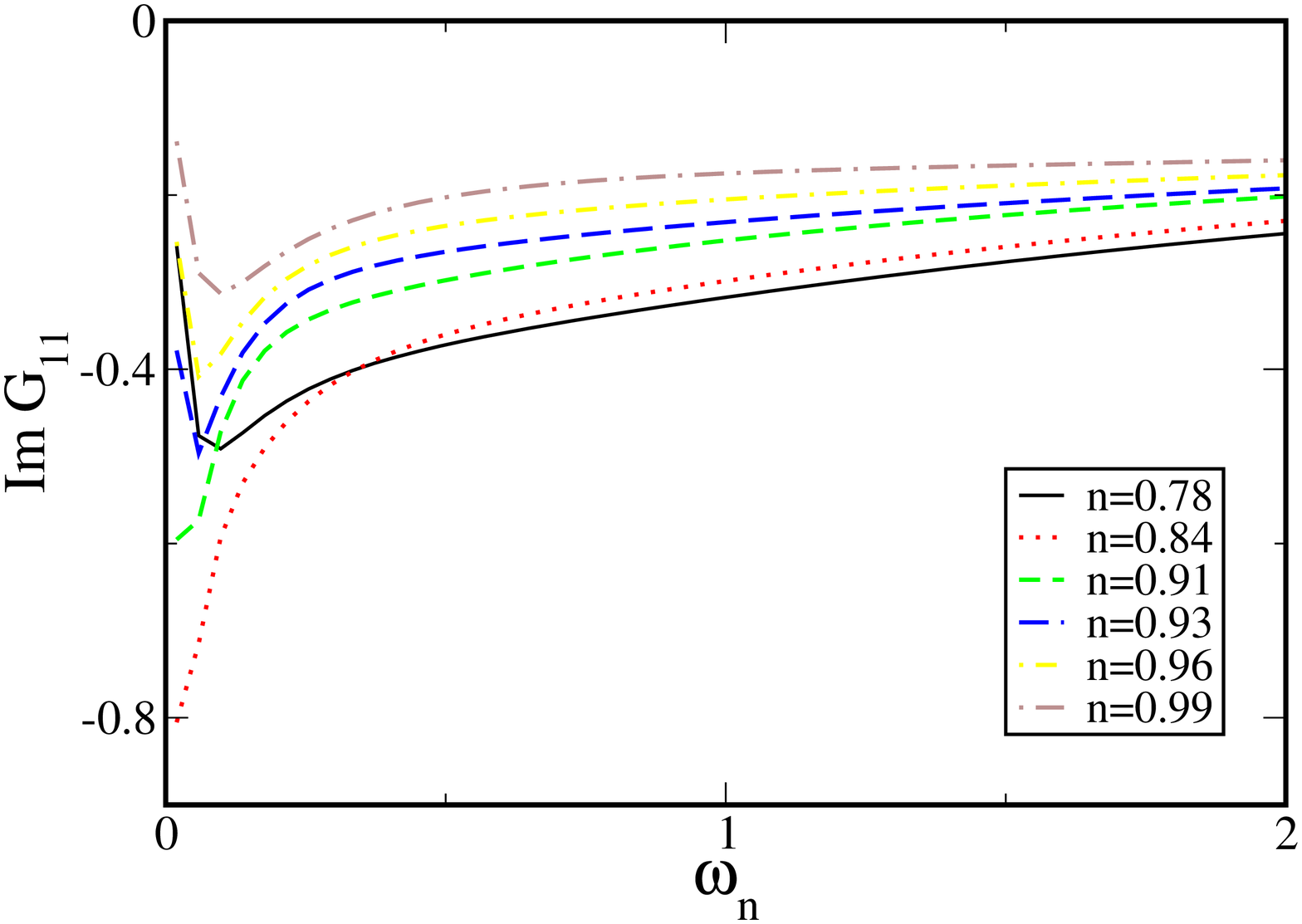}
\includegraphics[width=10.0cm,height=8.0cm,angle=-0] {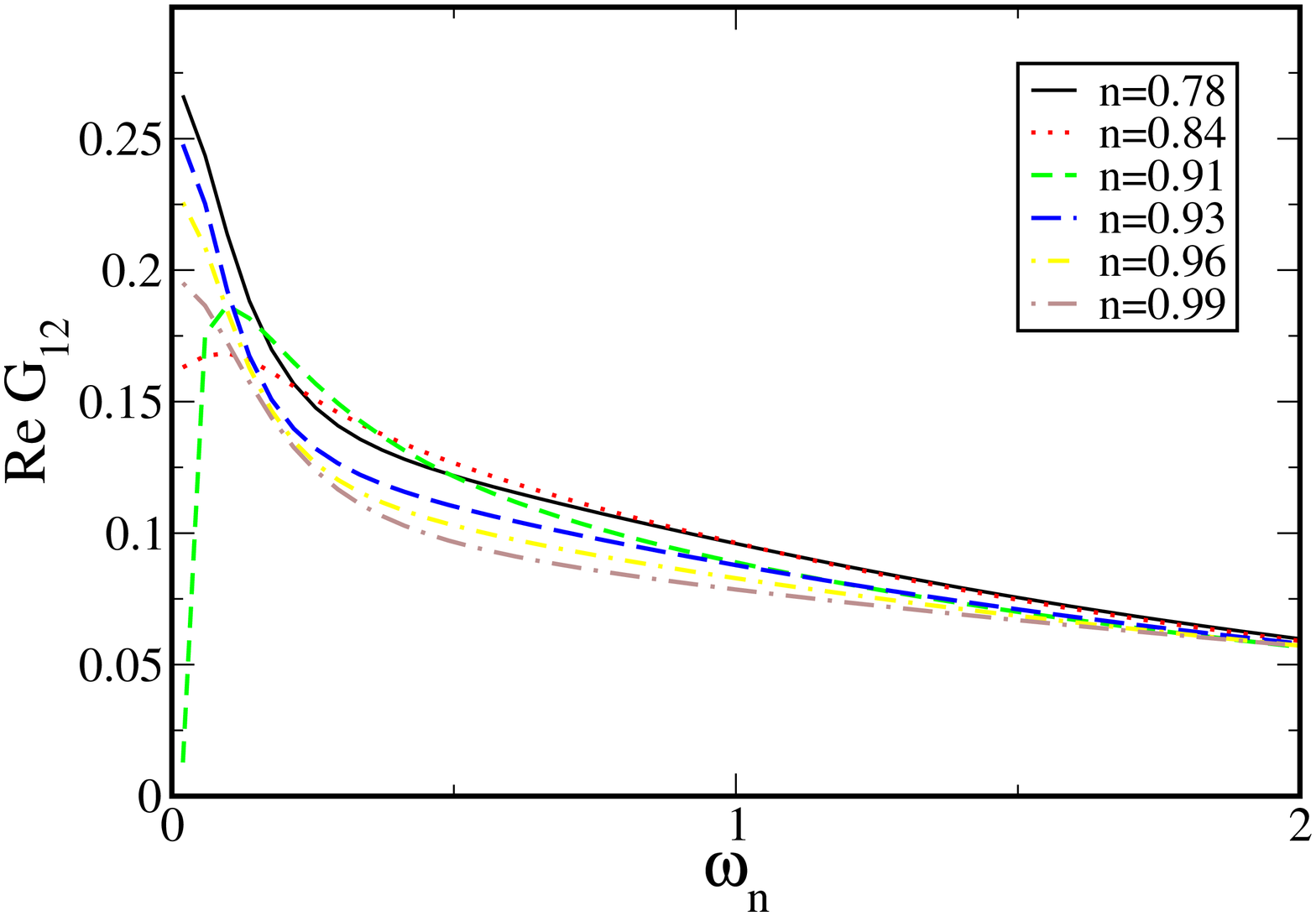}
\caption{Imaginary part of the local Green's function $\Im
G_{11}$ (top) and real part of the nearest neighbor Green's
function Re$G_{12}$ for $U/t=8.0$ and no frustration
$t^{\prime}=0$.} \label{G_U8}
\end{center}
\end{figure}
\begin{figure}[!htbp]
\begin{center}
\includegraphics[width=10.0cm,height=8.0cm,angle=-0] {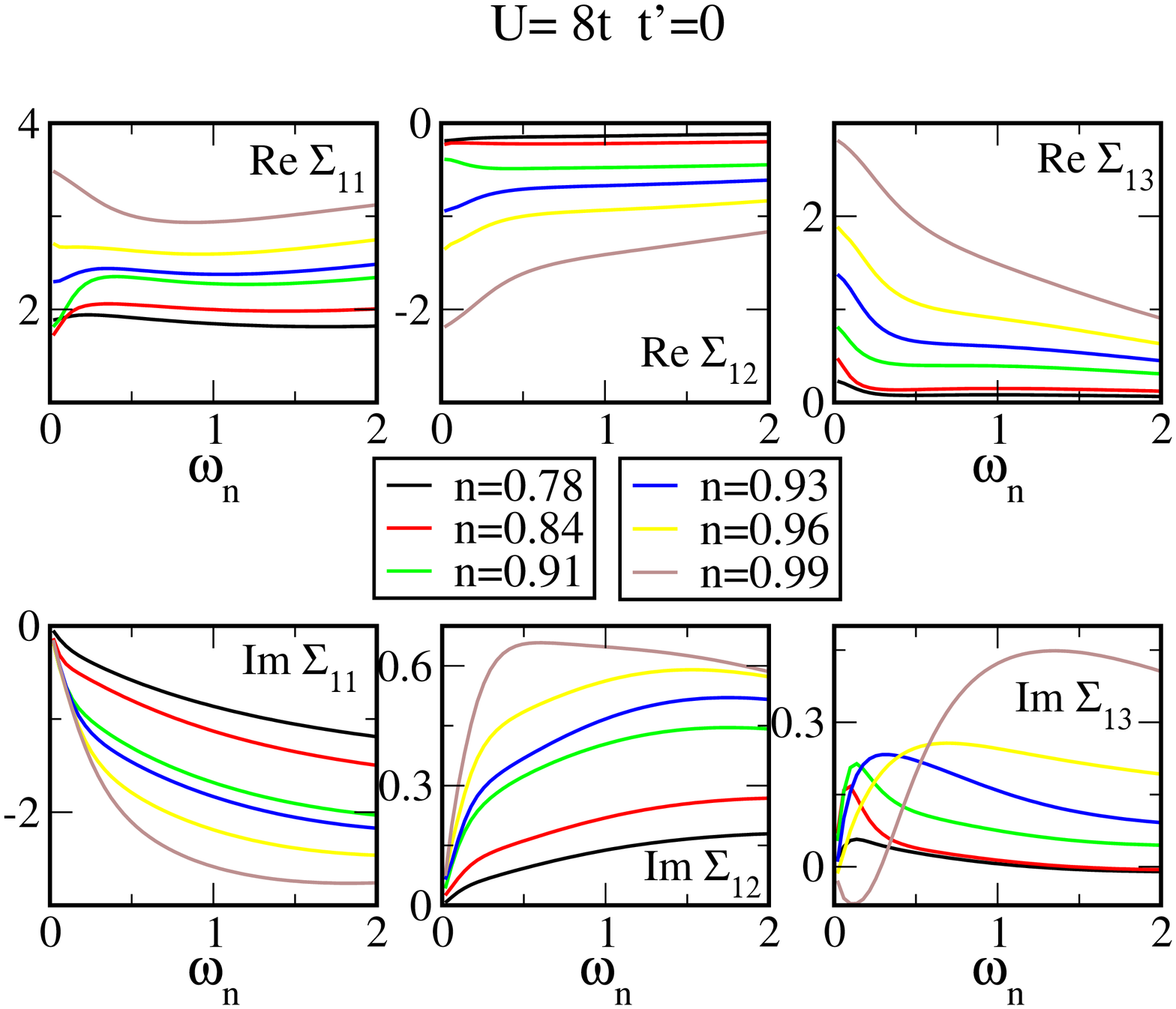}
\caption{Cluster self-energies $\Sigma_{\mu\nu}$ on the Matsubara
axis for $U/t=8.0$, no frustration $t^{\prime}=0$ and inverse
virtual temperature $\beta=160/t$.} \label{S_U8}
\end{center}
\end{figure}
\begin{figure}[!htbp]
\begin{center}
\includegraphics[width=10.0cm,height=8.0cm,angle=-0]
{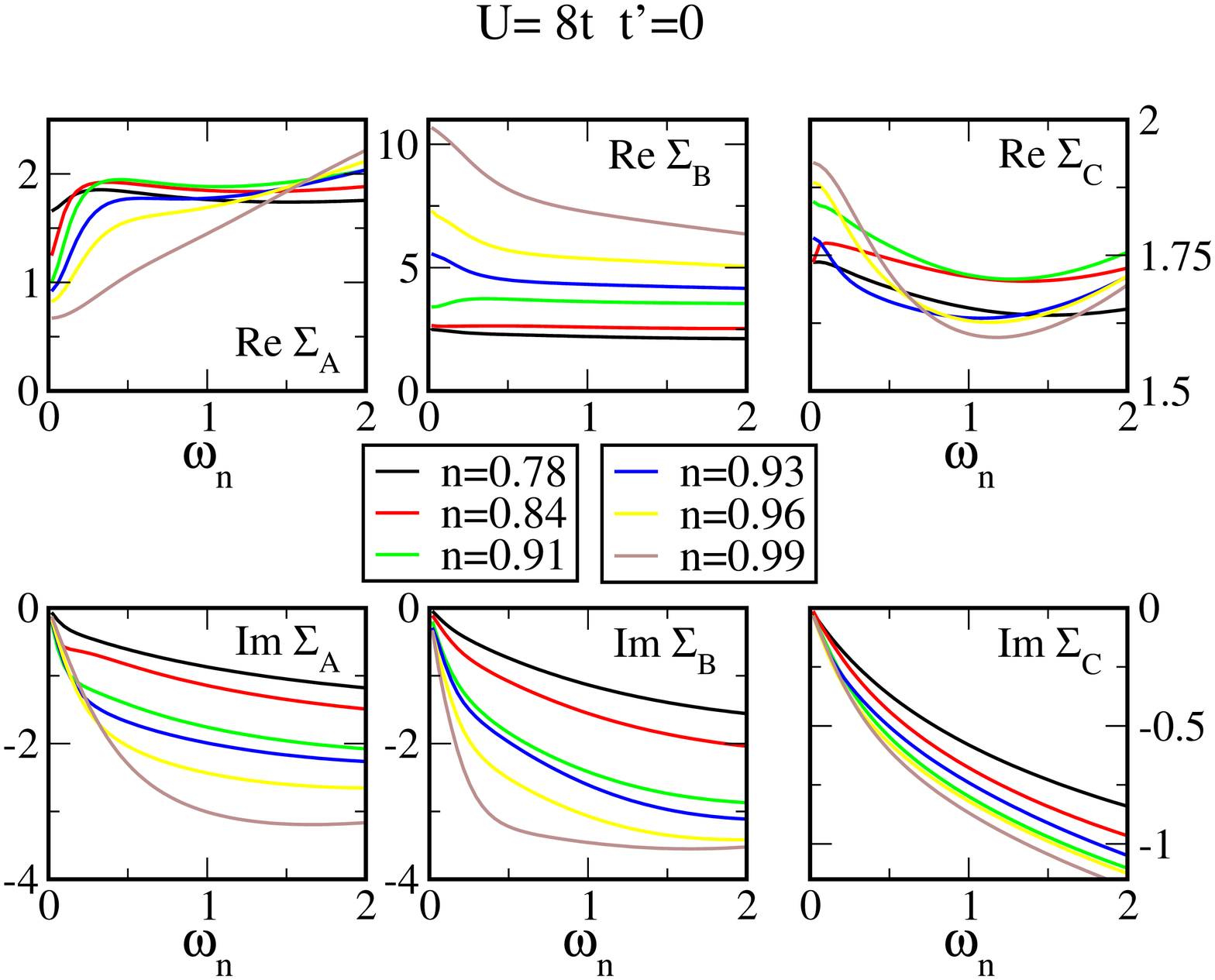}, \caption{Eigenvalues of the Cluster
self-energies $\Sigma_{A}$, $\Sigma_{B}$ and $\Sigma_{C}$ on the
Matsubara axis for $U/t=8.0$, no frustration $t^{\prime}=0$ and
inverse virtual temperature $\beta=160/t$.} \label{Sx_U8}
\end{center}
\end{figure}
\begin{figure}[!htbp]
\begin{center}
{\bf U=8t, t'=0 } \vspace{1cm}\\
\includegraphics[width=6.0cm,height=5.0cm,angle=-0]
{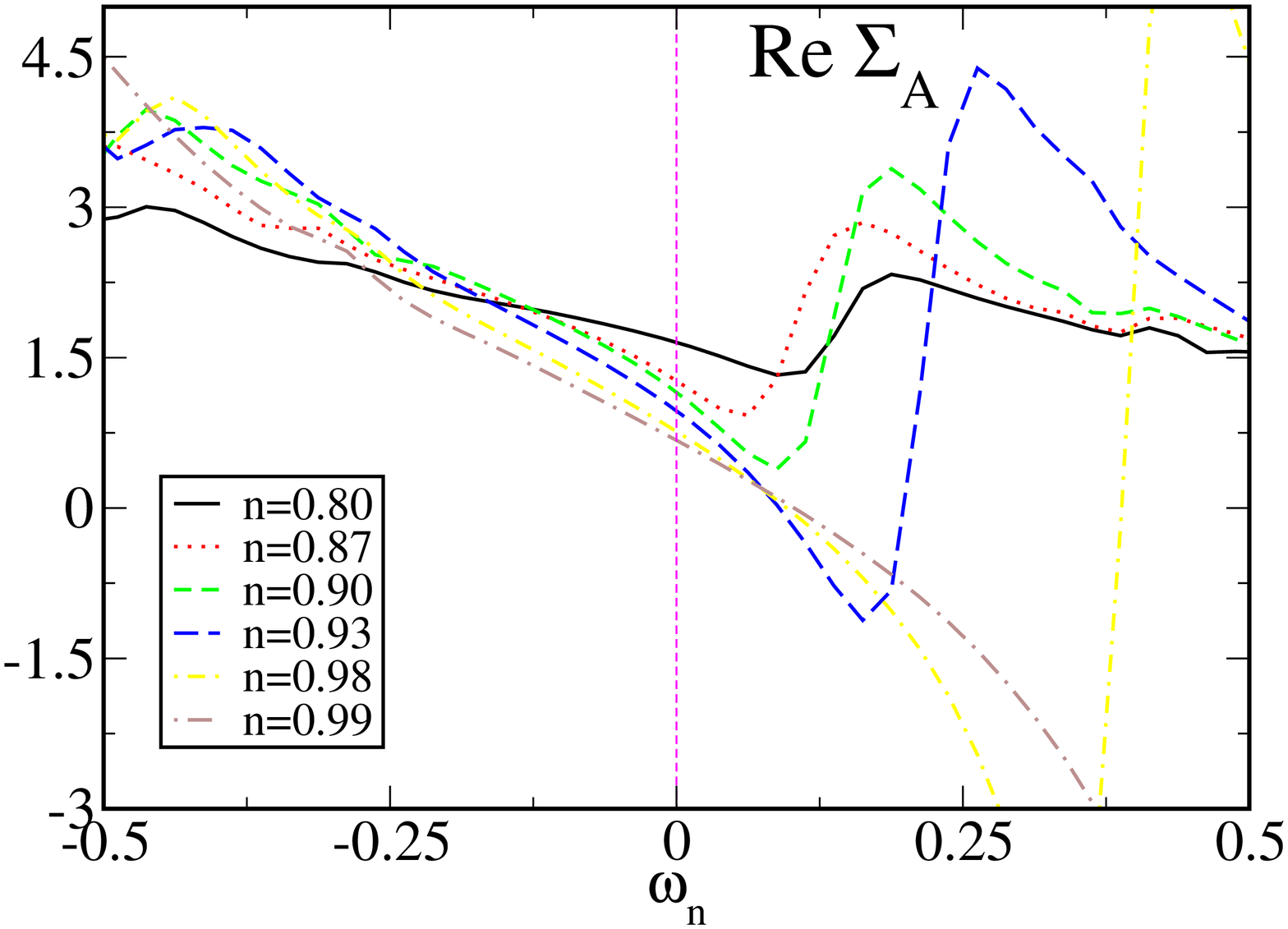}
\includegraphics[width=6.0cm,height=5.0cm,angle=-0]
{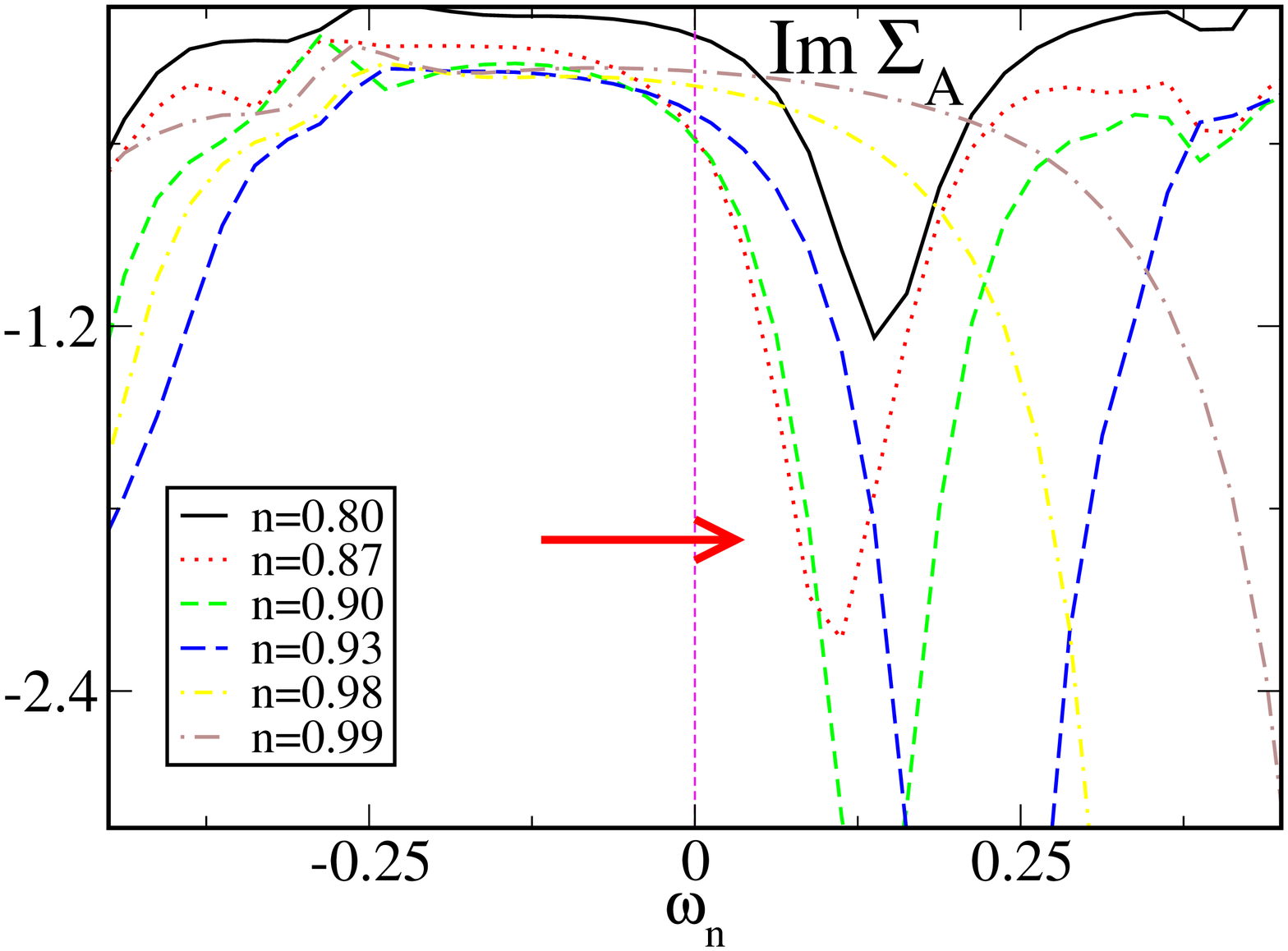} \\
\includegraphics[width=6.0cm,height=5.0cm,angle=-0]
{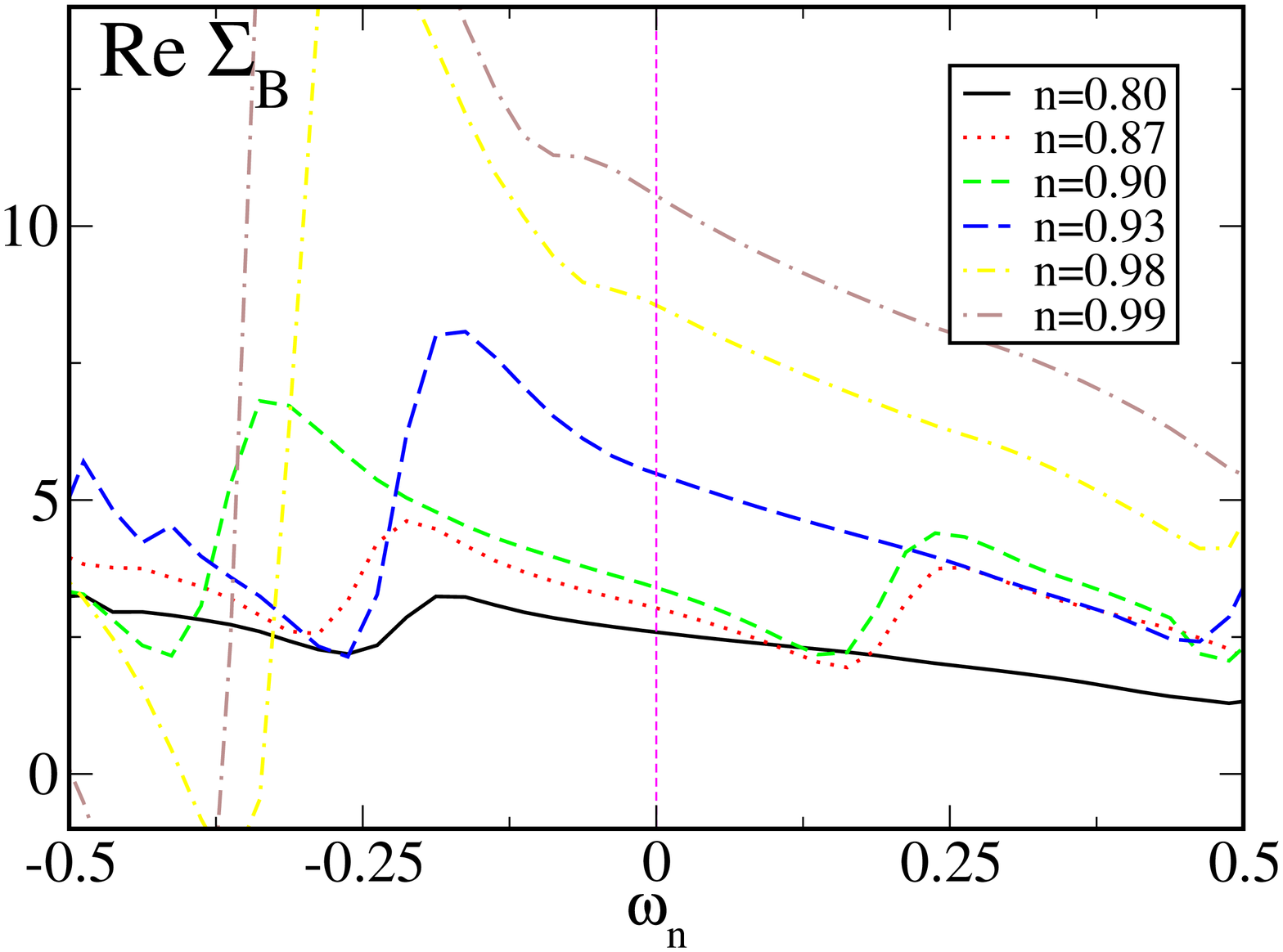}
\includegraphics[width=6.0cm,height=5.0cm,angle=-0]
{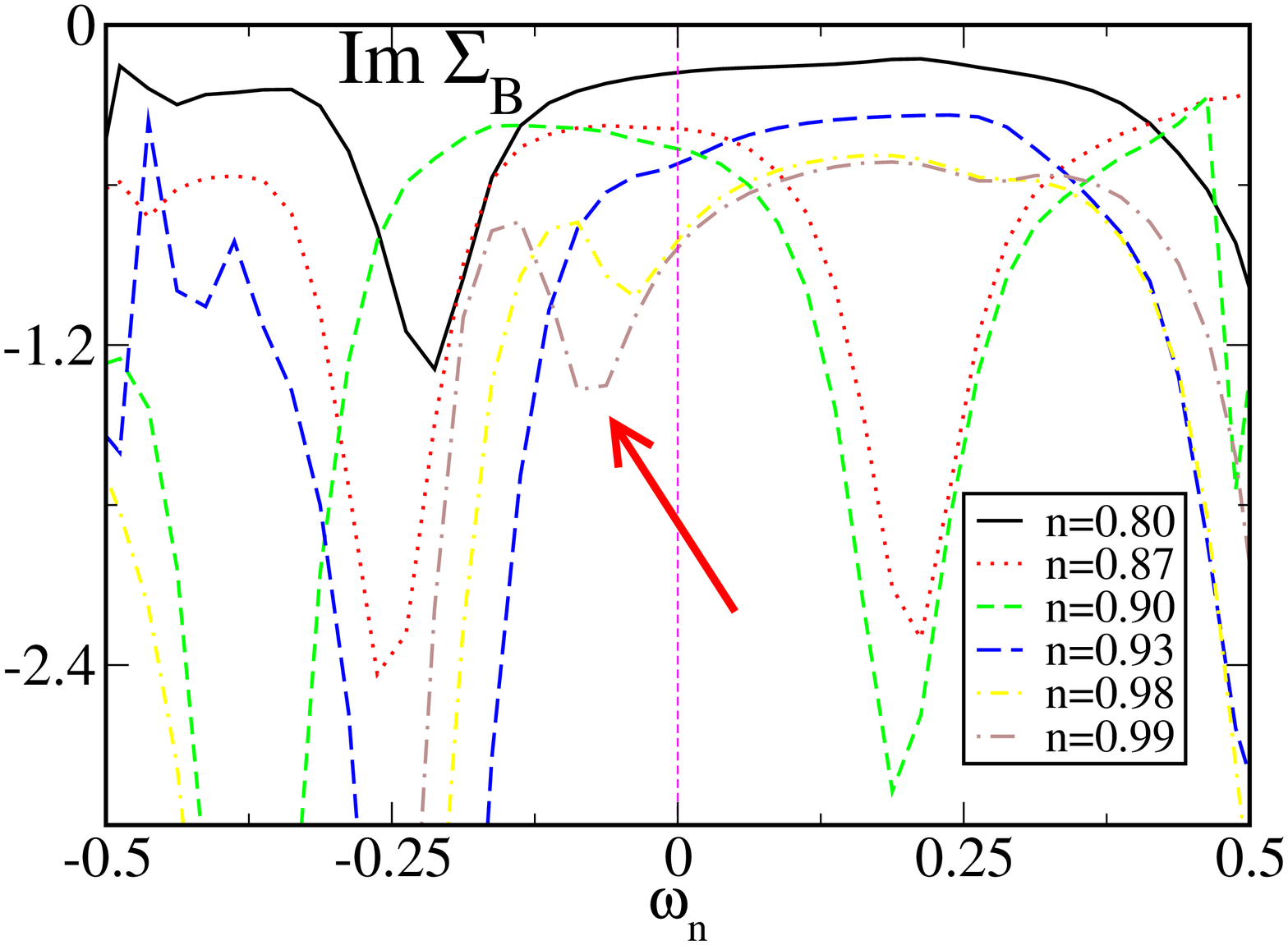} \\
\includegraphics[width=6.0cm,height=5.0cm,angle=-0]
{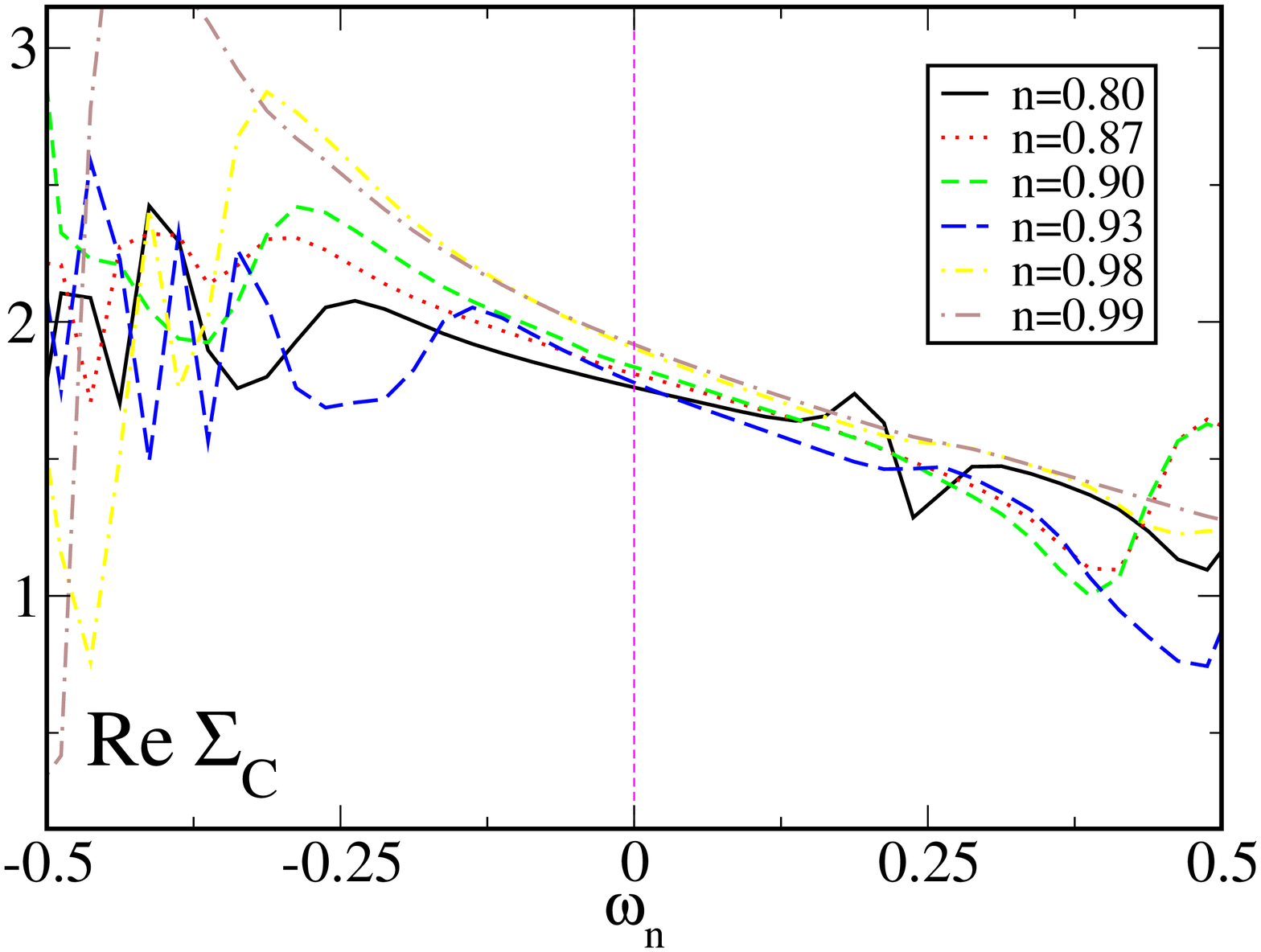}
\includegraphics[width=6.0cm,height=5.0cm,angle=-0]
{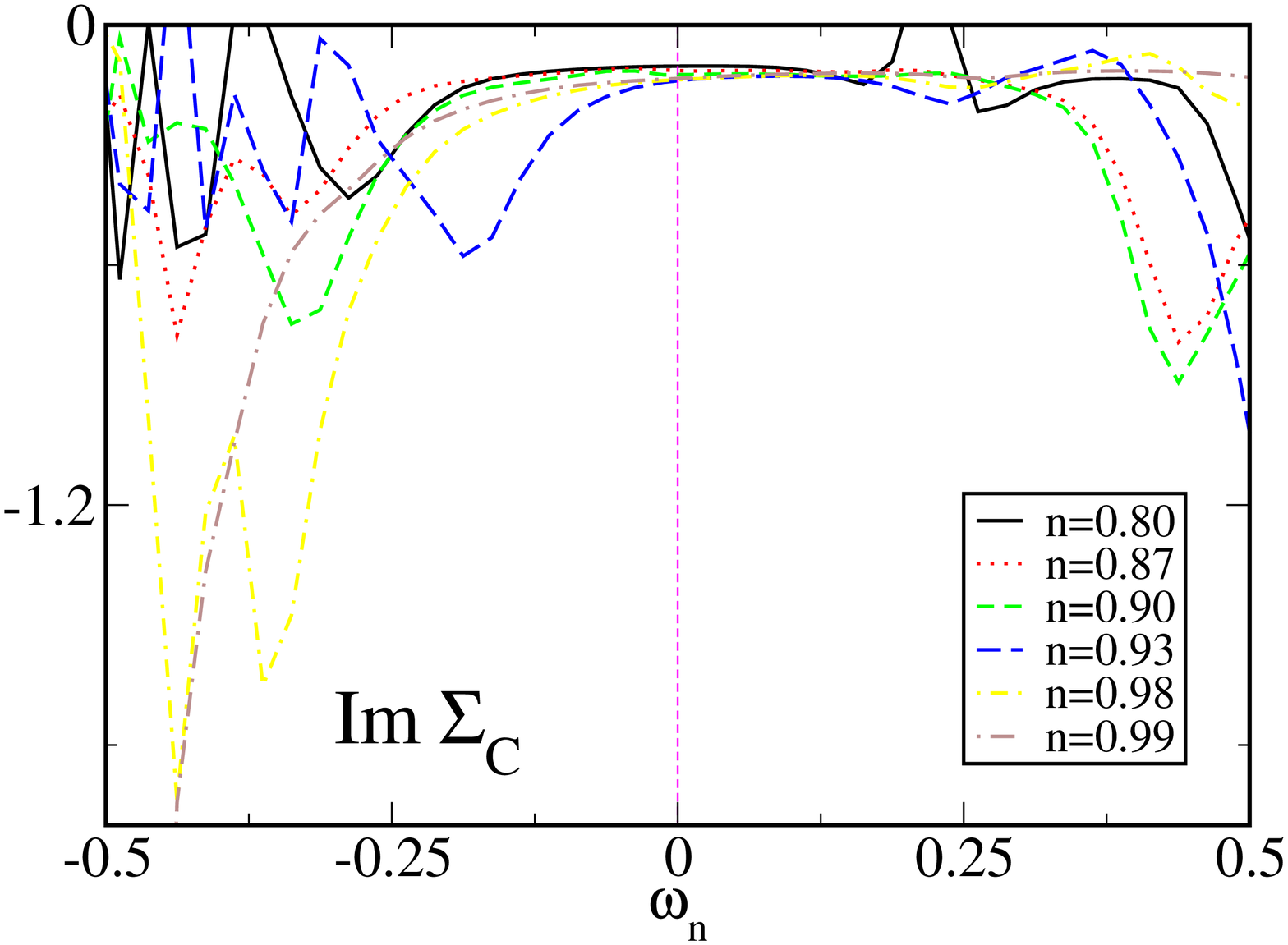}  \caption{Eigenvalues of the Cluster
self-energies $\Sigma_{A}$, $\Sigma_{B}$ and $\Sigma_{C}$ on the
real axis for $U/t=8.0$, no frustration $t^{\prime}=0$ and
inverse virtual temperature $\beta=160/t$.} \label{SABCret}
\end{center}
\end{figure}
We present here the case $U/t=8.0$ and $t'=0$. With this value of
the local interaction, relevant for cuprates superconductor
materials, we were able to use a very low virtual temperature on
the Matsubara axis. The inverse virtual temperature $\beta$ in the
data presented here is $\beta=160/t$ (energy units of
half-band-width $4t$). This allows us to better describe the low
energy physics, capturing the small frequency behaviour of the
Green's functions (displayed in Fig. \ref{G_U8} on the Matsubara
axis, where no artificial spreading in the analytic continuation
is required) and of the cluster self-energies $\Sigma_{ij}$ (Fig.
\ref{S_U8}) and the eigenvalues of the cluster-self-energy
$\Sigma_{X}$ (Fig. \ref{Sx_U8}). From $\Im \Sigma_{ij}$ in
particular (Fig. \ref{S_U8}) we can see that the extrapolation to
a FL behaviour in the cluster quantities for $\omega \rightarrow
0$, extracted previously in the strong coupling case $U= 16t$,
holds up to the low temperature reached in this case. In Fig.
\ref{SABCret} we also present the eigenvalues of the
cluster-self-energy on the real axis, which, though less clean
than those on the Matsubara axis, are easier to physically
interpret in comparison to a FL system. In Fig. \ref{mu_dens} we
present first the density of particles in the system versus the
chemical potential. We notice there are two lines of solutions
which coexist in a small region around 10\% doping, well before
the MT. Both phases are in a metal state. At this level of energy
resolution therefore, the CDMFT solution apparently detects a
first-order phase transition between two states with no evident
order parameter. It is difficult to be categorical on this point
and on the nature of the transition, as it is difficult to reach
this region accurately. The existence of two solutions may indeed
be a fake effect of CDMFT which tries to describe the total
behavior of the system with only 3 cluster-selfenergy parameters.
A possibility of a second order phase transition, maybe even a
quantum critical point (QCP) cannot be ruled out. Looking at the
cluster quantities, it is not trivial to detect the appearance of
the critical region. For example, the local propagator
$G_{11}(\omega_n)$ (in the top panel of Fig. \ref{G_U8} we show
the imaginary part on the Matsubara axis) and the eigenvalue of
the cluster-self-energies (Fig. \ref{Sx_U8}) show a systematic
and continuous behavior approaching the MT, as already described
for the $U=16t$, $t^{\prime}=\pm0.3$ case in section 5.1. An
irregular behavior is observed in $\Re G_{12}$, which decreases
for decreasing doping up to $10\%$, where it has a sudden drop to
zero. It then jumps back to a finite value for smaller doping,
and increases until the Mott insulator is reached. A singular
behavior , not observed in $U=16t\, t^{\prime}=\pm0.3$ case of
section 5.1 because of the too high virtual temperature
$\beta^{-1}=32$, is displayed also by the low energy part of the
nearest next-neighbor cluster-self-energy $\Im \Sigma_{13}$ (Fig.
\ref{S_U8}). As a matter of facts $\Im \Sigma_{13}$ grows
positive up to $10\%$ doping with an increasing positive slope
for $\omega_n\rightarrow 0$ and after passing this critical
point, it decreases, with the slope going to zero and eventually
changing sign close to the MT. Notice that the energy-scale of
this feature is small ($\sim 0.1t$). This suggest a scenario in
which approaching the critical doping ($10\%$) the system has the
tendency to break the FL (i.e. $\Im \Sigma_{13}$ does not go to
zero for $\omega_n\rightarrow 0$), and the way CDMFT describes
this effect is by providing a region of coexistence of two FL
"close to the breaking" point. This idea can also be supported by
looking at the eigenvalues of the cluster self-energies on the
real axis instead than on the Matsubara Fig. \ref{SABCret}. In a
FL we expect the Real part to cross the $\omega=0$ point with a
linear slope, the imaginary part has to be negative, in order to
preserve casuality, and have a local minimum at $\omega=0$. We see
that approaching the critical doping ($10\%$), either from the
overdoped and the underdoped side, $-\Im \Sigma_A$ has a maximum
in the self-energy which moves close to $\omega=0$, i.e. it moves
toward breaking the FL (phenomenon indicated in Fig.
\ref{SABCret} by a arrow, top-right panel). A similar effect
(also indicated by a arrow, right panel in the second row) takes
place in $-\Im \Sigma_B$ at the MT point. Therefore, the region
around $k=(0,\pi) [(\pi,0)]$, labeled A, drives the instability
at $10\%$ doping, which hints to a breaking of the simple FL
picture (and the effect taking place in $k=(0,\pi) [(\pi,0)]$
strongly suggest the appearance of a pseudogap, as we will see in
the following), while, as already pointed out in sections 5.1 and
5.2, the region around $k=(\pi,\pi)$, labeled B, drives the system
into the Mott transition. The $M$-scheme scenario is exciting,
proposing a critical $10\%$ point which precedes the well-known
Mott critical point at zero doping. The former is in some way
connected to an instability appearing in the $k=(0,\pi)
[(\pi,0)]$ A-region, but we cannot unfortunately access this
point close enough with the present technique to make conclusive
statements.
We therefore concentrate our study on the over-doped region and at
the small doped region around this critical area.
\begin{figure}
\begin{center}
\includegraphics[width=12.0cm,height=9.0cm,angle=-0]{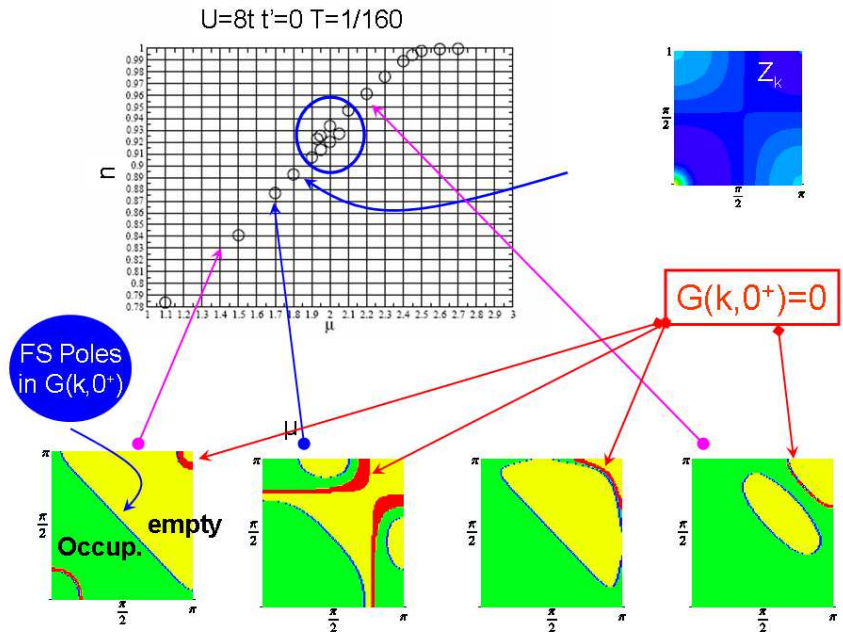}
\caption{Renormalized energy, $r({\bf k})$, (upper panels) and
spectral function, $A({\bf k})$, (lower panels) for the 2D
Hubbard model with $U=8t$ and $T=0$. The color code for the upper
panels is: green  ($r<0$), blue  ($r=0$), yellow ($r>0$), red
($r\rightarrow\infty$).} \label{FIGslide}
\end{center}
\end{figure}
\begin{figure}
\begin{center}
\includegraphics[width=0.5\textwidth]{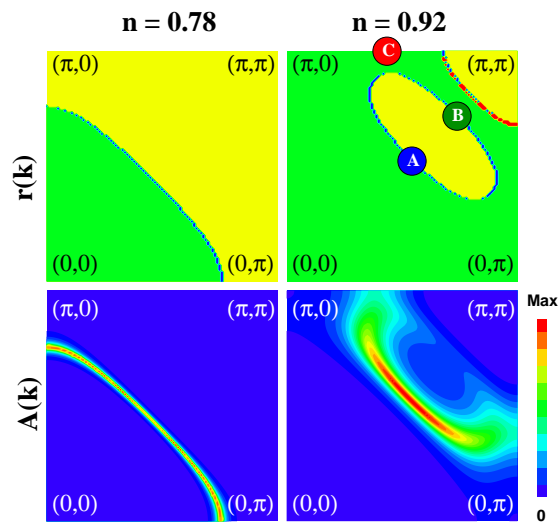}
\caption{Renormalized energy, $r({\bf k})$, (upper panels) and
spectral function, $A({\bf k})$, (lower panels) for the 2D
Hubbard model with $U=8t$ and $T=0$. The color code for the upper
panels is: green  ($r<0$), blue  ($r=0$), yellow  ($r>0$), red
($r\rightarrow\infty$).} \label{FIG2}
\end{center}
\end{figure}
To illustrate this, we show in \ref{FIGslide} how the
renormalized energy, $r({\bf k})= \varepsilon_k+ \mu
-\Sigma_{\epsilon_k}$, and the spectral function $A({\bf k},
\omega=0) = -(1/\pi)\, \Im G({\bf k},0)$ evolve as a function of
doping. For a better comparison, on the top of this figure we
show again the density $n$ vs chemical potential $\mu$ plot,
enlightening in a blue circle the region of co-existence around
$10\%$ doping, and the point where the quasiparticle residuum
$Z_k$ has the collapse to zero in the cold region. The lower part
of the figure shows the first quadrant of the BZ for various
values of doping. The blue lines represent poles in the Green's
function for which $r({\bf k})=0$, i.e. the FS. The green region
is the $k$-space occupied by electrons for which $r({\bf k}) < 0$,
the yellow is the empty region, $r({\bf k}) > 0$. Notice that
already at $\sim 17 \%$ doping (right side), besides the FS (blue
line) regions of infinite self-energy (red-line) appear. These
correspond to zeroes in the Green's function, and are absent at
high doping, as evidenced in the following picture \ref{FIG2},
where a high value of doping ($n=0.78$) is displayed (left panel).
At this point the line of zeros is far from the FS, therefore it
does not affect the physics of the system. However, approaching
the $10\%$ doping, the lines of zero approach the FS more and more
until touching and breaking it. At this point we are in the
critical region, which is not well described by our calculation.
So it is difficult to state in detail how the FS modifies. The
result of passing trough this critical point however is the most
left quadrant of Fig. \ref{FIGslide}, which shows a FS with a
different topology: a pocket centered around
$(\frac{\pi}{2},\frac{\pi}{2})$. The critical point can be seen
therefore like a topological phase transition of the FS, arising
from the appearance of zeros in the Green's function.
Although, as we said, we cannot describe well the critical point,
we can get however insight on the nature of the phase resulting
from this transition. In Fig. \ref{FIG2} we display the
renormalized energy, $r({\bf k})$, and the spectral function
$A({\bf k}, \omega=0) = -(1/\pi)\, \Im G({\bf k},0)$ for two
values of doping outside the critical region. For $n=0.78$ (left
panels) we have a large electron-type FS (blue line in the
$r({\bf k})$ panel, a pole for the Green's function) separating
the occupied region of the Brillouin zone (green), from the
unoccupied region (yellow). The Fermi surface can also be traced
in the $A({\bf k})$ panel as the maximum of the spectral function
(red colors). This what is expected from the FL picture. On the
other hand, for $n=0.92$, after we passed trough the critical
point, a qualitatively different picture emerges in the topology
of the FS . As we said, the FS (blue line) is now represented by
a hole pocket and, in addition, we have a line of zeros of the
Green's function (red line) close to the $(\pi,\pi)$ region of
the Brillouin zone. Furthermore, there is no one-to-one
correspondence between the Fermi surface and the maximum of the
spectral function. This behavior has two causes:
\begin{itemize}
\item the proximity of a zero line for the Green's function (the
red one) suppresses the weight of the quasiparticle, so that in
correspondence of the FS we do not observe a high spectral weight.
\item for k-points corresponding to $r({\bf k}) \neq 0$ the quasiparticles
are pushed away from $\omega=0$ and a pseudogap opens at the Fermi
level in the region close to the $(0,\pi)$ and $(\pi,0)$ points.
\end{itemize}
Several works \cite{psgap, psgap1} have established that a
pseudogap is present in the cluster solution of DMFT. We show this
explicitly in Fig. \ref{FIG3} by comparing the low frequency
dependence of the spectral function in three different points of
the Brillouin zone, marked by A, B and C in Fig. \ref{FIG2}.
Notice the suppression of the zero frequency peak at point B, due
to the proximity to the blue line of zeroes of the Green's
function and possibly not detected in ARPES experiments, and the
frequency shift $\delta = - 0.05t$ of the peak at point C, which
is the pseudogap effect . The $M$-approach provides a simple
interpretation of this effect, as observed in photo-emission
experiments, in terms of the emergence of infinite self-energy
lines (i.e zeroes of the Green's function). In Fig. \ref{Akw} we
show along the usual path in the first quadrant of the BZ
$(0,0)\rightarrow (\pi,\pi)\rightarrow (0,\pi)\rightarrow (0,0)$
the spectral function as function of frequency $\omega$, the Fermi
level is set at $\omega=0$. In ARPES experiments the $\omega>0$ is
not accessible, as it is in a theoretical calculation. In the
first row we show a large energy scale ($\omega_{max}\sim 8 t$),
while in the second row we have the same graphs in a more detailed
energy scale ($\omega_{max}\sim 2 t$). At the density $n=0.88$
just before the critical region at $10\%$ doping we can see that
indeed the spectral functions crosses the Fermi level twice, once
close to $(\pi/2,\pi/2)$ and the second close to $(0,\pi)$ with
well defined quasiparticle peaks, even if around $(0,\pi)$ the
peak is "less dispersive" as observed in the previous section for
the $U= 16 t$ results. For a density $n=0.95$ instead, there is
still a quasiparticle-like peak in correspondence of the region
around $(0,\pi)$, but this does not cross the Fermi level anymore,
but rather remains tangent to it at negative energy equal to the
$\delta=-0.05 t$ of peak C in $(0,\pi)$ as compared to peak A in
$(\pi/2,\pi/2)$. This is the real pseudogap, and the
correspondence of this behavior observed in the ARPES experimental
data of Fig. \ref{Marshall} (in particular at panels C1 going from
$(0,0)\rightarrow (\pi,\pi)$ and C2 going from $(0,0)\rightarrow
(0,\pi) \rightarrow (\pi,\pi)$) is quite impressive. We finally
remark that this pseudogap is not the same feature we studied in
\cite{bumsoo}, where the pseudogap observed was rather the higher
energy feature we see in curve C of Fig. \ref{FIG3} at energy
$\sim 0.5 t$ and labeled IR in Fig. \ref{Akw}, which is truly
present around the hot point $(0,\pi)$, but, being at positive
energies, does not appear in ARPES measure. It contributes
instead as a depression in the integrated DOS close to the Fermi
level.

\begin{figure}
\begin{center}
\includegraphics[width=0.4\textwidth]{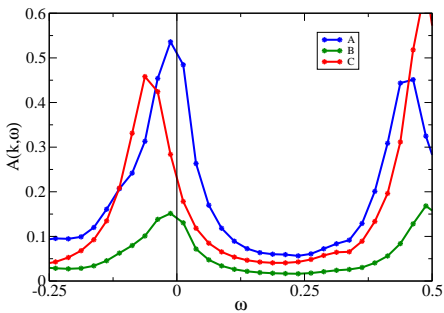}
\caption{Frequency dependence of the spectral function for three
points in the Brillouin zone marked by A, B, and C in Fig.
\ref{FIG2}.} \label{FIG3}
\end{center}
\end{figure}
\begin{figure}[!!t]
\begin{center}
\includegraphics[width=14cm,height=6cm,angle=-0] {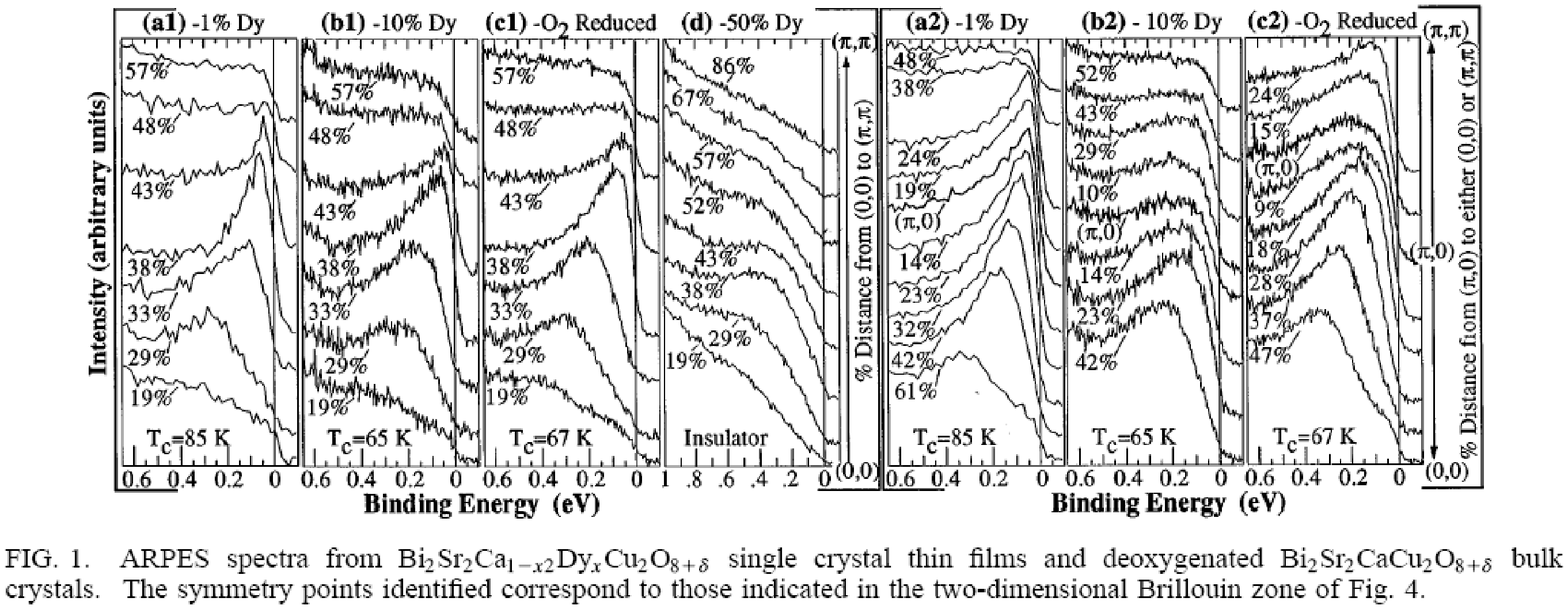}
\caption{Plot taken from D.S Marshall {\it at al.}, Phys. Rev.
Lett 76, 4841 (1996). }\label{Marshall}
\end{center}
\end{figure}
\begin{figure}[!!p]
\includegraphics[width=15.0cm,height=13.0cm,angle=-0] {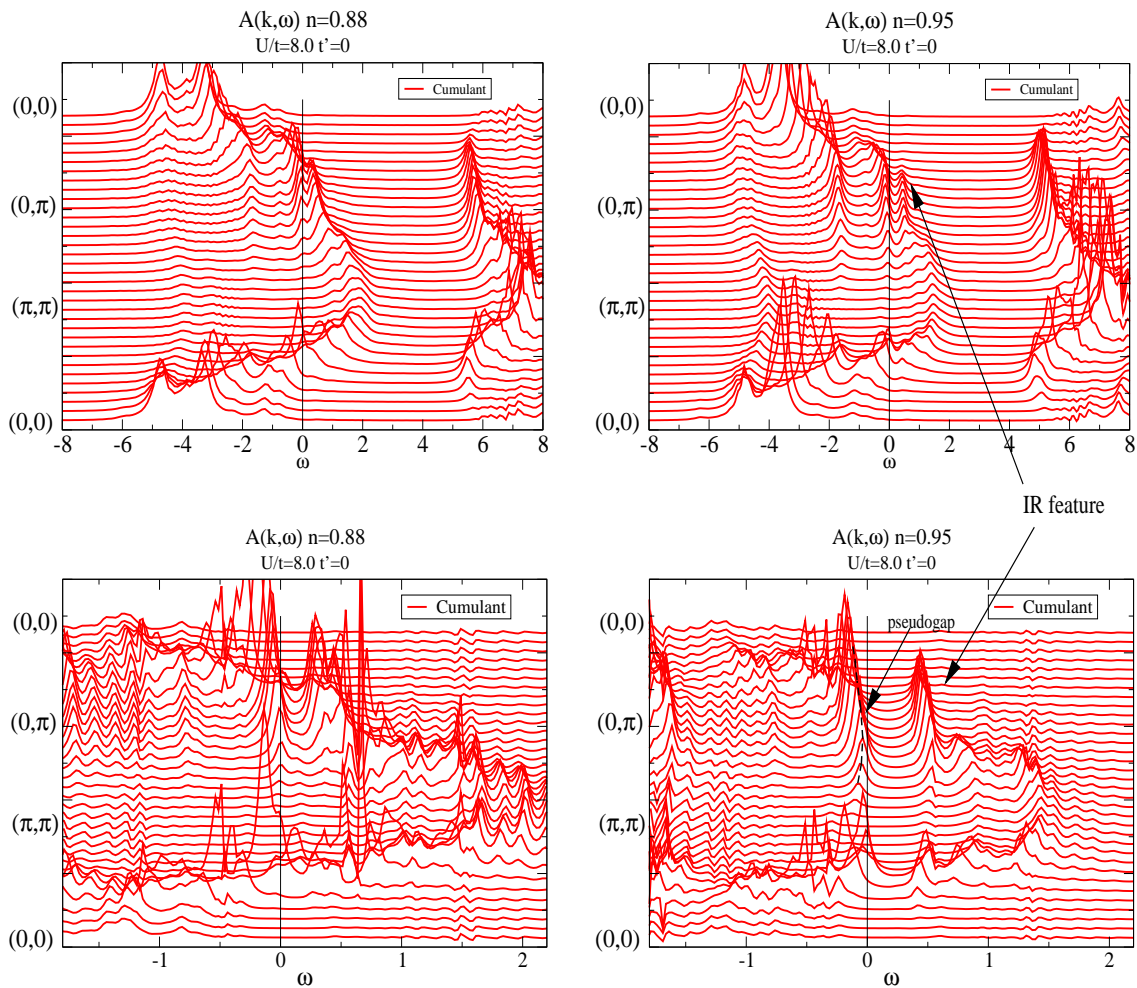}
\caption{Spectral function $A(k,\omega)$ along the path
$(0,0)\rightarrow (\pi,\pi)\rightarrow(0,\pi)\rightarrow(0,0)$ in
the first quadrant of the Brillouin Zone. The red curves are
calculated extracting onto the lattice the Cumulant $M$ from the
cluster./ In the left panel the density is $n=0.85$, the state is
at the left side of the coexistence region. In the right panel the
density of the system is $n=0.95$, the system is at the right of
the coexistence region. The bottom panels are details at the
Fermi level: notice that for $n=0.95$ in the region close to
$(0,\pi)$ the quasiparticle peak passes tangent, {\it but without
touching} the Fermi level, it's the pseudogap!
  } \label{Akw}
\end{figure}
\begin{figure}[!!htbp]
\begin{center}
\includegraphics[width=10.0cm,height=7.0cm,angle=-0] {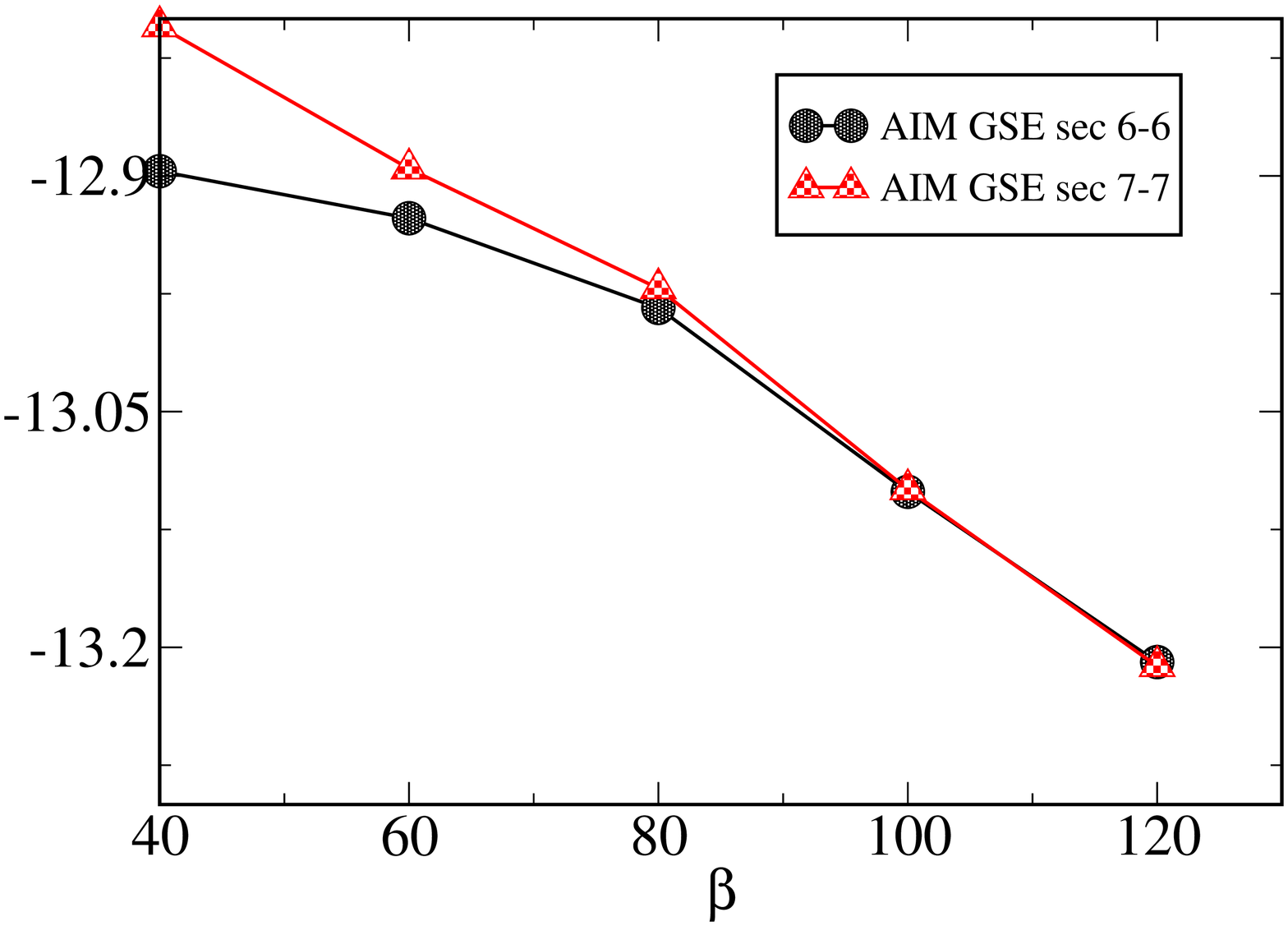}
\end{center} \caption{Ground state energy in two different but close
paericle-sectors, $(n_\uparrow,n_\downarrow)= (6,6)$ and $(7,7)$,
of the Anderson impurity model as a function of the inverse
virtual temperature $\beta$. The system parameters are $U=12$,
$t^{\prime}=0$, doping $10\%$.} \label{level_crosssing}
\end{figure}
Therefore the strong coupling $M$-CDMFT point of view suggests
that the short-range correlation, which play a key role in the
physics close to the Mott transition, can be naturally described
by finite range cumulants $M$ rather approximately in a cluster
scheme, than the self-energy. Consequently, the lattice cumulant
$M_{k}$, rather than the self-energy, is a smoothly varying
function of $k$ resulting in the appearance of surfaces of zeros
of the Green's function for small values of doping. The existence
of these lines is directly related to a change in the topology of
the FS and the opening of a pseudogap in the spectral function at
the Fermi energy. Remarkably,  the lines of poles of the self
energy appear first far from the Fermi surface. This is a strong
coupling instability which has no weak coupling precursors on the
Fermi surface. Our results suggest an interesting scenario: if the
evolution in Figure \ref{FIG2} from large to small doping is
continuous (rather than the first order transition we observe with
this CDMFT calculation), it has to go through a critical point
where the topology of the Fermi surface (and perhaps that of the
lines where the self energy is infinite) changes. This topological
change and its possible connection to an underlying critical point
at finite doping in the cuprate phase diagram deserves a deeper
investigation with
this techniques. 

Finally, we can try to connect the technical difficulties in
accessing the critical region (blue-circled area in the $n$ vs
$\mu$ plot of Fig. \ref{FIGslide}) with the underlying physics.
The impossibility of obtaining low virtual temperature
$\beta^{-1}$ results is an indication of the disappearance of a
the CDMFT solutions. To obtain insight on how this takes place,
we can start from a high temperature (low $\beta$) result and
gradually lower the temperature, until the last result is
obtained. This was done in Fig. \ref{level_crosssing}, where we
display the ground-state energy of two different particle-sectors
$(n_\uparrow,n_\downarrow)= (6,6)$ and $(7,7)$ as a function of
the inverse virtual temperature $\beta$. At low $\beta$, the
CDMFT-solution is determined in the sector with the lowest
ground-state energy [i.e. $(n_\uparrow,n_\downarrow)= (6,6)$].
However, as we approach the last temperature which produces a
stable CDMFT-solution ($\beta\sim 120$), the ground-state energy
of the two close particle sectors $(n_\uparrow,n_\downarrow)=
(6,6)$ and $(7,7)$ gets closer and closer, in the limit
degenerate. This strongly indicates that a phase transition is
taking place in this point as a function of temperature, and the
high-temperature NS-solution disappear. We are not able to state
if a different NS solution, maybe non-FL, will appear at lower
temperatures, or if a NS has in fact totally disappeared. We will
see in the next chapter that this critical point corresponds to
the maximum in the dome of the high-temperature SC
order-parameter. The possibility that the NS solution disappears
replaced by the SC solution is mostly appealing, and promises a
SC state which is the natural low-temperature continuation of an
anomalous NS. This idea is in line with the findings on the
momentum scattering differentiation, presented in section 5.1 and
5.2, the formation of hot/cold spots, whose position is inverted
from hole to electron doped systems, with the topological change
in the FS born out from a critical region that brings to the
formation of the pseudogap around $(0,\pi)$ and $(\pi,0)$ in the
underdoped side. All these NS-properties seem well fit to a
d-wave SC state for the hole-doped cuprates, as we will see in the
next chapter.

\chapter{ Superconductive Phase }
In this chapter we investigate the superconductive phase (SC). In
weakly correlated materials the superconducting state is well
described in the frame work of the BCS theory \cite{BCS}. In the
strongly correlated systems however the description of this
phenomenon is still one of the most challenging open problems in
condensed matter physics. We will present the results we can
achieve with CDMFT, which, as seen in the previous chapter, has
been able to provide new insights in the non-trivial NS of the
Hubbard Model in two dimension. In particular, we will try to
clarify the link between the SC state and the peculiarity of the
Mott transition, "the Mottness", and its relation with the NS
properties. The idea that the H-T$_{C}$ superconductive mechanism
is related to the Mottness of the cuprates can be traced back to
the Resonating Valence Bond (RVB) proposed by P.W. Anderson
\cite{anderson}. In this theory the insulating phase is a liquid
of pre-formed Cooper pairs, ready to become a superconductor once
doping is added into the system. Later slave boson studies
\cite{liu} refined this idea, showing that the super-exchange
interaction naturally selects a d-wave supeconductivity (dSC) and
that gaped one-particle excitation survive in the NS above the
critical temperature (the precursor of a pseudogap state). Further
investigation with variational approaches \cite{Zhang98} and more
extensive numerical RVB approaches \cite{Anderson:04} have
confirmed that many properties of cuprates can be qualitatively
described with these methods in the framework of Hubbard-like
models. We now extend the CDMFT-study to the SC state of the two
dimensional Hubbard model, adopting once again the 2X2 plaquette
as cluster-impurity. This is the minimal spatial arrangement
required in order to describe the short range interaction between
particles and the formation of a d-wave SC order parameter.

\section{CDMFT equations for a d-wave superconducting state}
In order to investigate the dSC phase within CDMFT we write an
effective action containing a Weiss dynamical field $\hat{\cal
G}_{o}$ with both normal (particle-hole) and anomalous
(particle-particle) components:
\begin{equation}
S_{\rm eff}= \int_{0}^{\beta }d\tau d\tau'\Psi_{c }^{\dagger
}(\tau) \left[ \hat{\cal G}_{o}^{-1} \right] \Psi_{c}(\tau')+ U
\sum_{\mu }\int_{0}^{\beta }d\tau \, n_{\mu \uparrow }n_{\mu
\downarrow }. \nonumber
\end{equation}
Here, for the case of a $2\times2$ plaquette, we indicate the
Nambu spinor,
$$\Psi_{c}^{\dagger } \equiv
(c_{1\uparrow}^{\dagger},\dots,c_{4\uparrow}^{\dagger},c_{1\downarrow},\dots,c_{4\downarrow})$$
$\mu$ labels the degrees of freedom inside the cluster.
Physically this action corresponds to a cluster embedded in a
self-consistently determined bath with SC correlations. Given the
effective action with a starting guess for the Weiss field
$\hat{\cal G}_{o}$, we compute the cluster propagator $\hat G_c$
and the cluster self energy $\hat \Sigma_c = \hat {\cal G}_o^{-1}
- \hat G_c^{-1}$.  Here,
\begin{equation}
\hat G_c\left( \tau ,\tau ^{\prime }\right) =\left(
\begin{array}{cc}
\hat G_{\uparrow }\left( \tau ,\tau ^{\prime }\right) & \hat
F\left( \tau ,\tau
^{\prime }\right) \\
\hat F^{\dagger}(\tau ,\tau ^{\prime }) & -{\hat G}_{\downarrow
}\left( \tau ^{\prime },\tau \right)
\end{array}
\right) \label{nambugreen}
\end{equation}
is a 8 X 8 matrix for a 2X2-impurity-cluster,
$$G_{\mu\nu,\sigma}\equiv -T \langle
c_{\mu\sigma}(\tau)c^{\dagger}_{\nu\sigma}(0)\rangle$$ and
$$F_{\mu\nu}\equiv -T \langle
c_{\mu\downarrow}(\tau)c_{\nu\uparrow}(0)\rangle$$ are the normal
and anomalous Green's functions respectively. Once again using the
CDMFT self-consistency condition \ref{self-consistency},
\begin{equation}
\hat {\cal G}_{o}(i\omega_n)^{-1}= \left[\sum_K \hat
G(K,i\omega_n)\right]^{-1} + \hat \Sigma_c(i\omega_n)
\label{selfcon}
\end{equation}
we recompute the Weiss field and iterate until convergence. The
self-consistency condition is here expressed by a 8X8 superlattice
local Green's function (as there are the off-diagonal anomalous
terms)
$$\hat
G(K,i\omega_n)=\left[i\omega_n + \mu - \hat t(K) - {\hat
\Sigma_c}(i\omega_n)\right]^{-1} $$ with $\hat t(K)$ the Fourier
transform of the superlattice hopping matrix with appropriate sign
flip between propagators for up and down spin and the sum over $K$
is performed over the reduced Brillouin zone of the superlattice.
To solve the cluster impurity problem represented by the effective
action above, we use, as usual, an Anderson cluster-impurity
Hamiltonian with a special superconductive bath:
\begin{eqnarray}
H_{\rm imp}&=&\sum_{\mu \nu \sigma}E_{\mu \nu \sigma}c_{\mu
\sigma }^{\dagger }c_{\nu \sigma}+ \sum_{m \sigma}\epsilon _{m
\sigma}^{\alpha}a_{m\sigma }^{\dagger\alpha
}a_{m\sigma }^{\alpha} \nonumber \\
&&+\sum_{m\mu \sigma} V_{m\mu \sigma}^{\alpha}a_{m\sigma }^{\dagger\alpha } (c_{\mu \sigma} + {\rm h.c.}) + U \sum_{\mu} n_{\mu\uparrow}n_{\mu\downarrow} \nonumber \\
&& + \sum_{\alpha} \Delta^{\alpha}(a_{1\uparrow}^{\alpha} a_{2\downarrow}^{\alpha} - a_{2\uparrow}^{\alpha} a_{3\downarrow}^{\alpha} + a_{3\uparrow}^{\alpha} a_{4\downarrow}^{\alpha} - a_{4\uparrow}^{\alpha} a_{1\downarrow}^{\alpha} \nonumber \\
&& + a_{2\uparrow}^{\alpha} a_{1\downarrow}^{\alpha} -
a_{3\uparrow}^{\alpha} a_{2\downarrow}^{\alpha} +
a_{4\uparrow}^{\alpha} a_{3\downarrow}^{\alpha} -
a_{1\uparrow}^{\alpha} a_{4\downarrow}^{\alpha}+ h.c.).\nonumber
\end{eqnarray}
Here $\mu,\nu=1, ..., N_c$ label the sites in the cluster and
$E_{\mu \nu \sigma}$ represents the hopping and the chemical
potential within the cluster. The energy levels in the bath are
grouped into multiples of the cluster size ($N_c=4$) with the
labels $m=1,\cdots,N_c$ and $\alpha=1,2$ such that we have 8 bath
energy levels $\epsilon _{m\sigma}^{\alpha}$ coupled to the
cluster via the hybridization matrix $V_{m\mu\sigma}^{\alpha}$.
Using lattice symmetries we take $V_{m\mu\sigma}^{\alpha}\equiv
V^{\alpha}\delta_{m\mu}$ and $\epsilon^{\alpha}_{m\sigma}\equiv
\epsilon^{\alpha}$. This is similar to the reduced parametrization
introduced in chapter 4.5, in order to target a first guess in the
NS case. This time, however, we added particle-particle
destruction and creation terms through the coupling
$\Delta^{\alpha}$, which represents the amplitude of SC
correlations in the bath. In this case no static mean-field order
parameter acts directly on the cluster sites\cite{Poilblanc}.
$\epsilon^{\alpha}$, $V^{\alpha}$ and $\Delta^{\alpha}$ are
determined by imposing the self-consistency condition in
Eq.~\ref{selfcon} using a the usual conjugate gradient
minimization algorithm with the distance function $f$ introduced
in chapter 3.4.1 that emphasizes the lowest frequencies of the
Weiss field \cite{marce}. With the dSC order parameter defined as
$\psi _{ij}=\left\langle c_{i\downarrow
}c_{j\uparrow}\right\rangle $ we consider $d$-wave singlet pairing
($\psi =\psi_{12}=-\psi _{23}=\psi_{34}=-\psi_{41}$).

\section{An anomalous superconductive state}
\begin{figure}[tbp]
\begin{center}
\includegraphics[angle=270,width=8cm]{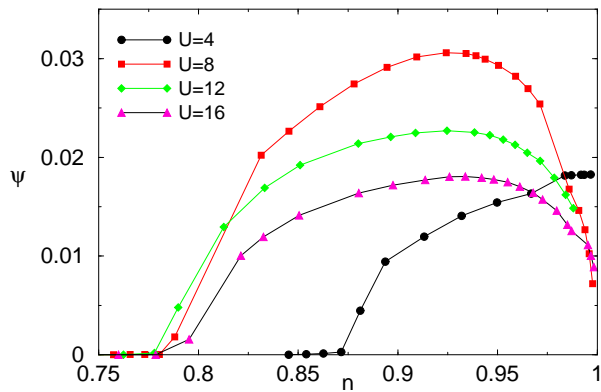}
\caption{SC order parameter $ \psi= < c_\mu c_\nu > $ as a
function of filling $n$ and onsite Coulomb repulsion $U$, $t'=0$.}
\label{ordpar}
\end{center}
\end{figure}
We first observe that in our zero-temperature calculation the
CDMFT-solution of the 2-dimensional Hubbard Model does support a
d-wave superconducting state. The dSC order parameter $\psi$ is
shown in Fig.~\ref{ordpar} as a function of the density $n$ for
$t'=0$ and different values of on-site Coulomb repulsion $U$. As
$U$ is increased to $8t$ the order parameter acquires a maximum
around $\delta=0.1$ (optimal doping) with a suppression above and
below this doping. On the overdoped region, dSC disappears at a
doping that is comparable with DCA and VCPT but smaller than the
first case \cite{Jarrell:2000} and larger than the second
\cite{Senechal:2005}. $\psi$ shows a clearly different behavior
for large and small interaction $U$, which can be ascribed to the
proximity to a Mott-insulating state. For $U=4t$, $\psi$ increases
as $n\rightarrow 1$, in this case the half-filling insulating
state is mostly an AF band-insulator. On the other hand, for
$U=8t-12t$, the half-filled insulating state is a Mott-insulator
even in the absence of AF. In this case, the superconducting state
senses the proximity to the Mott state, and $\psi$ goes to zero as
the doping disappears because of the progressive localization of
quasiparticles and hence, despite we observe stronger dSC Weiss
field ($\Delta$) in this region, there is a suppression of the dSC
order parameter. This suppression is not seen for $U=4t$ because
in that case it is the insulating gap that emerges from AF
(Slater) correlations only at half-filling that kills dSC
\cite{Kyung:2003}. The maximum of the order parameter increases as
$U$ goes from $4t$ to $8t$ but then decreases as $U$ changes to
$12t$ and then $16t$. This clearly signals that dSC is strongest
at intermediate coupling and that at strong coupling the dSC order
parameter scales as the magnetic exchange coupling $J=4t^2/U$, as
also found in early slave boson studies \cite{liu} and in other
cluster approaches (like VCPT \cite{Senechal:2005}). The value of
optimal doping is nearly independent of $U$ in the intermediate to
strong coupling regime.
\subsection{Discrepancy between order parameter and anomalous gap}
\begin{figure}
\begin{center}
\includegraphics[angle=270,width=12cm]{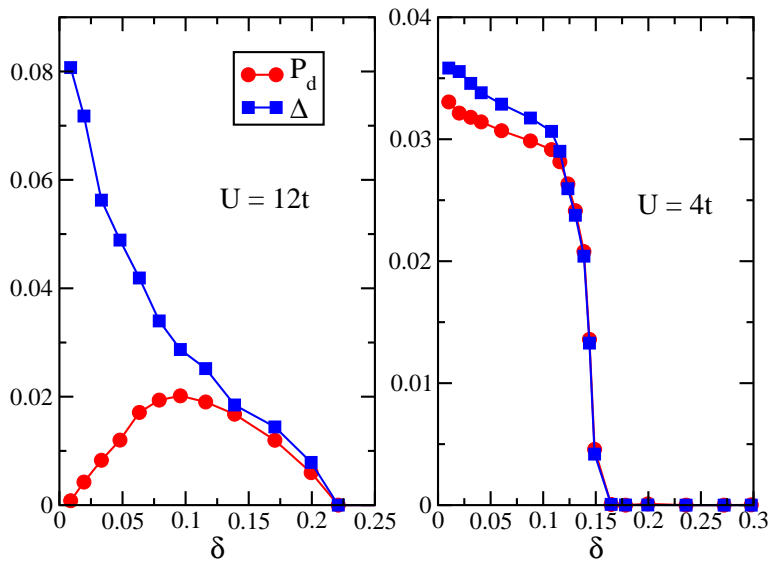}
\caption{ $d$-wave order parameter $P_d= < c_\mu c_\nu > $ and
superconducting gap $\Delta$ as a function of doping for
weak-coupling $U/t = 4$ (right) and strong-coupling $U/t = 12$
(left). } \label{gap}
\end{center}
\end{figure}
Another indication of the anomalous nature of the
superconductivity in the strong correlated regime is given by the
behavior of the anomalous gap $\Delta$ (Fig.~\ref{gap}), which we
computed from the density of states. The latter is obtained from
the lattice Green's function which we can calculate with one of
the three periodizing schemes ($\Sigma$,$M$ or $G$-schemes).
$\Delta$ monotonically increases as $n\rightarrow 1$, even when
the order parameter $\psi$ is decreasing to zero (left panel of
Fig.\ref{gap} where the case $U= 12t$ is displayed), in clear
contrast with the prediction of BCS theory, where this two
quantities are proportional, as instead takes place in the weak
interacting regime (right panel of Fig.\ref{gap} where the case
$U= 4t$ is displayed). This behavior is reminiscent of the cuprate
superconductors, where the $T=0$ superconducting gap scales
proportionally to the pseudogap as the doping $\delta\rightarrow
0$, increasing instead of decreasing like the critical temperature
$T_{C}$. It is also in accordance with experimental observation on
the cuprates, e.g. thermal conductivity experiments
\cite{thermal}.
These results in the strong coupling regime and the behavior of
the maximum in the dome of the order parameter, which varies
inversially proportional to U, support the super-exchange pairing
hypothesis. This shows that the Hubbard Model presents underdoped
and overdoped regions which reflect the properties of the
corresponding phases in the cuprates. The magnitude of the gap at
optimal doping is estimated to be around $30meV$, in good
agreement with experimental estimates \cite{thermal} for the
cuprates, if we take a reasonable value for $t$ of $300meV$.
\subsection{Disappearance of the quasiparticle spectral weight at
the MT}
\begin{figure}[!htb]
\begin{center}
\includegraphics[angle=270,width=12cm]{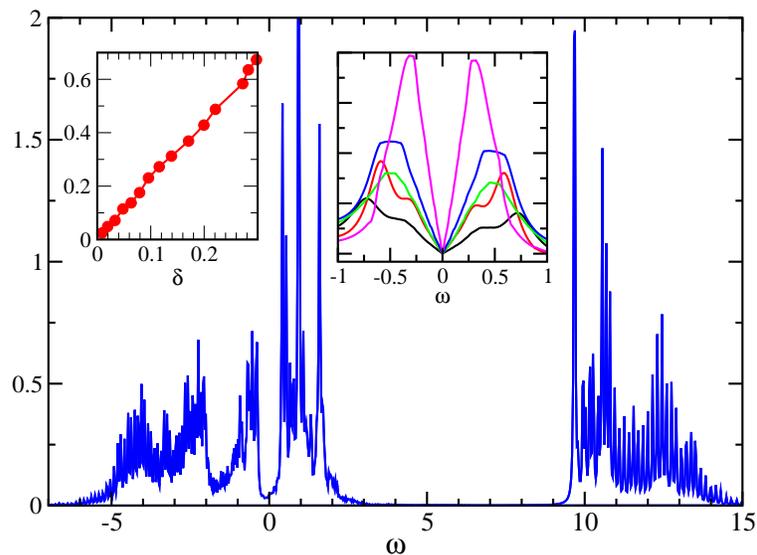}
\caption{ The local density of states $\rho_0(\omega)$ in the
superconducting phase for $U/t =12$. Notice the presence of high
and low energy features, and the strong particle-hole asymetry. b)
Integrated spectral weight of the low-energy feature as a function
of doping. The vanishing of the low energy spectral weight signals
the approach to the Mott insulator. c) Low energy spectral
function vs energy. The large doping curves have a typical d-wave
shape, and more involved structures appear for low-doping. Notice
the similarity of the particle hole asymmetry with the
experimental results of Ref. (16)
 } \label{spettro}
\end{center}
\end{figure}
In order to better understand how the quasiparticles disappear in
approaching the MT, bringing to the decreasing of the dSC order
parameter, we can observe the local density of states
$\rho(\omega) = -1/\pi \Im G_{loc}(\omega)$ ($G_{loc}$ being the
local Green's function), a quantity which can be measured
accurately  on the surface of a material by Scanning Tunneling
Microscopy (STM). In proximity to the Mott insulator, most of the
spectral weight lies in the high-energy Hubbard bands, but there
is also a low-energy feature in correspondence of the Fermi level,
taken as the reference zero-energy. In Fig.\ref{spettro}, we plot
$\rho(\omega)$ from ED-CDMFT (the discretization inherent to the
ED approach results in the spikes clearly visible in figure) for
$\delta \simeq 0.08$. In Fig. \ref{spettro}-(b) we plot the
integrated spectral weight of the low-energy feature as the doping
is reduced to zero. Approaching half-filling, spectral weight is
transferred from the low-energy feature to the Hubbard bands. The
low-energy feature completely disappears as the Mott insulator is
reached. The disappearance of mobile low-energy carriers, is an
unequivocal feature of the approach to the Mott transition. This
is ultimately the cause of the disappearance of superconductivity
at the Mott boundary, rather than the vanishing of the attraction.
In Fig. \ref{S-ano}(c) we focus on the frequency dependence of
the low-energy feature. A first evident observation is the
asymmetry between positive and negative frequencies. Such an
asymmetry has been first discussed by Anderson and Ong, who
noticed it in the STM spectra, and ascribed it to the proximity
to the Mott insulator\cite{asymmetry}. On top of that, we
identify a clear evolution from the underdoped to the overdoped
physics. In all cases the low-energy spectrum is what we can
expect for a d-wave superconductor, with zero weight only at
$\omega=0$ and a linear behavior for small frequencies. As the
doping is reduced $\rho(\omega)$ develops a more complicated
shape of the low-frequency profile of the spectrum, where  the
linear behavior is only limited to very low frequency, and a
second feature appears at larger frequency.
The behavior we have just described  shares many similarities with
what is observed in Bi2212 cuprate\cite{stm}.
\subsection{Nature of the pairing}
\begin{figure}[htb]
\begin{center}
\includegraphics[angle=0,width=12cm]{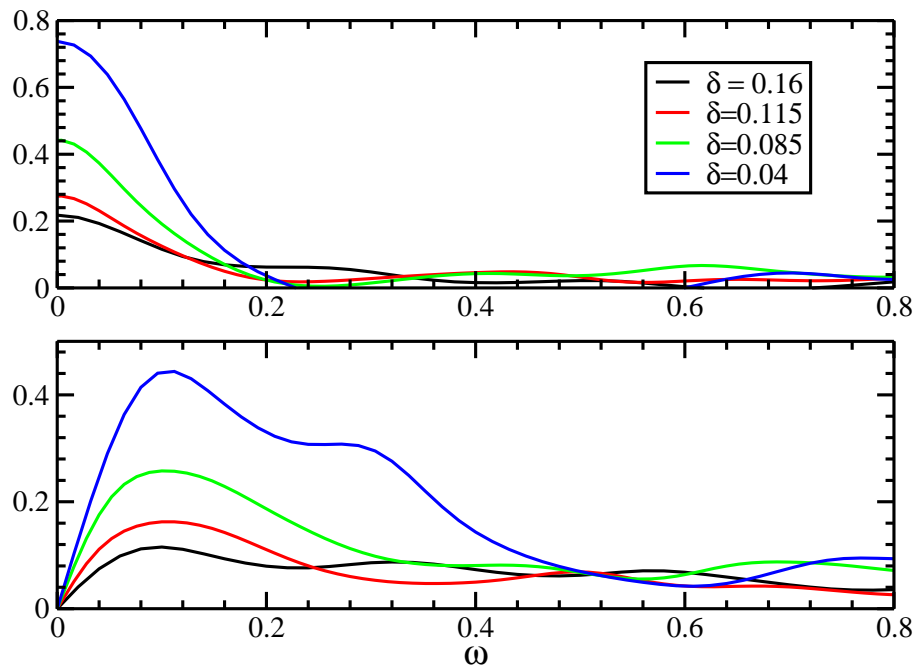}
\caption{Real and imaginary part of the anomalous at low energy
as a function of frequency for different doping levels. Notice
that the scale over which the pairing is attractive is
independent of doping. The real part of the self energy goes to a
very small value beyond a scale of order $J$ indicating that
retardation plays a small role in this strongly correlated
limit.} \label{S-ano}
\end{center}
\end{figure}
We want now to get some insight in the pairing attraction,
considering the frequency dependence of the anomalous self-energy
in the d-wave channel $S_d(\omega)$. This quantity, as well as any
other dynamical observable, is directly obtained within ED-CDMFT,
as opposed to many other approaches, where the dynamics is hardly
accessible. In the classic Migdal-Eliashberg theory, the imaginary
part of the anomalous self-energy is peaked at the frequencies of
the  phonons which mediate the pairing, while the real part
changes sign around the same frequency, testifying that the
attraction only leaves up to the phonon energy scales, while high
energies are dominated by Coulomb repulsion. In Fig. \ref{S-ano}
we plot the evolution of this quantity with doping in our
strong-coupling Hubbbard model. Remarkably, we do not find sharp
features, but rather a broad continuum which lives  up to a scale
of the order of the AFM superexchange coupling, indicating the
magnetic origin of the pairing. Accordingly, the real part has
less sharp behavior than in ordinary superconductors, but it is
attractive up to a scale of order $J$, after which it is
substantially zero, indicating a little role of retardation
effects. Remarkably, the characteristic energy scale where the
anomalous self-energy changes is weakly doping-dependent.
\subsection{The superconducting phase as "cure" for a ill normal state}
\begin{figure}[htb]
\begin{center}
\includegraphics[angle=0,width=7.0cm]{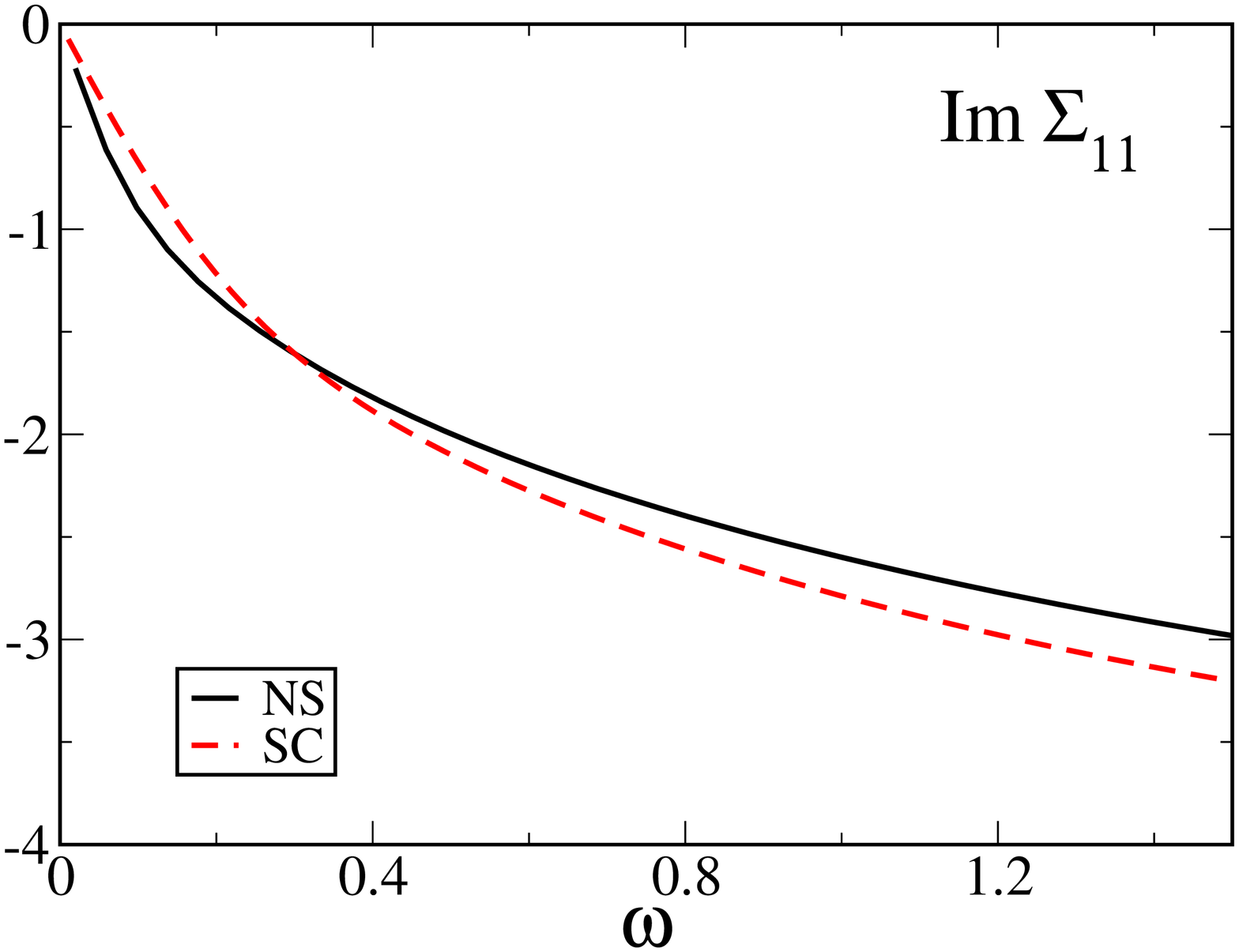}
\includegraphics[angle=0,width=7.0cm]{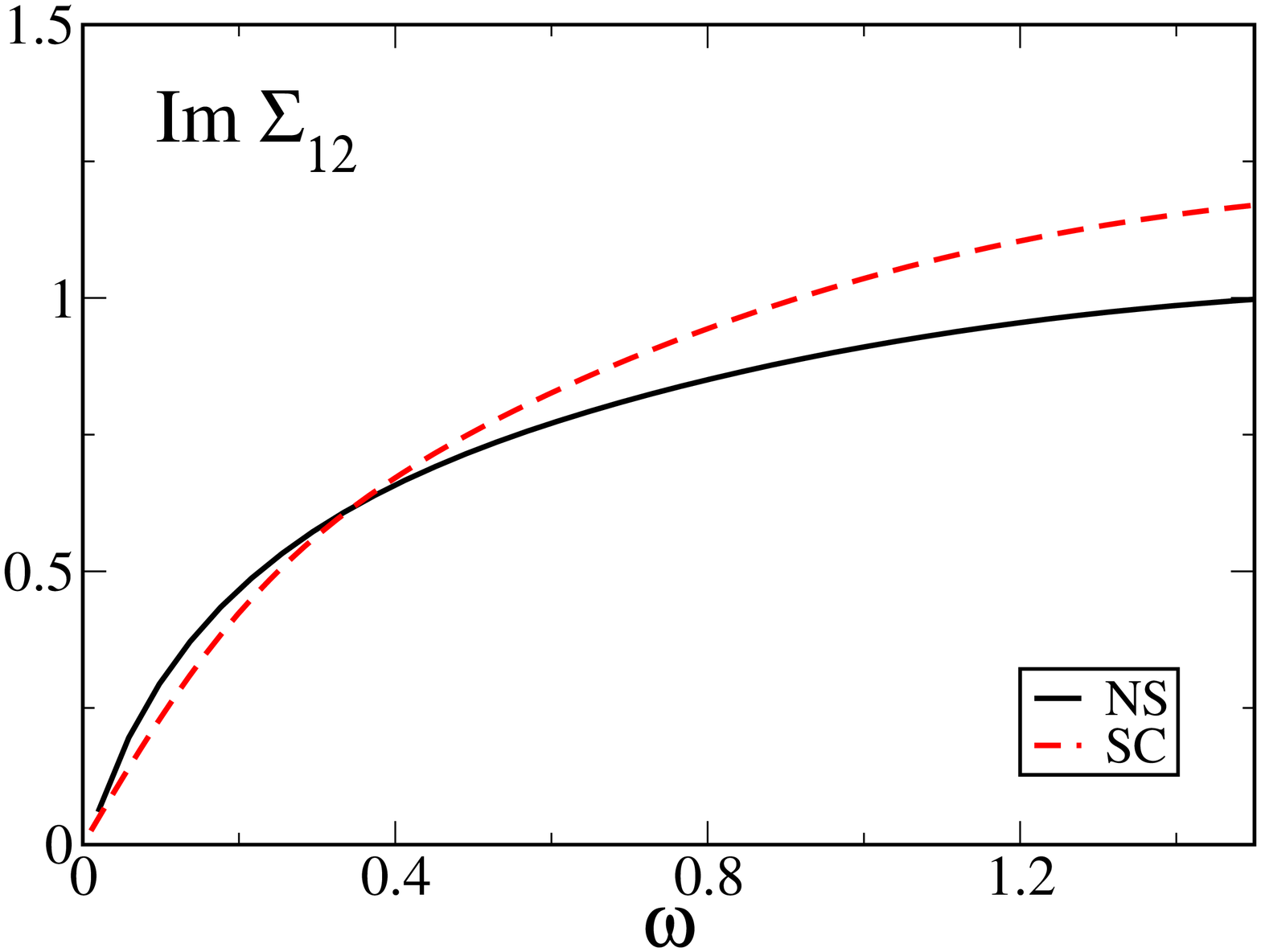}
\includegraphics[angle=0,width=7.0cm]{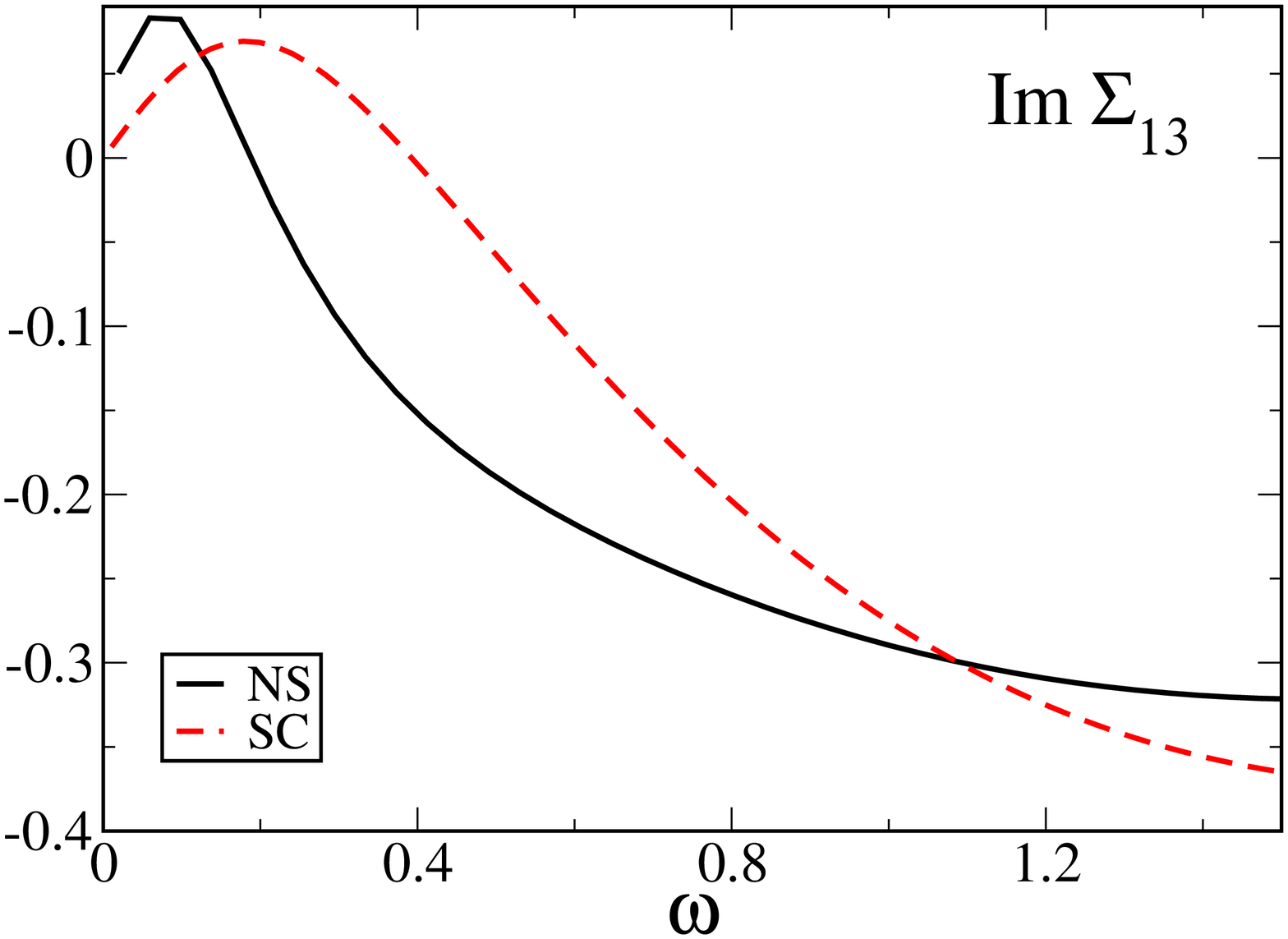}
\caption{Imaginary parts of the normal-cluster components of the
self-energy for $U= 12t$, $t^{\prime}=0$ and $10\%$ doping. The
normal state self-energies (black continuous line) are compared
with the superconducting-state ones (red dashed line). Notice the
enhanced FL-like behavior of the nearest-next-neighbor component
$\Im\Sigma_{13}\rightarrow 0$ for $\omega\rightarrow 0$ in the SC
state. } \label{Scluster-ano}
\end{center}
\end{figure}
\begin{figure}[htb]
\begin{center}
\includegraphics[width=4.5cm,angle=270]{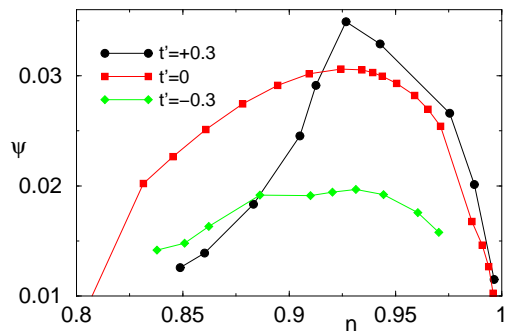}
\caption{dSC order parameter as a function of $n$ for various
values of t' (frustration).} \label{frust}
\end{center}
\end{figure}
It is useful to compare the dSC solutions with the corresponding
NS ones. We have observed in the previous chapter that,
approaching the MT, the NS quasiparticles present the formation of
large scattering rates in some regions of the $k$-space, mainly
around the $(\pi,\pi)$ point in the first quadrant of the BZ (at
least for the energy resolution used in this case, $\beta^{-1}=
128$ in half-bandwidth units). This was at the origin of the
momentum-space differentiation effect and the set up of an
"anomalous" FL regime. Under the $M$-scheme viewpoint, the
anomalies of this regime were enhanced with respect to the
$\Sigma$-scheme viewpoint, displaying the formation of lines of
zeroes of the one-particle Green's function in the $k$-space,
accompanied by a radical change in the topology of the FS. The
regime was nicely linked to the PG phase observed in experimental
ARPES results. Zeroes in the Green's function are related to zeros
in the lattice-cumulant $M$, or in other words, to infinities of
the lattice-self-energy $\Sigma_k$. Even if the relation of the
lattice quantities with the cluster ones obtained as direct output
of the CDMFT calculation is indeed $\Sigma$-$M$-$G$-scheme
dependent, nevertheless we expect to observe an effect of the
diverging behavior of $\Sigma_k$ in certain points of the momentum
mirrored in the cluster-self-energies. Rather than in the
eigenvalues of the cluster-self-energy matrix, this is most
evident in the cluster-self-energies , whose imaginary part we
display in Fig. \ref{Scluster-ano} for the intermediate coupling
case $U/t=12$, $t^{\prime}=0.0$ and 8\% doping, and for a very low
virtual temperature $\beta^{-1}= 640$ (in half-bandwidth units).
In spite at very low energies $\omega \rightarrow 0$ all the
cluster self-energies extrapolate to zero, the
next-nearest-neighbor self-energy $\Sigma_{13}$ present a very
sharp low low energy feature, with a turning down very steep only
at the last points of the Matsubara axis. This contributes to
create in the $k$-space a strongly scattering quasiparticle with a
small residuum $Z$, or, said in the $M$-language, to the formation
of the self-energies-divergencies. We super-impose, for the same
values of the Hubbard parameters, the normal-cluster-self-energies
obtained from a SC solution. The effect of superconductivity is
such as to "normalize" the cluster-self-energies, which are better
behaved in terms of regular quasiparticles. The slopes of the
$\Im\Sigma$ for $\omega\rightarrow 0$ are going to zero more
linearly, most evidently in the next-nearest-neighbor case
$\Im\Sigma_{13}$. In other words it means that quasiparticles are
better defined in the dSC phase than in its parent NS, in analogy
with ARPES experiments on the cuprates \cite{quasiparticlesarpes}.
This suggests that the origin of the dSC could be interpreted as
the natural low-energy cure to an anomalous-FL NS.

As we did in the previous chapter for the NS, we now add a
frustrating next-nearest neighbor (diagonal) hopping $t^{\prime}=
\pm 0.3$ to the system. In Fig.~\ref{frust} we plot the dSC order
parameter $\psi$ as a function of doping for various values of
the $t^{\prime}$. In the underdoped region, $\psi$ grows with
increasing $t'$, as it also was found in previous Dynamical
Cluster Approximation (DCA) calculations\cite{Jarrell:2000}. This
is un-expected if compared with experimental observation and band
structure studies of the cuprates \cite{pavarini}. A possible
explanation could the proposal advanced in previous VCPT
(Variational Cluster Perturbation Theory) studies
\cite{Senechal:2005,lich} suggesting that dSC may survive the
onset of AF on the electron-doped side.
\subsection{$k$-dependent quantities}
To try better clarifying this point, we need to extract the
lattice quantities which can be directly connected to
experimental observation, giving a physical interpretation to the
results. As explained in chapter 4, this is achieved trough a
periodizing scheme, able to interpolate three cluster degree of
freedoms (the cluster self-energies or the cluster cumulants)
onto $k$-dependent quantities. In studying the NS in chapter 4, we
adopted three of such schemes, the $\Sigma$, $M$ and $G$-schemes,
which now we extend to the case of a superconductive state, where
we have a non-zero pair-correlation function:
\begin{equation}
F(k,\tau)=\, -T \, \langle \, c_{k \uparrow} c_{-k \downarrow} \,
\rangle \label{F_pair}
\end{equation}
This is better described in the Nambu-spinor notation:
$$
\Psi^{\dagger}_{k}=\, \left( \, c^{\dagger}_{k\uparrow}, \,
c_{k\downarrow} \, \right)\label{spinor1}
$$
and the matrix formulation of the one-particle Green's function:
\beqa \label{Gk-nambu}
 \hat{G}(k,\tau) & =\,-T \, \langle \,
\Psi_{k}(\tau)\,\Psi^{\dagger}_{k}(0) \, \rangle\cr
 & \cr
 & = \, \begin{pmatrix}
  G_{\uparrow}(k,\tau) & F(k,\tau) \\
  F(k,\tau)^{*}        & -G_{\uparrow}(-k,-\tau)
\end{pmatrix}
\eeqa The kinetic part of the Hamiltonian is simply re-written as:
\begin{equation}
{\cal H}_{kin}= \,-t\, \sum_{\langle ij \rangle}\,
\Psi^{\dagger}_{i}\, \hat{\sigma}_{3}\, \Psi_{j}
\end{equation}
where $\hat{\sigma}_{3}$ represents the Pauli matrix. A matrix
self-energy is associated to the matrix Green's function:
\begin{equation}
\hat{G}^{-1}(k,\omega)=\, \imath\omega\, \hat{1}+
\left(\,\mu-\varepsilon_{k} \, \right)\,\hat{\sigma}_{3}-
\hat{\Sigma}_{k}(\imath\omega)
\end{equation}
which contains normal $\Sigma_k$ as well as anomalous
$\Sigma^{A}_k$ contributions:
\begin{equation}
\hat{\Sigma}_{k}(\imath\omega)=\, \begin{pmatrix}
  \Sigma_{k}(\imath\omega)      &  \Sigma^{A}_{k}(\imath\omega)\\
   \Sigma^{A}_{k}(\imath\omega) & -\Sigma_{k}(\imath\omega)^{*}
\end{pmatrix}
\end{equation}
Naturally, this can be re-formulated in terms of the cumulant
$M_k$ :
\begin{equation}
\hat{G}^{-1}(k,\omega)=\, \hat{M}^{-1}_{k}(\imath\omega)+
\left(\,\mu-\varepsilon_{k} \, \right)\,\hat{\sigma}_{3}
\end{equation}
which contains normal $M_k$ as well as anomalous $M^{A}_k$
contributions:
\begin{equation}
\hat{M}_{k}(\imath\omega)=\, \begin{pmatrix}
  M_{k}(\imath\omega)      &  M^{A}_{k}(\imath\omega)\\
  M^{A}_{k}(\imath\omega) & -M_{k}(\imath\omega)^{*}
\end{pmatrix}
\end{equation}
Once again, in the $\Sigma$-scheme we periodize the
cluster-self-energy, normal and anomalous: \beqa
\Sigma_{k}(\omega)= &\,\Sigma_{11}(\omega)+
\Sigma_{12}(\omega)\,\left( \cos k_{x}+ \cos k_{y} \right)+
\Sigma_{13}(\omega) \cos k_{x} \cos k_{y}\cr
\Sigma^{A}_{k}(\omega)= &\, \Sigma^{A}_{12}(\omega)\,\left( \cos
k_{x}- \cos k_{y} \right) \eeqa and in the $M$-scheme the
cluster-cumulant : \beqa M_{k}(\omega)= &\,M_{11}(\omega)+
M_{12}(\omega)\,\left( \cos k_{x}+ \cos k_{y} \right)+
M_{13}(\omega) \cos k_{x} \cos k_{y}\cr M^{A}_{k}(\omega)= &\,
M^{A}_{12}(\omega)\,\left( \cos k_{x}- \cos k_{y} \right) \eeqa
In the $G$-scheme we again periodize the super-lattice Green's
function on the full lattice like in formula \ref{Gk}, but this
time the superlattice Green's function is doubled in size ($2
N_{c}X 2 N_{c}$), considering the superconductive pairing
correlations (formula \ref{F_pair}) in the blocks out of diagonal,
according to the super-cluster spinors $\Psi_{c}$ notation
introduced in formula \ref{nambugreen}:
\begin{eqnarray}
\label{Gk2} \hat{G}(k,\omega)=\, \frac{1}{N_{c}}\,
\sum_{\mu,\nu=\,1}^{N_{c}}\, e^{\imath k\mu}\,
 \left[
\mathbf{\hat{1}}\,\omega+ \hat{t}^{c}_{k}- \hat{\Sigma}
\right]^{-1}_{\mu \nu }\, e^{-\imath k\nu}
\end{eqnarray}
where $\hat{G}(k,\omega)$ is $2X2$ Nambu's notation Green's
function (\ref{Gk-nambu}) ,$\hat{1}$ is the $2 N_{c}X 2 N_{c}$
identity matrix, $$\hat{t}^{c}_{k}= \, \begin{pmatrix}
   \hat{t}^{}_{k}& 0  \\
   0             &-\hat{t}^{}_{k}
\end{pmatrix}$$
$\hat{t}^{}_{k}$ is the $N_{c}X N_{c}$ cluster-hopping matrix and
$\hat{\Sigma}$ the super-cluster self energy (\ref{selfcon}).
\begin{figure}[htb]
\begin{center}
{\bf U= 12t, t$^{\prime}$=-0.3 t, n=0.87 } \vspace{0.5cm}\\
$\Sigma$-scheme\\
\includegraphics[width=9.0cm,height=3.0cm,angle=-0] {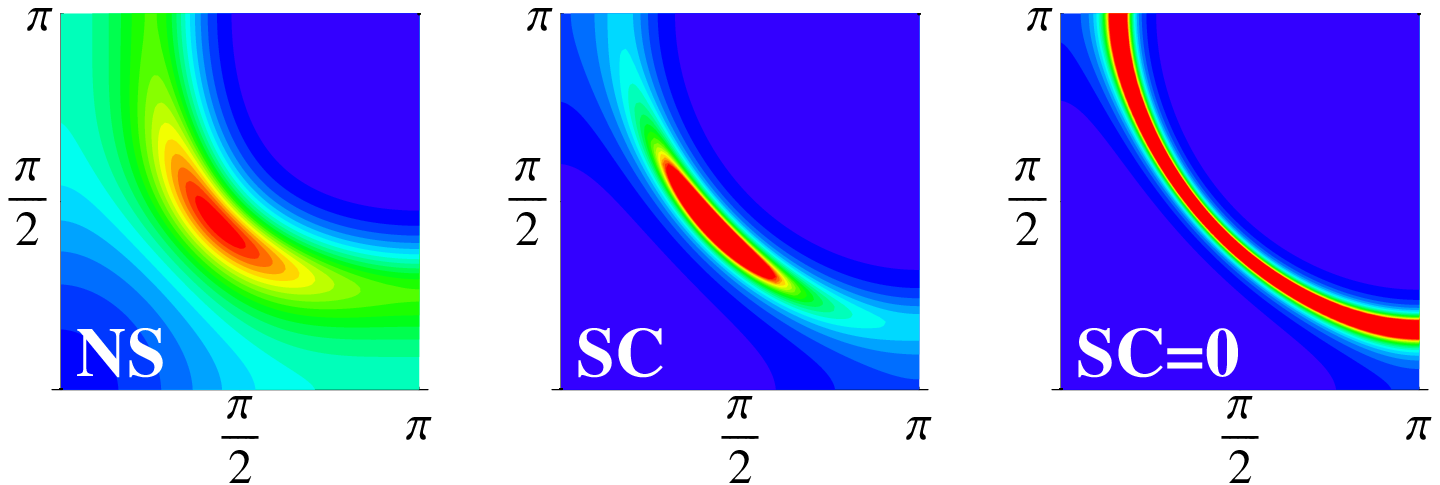}\\
$M$-scheme \\
\includegraphics[width=9.0cm,height=3.0cm,angle=-0] {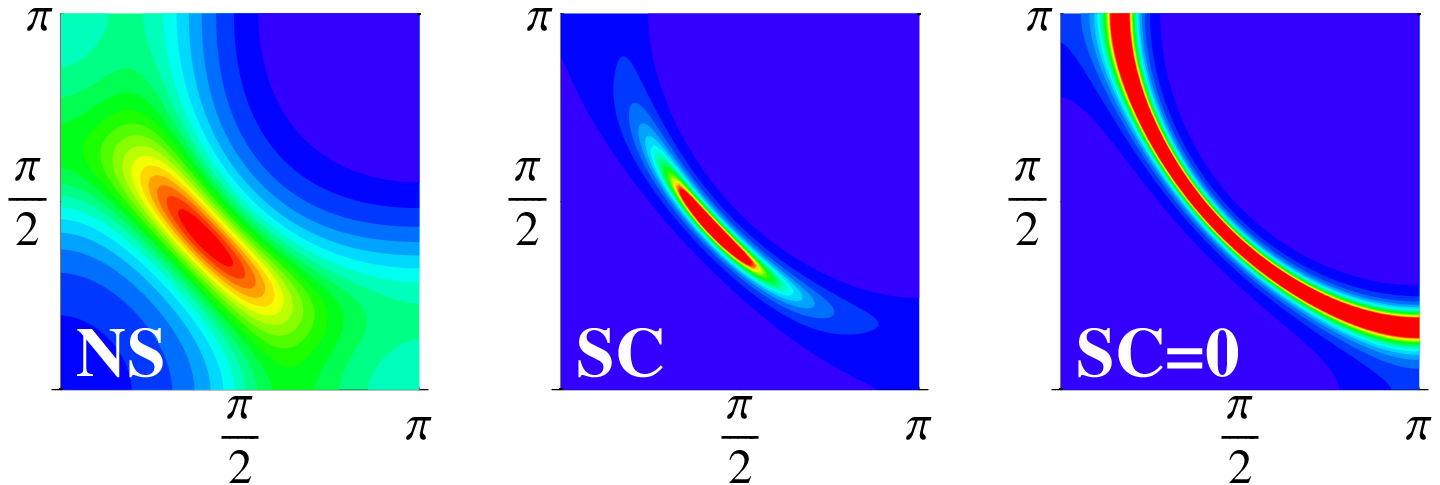}\\
$G$-scheme \\
\includegraphics[width=9.0cm,height=3.0cm,angle=-0] {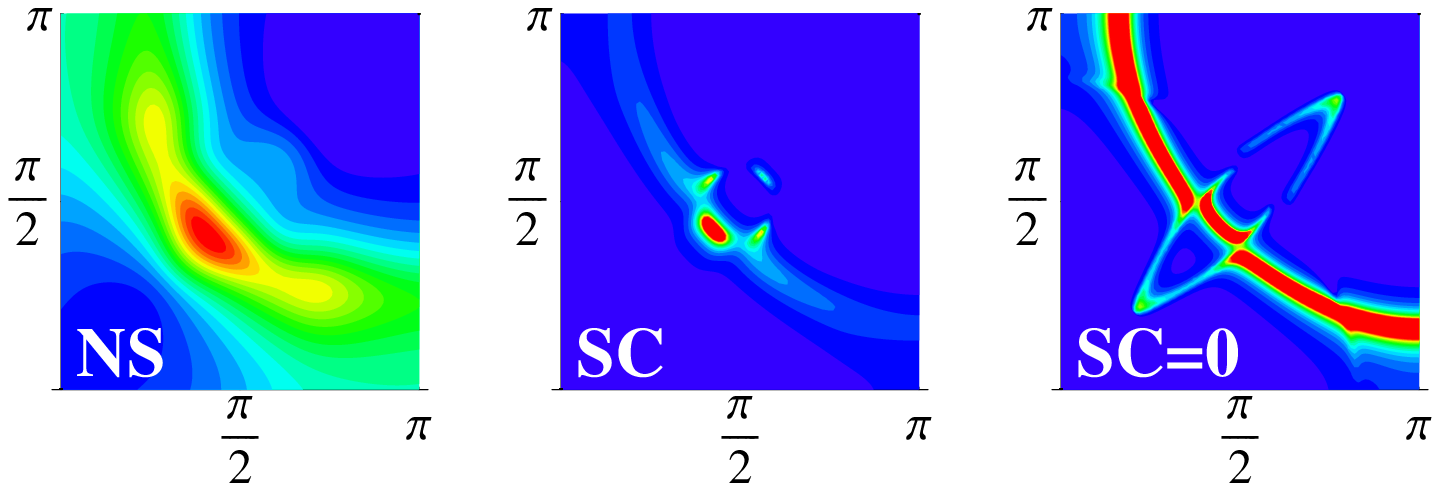}\\
\caption{Spectral function $A(k,\omega\rightarrow 0^{+})$ for the
case $U= 12t$, $t^{\prime}= -0.3t$ and density $n=0.87$. The
result of the three periodizing schemes are displayed: the
$\Sigma$-scheme in the first row, the $M$-scheme in the second,
the $G$-scheme in the third. In the first column of each row the
NS-result is presented, whose color scale is kept in the
following SC (second column) and SC=0 (third column) plot. The SC
is the superconducting state result, SC=0 is the normal
quasiparticle spectrum in the SC state obtained zeroing the
anomalous components from the SC result.}\label{spectrat1-0.3}
\end{center}
\end{figure}
\begin{figure}[htb]
\begin{center}
{\bf U= 12t, t$^{\prime}$= 0.0, n=0.93  } \vspace{0.5cm}\\
$\Sigma$-scheme\\
\includegraphics[width=9.0cm,height=3.0cm,angle=-0] {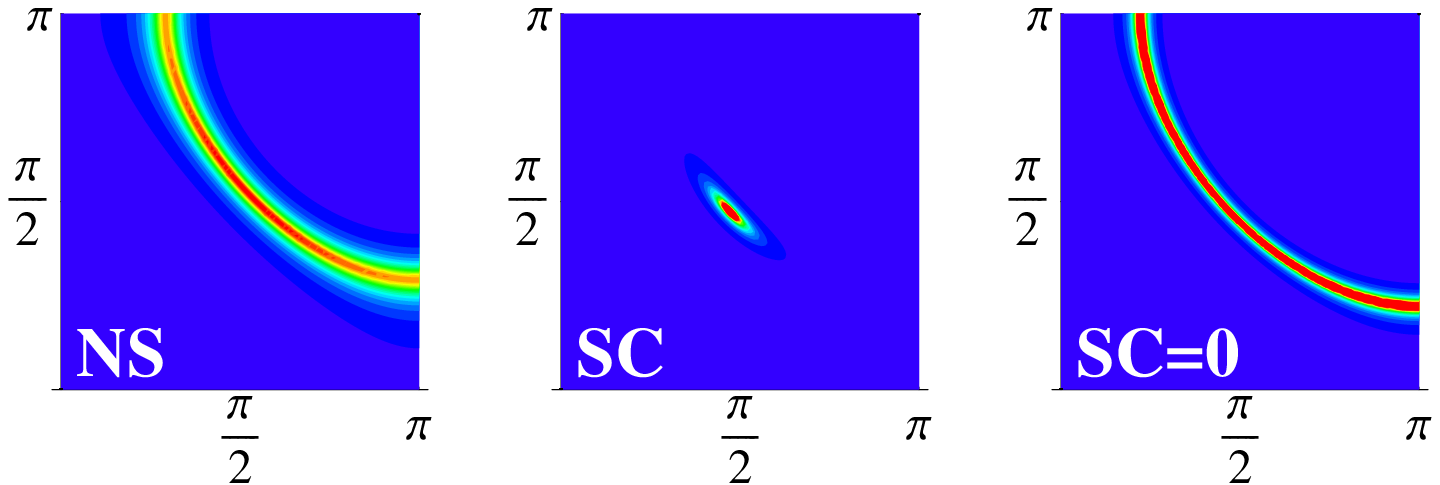}\\
$M$-scheme \\
\includegraphics[width=9.0cm,height=3.0cm,angle=-0] {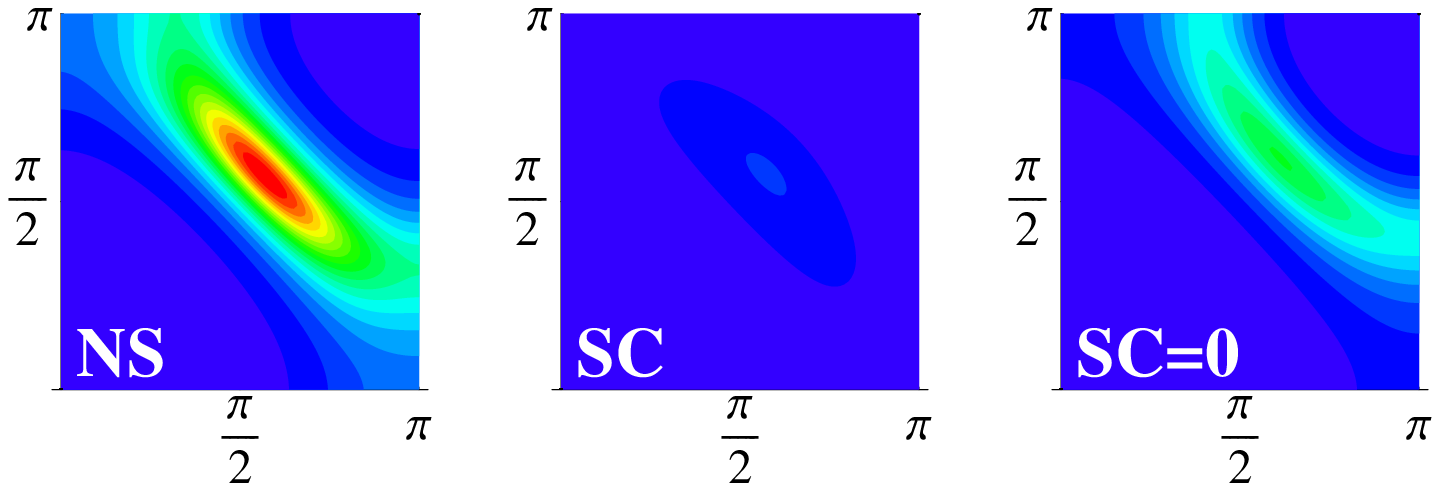}\\
$G$-scheme \\
\includegraphics[width=9.0cm,height=3.0cm,angle=-0] {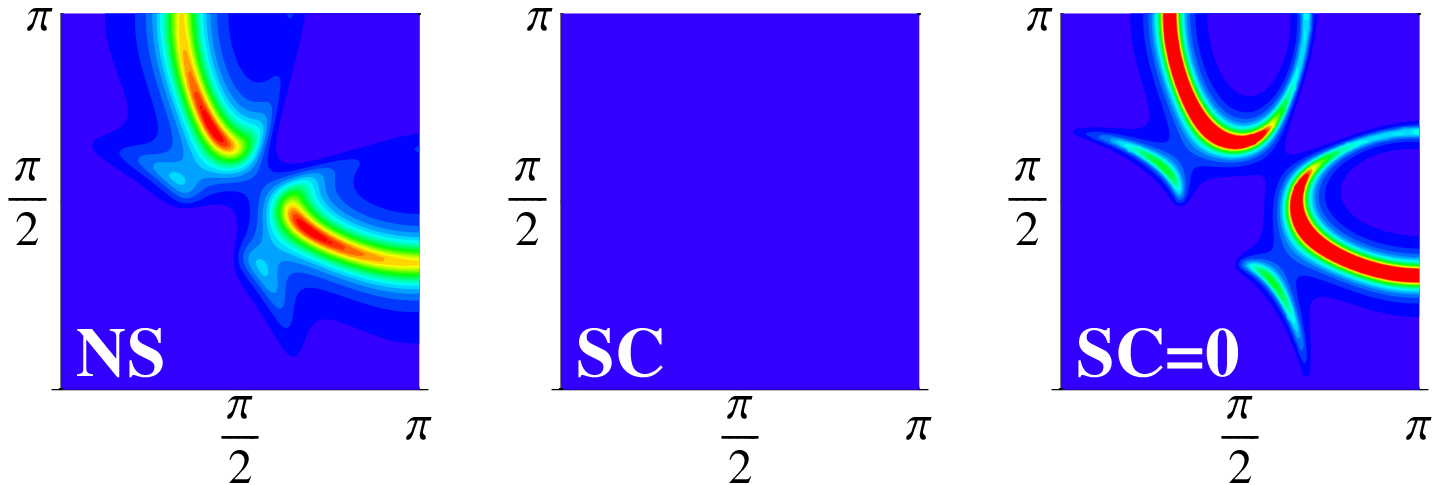}\\
\caption{Spectral function $A(k,\omega\rightarrow 0^{+})$ for the
case $U= 12t$, $t^{\prime}= 0.0$ and density $n=0.93$. The result
of the three periodizing schemes are displayed: the
$\Sigma$-scheme in the first row, the $M$-scheme in the second,
the $G$-scheme in the third. In the first column of each row the
NS-result is presented, whose color scale is kept in the
following SC (second column) and SC=0 (third column) plot. The SC
is the superconducting state result, SC=0 is the normal
quasiparticle spectrum in the SC state obtained zeroing the
anomalous components of the SC state.}\label{spectrat1-0}
\end{center}
\end{figure}
We plot in Fig.s \ref{spectrat1-0.3}, \ref{spectrat1-0} the
spectral density $A(k,\omega=0^{+})$ for a doping corresponding
to the critical temperature and $U= 12t$ in the first quadrant of
the BZ for the two cases $t^{\prime}= -0.3t$ (relevant for
cuprate materials Fig. \ref{spectrat1-0.3}), and $t^{\prime}=
0.0$ (Fig. \ref{spectrat1-0.3}) showing the 3 periodizing schemes
(the $\Sigma$-scheme on the top row, the $M$-scheme in the middle
row and the $G$-scheme in the bottom). In the first column we
display $A(k,\omega=0^{+})$ for the normal state (NS)
corresponding to the same critical density. These plots are
chosen to fix a reference scale for the color scale in the
spectral density plots for the superconducting state (SC) in the
second and third column, i.e. reading the plots horizontally the
color scale is absolute and fixed by the NS plot. In the third
column we extracted the quasiparticle spectrum underlying the SC
by artificially zeroing the anomalous part of the self-energy in
the SC solution (displayed in the second column).

\begin{figure}[!!tbh]
\begin{center}
\includegraphics[width=7.0cm,height=6.0cm,angle=-0] {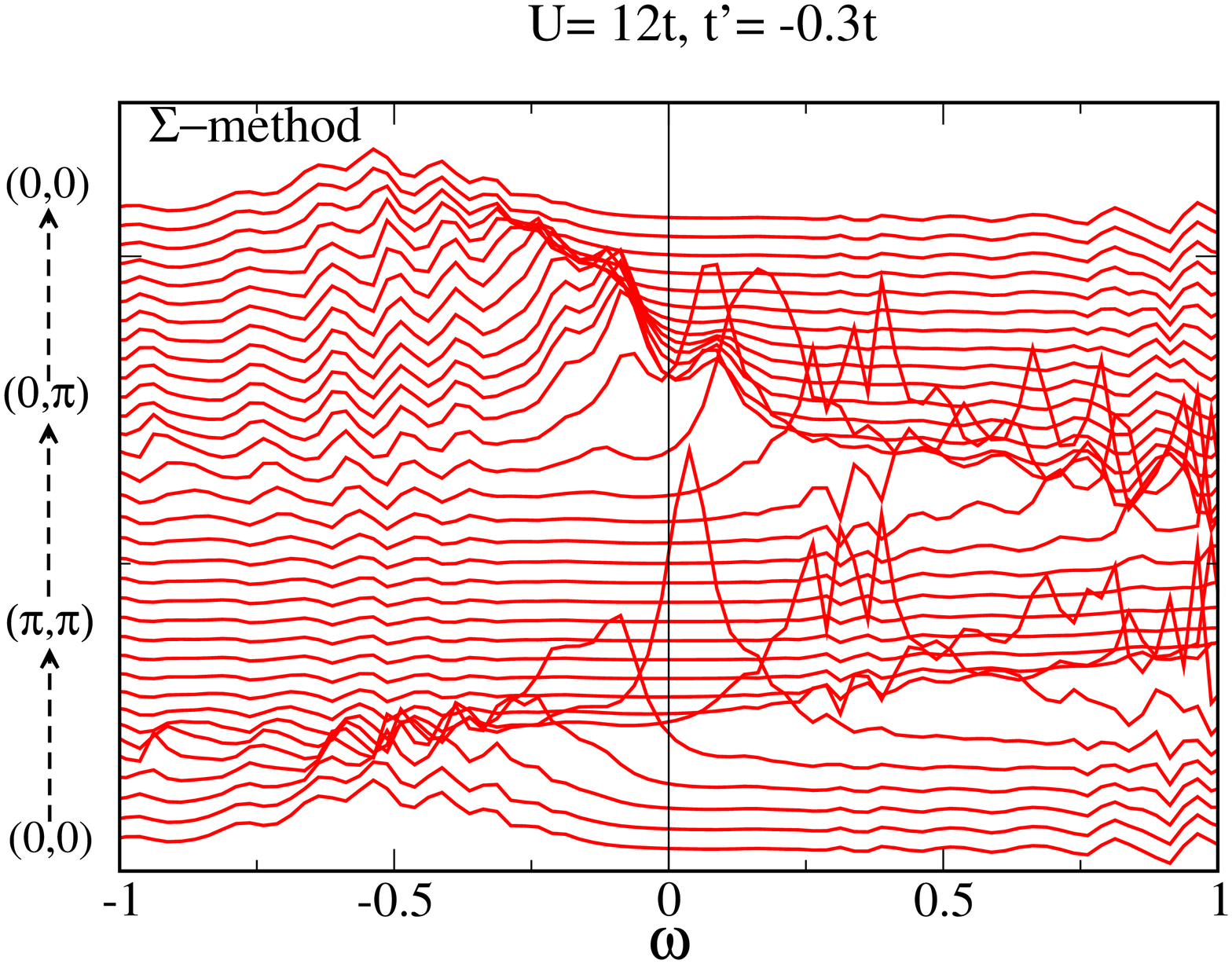}
\includegraphics[width=7.0cm,height=6.0cm,angle=-0] {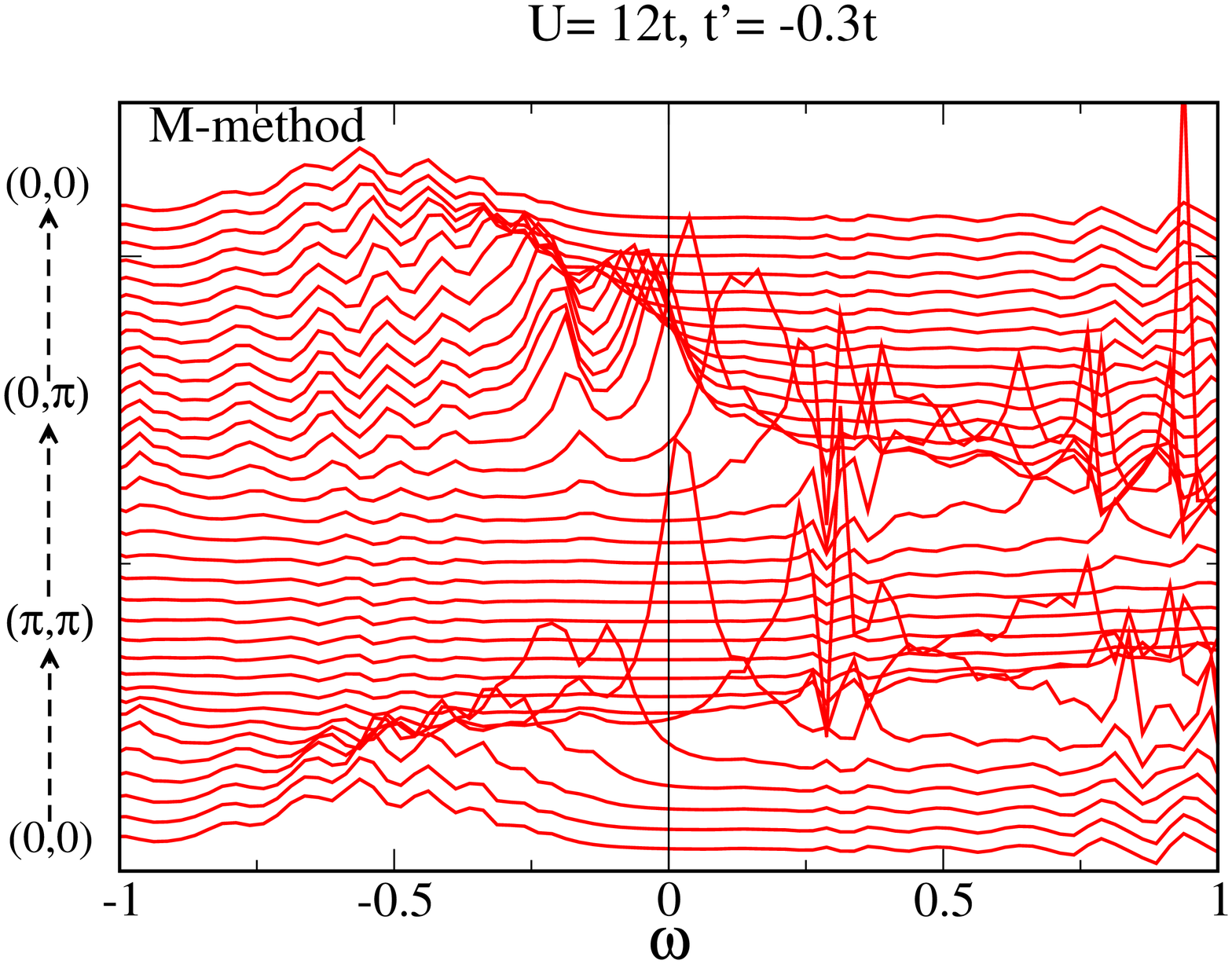}\vspace{0.5cm}\\
\includegraphics[width=7.0cm,height=6.0cm,angle=-0] {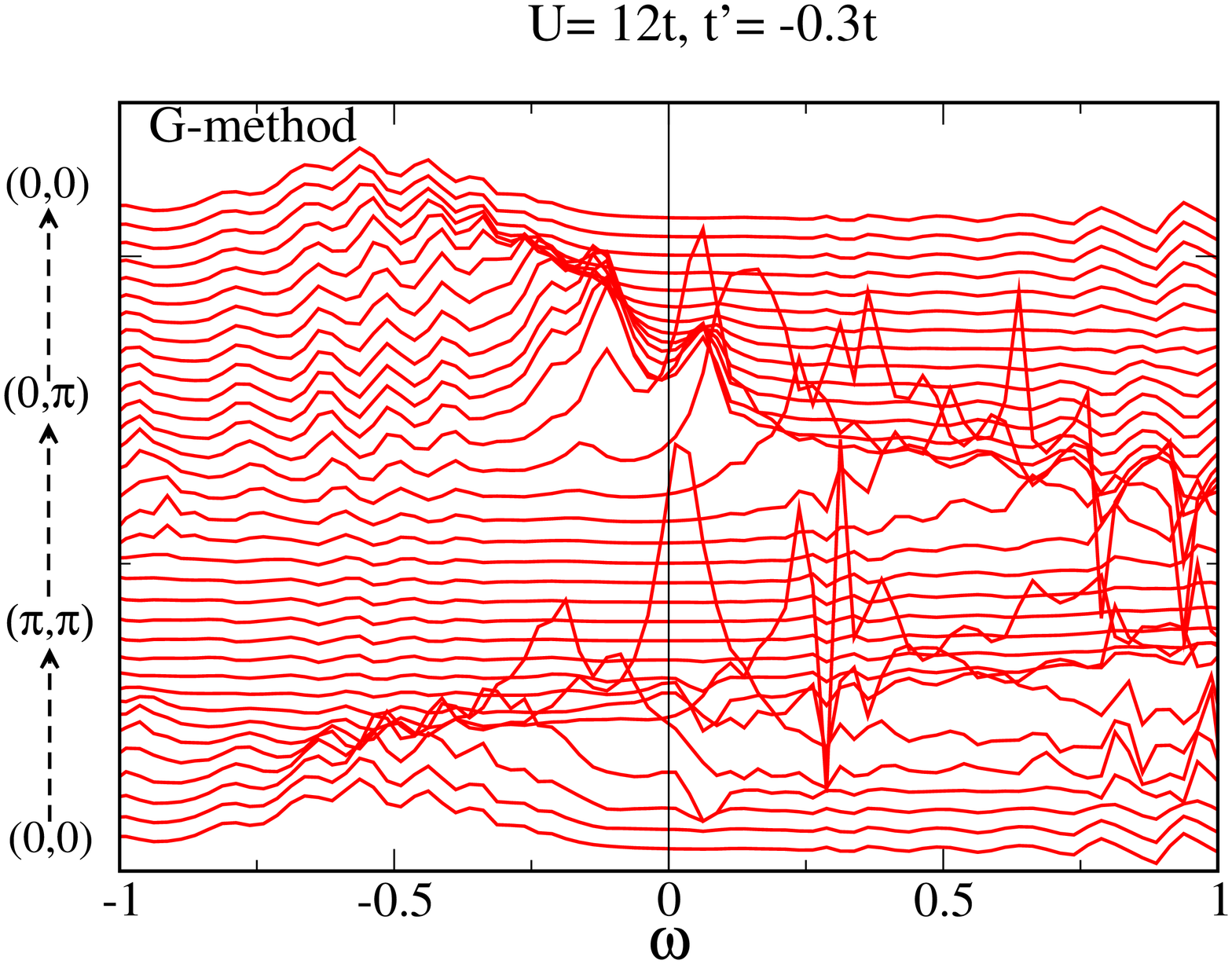}
\caption{Spectral function $A(k,\omega)$ along the path
$(0,0)\rightarrow (\pi,\pi)\rightarrow(0,\pi)\rightarrow(0,0)$ in
the first quadrant of the Brillouin Zone obtained with the 3
periodyizing schemes } \label{Akw_suprat1-0.3}
\end{center}
\end{figure}

In the case of $t^{\prime}= -0.3 t$ all the three methods produce
a consistent picture, where a dSC state is born out of a NS whose
momentum-dependent scattering gives rise to the cold spots
appearing in the $(\frac{\pi}{2},\frac{\pi}{2})$ region in the
first quadrant of the BZ, the right place for the nodal gapless
spectrum of a d-wave superconductor. In this sense we state the
dSC state is a cure to an instability already evident in the NS
and which originates the hot/cold spot modulation. The observation
made in Fig. \ref{Scluster-ano} that the dSC state rebuilds a more
FL system, originating cluster self-energies whose $\Im
\Sigma_{ij}\rightarrow 0$ more linearly (this above all in the
nearest next-neighbor component) is here enforced by the third
column graph (labeled $SC=0$), once again this is consistent in
all the three periodizing schemes. As mentioned above, this plot
has been obtained by switching off superconductivity in the SC
solution by simply putting to zero the anomalous components. It is
evident how a full homogeneous FS is re-constructed equivalent in
shape to the NS one (first column) but with a less $k$-dependent
modulation in the spectral intensity, i.e. the one particle
spectrum underlying the SC state is more regularly a FL than its
corresponding NS. In Fig. \ref{Akw_suprat1-0.3} we show the
spectral function $A(k,\omega)$ displayed in the path
$(0,0)\rightarrow (\pi,\pi)\rightarrow (0,\pi)\rightarrow (0,0)$
of the first quadrant of the BZ as a function of the energy
$\omega$, obtained in the SC system with the three periodyizing
schemes. As observed in experimental ARPES data \cite{damascelli}
\cite{campuzano}, a FL quasiparticle peak is observed around the
$k=(\pi/2,\pi/2)$ nodal point, even more promounced than in the
NS, while a superconducting gap opens in the pseudogap region
around $k=(0,\pi)$ point, as required in a d-wave shaped SC gap.
This enforces the idea that the dSC state is derived naturally
from the pseudogap NS in the hole-doped case, and it represents
the natural low-tempearture elongation of the pseudogap
instability at lower temperatures. The strong similarity of these
graphs in all the 3 periodizing schemes considered supports the
robustness of this result.

We now show in Fig. \ref{spectrat1-0} the case $t^{\prime}=0$.
This is qualitatively different from the previous one, and the
scenario more complex. In the NS already (first column) the three
periodizing schemes give a non-uniform answer: in the
$\Sigma$-scheme a weak modulation on the FS appears, with more
spectral weight around $(\frac{\pi}{2},\frac{\pi}{2})$, in the
$M$-scheme the modulation is more enhanced with a clear cold spot
forming around $(\frac{\pi}{2},\frac{\pi}{2})$ (but the total
intensity, which is not explicitly shown in this graph, is very
much reduced, being $\frac{1}{5}$ of the one resulting from the
$\Sigma$ and $G$-schemes), in the $G$-scheme spectral weight
disappears around $(0,\pi)$ and $(\pi,0)$ regions of the first
quadrant as well as around $(\frac{\pi}{2},\frac{\pi}{2})$. More
instabilities which (at least for the $M$ and $G$ schemes) push
away spectral weight from $(\frac{\pi}{2},\frac{\pi}{2})$ seems to
be taking place. Indeed, for the $M$ and $G$ schemes the
$t^{\prime}=0$ system is more connected to the
electron-like-shaped FS, presenting (as seen in chapter 5) a
dramatic change in the topology of the FS in correspondence of an
hypothetical quantum critical point. The latter was instead washed
away under a $\Sigma$-scheme viewpoint. We therefore see that in
the $\Sigma$-scheme (first row Fig. \ref{spectrat1-0}) the
scenario proposed is the same as in the case $t^{\prime}=-0.3t$,
where a d-wave SC was arising from a cold spot centered around the
$(\frac{\pi}{2},\frac{\pi}{2})$ region. Switching off SC (third
column labeled $SC=0$) re-builds a uniform FL-like FS. In the
$M$-scheme (second row) the spectrum totally gaps in the SC state,
and the nodal quasiparticle around the
$(\frac{\pi}{2},\frac{\pi}{2})$ region disappears too. Restoring
the NS from the SC (third column) partially re-insert spectral
weight in $(\frac{\pi}{2},\frac{\pi}{2})$, but strongly reduced as
compared to the NS one. In this case the simple FL picture of the
$\Sigma$-scheme is not recovered. We encounter some similarities
in the $G$-scheme result (third row). Here, as was stated, already
the NS is not a regular FL. Similar results for the electron-doped
side were obtained from VCPT with AF and dSC coexisting as well as
from CPT with no long range order when $U$ is larger than $6t$
\cite{Senechal:2005, Dahnken:2005}. Here, according to this
viewpoint,  short-range AF correlations would be sufficient to
suppress weight at $(\frac{\pi}{2},\frac{\pi}{2})$. The SC (second
column) state is once again, like in the $M$-scheme fully gaped.
Restoring the normal state re-builts the NS FS, however spectral
weight is not recovered in $(\pi/2,\pi/2)$. This, together with
the behaviour observed in the $M$-scheme, may indicate that in
these two schemes the instability responsible for the
disappearance of weight in $(\frac{\pi}{2},\frac{\pi}{2})$ is a
mechanism not directly connected with the dSC mechanism.

\addcontentsline{toc}{chapter}{Conclusions} 



\chapter*{{\huge Conclusions}}

In this thesis we have studied a strongly correlated electron
model, the two dimensional Hubbard Model, which is considered to
embody the essential physics of high temperature superconducting
materials (chapter 1). Strongly interacting many-body systems are
difficult to tackle because it is not possible to apply standard
techniques like perturbation theory. In this work we have used an
extension of Dynamical Mean Field Theory (DMFT) \cite{bibble}. In
the past decade, DMFT has proved a powerful tool to access
strongly interacting many-body models. In spite of its successes
however, DMFT is not able to accurately describe the physics of
the cuprate materials as, being a local theory exact only in the
infinite dimensional limit, misses the short ranged (i.e. momentum
space dependent) spatial correlations, which experimentally showed
to be an essential ingredient in the cuprate physics
\cite{damascelli}\cite{campuzano} (chapter 2). Moreover, DMFT by
construction does not allow one to study phases with a definite
spatial arrangement of the order parameter, like for example
d-wave superconductivity. We have therefore:
\begin{itemize}
  \item Developed an extension of DMFT, the Cellular Dynamical
  Mean Field Theory (CDMFT), which extends the single site
  character of DMFT to a cluster of sites, allowing for a first
  momentum space description of physical properties (chapter 3).
  The numerical technique used to solve an effective associate
  Anderson Impurity Model is the Lanczos algorithm, which allows
  to virtually access the zero temperature physics.
  \item Benchmarked the technique with exact known results in one
  dimension, with previous Quantum Montecarlo studies, set up
  the right implementation of the method and extracted
  physically interpretable results (chpater 4).
  \item Applied the method to the two-dimensional Hubbard Model,
   including the effect of a next-nearest neighbor hopping. We
   studied the density-driven Mott metal-insulator (MT) transition
   in the hole-doped and electron-doped systems (chapter 5). Our
   findings can be summarized as follows:
    \begin{enumerate}
    \item Approaching the MT point the Fermi Surface (FS) is strongly renormalized
    by interaction, contrary to the common band-structure approach
    which fix the shape of the FS in the MT description.
    \item In the region preceding the MT, we observe the phenomenon of momentum-space
    differentiation of scattering properties, as found in angle
    photo-emission experiments (ARPES)
    experiments \cite{damascelli}\cite{campuzano}. In particular
    in the hole-doped systems a
    cold (low scattering) spot is formed around the point
    $k=(\frac{\pi}{2},\frac{\pi}{2})$ in the first quadrant of the
    Brillouin Zone (BZ), while high scattering regions (hot spots) develop
    around the points $k=(0,\pi),(\pi,0)$. This state can be
    smoothly connected
    to a d-wave superconducting state (dSC), where a
    nodal quasiparticle survives in $k=(\frac{\pi}{2},\frac{\pi}{2})$.
    In the electron-doped systems instead, the cold/hot spot
    position in the $k$-space is switched. This state cannot be
    easily connected with a d-wave SC.
    \item Enforcing an artificially strong frustration into the
    system ($t^{\prime}= +0.9t$), we have shown that the
    phenomenon of momentum-space modulation is indeed associated
    with the proximity to the Mott transition rather than to
    long-ranged antiferromagnetic (AF) correlations of a insulating parent
    ordered state.
    \item We have found hints of the presence of a quantum
    critical instability which preludes the formation of the phase
    with momentum-modulated properties (hot/cold spot formation).
    This quantum critical region is associated with the appearance of zeroes of the
    one particle Green's function in the
    momentum-space and with a dramatic change in topology of the FS.
    In this phase a depression of spectral weight forms in
    correspondence of the Fermi level in the hot regions of
    momentum space: this is the pseudogap phase observed also
    in ARPES experiments (\cite{damascelli}\cite{campuzano}).
    \end{enumerate}
\item finally applied CDMFT to study a dSC-state (chapter 6). We have found
that at zero temperature the two-dimensional Hubbard model is
capable of sustaining d-wave superconducting state, which present some anomalies
compared with a standard Bardeen-Cooper-Shifferer (BCS) superconductor:
    \begin{enumerate}
    \item the order parameter and the superconductive gap do not
    scale together, rather the superconductive gap increases with
    decreasing doping while the order parameter decreases to zero
    as the critical temperature T$_{C}$. This is in major contrast
    with the BCS predictions where these two quantities are
    proportional.
    \item The disappearance of SC is due to the disappearance of
    quasiparticles approaching the MT, rather than to weakening
    of the pairing interaction.
    \item The pairing mechanism originates from the magnetic
    super-exchange $J\sim \frac{1}{U}$, as shown by the maximum value of the order
    parameter that scales like $J$ and by the frequency-dependence of the anomalous
    self-energy, which give insights into the range of the paring
    attraction (again of the order of $J$ and roughly doping independent).
    \end{enumerate}
We finally conjecture that in the hole-doped system the d-wave nature
of the superconducting state is naturally born out from its anomalous
pseudogap normal state. We have shown that the momentum space modulated
FS of the NS is uniformly restored in the normal-quasiparticle spectrum of
the corresponding dSC state, which in this way "cures" the instability of the NS.
In the electron-doped systems instead the scenario is more complicated, where other
instabilities (like for example short ranged AF correlations) may play a role in determining
a non d-wave SC gap, i.e. NS and dSC states may not be smoothly connected.
Further investigations are in this case required.
\end{itemize}

\appendix

\chapter{The tJ Model}

In Chapter 1 we presented a Hartree-Fock mean field treatment of
the Hubbard hamiltonian which is able to take into account the
high-energy charge excitation. This is responsible for the
splitting of the conduction band in an upper and lower part,
separated by a gap of the order of the on-site repulsion U. We
then showed that a Heisenberg Hamiltonian, introduced as the
strong repulsion limit of the Hubbard model, is able to describe
the low-energy collective spin-modes also deriving from the
on-site repulsion but not encompassed into the mean-field
approach. In this appendix we want now to extend the Heisenberg
treatment to the doped case. The basic idea is to be able to
describe the physics of few charge carriers immersed in an AF
background, taking into account the forbidden double occupancy as
well as the collective motion of the spin degrees of freedom. To
this purpose, we define the interaction part of the Hamiltonian
$H_{0}$ and the kinetic part $H_{1}$, which we consider as the
perturbing term. We seek for the effective Hamiltonian defined
into the subspace $\Xi$ of forbidden double occupation for
$H_{0}$, corresponding to the zero eigenvalue. The difference with
the previous section is that the states called $\phi_{0}$, which
specify the positions of the electrons in the lattice, have now
some holes corresponding to empty sites. Applying $H_{1}$ to
$\phi_{0}$ creates a superposition of states which are not all
double occupied. Said $|\Psi_{0}\rangle \in \Xi$ a generic linear
combination of $|\phi_{0}\rangle$, we have that $H_{1}
|\Psi_{0}\rangle$ has a non-zero component in $\Xi$. So, for
convenience, let's decompose $H_{1} |\Psi_{0}\rangle$ into two
components in $\Xi$ and in $\Xi^{\perp}$, introducing the
projecting-operators onto these subspaces, respectively $P_{0}$ e
$P_{1}= 1-P_{0}$. Given then $H=H_{0}+\lambda H_{1}$ and $|\Psi
\rangle= |\Psi_{0} \rangle +\lambda |\Psi_{1} \rangle $, where,
without loss of generality, $| \Psi_{1} \rangle \in \Xi^{\perp}$,
we write the eigenvalues equation $ H|\Psi \rangle= E |\Psi
\rangle$, and we have: \beq \lambda H_{0} |\Psi_{1}+ \lambda H_{1}
|\Psi_{0} \rangle+ \lambda^{2} H_{1} |\Psi_{1} \rangle= E (
|\Psi_{0} \rangle+ \lambda |\Psi_{1} \rangle. \label{1.9}\eeq
Let's multiply both terms by $\langle \Psi_{0}|$. Noticing that
the first term in the first side of the equation is zero, as
$H_{0} |\Psi_{1} \rangle \in \Xi^{\perp}$, it is: \beq \lambda
\langle \Psi_{0}| H_{1} |\Psi_{0}\rangle+ \lambda^{2} \langle
\Psi_{0}| H_{1} | \Psi_{1}\rangle= E. \label{1.10}\eeq If instead
we multiply the (\ref{1.9}) by  $\langle \Psi_{0}| H_{1}P_{1}$, we
have at the first order
$$\lambda \langle \Psi_{0}| H_{1}P_{1}H_{0}| \Psi_{1}\rangle+
\lambda \langle \Psi_{0}| H_{1}P_{1}H_{1}| \Psi_{1}\rangle= E
\langle \Psi_{0}| H_{1}P_{1}|\Psi_{0}\rangle,$$ and, as the second
side is zero given $P_{1}H_{1}|\Psi_{0} \rangle \in \Xi^{\perp}$
(in fact $H_{0}P_{1}H_{1} |\Psi_{0} \rangle= U P_{1}H_{1}
|\Psi_{0} \rangle$, as $P_{1}H_{1} |\Psi_{0} \rangle$ is a linear
combination of configurations with only one site double occupied),
we obtain \beq \langle \Psi_{0}| H_{1} |\Psi_{1}\rangle=
-\frac{1}{U} \langle \Psi_{0}| H_{1}P_{1}H_{1} |\Psi_{0}\rangle.
\label{1.11} \eeq We can plug the (\ref{1.11}) into (\ref{1.10}),
and finally obtain :
$$\langle \Psi_{0}| \lambda H_{1}- \frac{\lambda^{2}}{U} H_{1}P_{1}H_{1} |\Psi_{0}\rangle= E. $$
The effective Hamiltonian we want, which is called $tJ$, is then:
$$H_{tJ}= P_{0} (H_{1}-\frac{1}{U} H_{1}P_{1}H_{1}) P_{0}$$
with the operators $P_{0}$ limiting its action into the subspace
$\Xi$ only. Once again we can get a more representative expression
by introducing spin operators. The action of the operators $P_{0}$
e $P_{1}$ can be described by redefining the creation and
destruction operators. So given $|\varphi \rangle$ a generic state
in $\Xi$, we can write with good approximation:
$$c^{+}_{i\sigma}|\varphi\rangle=
P_{0}|\varphi\rangle+P_{1}|\varphi\rangle=
[(1-n_{i,-\sigma})c^{+}_{i\sigma}+n_{i,-\sigma}c^{+}_{i\sigma}] \
|\varphi \rangle .$$ Essentially the action of $c^{+}_{i\sigma}$
on $\Xi$ is described by $(1-n_{i,-\sigma})c^{+}_{i\sigma}$; we
can therefore create an electron with spin $\sigma$ on the site
$i$ only if there is not already an electron of opposite spin.
The term $P_{0}H_{1}P_{1}H_{1}P_{0}$ can be written $$ t^{2}
\sum_{ij,\sigma \sigma^{'}}^{}
(1-n_{i,-\sigma})c^{+}_{i\sigma}c_{j\sigma}
n_{j,-\sigma}n_{j,-\sigma^{'}}c^{+}_{j\sigma^{'}}c_{i\sigma^{'}}(1-n_{l,-\sigma^{'}})
.$$ We want to consider the case for which the doping $\delta<<
1$, so that at the first order in $\delta$ we can disregard the
three-site-hopping terms. As $S^{+}=
c^{+}_{\uparrow}c_{\downarrow} \hbox{ e } S^{-}=
c^{+}_{\downarrow}c_{\uparrow}$, after some simple passage:
$$H_{tj}=
-t \sum_{\langle ij \rangle,\sigma}^{} P_{0} [
c^{+}_{i\sigma}c_{j\sigma}+ h.c.]P_{0}+ J\sum_{\langle
ij\rangle}^{} ({\bf S_{i}S_{j}}-\frac{1}{4}n_{i} n_{j}).$$ This
Hamiltonian differs from Heisenberg's only in the kinetic term,
which appears because of doping. The second term describes the
spin-spin interaction due to the virtual hopping of the electrons
between occupied sites. Notice that in the half-filled case
($\delta= 0$) the kinetic term disappears.

\chapter{The Lanczos Algorithm}
\section{The Lanczos idea}
We briefly overview the numerical principles underlying the
numerical exact diagonalization method (Lanczos) that we used in
this work to solve the AIM associated with the original non local
problem. A complete and detailed explanation of this method, with
its virtues and defects, can be found for example in \cite{ANLA}.
The problem aims to find few eigenvalues and eigenvectors of a
large matrix simmetric $A$ which cannot be solved by direct
methods. The simplest problem consist for example in finding the
largest eigenvalue in absolute value, along with the corresponding
eigenvector, and the simple algorithm able to achieve this goal
is the power method:
\begin{eqnarray}
y_{i+1}=\, A\,x_{i} \nonumber \\
x_{i+1}=\, y_{i+1}/\parallel y_{i+1}
\parallel
\end{eqnarray}
If there exists only one eigenvalue of largest absolute value and
the starting $x_{1}$ it is not perpendicular to it, $x_{i}$
converges to the desired eigenvector. Notice that the main
calculation in the algorithm, the matrix multiplication
$A\,x_{i}$, can be considered as a black box with $x_{i}$ in input
and $A\,x_{i}$ in output. The procedure after $k$ steps builds
$x_{1}...x_{k}$ vectors which span the so called {\it Krylov
subspace} ${\cal K}_{k}(A,x_{1})=\, \{\, A x_{1},...,A^{k-1} x_{1}
\,\}$. In fact, rather then considering only $x_k$, a stronger
procedure consist in seeking the best linear combination
$\sum_{i=1}^{k}\, \alpha_i x_i$ in the Krylov subspace, the {\it
Ritz} vector, which best approximates the desired eigenvector. The
corresponding approximation to the desired eigenvalue is the so
called Ritz value and the proceudure is called the Rayleigh-Ritz
Method. In more precise terms, given the $n \times n$ orthogonal
matrix $Q= (Q_k,Q_{n-k})$, where with $Q_k$ we indicate the $k$
vector-columns (which in practice span the Krylov subspace) and
\begin{eqnarray*}
T=\, Q^{T}\, A \, Q
\end{eqnarray*}
Rayleigh-Ritz Method approximates the eigenvalues of $A$ by the
eigenvalues of $T_{k}=\,Q_k^{T}\, A \, Q_k$, the Ritz values, and,
given the eigendecomposition of $T_k=\, V\,\Lambda\,V^{T}$, the
corresponding eigenvectors approximations are the columns of
$Q_k\,V$, the Ritz vectors. In using the Lanczos algorithm to
compute $Q_k$, $T$ assumes the particular symmetric triangular
form:
\begin{eqnarray*}
T=\, \left(
\begin{array}{c|c}
  T_k    &  T^{T}_{ku}\\ \hline
  T_{ku} &  T_u
\end{array}
\right)=\quad \left(
\begin{array}{cccc|cccc}
  \alpha_1 \quad & \beta_1     &              &            &  &  &  &  \\
  \beta_1  \quad & \ddots           & \ddots            &            &  &  &  &  \\
                 & \ddots           & \ddots            &\beta_{k-1} &  &  &  &  \\
                 &             & \beta_{k-1}  &\alpha_k    &\beta_k  &  &  &  \\
           \hline 
&  &  &\beta_k  &  \alpha_{k+1} & \beta_{k+1}     &             &                \\
&  &  &         &  \beta_{k+1}  & \ddots               & \ddots           &                \\
&  &  &         &               & \ddots               & \ddots           &\beta_{n-1}     \\
&  &  &         &               &                 & \beta_{n-1}
&\alpha_n
\end{array}
\right)
\end{eqnarray*}
for which it is easy to compute eigenvalues and eigenvectors, and
the Rayleigh-Ritz procedure simplifies. The following theorem,
which we enunciate here without proof (for it we refer to
\cite{ANLA}) elucidates the advantages:\vspace{1cm}\\
{\it Let $T_k=\, V\,\Lambda\,V^{T}$ be the eigendecomposition of
$T_k$, where $V=[v_1...v_k]$ is orthogonal and $\Lambda=
\,\hbox{diag}(\theta_1...\theta_k)$. Then \begin{enumerate} \item
There are $k$ eigenvalues $\alpha_1, ..., \alpha_k$ of $A$ (not
necessarily the largest k) such that $|\theta_i-\alpha_i|\leq
\parallel T_{ku}\parallel= \beta_k$, the single (possibly) nonzero
entry in the upper right corner of $T_{ku}$. \item $\parallel
\,A\,(Q_k\,v_i)- (Q_k\,v_i)\,\theta_i \,
\parallel= \, \parallel T_{ku} v_i \parallel=\, \beta_k \,
|v_i(k)|$ where $v_i(k)$ is the $k^{th}$ (bottom) entry of $v_i$.
Thus, the difference between the Ritz value $\theta_i$ and some
eigenvalue $\alpha$ of $A$ is at most $\beta_k \, |v_i(k)|$, which
may be much smaller than $\beta_k$. Moreover this formula allow us
to compute the residual $\parallel \,A\,(Q_k\,v_i)-
(Q_k\,v_i)\,\theta_i \,\parallel$ cheaply without multiplying any
vector by $Q_k$ or by $A$. \item Without any further information
on the spectrum of $T_u$, we cannot deduce any useful error bound
on the Ritz vector $Q_k v_i$. However, if we know that the gap
between $\theta_i$ and any other eigenvalue of $T_k$ or $T_u$ is
at least $g$, then we can bound the angle $\varphi$ between $Q_k
v_i$ and a true eigenvector of $A$ by
\begin{eqnarray*}
\frac{1}{2} \, \sin 2\varphi \leq \, \frac{\beta_k}{g}
\end{eqnarray*}
\end{enumerate}}

In order to find the eigenvalues and the eigenstates of the
symmetric matrix $A$ the Lanczos procedure combines the Lanczos
power method to construct the Krylov subspace and the
Rayleigh-Ritz method: it first builds the orthogonal matrix $Q_k=
\,[q_1...q_k]$ of Lanczos vectors and approximates the
eigenvalues of $A$ by the Ritz eigenvalues of the tridiagonal
symmetric matrix $T_k=\, Q^T_k \, A\, Q_k$.
\begin{center}
{\bf The Lanczos algorithm:}
\end{center}
\begin{enumerate}
\item[] $q_1=\, b/\parallel b \parallel$, $\beta_0= 0$, $q_0= 0$
\item[] for $j=1$ to $k$
      \begin{enumerate}
      \item[] $z= A q_j$
      \item[] $\alpha_j= q^{T}_{j} z$
      \item[] $z= z- \alpha_j q_j- \beta_{j-1} q_{j-1}$
      \item[] $\beta_j= \parallel z \parallel$
      \item[] if $\beta_j=0$ exit
      \item[] $q_{j+1}= z/\beta_j$
      Compute eigenvalues, eigenvectors and error bounds of $T_j$
      \end{enumerate}
\item[] end for
\end{enumerate}
The algorithm is in principle able to provide the best $k$ Ritz
approximation to the eigenvalues and eigenvectors of $A$, though
the largest eigenvalues converge much faster than the others. The
starting vector should not however be nearly orthogonal to one of
the desired eigenvectors. If this is the case in fact, a
difficulty arises in the procedure wherein the eigenvector
orthogonal to $q_1$ is missing in the solution. Choosing $q_1$ at
random can prevent this kind of problem and we can always rerun
the Lanczos algorithm with a different random $q_1$ to provide
more "statistical" evidence that we did not miss any eigenvalues.
Another difficulty arises if there are degenerate eigenvalues in
the spectrum. In this case the Lanczos procedure is never able to
obtain both eigenvalues, and one of them will be missing.
Fortunately there are many applications where it is enough to
determine the value of the eigenvalue without knowing its
multiplicity. There are however procedure like the "Block" Lanczos
that are able, at least to some extend, to tackle degenerate
spectra and recover the multiplicity of the eigenvalues. We will
return later on this method.
\section{The Lanczos algorithm in floating point arithmetic}
We presented in the previous section an "ideal" Lanczos procedure
which did not take into account the roundoff, unavoidable in any
realistic calculation. At each Lanczos loop, the vector $z$ is by
construction orthogonal to $q_1$ through $q_{j-1}$. Roundoff
unfortunately destroys this orthogonality. However the algorithm
does not become totally unpredictable: it can be shown (Paige's
theorem \cite{ANLA}) that the $q_k$ lose orthogonality because
they acquire large components in the direction of the Ritz vectors
$y_{i,k}= Q_k v_i$ whose Ritz value $\theta_i$ have already
converged. This creates ghost copies of these Ritz vectors, i.e.
instead of a $T_k$ having one eigenvalue nearly equal to the
desired value $\lambda_i(A)$ it may have many eigenvalues nearly
equal to $\lambda_i(A)$. This is not dramatic if the multiplicity
of the eigenvalue it is not important in the problem studied, only
the convergence of interior eigenvalues is further delayed. But if
accurate multiplicities are important (as it is for example in
calculating partition functions), then it is extremely important
to keep the Lanczos vectors nearly orthogonal. One could for
example use a Lanczos algorithm with full orthogonalization,
imposing the $z$ vector orthogonal to each converged Ritz vector
at each Lanczos step:
\begin{eqnarray}
z= z- \alpha_j q_j- \beta_{j-1} q_{j-1} \quad \rightarrow \quad
\left\{
\begin{array}{c}
z=\, z- \sum_{i=1}^{j-1}\, (z^{T} q_i) q_i \\
z= z- \alpha_j q_j- \beta_{j-1} q_{j-1}
\end{array}
 \right.
\end{eqnarray}
This implementation does cure the problem, however it is very
expensive. One can easily check it costs $O(k^2 n)$ flops instead
of $O(k n)$ and requires $O(k n)$ space instead of $O(n)$: this
may be a too high prize to pay. Fortunately there is a middle
ground between no reorthogonalization and full reothogonalization
capable to get the best from both methods: the selective
reothogonalization.
\subsection{Selective reothogonalization}
Selective reothogonalization  exploits Paige's theorem which tells
us that the ghost-Ritz-vector develop in the direction of the
already converged Ritz vector. So we can simply monitor the error
bound $\beta_k |v_i(k)|$ at each step, and when it becomes small
enough (less than the converging precision $\epsilon$), the vector
$z$ in the inner loop of the Lanczos algorithm is orthogonalized
against $y_{i,k}$: $z= z- (y^{T}_{i,k} z) y_{i,k}$:
\begin{center}
{\bf The Lanczos algorithm with selective orthogonalization:}
\end{center}
\begin{enumerate}
\item[] $q_1=\, b/\parallel b \parallel$, $\beta_0= 0$, $q_0= 0$
\item[] for $j=1$ to $k$
      \begin{enumerate}
      \item[] $z= A q_j$
      \item[] $\alpha_j= q^{T}_{j} z$
      \item[] $z= z- \alpha_j q_j- \beta_{j-1} q_{j-1}$
      \item[] {\it Selective orthogonalization against Ritz
      vectors}:\\
      for all $i\leq k$ such that $\beta_k |v_i(k)|\leq \epsilon$\\
      $\qquad z= z- (y^{T}_{i,k} z) y_{i,k}$\\
      end for
      \item[] $\beta_j= \parallel z \parallel$
      \item[] if $\beta_j=0$ exit
      \item[] $q_{j+1}= z/\beta_j$
      Compute eigenvalues, eigenvectors and error bounds of $T_j$
      \end{enumerate}
\item[] end for
\end{enumerate}


\bibliographystyle{plain}
{
\renewcommand{\baselinestretch}{1}\huge\normalsize
\bibliography{thesis}

\newcommand{\PRB}{Phys. Rev. B}\newcommand{\PRL}{Phys. Rev.
  Lett}\newcommand{\NPB}{Nucl. Phys. B}\newcommand{\RMP}{Rev. Mod.
  Phys.}\newcommand{\ADV}{Adv. Phys.}
\begin{thebibliography}{10}

\bibitem{anderson}
P.~W. Anderson.
\newblock {\em Science}, 235:1196, 1987.

\bibitem{Anderson:04}
P.~W. Anderson, P.~A. Lee, M.~Randeria, T.~M. Rice, N.~Trivedi, and F.~C.
  Zhang.
\newblock {\em J. Phys. Condens. Matter}, 16:R755--R769, 2004.

\bibitem{asymmetry}
P.~W. Anderson and P.~Ong.
\newblock cond-mat/0405518.

\bibitem{Ando}
Y.~Ando and T.~Murayama.
\newblock {\em \PRB}, 60:6991, 1999.

\bibitem{AM}
N.W. Ashcroft and N.D. Mermin.
\newblock {\em Solid State Physics}.
\newblock Saunders College Publishing, 1976.

\bibitem{Takagi}
Takagi at~al.
\newblock {\em J. of Appl.Phys. (Lett.)}, 26:L123, 1987.

\bibitem{BCS}
J.~Bardeen, L.N. Cooper, and J.R. Schrieffer.
\newblock {\em Phys. Rev.}, 108(5):1175, 1957.

\bibitem{Bethe}
H.~Bethe.
\newblock {\em Zeitschrift f\"ur Physik}, 71:205, 1931.
\newblock English translation by V. Frederick in: D.C. Mattis {\it The
  many-body problem}, World Scientific, (1993), pg 689.

\bibitem{venky}
C.~J. Bolech, S.~S. Kancharla, and G.~Kotliar.
\newblock {\em \PRB}, 67:075110, 2003.

\bibitem{Caffarel94}
M.~Caffarel and W.~Krauth.
\newblock {\em \PRL}, 72:1545, 1994.

\bibitem{campuzano}
J.~C. Campuzano, M.~R Norman, and M.~Randeria.
\newblock {\em Physics of Superconductors II}.
\newblock K. H. Bennemann and J. B. Ketterson, 2004.
\newblock 167-273.

\bibitem{marce}
M.~Capone, M.~Civelli, S.~S. Kancharla, C.~Castellani, and G.~Kotliar.
\newblock {\em \PRB}, 69:195105, 2004.

\bibitem{Chien}
T.R. Chien, Z.Z. Wang, and N.P. Ong.
\newblock {\em Phys. Rev. Lett.}, 67:2088, 1991.

\bibitem{Pines}
P.~Nozi\'eres D.~Pines.
\newblock {\em The Theory of Quantum Liquids}, volume~I.
\newblock W.A. Benjamin, Inc., New York, 1966.

\bibitem{damascelli}
A.~Damascelli, Z.~X. Shen, and Z.~Hussain.
\newblock {\em \RMP}, 75:473, 2003.

\bibitem{ANLA}
JW~Demmel.
\newblock {\em Applied Numerical Linear Algebra}.
\newblock SIAM, Philadelphia, PA, 1997.

\bibitem{quasiparticlesarpes}
A.~Kaminski et~al.
\newblock {\em \PRL}, 84:1788, 2000.

\bibitem{Malinowski}
A.~Malinowski et~al.
\newblock cond-mat/0108360.

\bibitem{bumsoo}
B.~Kyung et~al.
\newblock cond-mat/0502565.

\bibitem{Kyung:2003}
B.~Kyung et~al.
\newblock {\em \PRB}, 68:174502, 2003.

\bibitem{Dahnken:2005}
C.~Dahnken et~al.
\newblock cond-mat/0504618.

\bibitem{thermal}
D.~G.~Hawthorn et~al.
\newblock cond-mat/0502273.

\bibitem{Senechal:2005}
D.~S\'en\'echal et~al.
\newblock {\em \PRL}, 94:156404, 2005.

\bibitem{pavarini}
E.~Pavarini et~al.
\newblock {\em \PRL}, 87:047003, 2001.

\bibitem{Biroli:2005}
G.~Biroli et~al.
\newblock {\em \PRB}, 71:037102, 2005.

\bibitem{Mesot}
J.~Mesot et~al.
\newblock {\em \PRB}, 63:224516, 2001.

\bibitem{KShen}
K.~M.~Shen et~al.
\newblock {\em Science}, 307:901, 2005.

\bibitem{Yamada}
K.~Yamada et~al.
\newblock {\em \PRB}, 40:4557, 1989.

\bibitem{Nucker}
N.~N\"ucker et~al.
\newblock {\em Zeitschrift f\"ur Physik B}, 67:9, 1987.

\bibitem{stm}
S.~H.~Pan et~al.
\newblock {\em Nature}, 413:282, 2002.

\bibitem{Jarrell:2000}
Th.~Maier et~al.
\newblock {\em \PRL}, 85:1524, 2000.

\bibitem{Fano}
G.~Fano, F.~Ortolani, and A.~Parola.
\newblock {\em \PRB}, 46:1048, 1992.

\bibitem{Forro}
L.~Forro, D.~Mandrus, C.~Kendziora, L.~Mihaly, and R.~Reeder.
\newblock {\em Phys. Rev. B}, 42:8704, 1990.

\bibitem{George-Kotliar92}
A.~Georges and G.~Kotliar.
\newblock {\em \PRB}, 45:6479--6483, 1992.

\bibitem{bibble}
A.~Georges, G.~Kotliar, W.~Krauth, and M.~Rozenberg.
\newblock {\em \RMP}, 68:13--125, 1996.

\bibitem{grilli}
M.~Grilli and G.~Kotliar.
\newblock {\em \PRL}, 64:1170, 1990.

\bibitem{haule}
K.~Haule, A.~Rosch, J.~Kroha, and P.~W\"olfle.
\newblock {\em \PRL}, 89:236402, 2002.

\bibitem{hubbard:64}
J.~Hubbard.
\newblock {\em Proc. Roy. Soc. London A}, 276:238, 1963.

\bibitem{Hybertsen}
M.~S. Hybertsen, M.~Schl\"uter, and N.~E. Christensen.
\newblock {\em \PRB}, 39:9028, 1989.

\bibitem{Hyrsch-Fye86}
J.E. Hyrsch and R.M. Fye.
\newblock {\em \PRL}, 56:2521, 1986.

\bibitem{Poilblanc}
D.~Poilblanc it~et al.
\newblock cond-mat/0202180 (2002).

\bibitem{Konsta}
Z.~Konstantinovic, Z.Z. Li, and H.~Raffy.
\newblock {\em \PRB}, 62:11989, 2000.

\bibitem{liu}
G.~Kotliar and J.~Liu.
\newblock {\em \PRB}, 38:R5142, 1988.

\bibitem{cdmft}
G.~Kotliar, S.Y. Savrasov, G.~Palsson, and G.~Biroli.
\newblock {\em \PRL}, 87:186401, 2001.

\bibitem{kotliar_P2day}
G.~Kotliar and D.~Vollhardt.
\newblock {\em Physics Today}, 57:53, 2004.

\bibitem{psgap1}
B.~Kyung, S.S. Kancharla, D.~S\'en\'echal, A.-M.S. Tremblay, M.~Civelli, and
  G.~Kotliar.
\newblock 2005.
\newblock cond-mat/0502565.

\bibitem{bumsoo06}
B.~Kyung, G.~Kotliar, and A.~M.~S. Tremblay.
\newblock cond-mat/0601271.

\bibitem{Lanczos}
C.~Lanczos.
\newblock {\em J. Res. Nat. Bur. Stand.}, 45:255, 1950.

\bibitem{lich}
A.~I. Lichtenstein and M.~I. Katsnelson.
\newblock {\em \PRB}, 62:R9283, 2000.

\bibitem{Lieb-Wu}
E.H. Lieb and F.Y. Wu.
\newblock {\em \PRL}, 20:1445, 1968.

\bibitem{psgap}
T.~Maier, M.~Jarreland~T. Pruschke, and J.~Keller.
\newblock {\em Eur. Phys. J. B}, 13:613, 2000.

\bibitem{McIntosh}
G.C. McIntosh and A.B. Kaiser.
\newblock {\em Phys. Rev. B}, 54:12569, 1996.

\bibitem{Metzner:89}
W.~Metzner and D.~Vollhardt.
\newblock {\em \PRL}, 62:324, 1989.

\bibitem{Obertelli}
S.D. Obertelli, J.R. Cooper, and J.L. Tallon.
\newblock {\em \PRB}, 46:14928, 1992.

\bibitem{bpk}
O.~Parcollet, G.~Biroli, and G.~Kotliar.
\newblock {\em \PRL}, 92:226402, 2004.

\bibitem{parcollet}
O.~Parcollet and A.~Georges.
\newblock {\em \PRB}, 59:5341, 1999.

\bibitem{PSK}
A.~Perali, M.~Sindel, and G.~Kotliar.
\newblock {\em Eur. Phys. J. B}, 24:487, 2002.

\bibitem{Pini}
D.~Pini.
\newblock Il modello di hubbard nella superconduttivit\'a ad alta temperatura.
\newblock Master's thesis, Universit\'a degli studi di Milano, Facolt\'a di
  Scienze MM FF NN, 1989.

\bibitem{Quijada}
M.A. Quijada, D.B. Tanner, R.J. Kelley, and M.~Onellion.
\newblock {\em Physica C}, 235-240:1123, 1994.

\bibitem{Rozenberg94}
M.J. Rozenberg, G.~Moeller, and G.~Kotliar.
\newblock {\em Mod. Phys, Lett. B}, 8:535, 1994.

\bibitem{senechal}
D.~Senechal, P.-L. Lavertu, M.-A. Marois, and A.-M.~S. Tremblay.
\newblock cond-mat/0410162.

\bibitem{Si94}
Q.M. Si, M.J. Rozenberg, G.~Kotliar, and A.E Ruckenstein.
\newblock {\em \PRL}, 72:2761, 1994.

\bibitem{Singh}
R.R.P. Singh.
\newblock {\em \PRB}, 39:9760, 1989.

\bibitem{tudor}
T.~D. Stanescu and G.~Kotliar.
\newblock cond-mat/0508302.

\bibitem{tudor-phillips}
T.~D. Stanescu and P.~Phillips.
\newblock {\em \PRL}, 91:049901(E), 2003.

\bibitem{tudorpr}
Tudor Stanescu.
\newblock Private communication.

\bibitem{Takagi2}
K.~Takagi, B.~Batlogg, H.L. Kao, J.~Kwo, R.J. Cava, J.J. Krajeski, and W.F.
  Peck.
\newblock {\em \PRL}, 69:2975, 1992.

\bibitem{Varma}
C.M. Varma and E.~Abrahams.
\newblock {\em \PRL}, 86:4652, 2001.

\bibitem{Zhang98}
F.C. Zhang, C.~Gross, T.M. Rice, and H.~Shiba.
\newblock {\em Supercond. Sci. Tech.}, 1:36, 1998.

\bibitem{Zhang}
Y.~Zhang, N.P. Ong, Z.A. Xu, K.~Krishana, R.~Gagnon, and L.~Taillefer.
\newblock {\em \PRL}, 84:2219, 2000.

\end{thebibliography}
}


\begin{vita}
\heading{Marcello Civelli} \vspace{25pt}
\begin{center}
{\bf Education} \end{center}
\begin{descriptionlist}{xxxxx-xxxxx} 
\item[2006] Ph. D. in Physics, Rutgers, The State University of
NJ, New Jersey, USA.
\item[2000] Italian "Laurea" in Physics, Universit\'a degli
Studi dell' Insubria, Como, Italy.
\end{descriptionlist}
\medskip
\begin{center}
{\bf Positions} \end{center}
\begin{descriptionlist}{xxxxx-xxxxx} 
\item[2003-2006] Graduate assistant, Department of Physics, Rutgers
University.
\item[2000-2003] Teaching assistant, Department of Physics, Rutgers
University.
\end{descriptionlist}
\medskip
\begin{center}
{\bf Publication List} \end{center}
\begin{descriptionlist}{xxxxx-xxxxx}
\item[2005]  S. S. Kancharla, M. Civelli, M. Capone, B. Kyung, D. S\'en\'echal, G. Kotliar and A.-M.S.
Tremblay, {\em Anomalous superconductivity in doped Mott
insulators}, cond-mat/0508205.
\item[2005]  B. Kyung, S. S. Kancharla, D. S\'en\'echal, A. -M. S. Tremblay, M. Civelli and G.
Kotliar, {\em Short-Range Correlation Induced Pseudogap in Doped
Mott Insulators}, cond-mat/0502565.
\item[2005] M. Civelli, M. Capone, S. S. Kancharla, O. Parcollet and G. Kotliar
{\em Dynamical Breakup of the Fermi Surface in a doped Mott
Insulator}, Phys. Rev. Lett. 95, 106402 (2005).
\item[2004] M. Capone, M. Civelli, S.S. Kancharla, C. Castellani and G.
Kotliar,{\em Cluster Dynamical Mean-Field Theory of the
density-driven Mott transition in the one-dimensional Hubbard
model}, Phys. Rev. B 69, 195105 (2004).
\end{descriptionlist}
\end{vita}

\end{document}